\documentclass[11pt]{article}

\usepackage[usenames,dvipsnames]{xcolor}
\usepackage[a4paper]{geometry}
\usepackage[english]{babel}
\usepackage{cite,enumerate,enumitem,booktabs,float,graphicx}
\usepackage[utf8]{inputenc}
\usepackage{amsmath,amssymb}
 
\usepackage{caption}
\usepackage{scalerel}
\usepackage{mathtools}
\usepackage{graphicx, nicefrac}
\usepackage{wrapfig}
\usepackage{xcolor}
\usepackage{braket}
\usepackage{comment}
\usepackage{color,soul}
\usepackage{yfonts}
\usepackage{subcaption}
\usepackage{cite}
\usepackage{amsfonts} 
\DeclareMathSymbol{\sm}{\mathbin}{AMSa}{"39}
\numberwithin{equation}{section}

\usepackage[T1]{fontenc}
\usepackage[utf8]{inputenc}
\usepackage{lmodern}

\makeatletter
\g@addto@macro\bfseries{\boldmath}
\makeatother

\usepackage{bm,amsmath,amssymb,tensor,mathtools}
\numberwithin{equation}{section}
\usepackage{nth}

\usepackage{tikz}
\usetikzlibrary{cd}

\usepackage{caption}
\usepackage{subcaption}

\usepackage[pdftex]{hyperref}
\definecolor{dark-blue}{rgb}{0.15,0.15,0.4}
\hypersetup{
  colorlinks,
  linkcolor={dark-blue},
  citecolor={blue}
}

\usepackage{enumitem}
\usepackage{amsthm}
\usepackage{array}

\usepackage{authblk}

\newcommand{\sL}{\mathcal{L}}
\newcommand{\E}{\mathcal{E}}
\newcommand{\T}{\mathcal{T}}
\newcommand{\RA}{R^{\scaleto{[-4]}{6pt}}}
\newcommand{\RB}{{R}^{\scaleto{[-2]}{6pt}}}
\newcommand{\RC}{R^{\scaleto{[0]}{6pt}}}
\newcommand{\RD}{R^{\scaleto{[2]}{6pt}}}

\definecolor{darkgreen}{rgb}{0.0, 0.4, 0.1}
\definecolor{darkyellow}{rgb}{0.75, 0.75, 0.1}

\title{\textbf{Towards a covariant framework for post-Newtonian expansions for radiative sources}
}
\date{}
\author{Jelle Hartong}
\author{J\o rgen Musaeus}
\affil{%
  School of Mathematics and Maxwell Institute for Mathematical Sciences,
  \protect\\
  University of Edinburgh,
  Peter Guthrie Tait Road,
  Edinburgh EH9 3FD, UK

}

\begin{document}

\maketitle
\thispagestyle{empty}

\begin{abstract}
\noindent

We consider the classic problem of a compact fluid source that behaves non-relativistically and that radiates gravitational waves. The problem consists of determining the metric close to the source as well as far away from it. The non-relativistic nature of the source leads to a separation of scales resulting in an overlap region where both the $1/c$ and (multipolar) $G$-expansions are valid. Standard approaches to this problem (the Blanchet--Damour and the DIRE approach) use the harmonic gauge. We define a `post-Newtonian' class of gauges  that admit a Newtonian regime in inertial coordinates. In this paper we set up a formalism to solve for the metric for any post-Newtonian gauge choice. Our methods are based on previous work on the covariant theory of non-relativistic gravity (a $1/c$-expansion of general relativity that uses post-Newton-Cartan variables). At the order of interest in the $1/c$ and $G$-expansions we split the variables into two sets: transverse and longitudinal. We show that for the transverse variables the problem can be reduced to inverting Laplacian and d'Alembertian operators on their respective domains subject to appropriate boundary conditions. The latter are regularity in the interior and asymptotic flatness with a Sommerfeld no-incoming radiation condition imposed at past null infinity. The longitudinal variables follow from the gauge choice. The full solution is then obtained by the method of matched asymptotic expansion. We show that our methods reproduce existing results in harmonic gauge to 2.5PN order.

\end{abstract}

\newpage
\tableofcontents
\newpage

\section{Introduction}

The post-Newtonian expansion is an expansion of general relativity assuming weak fields and slow motion. The expansion is almost as old as general relativity itself and has played a key role in our understanding of gravity. Its applications go as far back as the precession of the perihelion of Mercury. Currently, it plays  a key role in gravitational wave physics. In fact, one can argue that the demand for high accuracy predictions in gravitational wave physics has driven modern developments in post-Newtonian theory. One of the hurdles that had to be overcome was finding a way to glue together the physics of the slowly evolving system (for example some fluid with compact support) with that of the relativistic phenomenon of gravitational radiation that one observes far away from the source. The objective is to compute the metric both close and far away from the source. This problem has led to two different but equivalent approaches, namely the Blanchet-Damour approach (for a review see \cite{Blanchet2014}) and the direct integration of the relaxed Einstein equations (DIRE) approach (for a review see \cite{poisson2014gravity})\footnote{Other important approaches to post-Newtonian gravity include: the effective field theory methods reviewed in \cite{Levi:2018nxp}, the celestial mechanics for N-body systems \cite{Damour:1990pi}, and the Hamiltonian approach for compact binary systems \cite{PhysRevD.57.7274,Damour:2000kk}.}. Both approaches make use of the relaxed Einstein equations, which is a clever rewriting of Einstein gravity adapted to the harmonic gauge. Then through a separation of scales one is able to split spacetime into separate but overlapping regions for which different approximations are valid.

In recent times there has been a revival of work done in developing covariant non-relativistic expansions of gravity described in terms of Newton-Cartan type geometries plus relativistic corrections \cite{Tichy:2011te,VandenBleeken:2017rij,Hansen:2019pkl,VandenBleeken:2019gqa,HHO,Hartong:2023yxo}. For a review see \cite{Hartong:2022lsy}. In this covariant approach the non-relativistic expansion of gravity essentially takes place in tangent space in the limit in which the tangent space lightcones flatten ($1/c \rightarrow \infty$). This expansion is more general than the post-Newtonian expansion for two reasons. The first reason is that being `post-Newtonian' already presupposes that one is working in a gauge in which there is a Newtonian regime (this is not true in all gauge choices\footnote{The covariant formulation of Newtonian gravity is Newton-Cartan gravity of which Newtonian gravity is a gauge-fixed version.}). Secondly, the covariant $1/c$-expansion is not necessarily a weak field expansion. There is a regime called strong non-relativistic gravity that includes solutions such as a non-relativistic Schwarzchild geometry \cite{VandenBleeken:2017rij,VandenBleeken:2019gqa,Hansen:2019vqf}. It depends on what assumptions are made regarding the non-relativistic expansion of the matter fields whether one ends up with a weak or strong non-relativistic gravity regime \cite{HHO}. One of the purposes of this paper is to use the covariant non-relativistic gravity approach to find a systematic framework that allows us to perform post-Newtonian calculations in a more covariant manner. There are in some sense three increasingly challenging generalisations of the current state of the art (reviewed below) regarding post-Newtonian methods. The first layer (the scope of this paper) is to find a framework in which we can perform post-Newtonian calculations in any gauge that admits a Newtonian regime. The second layer of sophistication is to generalise this further to a framework that is properly covariant in the sense of some yet to be constructed post-Newton-Cartan theory, but still assuming weak fields. Finally, the ultimate aim is to develop methods that are based on post-Newtonian ideas but where the leading order theory is not Newtonian gravity but rather the strong non-relativistic gravity regime\footnote{To be clear, this aim can be achieved independently from the second layer/aim.} alluded to above.

With this work we intend to build a clear bridge between the covariant non-relativistic expansion and the post-Newtonian expansion which will serve multiple purposes. Firstly, it gives us a better understanding of the covariant $1/c$-expansion and what its capabilities as well as its limitations are. Secondly, this will provide us with a new framework for the post-Newtonian expansion that is able to improve upon certain aspect of the otherwise very well developed theory. In our endeavour to construct a more covariant approach to the post-Newtonian expansion we will also have to develop a more covariant framework for the post-Minkowskian expansion outside the source, which is necessary in order to describe radiating systems. 
We will set up a formalism that allows us to compute the metric close and far away from a radiating source for any gauge choice that admits a Newtonian regime and for which the vacuum is described in inertial coordinates.

This framework is of course not going to compete with the Blanchet-Damour or DIRE approach when it comes to the accuracy with which calculations have been performed in the harmonic gauge. However, a more covariant framework might make it easier to identify gauge-independent physics and develop intuition about the expansion. Furthermore, there might be advantages to working in other gauges, depending on the problem at hand.

Apart from developing the ingredients of a more covariant framework we also show how our approach works in the standard harmonic gauge (to show that the method works and to facilitate comparison with the literature) as well as in another gauge, that we refer to as the transverse gauge (see equation (\ref{eq:Cgauge})). The latter can be thought of as the GR version of the Coulomb gauge familiar from electromagnetism. In the companion paper \cite{companionpaper} we will report in more detail on how the post-Newtonian expansion works in that case.

\subsection{State of the art}
Here we give a very brief review of the Blanchet-Damour approach \cite{damour1983gravitational,Blanchet:1985sp,blanchet1987radiative,Blanchet:1989ki,Blanchet:1992br,PhysRevD.51.2559,LucBlanchet_1998,Damour:1990ji,PhysRevD.47.4392,PhysRevD.65.124020,PhysRevD.72.044024} as well as the DIRE approach \cite{Will:1996zj,Pati:2000vt,Pati:2002ux,Will:2005sn,Wang:2007ntb,Mitchell:2007ea}  which themselves build on a lot of previous work (see for example \cite{bonnor1961transport,bonnor1966gravitational,bonnor1959spherical,thorne1980multipole,10.1063/1.1665603,hunter1968double,anderson1975equations} or for a much more comprehensive list of references see \cite{Blanchet2014})  
that helped bridge the gap between the classic approach\footnote{In the classic approach the $1/c$-expansion is assumed to be valid everywhere, i.e. all the way up to infinity.} and modern day post-Newtonian theory. The basic post-Newtonian setup goes as follows. One assumes that the matter source is compact with some characteristic length scale, $l_c$, and characteristic time scale, $t_c$. Then one assumes slow motion $\frac{v_c}{c} \ll 1$ where $v_c := l_c/t_c$,  and through the virial theorem it then follows that the gravitational field strength is weak as well, $\frac{GM}{c^2 l_c} \sim \frac{v_c^2}{c^2} \ll 1$ where $M$ is the total mass. The post-Newtonian expansion has a limited region of validity, called the near zone, which is the part of the spacetime where retardation effects can be treated perturbatively, i.e. $r \ll \lambda_c= ct_c$. Outside of the near zone one has to rely on post-Minkowskian techniques, i.e. expansions in Newton's constant $G$. 

Both approaches are reliant on the harmonic gauge which can be expressed as 
\begin{align}
    \partial_\nu h^{\mu \nu} = 0.
\end{align}
where $h^{\mu \nu} = \eta^{\mu \nu} - \sqrt{-g} g^{\mu \nu}$ and $\mu =0,1,2,3$. In this gauge Einstein's equations can be rewritten as
\begin{align}
    \square h^{\mu \nu} = - \frac{16 \pi G}{c^4} \tau^{\mu \nu}, \label{RE}
\end{align}
where $\square \equiv - \frac{1}{c^2} \frac{\partial^2 }{\partial t^2} + \nabla^2$ is the flat-spacetime d'Alembertian and $\tau^{\mu \nu}$ depends on non-linear combinations of $h^{\mu \nu}$ and its derivatives as well as the energy-momentum tensor $T^{\mu \nu}$ of the matter source. Once $h^{\mu \nu}$ is determined one can derive the metric by simply solving $h^{\mu \nu} = \eta^{\mu \nu} - \sqrt{-g} g^{\mu \nu}$ for $g_{\mu\nu}$. One can then derive the matter EOM from the near zone metric or the waveform from the asymptotic behavior of the metric. Equation (\ref{RE}) is the starting point for both approaches but they differ in how they solve this equation.

\textbf{The Blanchet-Damour approach}
relies on the method of matched asymptotic expansions. One solves equation (\ref{RE}) in the exterior ($l_c <r $) of the source using a multipolar post-Minkowskian (MPM) expansion and one solves the equation in the near zone ($r \ll \lambda_c$) using a post-Newtonian expansion.
The two solutions are matched in the overlap region, fixing undetermined functions on both sides. The near zone solution takes the following form
\begin{subequations}
\begin{align}
    h^{\mu\nu} =& \frac{16 \pi G}{c^4} \square^{-1}_{\text{ret}}[\bar{\tau}^{\mu\nu}] - \frac{4G}{c^4} \sum_{l=0}^{\infty} \frac{(-)^l}{l!} \partial_L \left( \frac{\mathcal{R}^{\mu\nu}_L(t-r/c) - \mathcal{R}^{\mu\nu}_L (t+r/c)}{2r} \right)\,, \label{eq:BDPN}
    \\
    \square^{-1}_{\text{ret}}[\bar{\tau}^{\mu\nu}] :=& - \frac{1}{4 \pi} \sum_{m=0}^{\infty} \frac{(-)^m}{m!} \left( \frac{\partial}{c \partial t} \right)^m \mathcal{FP}\int d^3x' |x-x'|^{m-1} \bar{\tau}^{\mu\nu}(x',t)\,,
\end{align}
\end{subequations}
where the bar over $\tau^{\mu\nu}$ indicates that the source $\tau^{\mu\nu}$ has been  $1/c$-expanded. The index $L$ is a multi-index $i_1\cdots i_l$. Meanwhile, $\mathcal{FP}$ denotes a regularisation procedure to find the finite part of the integral. The functions $\mathcal{R}^{\mu\nu}_L(t-r/c)$ are fixed in the matching and are in general not analytic in $1/c$. However, to 2.5PN order these terms will be zero. The source $\tau^{\mu\nu}$ of course depends on $h^{\mu\nu}$ as well but only non-linearly and so \eqref{eq:BDPN} can be computed iteratively.

In the exterior zone $T^{\mu \nu} = 0$ and therefore $\tau^{\mu \nu}$ simply consists of nonlinear combinations of $h^{\mu\nu}$ and its derivatives. So, for the $G$-expansion
\begin{align}
    h^{\mu \nu} =& G h^{\mu \nu}_{(1)} + G^2h^{\mu \nu}_{(2)} + G^3h^{\mu \nu}_{(3)} +\mathcal{O}(G^4),
    \\
    \tau^{\mu \nu} =& G^2 \tau^{\mu \nu}_{(2)} + G^3 \tau^{\mu \nu}_{(3)} + \mathcal{O}(G^4)\,,
\end{align}
the leading order equation is simply $\square h_{(1)}^{\mu\nu} = 0$. This is then solved  making use of the past-stationarity condition $\{ \partial_t h^{\mu \nu} =0 | t\leq -T_0\}$ for some finite positive number $T_0$. The solution can be expressed as 
\begin{align}
    h^{00}_{(1)} =& - \frac{4}{c^2 } \sum_{l \geq 0} \frac{(-)^l}{l!} \partial_L \bigg[ \frac{1}{r} I_L (u)\bigg] + \partial_k \phi^k - \frac{1}{c}\partial_t \phi^0\,, \label{hK1}
    \\
    h^{0i}_{(1)} =& \frac{4}{c^3} \sum_{l \geq 1} \frac{(-)^l}{l!} \partial_{L-1} \bigg[ \frac{1}{r} \dot{I}_{iL-1}(u) + \frac{l}{l+1} \epsilon_{iab} \partial_a \Big( \frac{1}{r} J_{bL-1}(u) \Big) \bigg] + \partial_i \phi^0 - \frac{1}{c} \partial_t \phi^i\,,
    \\
    h^{ij}_{(1)} =& - \frac{4}{c^4} \sum_{l \geq 2} \frac{(-)^l}{l!} \partial_{L-2} \bigg[ \frac{1}{r} \ddot{I}_{ijL-2}(u) + \frac{2l}{l+1} \partial_a \Big(  \frac{1}{r} \epsilon_{ab(i} \dot{J}_{j)bL-2}(u) \Big) \bigg] \nonumber
    \\
    &+2 \partial^{(i} \phi^{j)} - \delta^{ij} \partial_\alpha \phi^\alpha\,, \label{hK2}
\end{align}
with
\begin{align}
    \phi^0 =& \frac{4}{c^3} \sum_{l\geq 0} \frac{(-)^l}{l!} \partial_L \bigg[ \frac{1}{r} W_L(u)\bigg]\,, \label{eq:MPMgauge1}
    \\
    \phi^i =& - \frac{4}{c^4} \sum_{l \geq 0} \frac{(-)^l}{l!} \partial_{iL} \bigg[ \frac{X_L(u)}{r} \bigg] - \frac{4}{c^4} \sum_{l \geq 1} \frac{(-1)^l}{l!} \partial_{L-1} \bigg[ \frac{Y_{iL-1}(u)}{r}  + \frac{l}{l+1} \epsilon_{iab} \partial_a \big( \frac{1}{r} Z_{bL-1}(u) \big)\bigg]\,, \label{eq:MPMgauge2}
\end{align}
where $I_L,J_L,W_L,X_L,Y_L,Z_L$ are undetermined STF (symmetric trace-free) tensors that will be fixed in the matching procedure in terms of multipole moments of the matter source. The resulting expression for $h_{(1)}^{\mu\nu}$ will then determine the source term $\tau^{\mu\nu}_{(2)}$ in the wave equation for $h^{\mu\nu}_{(2)}$, which itself enters in $\tau^{\mu\nu}_{(3)}$ and so on.

The full $n$th order solution can then be written as
\begin{align}
    h^{\mu\nu} =&  Gh^{\mu\nu}_{\text{hom}}+ \sum_{n=1}^\infty G^n\left( u^{\mu\nu}_{(n)} + v^{\mu\nu}_{(n)}  \right),
    \\
    u^{\mu\nu}_{(n)} =& \mathcal{FP}\int d^3x' \frac{\tau^{\mu\nu}_{(n)}(t-|x-x'|/c,x')}{|x-x'|}\, ,
\end{align}
where $v_{(n)}^{\mu\nu}$ is a specific homogeneous solution that is determined through $\partial_\mu v^{\mu\nu}_{(n)} = - \partial_\mu u^{\mu\nu}_{(n)}$. This is to ensure that $u^{\mu\nu}_{(n)} + v^{\mu\nu}_{(n)}$ forms a particular solution that fulfills the harmonic gauge condition. Finally, $h_{\text{hom}}^{\mu\nu}$ is the general solution to the homogeneous equation, which is given by taking $h_{(1)}^{\mu\nu}$ and adding correction to $I_L,\ldots,Z_L$ up to the desired order in $G$. For more details and in-depth analysis we refer the reader to the review paper \cite{Blanchet2014}.

In \textbf{the DIRE approach} the first step is to formally integrate (\ref{RE}) using the retarded Green's function
\begin{align}
    h^{\mu\nu} = \int d^3x' \frac{ \tau^{\mu\nu}(t-|x-x'|/c,x')}{|x-x'|}\,.
\end{align}
Then one splits up the integration domain in a near zone $\mathcal{N} = \{\vec{x} \in \mathbb{R}^3 | r<\mathcal{R}\}$ and a wave zone $\mathcal{W} = \{\vec{x} \in \mathbb{R}^3 | r>\mathcal{R}\}$, where by definition $\mathcal{R}$ is the boundary of the near zone. On then gets
\begin{align}
    &h^{\mu\nu} = h^{\mu\nu}_\mathcal{N} + h^{\mu\nu}_\mathcal{W}\,, 
    \\
    &h^{\mu\nu}_\mathcal{N} = \int_\mathcal{N} d^3x' \frac{ \tau^{\mu\nu}(t-|x-x'|/c,x')}{|x-x'|}, \qquad h^{\mu\nu}_\mathcal{W} = \int_\mathcal{W} d^3x' \frac{ \tau^{\mu\nu}(t-|x-x'|/c,x')}{|x-x'|}\,. \label{eq:Wave&Near_int}
\end{align}
In here $h^{\mu\nu}_\mathcal{N}$ and $h^{\mu\nu}_\mathcal{W}$ are each subject to different approximations depending on whether one is evaluating at a field point $x \in \mathcal{N}$ or $x \in \mathcal{W}$. This leads to four different integral equations that one solves iteratively. At leading order $\tau^{\mu\nu} = T^{\mu \nu}$  and thus $h^{\mu\nu}_\mathcal{W} = 0$. For the near zone integrations the following approximations are used
\begin{align}
    h^{\mu\nu}_\mathcal{N} =& \sum_{l=0}^\infty \frac{(-)^l}{l!c^l} \left( \frac{\partial}{\partial t} \right)^l \int_{\mathcal{N}}d^3x' \tau^{\mu\nu}(t,x') |x-x'|^{l-1}  \qquad \text{for}\ x \in \mathcal{N} \,,\label{eq:NNint}
    \\
    h^{\mu\nu}_\mathcal{N} =&\sum_{l=0}^\infty \frac{(-)^l}{l!} \partial_L \left[\frac{1}{r} \int_{\mathcal{N}} d^3x'\tau^{\mu\nu}(t-r/c,x') x'^L  \right]\qquad \text{for}\ x \in \mathcal{W} \,,\label{eq:NWint}
\end{align}
where the expression in (\ref{eq:NNint}) has been $1/c$-expanded and the expression (\ref{eq:NWint}) has been multipole expanded using that the field points are in the near and wave zone, respectively. Equation (\ref{eq:NWint}) then gives rise to the source terms for $ h^{\mu\nu}_\mathcal{W}$ at the second iteration. It follows from this that the source term, for $x \in \mathcal{W}$, is going to be a sum over terms of the generic form
\begin{align}
    \frac{1}{4\pi} \frac{f^{\mu\nu}_L(u) n^{\langle L \rangle}}{r^m} \label{eq:MultipolSource}
\end{align}
where $m$ is a positive integer.
Using this the wave zone integrals can be written as follows (given here for just one generic term in (\ref{eq:MultipolSource}), but in actuality one would have to sum over multiple contributions of this type)
\begin{align}
    h^{\mu\nu}_\mathcal{W} =&\frac{n^{\langle L \rangle}}{r} \bigg[\int_0^{\mathcal{R}}ds f^{\mu\nu}_L(u-2s/c) A(s,r)  + \int_{\mathcal{R}}^\infty ds f^{\mu\nu}_L(u-2s/c) B(s,r)  \bigg] \quad \text{for}\ x \in \mathcal{W}\,, \label{eq:WWext}
    \\
    h^{\mu\nu}_\mathcal{W} =& \frac{n^{\langle L \rangle}}{r} \bigg[\int_{\mathcal{R}-r}^{\mathcal{R}}ds f^{\mu\nu}_L(u-2s/c) A(s,r)  + \int_{\mathcal{R}}^\infty ds f^{\mu\nu}_L(u-2s/c) B(s,r)  \bigg]\quad \text{for}\ x \in \mathcal{N}\,, \label{eq:WNext}
\end{align}
where $u = t-r/c$ and
\begin{align}
    A(s,r) := \int_{l_c}^{r+s} dr' \frac{P_l(\xi)}{r'^{(m-1)}}\,, \qquad B(s,r) := \int_s^{r+s} dr' \frac{P_l(\xi)}{r'^{(m-1)}}\,.
\end{align}
In here $P_l$ denotes the Legendre polynomial of degree $l$ and $\xi = (r+2s)/r-2s(r+s)/(rr')$. The functions $A(s,r)$ and $B(s,r)$ can be computed explicitly for a given $l$ and $m$. The integrals over $s$ are done by making continual use of integration by parts while throwing away terms that depend explicitly on the cut-off (these will be cancelled by similar boundary terms coming from (\ref{eq:NWint}) and (\ref{eq:NNint})).

Going beyond the second iteration the source term in the wave zone, $\tau^{\mu\nu}$, will be constructed out of non-linear combination of both (\ref{eq:WWext}) and (\ref{eq:NWint}) as well as their derivatives. Most of these terms will be on the form of (\ref{eq:MultipolSource}), but if one goes to high enough order $\log r$-terms will appear and then (\ref{eq:WWext}) and (\ref{eq:WNext}) no longer hold and one has to return to (\ref{eq:Wave&Near_int}). For a slightly different form of (\ref{eq:WWext}) and (\ref{eq:WNext}), and a more in-depth description see \cite{Pati:2000vt}.

\subsection{Statement of the problem}

Given a perfect fluid source with compact support the goal is to devise a computational scheme that is able to perturbatively compute the metric both near the source as well as far away from it (and in principle in the intermediate region). The source is assumed to behave non-relativistically so that the characteristic velocity is much smaller than the speed of light leading to a separation of scales $l_c\ll\lambda_c=t_c c$. The method should allow us to compute the metric in any gauge that admits a Newtonian regime for the near zone metric. Furthermore, we assume that the metric is asymptotically flat in inertial coordinates with Sommerfeld no-incoming radiation conditions imposed at past null infinity. This framework must include a suitably covariant framework for the multipolar post-Minkowskian expansion as this is necessary to capture the radiative effects. In this paper we construct this framework and test that it produces the correct results for the metric in harmonic gauge to 2.5PN order. In \cite{companionpaper} we show how the method works in transverse gauge.

We restrict ourselves to solving the post-Newtonian metric for a compact perfect fluid source. However, there exists a method of extracting the equations of a compact binary system from those of the perfect fluid \cite{Pati:2002ux}. This involves treating the bodies as small (compared to their separation), spherical, non-rotating balls of fluid. Doing this of course adds a whole extra layer of complication which is beyond the scope of this paper.

Additionally, Since we restrict ourselves to 2.5PN order we do not have to deal with tail terms that will eventually show up in the near zone and that signal a breakdown of the $1/c$ Taylor expansion. In order to fix this one needs to include $\log c$-terms \cite{Blanchet:1985sp,PhysRevD.25.2038}. We leave their incorporation for future work. 

\subsection{Summary of results}

In this paper we present a $1/c$-expansion approach to the post-Newtonian expansion that applies to any post-Newtonian gauge. By a post-Newtonian gauge we mean a gauge choice for which the metric admits a Newtonian regime. More concretely, these are gauge choices for which we can write the metric as $g_{\mu\nu}=\eta_{\mu\nu}+h_{\mu\nu}$ where $\eta_{\mu\nu}$ corresponds to the Minkowski metric in inertial coordinates and where there is a region of spacetime where the metric is Newtonian plus corrections. 

We start by working out the metric in the near zone, defined by $r\ll \lambda_c$, using the covariant $1/c$-expansion. Then we solve the metric in the exterior zone $r>l_c$ using a multipolar $G$-expansion that works for the same class of post-Newtonian gauge choices as used for the $1/c$-expansion. Finally, we match the two expansions in the overlap region. In both cases the general principle is to first expand the equations, split the variables into transverse and longitudinal variables, solve for the former and fix the latter by applying a gauge condition. We then  integrate the $1/c$ and $G$ expanded Einstein equations subject to appropriate boundary conditions and match them in the overlap region.

The covariant $1/c$-expansion starts by expressing the metric in pre-non-relativistic variables $T_\mu$ and $\Pi_{\mu \nu}$ as
\begin{align}
g_{\mu \nu} = - c^2 T_\mu T_\nu + \Pi_{\mu \nu}\,, 
\end{align}
where $\Pi_{\mu \nu}$ has signature $(0,1,1,1)$. The choice of $T_\mu$ and $\Pi_{\mu \nu}$ is however not unique and is subject to local Lorentz boost transformations. We use this freedom to set $\Pi_{it} =0$ (where $i=1,2,3$ is a spatial index), in which case we get
\begin{equation}
    ds^2=-c^2\left(T_\mu dx^\mu \right)^2+\Pi_{ij}dx^idx^j\,.
\end{equation}
The fields $T_\mu$ and $\Pi_{ij}$ are assumed to be analytic in $1/c$, which is valid to the order we are interested in\footnote{This will eventually break down at higher order but it can be fixed by including $\log c$ terms in the expansion.} which is 2.5PN. The $1/c$-expansions of $T_\mu$ and $\Pi_{ij}$ are then given by
\begin{align}\label{eq:postNgauge}
    T_\mu &=\tau_\mu+ \frac{1}{c^2}\tau_t^{(2)}+\sum_{n=4}^\infty \frac{1}{c^n} \tau_\mu^{(n)}, \qquad &\Pi_{ij} &=h_{ij}+  \sum_{n=2}^\infty \frac{1}{c^n} h_{ij}^{(n)}\,,
\end{align}
where we used that the 0.5PN metric (and the term in $T_\mu$ at order $1/c$) can always be gauged away. Since we use inertial coordinates for the vacuum we have
\begin{align}
    h_{\mu \nu} = \delta_{ij} \delta^i_\mu \delta^j_\mu\,, \qquad \tau_\mu = \delta_\mu^t\,. \label{eq:flat}
\end{align}
It then follows that $\tau^{(2)}_\mu = -U \delta_\mu^t$ with $U$ being the Newtonian potential. In order to construct the $\frac{n}{2}$PN metric one needs to determine $T_\mu$ to $\tau^{(n+2)}_\mu$ and $\Pi_{ij}$ to $h_{ij}^{(n)}$. We expand Einstein's equation in $1/c$, and apply the following decomposition of the post-Newtonian variables
\begin{eqnarray}
h_{ij}^{(n)} & = & h_{ij}^{(n)}(\text{TT})+\partial_i L^{(n)}_j+\partial_j L^{(n)}_i+\frac{1}{3}\delta_{ij}H^{(n)}\,,\\
\tau_i^{(n+2)} & = & M_i^{(n)}(\text{T})-\partial_t L_i^{(n)}-\partial_i N^{(n)}\,,\\
\tau_t^{(n+2)} & = & M_t^{(n)}-\partial_t N^{(n)}\,,
\end{eqnarray}
where $H^{(n)}=h_{kk}^{(n)}-2\partial_k L_k^{(n)}$ and where T denotes that the field is transverse and TT that it is transverse traceless. This leads to
\begin{eqnarray}
\hspace{-1cm}\partial^2 H^{(n)} & = & \frac{3}{4}S^{(n)}_{ii}\,,\\
\hspace{-1cm}\partial^2 h_{ij}^{(n)}(\text{TT}) & = & S^{(n)}_{ij}-\frac{1}{4}\delta_{ij}S^{(n)}_{kk}-\frac{1}{3}\partial_i\partial_j H^{(n)}\,,\\
\hspace{-1cm}\partial^2 M_i^{(n)}(\text{T}) &  = & S^{(n)}_i+\frac{2}{3}\partial_t\partial_i H^{(n)}\,,\\
\hspace{-1cm}\partial^2 M_t^{(n)} & = & S^{(n)}+\frac{1}{2}\partial_t^2 H^{(n)}+\partial^2 L_i^{(n)}\partial_i\tau_t^{(2)}-\frac{1}{6}\partial_i  H^{(n)}\partial_i\tau_t^{(2)}+h_{ij}^{(n)}\partial_i\partial_j\tau_t^{(2)}\,,
\end{eqnarray}
where $S^{[n]},S_i^{[n]}$ and $S_{ij}^{[n]}$ only depend on the fluid matter variables (which are also $1/c$ expanded) and lower order fields $h^{(k)}_{ij}$ and $\tau^{(k+2)}_\mu$ with $k<n$. 
These equation can be rewritten in the form of simple Poisson type equations. 
\begin{eqnarray}
\partial^2 H^{(n)} & = & \frac{3}{4}S^{(n)}_{ii}\,, \label{eq:H^nPoisson}\\
\partial^2 \left(M_i^{(n)}(\text{T})
-\frac{1}{3}x^i\partial_t H^{(n)}\right) &  = & S^{(n)}_i-\frac{1}{4}x^i\partial_t S^{(n)}_{jj}\,,
\end{eqnarray}
\begin{eqnarray}
    \hspace{-1cm}&&\partial^2\left(h^{(n)}_{ij}(\text{TT})+\frac{1}{12}\left[x^i\partial_j H^{(n)}+x^j\partial_i H^{(n)}-\frac{2}{3}\delta_{ij}x^k\partial_k H^{(n)}\right]\right)=\nonumber\\
    \hspace{-1cm}&&S^{(n)}_{ij}-\frac{1}{3}\delta_{ij}S^{(n)}_{kk}+\frac{1}{16}\left(x^i\partial_j S^{(n)}_{ll}+x^i\partial_j S^{(n)}_{ll}-\frac{2}{3}\delta_{ij}x^k\partial_k S^{(n)}_{ll}\right)\,,\\
\hspace{-1cm}&&\partial^2\left( M_t^{(n)} -\frac{1}{12}r^2\partial_t^2 H^{(n)}+\frac{1}{2}x^i\partial_t M_i^{(n)}(\text{T})\right) = \nonumber\\
\hspace{-1cm}&& S^{(n)} -\frac{1}{16}r^2\partial_t^2 S_{ii}^{(n)}+\frac{1}{2}x^i\partial_tS_i^{(n)}-\frac{1}{6}\partial_i  H^{(n)}\partial_i\tau_t^{(2)}+\partial^2 L_i^{(n)}\partial_i\tau_t^{(2)}+h_{ij}^{(n)}\partial_i\partial_j\tau_t^{(2)}\,.
\end{eqnarray}

At this point we have still not applied any gauge condition, except for what we assumed in \eqref{eq:postNgauge} to get Newtonian gravity. Thus, we see that for any post-Newtonian gauge the field equations all reduce to Poisson type equations and the gauge freedom is stored in the longitudinal fields $L_i^{(n)}$ and $N^{(n)}$ which are determined through an appropriate gauge choice. The latter is an important addition to the list of equation because the source terms depend on $L_i^{(k)}$ and $N^{(k)}$ for $k<n$.

The Poisson equations are formally solved using a regularised Poisson integral to which we add the most general harmonic function that is regular in the near zone. Take for example equation (\ref{eq:H^nPoisson}), the solution to this would be given by
\begin{equation}
    H^{[n]}=-\frac{3}{16\pi}\int_{\Omega_{R_\star}} d^3 x'\frac{S^{(n)}_{ii}(t,x')}{\vert x-x'\vert}+ \sum_{l=0}^\infty \mathcal{F}_{L} x^{L}\, ,
\end{equation}
where $L = i_1,\ldots, i_l$ and where $\mathcal{F}_{L}$ is completely symmetric and tracefree in all its indices. The coefficients $\mathcal{F}_{L}$ characterise our ignorance about boundary conditions imposed outside the near zone and they will be fixed in the matching process.

For the exterior zone metric we perform a post-Minkowskian or $G$-expansion
\begin{align}
    g_{\mu \nu} = \eta_{\mu \nu} + G h_{\mu \nu}^{[1]} +  G^2 h_{\mu \nu}^{[2]} + \cdots\,,
\end{align}
and we will use $x^0=ct$. The vacuum Einstein equations for $h_{\mu \nu}^{[n]}$ can be written as 
\begin{align}\label{eq:VacE}
    - \square h_{\mu \nu}^{[n]} + \eta^{\rho\sigma}\left(2 \partial_\rho \partial_{(\mu} h^{[n]}_{\nu)\sigma}- \partial_\mu \partial_\nu h^{[n]}_{\rho\sigma}\right) = \tau^{[n]}_{\mu \nu}\,,
\end{align}
where $\tau^{[n]}_{\mu \nu}$ only depends on products of lower order fields $h^{[n-1]}_{\mu \nu},\ldots,h^{[1]}_{\mu \nu}$ and their derivatives.  Similarly to what we did in the near zone, we then make a decomposition of $h^{[n]}_{\mu \nu}$ in terms of transverse and longitudinal fields 
\begin{eqnarray}
    h_{ij}^{[n]} & = & h_{ij}^{[n]}(\text{TT})+\partial_i L_j^{[n]}+\partial_j L_i^{[n]}+\frac{1}{3}\delta_{ij} H^{[n]}\,,\\
    h_{0i}^{[n]} & = & -M_i^{[n]}(\text{T})+\partial_0 L_i^{[n]}+\partial_i N^{[n]}\,,\\
    h_{00}^{[n]} & = & -2 M^{[n]}_0+2\partial_0 N^{[n]}\,,
\end{eqnarray}
where
\begin{equation}
    H^{[n]}=h_{ii}^{[n]}-2\partial_i L_i^{[n]}\,.
\end{equation}
The equations \eqref{eq:VacE} are then given by 
\begin{eqnarray}
    \partial^2 H^{[n]} & = & -\frac{3}{4}\left(\tau_{00}^{[n]}+\tau_{kk}^{[n]}\right)\,,\\
    \partial^2 M_0^{[n]} & = & \frac{1}{2}\partial_0^2 H^{[n]}+\frac{1}{2}\tau_{00}^{[n]}\,,\\
    \partial^2 M_i^{[n]}(\text{T}) & = & \frac{2}{3}\partial_0\partial_i H^{[n]}+\tau_{0i}^{[n]}\,,\\
    -\square h_{ij}^{[n]}(\text{TT}) & = & -2\partial_0\partial_{(i}M_{j)}^{[n]}(\text{T})+2\partial_{\langle i}\partial_{j\rangle}M_0^{[n]}+\frac{1}{3}\partial_{ i}\partial_{j} H^{[n]}+\tau^{[n]}_{\langle ij\rangle}\,,
\end{eqnarray}
which can be rewritten as 
\begin{eqnarray}
\partial^2 H^{[n]} & = & -\frac{3}{4}\left(\tau_{00}^{[n]}+\tau_{kk}^{[n]}\right)\,,\\
    \partial^2\left(M_0^{[n]}-\frac{r^2}{12}\partial_0^2 H^{[n]}+\frac{x^i}{2}\partial_0 M^{[n]}_i(\text{T})\right) & = & \frac{1}{2}\tau^{[n]}_{00}+\frac{r^2}{16}\partial_0^2\left(\tau^{[n]}_{00}+\tau^{[n]}_{kk}\right)+\frac{x^i}{2}\partial_0\tau^{[n]}_{0i}\,,\\
    \partial^2\left(M^{[n]}_i(\text{T})-\frac{1}{3}x^i\partial_0 H^{[n]}\right) & = & \tau_{0i}^{[n]}+\frac{x^i}{4}\partial_0\left(\tau^{[n]}_{00}+\tau^{[n]}_{kk}\right)\,,
\end{eqnarray}
\begin{equation}
     -\square h_{ij}^{[n]}(\text{TT}) = -2\partial_0\partial_{(i}M_{j)}^{[n]}(\text{T})+2\partial_{\langle i}\partial_{j\rangle }M_0^{[n]}+\frac{1}{3}\partial_{\langle i}\partial_{j\rangle} H^{[n]}+\tau^{[n]}_{\langle ij\rangle}\,.
\end{equation}
If we differentiate the latter equation with respect to $x^0$ twice we can rewrite it further to
\begin{eqnarray}
    &&\square\left(\partial_0^2 h^{[n]}_{ij}(\text{TT})+\partial_i\partial_0 M^{[n]}_j(\text{T})+\partial_j\partial_0 M^{[n]}_i(\text{T})-2\partial_i\partial_j M^{[n]}_0+\frac{1}{3}\delta_{ij}\partial_0^2 H^{[n]}\right)=\nonumber\\
    &&-\partial_0^2\tau^{[n]}_{ij}+\partial_0\partial_i\tau^{[n]}_{0j}+\partial_0\partial_j\tau^{[n]}_{0i}-\partial_i\partial_j\tau^{[n]}_{00}\,.
\end{eqnarray}
Thus, we see that the problem boils down to inverting the d'Alembertian and the Laplacian operators in the exterior zone. Again this holds for any post-Newtonian gauge. In solving for these equations we of course also need to apply boundary conditions. These are asymptotic flatness as well as Sommerfeld's no-incoming radiation condition at past null infinity.

More concretely, the homogeneous solution to these equations can be found in equations \eqref{eq:finalhomogsol2}--\eqref{eq:finalhomogsol4} and \eqref{eq:intermediatesol_hTT2}. For the particular solution to the sourced equations we need to invert the Laplacian and the d'Alembertian in the exterior zone. The boundary conditions are such that $H^{[n]}$, $M_0^{[n]}$ and $M^{[n]}_i(\text{T})$ are $\mathcal{O}(r^{-1})$ for large $r$ and $h^{[n]}_{ij}(\text{TT})$ obeys the Sommerfeld no-incoming radiation condition at past null infinity, which is the statement that
\begin{equation}
    \lim_{\overset{r\rightarrow\infty}{v=\text{cst}}}\partial_v\left(rh_{ij}^{[n]}(\text{TT})\right)=0\,,
\end{equation}
where $v=t+r/c$ (advanced time). 

The particular solution for $h_{ij}^{[n]}(\text{TT})$ can be obtained by using a retarded Green's function that is well-defined in the exterior zone. Again the longitudinal fields are fixed by an appropriate post-Newtonian gauge choice. These are important as they are part of the matching process with the near zone solution and because they appear in the sources for the higher order $G$ equations of motion. 

Once we have obtained the most general solution in both the near zone and in the exterior zone we apply the matching condition that is very reminiscent of what is done in the Blanchet--Damour approach in harmonic gauge, i.e. we require that in the overlap region we have
\begin{align}
    \mathcal{M} \left(g_{\mu \nu}^\mathcal{N} \right) = \mathcal{C}\left( g_{\mu \nu}^\mathcal{E} \right) \,,
\end{align}
where $\mathcal{C}$ indicates the operation of $1/c$-expanding the exterior zone metric, $g_{\mu \nu}^\mathcal{E}$, and $\mathcal{M}$ indicates the operation of multipole expanding the near zone metric, $g_{\mu \nu}^\mathcal{N}$.

\subsection{Outline of the paper}

This paper is organised as follows. In section \ref{ref:sec2} we review the covariant $1/c$-expansion of GR. This leads to a formulation of Einstein's equations in terms of what are called pre-non-relativistic variables. In section \ref{sec:Newton} we continue our review of non-relativistic gravity by spelling out the conditions under which the theory has a Newtonian gravity description which informs us later about the class of gauge choices we can make. Sections 4 and 5 constitute the first main part of the paper. In section \ref{sec:genPN} we outline the general structure of the $1/c$-expansion of the Einstein equations to any order in $1/c$ and we give the explicit equations to 2.5PN order in any post-Newtonian gauge. The details of the latter result are discussed in appendix \ref{app:1/cexpGR}. In section \ref{sec:covGexp} we essentially do the same for the $G$-expansion. We decompose the metric at a certain order in $G$ into transverse and longitudinal components. We then show how the $G$ expanded Einstein equations can be solved for the transverse components at any order in $G$ and how the gauge choice fixes the longitudinal components. We furthermore discuss the issue of asymptotic boundary conditions for the different components of the metric. In the case of the transverse gauge we solve for the homogeneous part of the $G$-expanded Einstein equations explicitly and we derive a useful parametrisation of the homogeneous part of the harmonic gauge metric. In sections \ref{sec:NZto1.5} and \ref{sec:NZmetric2.5} as well as appendix \ref{app:conservation} we then focus our attention entirely on the harmonic gauge and show that our methods lead to the known 2.5PN near zone metric. In appendix \ref{sec:Match_inhom_sol} we discuss the solution for the exterior zone metric and its matching onto the 2.5PN near zone metric. In appendix \ref{app:conventions} we collect our conventions. Appendix C is a review of the multipole expansion of solutions to the free wave equation in Cartesian coordinates (for the sake of keeping the paper as self-contained as possible).

\section{The covariant $1/c$-expansion}\label{ref:sec2}

In this section we begin our exposition of the covariant $1/c$-expansion of GR also known as non-relativistic gravity (for a review see \cite{Hartong:2022lsy}). Ultimately, we want to make contact with the post-Newtonian approximation, but  before doing so we will briefly recap some results from \cite{HHO} (which was based in part on the earlier works\footnote{For other works on non-relativistic gravity see \cite{Tichy:2011te,Hansen:2019vqf,VandenBleeken:2019gqa,Hansen:2019svu,Hansen:2020wqw,Ergen:2020yop,Wolf:2023xrv}.} \cite{Hansen:2019pkl,VandenBleeken:2017rij,Dautcourt:1996pm}). We will deviate from this reference in two important ways. First of all, in \cite{HHO} they consider a $1/c^2$-expansion of Einstein gravity. However, in order to reproduce the half integer PN orders we will need to include odd powers in $1/c$ for our non-relativistic expansion. The second deviation comes from the fact that we will be doing an expansion of Einstein's field equations rather than the Einstein-Hilbert action. We choose to do this, as it reduces the amount of computation needed, which is very valuable when going to high orders. 

\subsection{Pre-nonrelativistic variables}

We state our conventions in appendix \ref{app:conventions}. The first task will be to formulate Einstein's field equations in terms of what are known as pre-non-relatvistic (PNR) variables. We can always write the metric $g_{\mu\nu}$ in terms of vielbeine $T_\mu$ and $\mathcal{E}^a_\mu$ as 
\begin{equation}
    g_{\mu\nu}=-c^2T_\mu T_\nu+\delta_{ab}\mathcal{E}^a_\mu \mathcal{E}^b_\nu\,, 
\end{equation}
where $a,b=1,2,3$ are spatial tangent space indices. We have introduced a speed of light so that $T_\mu dx^\mu$ has dimensions of time and we will denote $x^\mu=(t,x^i)$ (only in section \ref{sec:covGexp} will we use the notation $x^0=ct$). This helps with the covariant formulation in the non-relativistic domain. The PNR variables are $T_\mu$ and  $\Pi_{\mu\nu}=\delta_{ab}\mathcal{E}^a_\mu \mathcal{E}^b_\nu$ which is a symmetric tensor with signature $(0,1,1,1)$. The variables $(T_\mu,\mathcal{E}^a_\mu)$ form an invertible square matrix whose inverse $(T^\mu,\mathcal{E}_a^\mu)$ follows from the completeness and orthogonality conditions given by
\begin{align}
T_\mu \E^\mu_a = 0\,, \quad T^\mu \E_\mu^a = 0\,, \quad T_\mu T^\mu = -1\,, \quad \E^\mu_a \E^b_\mu = \delta^b_a\,, \quad \E^\mu_a \E^a_\mu = \delta_\nu^\mu + T^\mu T_\nu\,. \label{completeness1}
\end{align}
The metric and its inverse are thus 
\begin{align}
g_{\mu \nu} = - c^2 T_\mu T_\nu + \Pi_{\mu \nu}\,, \label{metric 1}\\
g^{\mu \nu} = -\frac{1}{c^2} T^\mu T^\nu + \Pi^{\mu \nu}\,, \label{metric2}
\end{align}
where $\Pi^{\mu \nu} = \E^\mu_a \E^\nu_b \delta^{ab}$.

So far everything is fully general. The theory of non-relativistic gravity (i.e. the $1/c$-expansion of GR) relies on the following important assumption: $T_\mu$ and $\Pi_{\mu\nu}$ admit Taylor series expansions in $1/c$. This assumption is known to breakdown in post-Newtonian calculations when tail terms start appearing in which case we need to consider expansions in $c^{-n}(\log c)^m$. This happens at higher post-Newtonian orders than considered in this work (we restrict our attention to 2.5PN order) and so we will not consider this important possibility. We refer to \cite{Blanchet2014} for more details.

Next, we will formulate Einstein's field equations in terms of the variables $T_\mu$ and $\Pi_{\mu\nu}$. We are specifically interested in the PNR version of the trace-reversed Einstein equations
\begin{align}
    R_{\mu \nu} = \frac{8\pi G}{c^4} \mathcal{S}_{\mu\nu}\,, \label{EE}
\end{align}
where we defined 
\begin{equation}\label{EE2}
    \mathcal{S}_{\mu\nu}= \T_{\mu \nu} - \frac{1}{2} g_{\mu \nu} \T\,,
\end{equation}
with $\T_{\mu \nu}$ the energy-momentum tensor (and $\T$ its trace).
Therefore, the main task is to rewrite the Ricci tensor, $R_{\mu \nu}$, in terms of PNR variables.

We know that the Ricci tensor can be expressed in terms of the Levi-Civita connection as
\begin{align}
    R_{\mu \nu} = \partial_{\sigma} \Gamma^{\sigma}_{\mu \nu} - \partial_{\mu} \Gamma^{\sigma}_{\sigma \nu} + \Gamma^{\sigma}_{\sigma \lambda} \Gamma^{\lambda}_{\mu \nu} - \Gamma^{\sigma}_{\mu \lambda} \Gamma^{\lambda}_{\sigma \nu}\,. \label{RelRicci}
\end{align}
So, first of all, we will have to work out the PNR version of the Levi-Civita connection. We know that
\begin{align}
    \Gamma^\rho_{\mu \nu} = \frac{1}{2} g^{\rho \sigma} \big( \partial_{\nu} g_{\mu \sigma} + \partial_{\mu} g_{\nu\sigma} - \partial_{\sigma} g_{\mu \nu}\big)\,,
\end{align}
so from equations (\ref{metric 1}) and (\ref{metric2}) we find that
\begin{align}\label{expGamma}
\Gamma^\rho_{\mu \nu} = c^2 {W^\rho_{\mu \nu}} + C^\rho_{\mu \nu} + S^\rho_{\mu \nu} + c^{-2} {V^\rho_{\mu \nu}}\,,
\end{align}
where we have defined
\begin{align}
{W^\rho_{\mu \nu}} := & \frac{1}{2} T_\mu \Pi^{\rho \sigma} (\partial_\sigma T_\nu - \partial_\nu T_\sigma) + \frac{1}{2}T_\nu \Pi^{\rho \sigma} (\partial_\sigma T_\mu - \partial_\mu T_\sigma)\,, \label{C(-2)}\\
C^\rho_{\mu \nu}  :=& - T^\rho \partial_\mu T_\nu + \frac{1}{2} \Pi^{\rho \sigma} (\partial_{\nu} \Pi_{\mu \sigma} + \partial_{\mu} \Pi_{\nu\sigma} - \partial_{\sigma} \Pi_{\mu \nu})\,,\label{C-conn}\\
S^\rho_{\mu \nu} :=& \frac{1}{2} T^\rho (\partial_\mu T_\nu - \partial_\nu T_\mu - T_\mu \sL_T T_\nu - T_\nu \sL_T T_\mu)\,, \label{C(0)}\\
{V^\rho_{\mu \nu}} := & \frac{1}{2} T^\rho \mathcal{L}_{T} \Pi_{\mu \nu}\,.\label{C(2)}
\end{align}
In here $\mathcal{L}_T$ denotes the Lie derivative along $T^\mu$.
We note that with the exception of \eqref{C-conn} all objects are tensorial. We refer to \eqref{C-conn} as the $C$-connection. Its LO expansion in $1/c$ gives us a useful connection that can be used in the covariant formulation of Newtonian gravity \cite{HHO,Hartong:2022lsy}. We note that the $C$-connection is not symmetric and so contains torsion. We stress that this is merely a reformulation of GR in terms of a torsionful and non-GR metric compatible connection whose features are chosen such that it gives us a useful Newton--Cartan connection when expanding in $1/c$.

We now insert the expression for the PNR Levi-Civita connection into equation (\ref{RelRicci}) and find that
\begin{align}\label{eq:PNRdecomRicci}
    R_{\mu \nu} = c^4 \RA_{\mu \nu} + c^2 \RB_{\mu \nu}+  \RC_{\mu \nu}+ c^{-2} \RD_{\mu \nu}\,,
\end{align}
with
\begin{align}
\RA_{\mu \nu} &= \frac{1}{4} T_\mu T_\nu \Pi^{\alpha \beta} \Pi^{\rho \sigma} T_{\alpha \rho} T_{\beta \sigma}\,, \label{ER-4}\\
\RB_{\mu \nu} &= \overset{(C)}{\nabla}_\sigma {W^\sigma_{\mu \nu}} +{W^\sigma_{\mu \nu}} S^\lambda_{\lambda \sigma} - {W^\sigma_{\mu \lambda}} S^\lambda_{\sigma \nu} -{W^\sigma_{\nu\lambda }} S^\lambda_{\sigma\mu } \,,\label{ER-2}
\\
\RC_{\mu \nu} &= \overset{(C)}{R}_{\mu \nu} - {W^{\sigma}_{\mu \lambda}} {V^{\lambda}_{\sigma \nu}} - {W^{\sigma}_{\nu \lambda}} {V^{\lambda}_{\sigma \mu}}  - \overset{(C)}{\nabla}_{\mu}  S^{\sigma}_{\sigma \nu} + \overset{(C)}{\nabla}_{\sigma} S^{\sigma}_{\mu \nu} -2C^{\lambda}_{[\mu \sigma]}S^{\sigma}_{\lambda \nu}\,, \label{check}
\\
\RD_{\mu \nu} &= \overset{(C)}{\nabla}_\sigma {V^\sigma_{\mu \nu}}\,, \label{ER2}
\end{align}
where we defined 
\begin{equation}
T_{\mu \nu} = \partial_\mu T_\nu - \partial_\nu T_\mu\,,    
\end{equation}
and where the overscript $(C)$ means that the object in question has been computed with respect to the $C$-connection \eqref{C-conn}. The expression for $\overset{(C)}{R}_{\mu \nu}$ is given by
\begin{equation}
    \overset{(C)}{R}_{\mu \nu} = \partial_{\sigma} C^{\sigma}_{\mu \nu} - \partial_{\mu} C^{\sigma}_{\sigma \nu} + C^{\sigma}_{\sigma \lambda} C^{\lambda}_{\mu \nu} - C^{\sigma}_{\mu \lambda} C^{\lambda}_{\sigma \nu}\,,
\end{equation}
which is not symmetric in $\mu$ and $\nu$ due to the fact that $C^\rho_{\mu\nu}$ has torsion.

We have now dealt with the left-hand side of equation (\ref{EE}). However, it will be convenient to rewrite the right-hand side of (\ref{EE}) as well, since we are generally going to be working with $\T^{\mu \nu}$ rather than $\T_{\mu \nu}$. For example, for the trace $\T =T^{\mu}{}_\mu$ we find that
\begin{align}
    \T = -c^2 T_\alpha T_\beta \T^{\alpha \beta} + \Pi_{\alpha \beta} \T^{\alpha \beta}\,.
\end{align}
We can also express $\T_{\mu \nu}$ in terms of $\T^{\mu \nu}$ in which case we get
\begin{align}
    \T_{\mu \nu} = c^4 T_\mu T_\alpha T_\nu T_\beta \T^{\alpha \beta} - c^2 (T_\mu T_\alpha \Pi_{\nu \beta} + T_\nu T_\beta \Pi_{\mu \alpha}) \T^{\alpha \beta}
    + \Pi_{\mu \alpha} \Pi_{\nu \beta} \T^{\alpha \beta}\,.
\end{align}
This leads us to our final expression for the PNR version of Einstein's equations
\begin{align}
    \sum_{n=0}^3 c^{(4-2n)} R_{\mu \nu}^{[-4+2n]} = 4 \pi G \Big[& T_\mu T_\nu T_\alpha T_\beta - \frac{1}{c^2}(2T_\mu T_\alpha \Pi_{\nu \beta} + 2T_\nu T_\beta \Pi_{\mu \alpha} - T_\mu T_\nu \Pi_{\alpha \beta} \nonumber
    \\
    &-T_\alpha T_\beta \Pi_{\mu \nu}) + \frac{2}{c^4} \Pi_{\mu \alpha} \Pi_{\nu \beta} - \frac{1}{c^4} \Pi_{\mu \nu} \Pi_{\alpha \beta} \Big] \mathcal{T}^{\alpha \beta}\,. \label{PNR Einstein}
\end{align}
This equation may seem daunting but it is the $1/c$-expansion we are interested in and when performing that expansion this will prove to be a useful starting point.

\subsection{Notation and basic identities}
The basics objects we are going to be expanding are $T_\mu$, $\Pi_{\mu \nu}$, $T^\mu$ and $\Pi^{\mu \nu}$. Since we are planning to go to high orders in the long run, we introduce the following notation for the expansion of the PNR fields
\begin{align}
    T_\mu &=\tau_\mu+ \sum_{n=1}^\infty \frac{1}{c^n} \tau_\mu^{(n)}, \qquad & T^\mu &=v^\mu+ \sum_{n=1}^\infty \frac{1}{c^n} v^\mu_{(n)}\,,
    \\
    \Pi_{\mu \nu} &=h_{\mu\nu}+  \sum_{n=1}^\infty \frac{1}{c^n} h_{\mu \nu}^{(n)}, \qquad & \Pi^{\mu \nu} &=h^{\mu\nu}+  \sum_{n=1}^\infty \frac{1}{c^n} h^{\mu \nu}_{(n)}\,.
\end{align}
The leading order (LO) geometry is of Newton--Cartan type and is described by $\tau_\mu$ and $h_{\mu\nu}$. For ease of notation in some expressions below we will sometimes denote the LO objects $\tau_\mu$ with a $(0)$ superscript, i.e. $\tau_\mu=\tau^{(0)}_\mu$, and similarly for $h_{\mu\nu}$, $v^\mu$ and $h^{\mu\nu}$.

Now, the variables above are not all independent. They are related through the completeness/orthogonality relations that we know from GR, equation \eqref{completeness1}. These relations given in terms of $T_\mu$ and $\Pi_{\mu\nu}$ read
\begin{align}
    T_\mu T^\mu = -1\,, \qquad T_\mu \Pi^{\mu \nu} = T^\mu \Pi_{\mu \nu} = 0\,,\qquad \Pi^{\mu \rho} \Pi_{\rho \nu} = \delta^\mu_{ \nu} + T^\mu T_\nu\,. \label{Completeness}
\end{align}
These hold order by order in the $1/c$-expansion. At leading order we simply get
\begin{align}
    \tau_\mu v^\mu = -1\,, \qquad
    \tau_\mu h^{\mu \nu} = v^\mu h_{\mu \nu} = 0\,, \qquad  h^{\mu \rho} h_{\rho \nu} = \delta^\mu_\nu + v^\mu \tau_\nu\,. \label{LOcompleteness}
\end{align}

For every subsequent order we get a new set of constraints from equation (\ref{Completeness}). At the $N$th order in $1/c$ (for $N \geq 1$) we get
\begin{align}
 \sum_{n=0}^{N} v^\mu_{(n)} \tau_\mu^{(N-n)}&=0\,, \qquad&   \sum_{n=0}^{N} v^\nu_{(n)} h_{\nu \mu}^{(N-n)} &= 0\,, \label{Complete1}
 \\  
 \sum_{n=0}^{N} h^{\mu \nu}_{(n)} \tau_\nu^{(N-n)} &= 0\,, \qquad& \sum_{n=0}^{N} h^{\mu \rho}_{(n)} h_{\rho \nu}^{(N-n)}- \sum_{n=0}^{N} v^\mu_{(n)} \tau_\nu^{(N-n)}&= 0\,. \label{Complete2}
\end{align}
We can use these equations to express $h^{\mu \nu}_{(N)}$ and $v^{\mu}_{(N)}$ in terms of $h^{\mu \nu}$, $v^\mu$, $h_{\mu \nu}^{(n)}$ and $\tau_{\mu}^{(n)}$ (for $n \leq N$). From equations (\ref{Complete1}) and (\ref{Complete2}) we can derive the following expressions for $h^{\mu \nu}_{(N)}$ and $v^{\mu}_{(N)}$ 
\begin{align}
    v^{\mu}_{(N)} &= v^{\mu} \sum_{n=0}^{N-1} v^\sigma_{(n)} \tau_\sigma^{(N-n)} - h^{\mu \sigma} \sum_{n=0}^{N-1} v^\nu_{(n)} h_{\nu \sigma}^{(N-n)}\,, \label{inverse1}
    \\
    h^{\mu \nu}_{(N)} &=   h^{\mu \sigma} \sum_{n=0}^{N-1} \left( v^\nu_{(n)} \tau_{\sigma}^{(N-n)} - h^{\nu\rho}_{(n)} h_{\rho \sigma}^{(N-n)} \right) + v^\mu \sum_{n=0}^{N-1}  h^{\nu \sigma}_{(n)}  \tau_{\sigma}^{(N-n)}\,.
    \label{inverse2}
\end{align}
We can solve these equations iteratively, starting from $N=1$ and working our way up to the desired order.

It is clear that these expressions get messy very quickly. In practice, however, one determines the metric at a certain order in $1/c$ before going to the next order. It then often happens (especially at low orders) that certain components at a given order in $1/c$ will be zero which simplifies the expressions for the inverse objects at higher orders considerably compared to the general result.  It is therefore not very useful to compute the higher order contributions to the inverse objects without knowing anything about $\tau_\mu^{(n)}$ and $h_{\mu\nu}^{(n)}$ at lower orders.

\subsection{Gauge transformations}
Since we are working with a covariant theory of non-relativistic gravity, gauge transformations are going to play a crucial role. In order to describe the most general non-relativistic gauge transformation, we must first study the gauge transformations of our PNR variables $T_\mu$ and $\Pi_{\mu \nu}$. Because we have split the metric into $T_\mu$ and $\Pi_{\mu \nu}$, we are allowed to perform local Lorentz boosts that transform $T_\mu$ and $\Pi_{\mu \nu}$ into each other while leaving the metric invariant. Apart from that the only other gauge transformations that act on $T_\mu$ and $\Pi_{\mu \nu}$ are diffeomorphisms.

The action of the gauge transformations on $T_\mu$ and $\Pi_{\mu \nu}$ are thus given by
\begin{align}
    \delta T_\mu &= \mathcal{L}_{\Xi} T_\mu + c^{-2} \Lambda_\mu\,, \label{GaugeT}
    \\
    \delta \Pi_{\mu \nu} &= \mathcal{L}_{\Xi} \Pi_{\mu \nu} +  T_\mu\Lambda_\nu  + T_\nu\Lambda_\mu\,, \label{GaugePi}
\end{align}
where $\Xi^\mu$ is a vector field generating diffeomorphisms and where $\Lambda_\mu=\Lambda_b \mathcal{E}^b_\mu$ is any 1-form for which $T^\mu\Lambda_\mu=0$. The transformations with local parameter $\Lambda_b$ correspond to local Lorentz boost transformations.

The next step is to expand both sides of equations (\ref{GaugeT}) and (\ref{GaugePi}). We will assume that the gauge parameters are real analytic in $1/c$ in order that the gauge transformed objects $T_\mu$ and $\Pi_{\mu\nu}$ admit a Taylor series in $1/c$. We thus consider the following expansions
\begin{align}
    \Xi^\mu &= \xi^\mu_{(0)} + \frac{1}{c} \xi_{(1)}^\mu + \frac{1}{c^2} \xi_{(2)}^\mu +\cdots\,,
    \\
    \Lambda_\mu &= \lambda_\mu^{(0)} + \frac{1}{c} \lambda^{(1)}_\mu + \frac{1}{c^2} \lambda^{(2)}_\mu +\cdots\,.
\end{align}
For the LO gauge transformations we will write $\xi^\mu=\xi^\mu_{(0)}$ and $\lambda_\mu=\lambda_\mu^{(0)}$.

Starting with equation (\ref{GaugeT}), we find that the most general gauge transformations for $\tau_\mu$, $\tau_\mu^{(1)}$, $\tau_\mu^{(2)}$, $\tau_\mu^{(3)}$ and $\tau_\mu^{(N)}$ are given by
\begin{align}
    \delta \tau_\mu &= \mathcal{L}_{\xi} \tau_\mu\,, \label{gaugetau}\\
    \delta \tau^{(1)}_\mu &= \mathcal{L}_{\xi_{(1)}} \tau_\mu + \mathcal{L}_{\xi} \tau_\mu^{(1)}\,, \label{gaugetau1} \\
    \delta \tau^{(2)}_\mu &= \mathcal{L}_{\xi_{(2)}} \tau_\mu + \mathcal{L}_{\xi_{(1)}} \tau_\mu^{(1)}
    + \mathcal{L}_{\xi} \tau_\mu^{(2)} + \lambda_\mu\,, \label{gaugetau2}
    \\
    \delta \tau^{(3)}_\mu &= \mathcal{L}_{\xi_{(3)}} \tau_\mu + \mathcal{L}_{\xi_{(2)}} \tau_\mu^{(1)}
    + \mathcal{L}_{\xi_{(1)}} \tau_\mu^{(2)} + \mathcal{L}_{\xi} \tau_\mu^{(3)} + \lambda_\mu^{(1)}\,, \label{gaugetau3}
    \\
    \delta \tau^{(N)}_\mu &= \sum^N_{n=0} \mathcal{L}_{\xi_{(n)}} \tau_\mu^{(N-n)} +\lambda_\mu^{(N-2)}\,, \label{gaugetauN}
\end{align}
where the condition $T^\mu\Lambda_\mu=0$ implies that $\lambda_\mu^{(N)}$ obeys
\begin{equation}
    \sum_{n=0}^{N}v^\mu_{(n)}\lambda_\mu^{(N-n)}=0\,.
\end{equation} 
In the case of (\ref{GaugePi}) we find that the gauge transformations of $h_{\mu \nu}$, $h_{\mu \nu}^{(1)}$ and $h_{\mu \nu}^{(N)}$ are given by 
\begin{align}
    \delta h_{\mu \nu} &= \mathcal{L}_{\xi} h_{\mu \nu} + 2  \tau_{(\mu} \lambda_{\nu)}\,, \label{gaugeh}
    \\
    \delta h_{\mu \nu}^{(1)} &= \mathcal{L}_{\xi_{(1)}} h_{\mu \nu} + \mathcal{L}_{\xi} h_{\mu \nu}^{(1)} + 2 \tau_{(\mu} \lambda^{(1)}_{\nu)} + 2  \tau_{(\mu}^{(1)}\lambda_{\nu)}\,, \label{gaugeh1}
    \\
    \delta h_{\mu \nu}^{(N)} &= \sum_{n=0}^N \mathcal{L}_{\xi_{(n)}} h_{\mu \nu}^{(N-n)} + 2\sum_{n=0}^N\tau_{(\mu}^{(n)}\lambda_{\nu)}^{(N-n)}\,. \label{gaugehN}
\end{align}

If we go back to \eqref{GaugeT} and \eqref{GaugePi} we see that we can fix the local Lorentz transformations entirely by setting $\Pi_{it}=0$. In this gauge we have $\Pi_{tt}=0$ for otherwise the signature would not be $(0,1,1,1)$, i.e. the determinant of $\Pi_{\mu\nu}$ is zero while $\text{det}\,\Pi_{ij}$ is nonzero. In this gauge we also have that $T^i=0$ which follows from $T^\mu\Pi_{\mu\nu}=0$ for $\nu=j$. Hence, the condition $T^\mu\Lambda_\mu=0$ implies that $\Lambda_t=0$. The condition $T^\mu T_\mu=-1$ with $T^i=0$ tells us that $T^t$ must be nonzero since $T^\mu$ is non-vanishing and thus that $T_t\neq 0$. Explicitly, when we take $\Pi_{it}=0$ we obtain for the inverse objects
\begin{equation}
    T^t=-\frac{1}{T_t}\,,\qquad T^i=0\,,\qquad \Pi^{ti}=-\frac{1}{T_t}T_j\Pi^{ij}\,,\qquad\Pi^{tt}=\frac{1}{T_t^2}T_i T_j\Pi^{ij}\,,
\end{equation}
where $\Pi^{ij}$ follows from 
\begin{equation}
    \Pi^{ik}\Pi_{kj}=\delta^i_j\,.
\end{equation}

The residual gauge transformations of the gauge choice $\Pi_{ti}=0$ follow from setting $\delta\Pi_{ti}=0=\mathcal{L}_\Xi \Pi_{ti}+T_t\Lambda_i$ which tells us that $\Lambda_i$ is entirely fixed and given by
\begin{equation}
    \Lambda_i=-\frac{1}{T_t}\Pi_{ij}\partial_t\Xi^j\,.
\end{equation}
Using this result together with \eqref{GaugeT} and \eqref{GaugeT} we see that the residual gauge transformations act on $T_\mu$ and $\Pi_{ij}$ as follows 
\begin{eqnarray}
    \delta T_t & = & \Xi^\rho\partial_\rho T_t+T_\rho\partial_t\Xi^\rho\,,\\
    \delta T_i & = & \Xi^\rho\partial_\rho T_i+T_\rho\partial_i\Xi^\rho-\frac{1}{c^2}\frac{1}{T_t}\Pi_{ij}\partial_t\Xi^j\,,\\
    \delta \Pi_{ij}& = & \Xi^\rho\partial_\rho\Pi_{ij}+\Pi_{kj}\partial_i\Xi^k+\Pi_{ik}\partial_j\Xi^k-\frac{1}{T_t}T_i\Pi_{jk}\partial_t\Xi^k-\frac{1}{T_t}T_j\Pi_{ik}\partial_t\Xi^k\,.
\end{eqnarray}

In this gauge the metric is parameterised as
\begin{equation}\label{eq:KSmetric}
    ds^2=-c^2\left(T_t dt+T_i dx^i\right)^2+\Pi_{ij}dx^idx^j\,.
\end{equation}
This is the metric in Kol--Smolkin (KS) parameterisation \cite{Kol:2010si,Kol:2007bc}. We will refer to the choice $\Pi_{it}=0$ as the KS gauge. Alternatively, we could have fixed the local Lorentz transformations by setting $T_i=0$. This would have led to the metric in Arnowitt-Deser-Misner (ADM) parameterisation. See \cite{Elbistan:2022plu} for more information about these two choices in relation to $1/c$ and $c$-expansions of GR. We prefer the KS parameterisation because then the nonzero components of $\Pi_{\mu\nu}$ form a 3-dimensional invertible tensor $\Pi_{ij}$. We will henceforth always take $\Pi_{ti}=0$.

\subsection{The perfect fluid in non-relativistic gravity}\label{subsec:PFvariables}

In this paper we are going to work with a perfect fluid (with compact support) as our matter source. The energy-momentum tensor for a perfect fluid is given by
\begin{align}
    \T^{\mu \nu} = \frac{E+P}{c^2} U^\mu U^\nu + P\Pi^{\mu \nu} - \frac{1}{c^2}P T^\mu T^\nu\,. \label{PNREMT}
\end{align}
where $E$ is the relativistic internal energy density, $P$ is the pressure and $U^\mu$ is the four-velocity that is normalised such that 
\begin{align}
    g_{\mu \nu} U^\mu U^\nu = - c^2\,. \label{VeloNorm}
\end{align}
Using \eqref{metric 1} this can be solved for $T_\mu U^\mu$ by writing this as
\begin{equation}\label{eq:TU}
    \left(T_\mu U^\mu\right)^2=1+\frac{1}{c^2}\Pi_{\mu\nu}U^\mu U^\nu\,.
\end{equation}

Since we expand the metric in even and odd powers of $1/c$ it is inevitable that we also have to include even and odd powers in the expansion of the fluid variables. The even powers are of course the dominant ones that correspond to the 0PN, 1PN etc sources. It turns out that at low orders the odd powers in $1/c$ in the metric are either zero or pure gauge. Our approach to expanding the fluid variables in $1/c$ is to assume this to be an even power series expansion until that assumption breaks down. This breakdown can be seen by studying the fluid conservation equations (the $1/c$-expansion of the covariant constancy of the fluid's energy-momentum tensor) at each PN order and to ensure that each non-trivial PN order has its own set of fluid variables to avoid unphysical constraints on the solution\footnote{If we have a non-trivial equation at a given order in the expansion of the fluid conservation equations and the fields appearing in said equation are all lower order fields that have already been determined at previous orders that equation would appear as a constraint on these lower order fields. This would be unwanted and simply a consequence of not having introduced the appropriate coefficients in the $1/c$-expansion of the fluid variables.}. In this way it turns out that we need odd powers in $1/c$ in the expansion of the fluid variables for the first time at 2.5PN in the expansion of the fluid equation. Odd terms break time-reversal symmetry and this is related to the well-known fact that the fluid starts to dissipate at 2.5PN due to the emission of gravitational waves.

We expand the energy density $E$, pressure $P$ and 3-velocity $U^i$ in powers of $1/c^2$, until we get to 2.5PN. In order to recover the Newtonian limit we need to  assume that $E$ starts at order $c^2$ and that $P$ starts at order $c^0$. We therefore have the following expansion
\begin{align}
    E &= c^2 E_{(\sm2)}  + E_{(0)} + \frac{1}{c^2} E_{(2)}  +  \frac{1}{c^3} E_{(3)}+\mathcal{O} (c^{-4})\,,  \label{eq:E_2.5}  \\
    P &= P_{(0)} + \frac{1}{c^2} P_{(2)}+  \mathcal{O} (c^{-4})\,,\label{eq:P_2.5}\\
    U^i &= v^i + \frac{1}{c^2} v^i_{(2)}+  \mathcal{O}(c^{-4})\,.\label{eq:Ui_2.5}
\end{align}
We will always assume that $E_{(\sm2)}> 0$. At 0PN, i.e. order $c^0$ in the expansion of the fluid conservation equations the fluid variables are $E_{(\sm2)}$, $P_{(0)}$ and $v^i$. However, at 0PN the metric only features $E_{(\sm2)}$. The 2.5PN fluid variables are $E_{(3)}, P_{(5)}$ and $v_{(5)}^i$. However, since our goal is to work up to 2.5PN in the metric we will only need $E_{(3)}$ of these variables.

The 4-velocity $U^\mu$ is a constrained variable. The time component $U^t$ follows from \eqref{eq:TU} which in the KS gauge becomes
\begin{equation}\label{eq:Unorm}
    \left(T_\mu U^\mu\right)^2=1+\frac{1}{c^2}\Pi_{ij}U^i U^j\,.
\end{equation}
Hence, the expansion of $U^t$ follows from the expansion of the PNR variables and $U^i$. At LO we have 
\begin{equation}\label{eq:LOfluidu_1}
 U^\mu=u^\mu+\mathcal{O}(c^{-1})\,,   
\end{equation}
and \eqref{eq:Unorm} tells us that 
\begin{equation}\label{eq:LOfluidu_2}
\tau_\mu u^\mu=1\,.    
\end{equation}

\section{The Newtonian order}\label{sec:Newton}

In this section, we set the stage for the post-Newtonian expansion by discussing how the Newtonian limit of GR comes about in the non-relativistic gravity framework reviewed in the previous section. 

The general covariant framework that describes Newtonian gravity is Newton--Cartan gravity. Newtonian gravity is a gauge-fixed version of that setting (for details see for example the review paper \cite{Hartong:2022lsy})\footnote{In Newton-Cartan gravity it is perfectly possible to choose a gauge in which the Newtonian potential is zero whilst still being able to describe the same physics as we observe in Newtonian gravity \cite{Kapustin:2021omc}.}. A post-Newtonian framework therefore necessarily has to be consistent with the same gauge fixing that is done in Newton-Cartan gravity to obtain the Newtonian description. In this section we will show how this gauge fixing works in the framework of non-relativistic gravity that was introduced in the previous section. One of the main purposes of this paper is to set up a framework for post-Newtonian calculations that is not tied to a particular gauge choice such as the harmonic gauge. However, the very fact that we want to be {\it{post-Newtonian}} means that we inevitably have to restrict ourselves to those gauge choices that are compatible with a Newtonian viewpoint. It would be interesting to develop techniques to study what one might call post-Newton-Cartan gravity which would then have to be a fully covariant version of what we present here and of what has been done elsewhere.

Finally, we end this section by discussing the 0.5PN order (which is trivial) in this non-relativistic gravity framework.

\subsection{Absolute time and Newtonian gravity}

We start our discussion by showing how a perfect fluid with $E = \mathcal{O}(c^2)$ and $P = \mathcal{O}(c^0)$ gives rise to a non-relativistic spacetime with absolute time at leading order in $1/c$. We start with Einstein's field equations, which we have written in PNR form in equation (\ref{PNR Einstein}). To leading order equation (\ref{PNR Einstein}) becomes
\begin{align}
    \frac{1}{4} \tau_\mu \tau_\nu h^{\alpha \beta} h^{\rho \sigma} \tau_{\alpha \rho} \tau_{\beta \sigma} =0\,, \label{L-EOM}
\end{align}
where we defined
\begin{equation}
    \tau_{\mu \nu} = \partial_\mu \tau_\nu - \partial_\nu \tau_\mu\,.
\end{equation}
Equation \eqref{L-EOM} is simply the leading order expansion of $R_{\mu \nu}^{[-4]}$ which is set to zero because there is no term on the RHS of (\ref{PNR Einstein}) that is of order $c^4$. We see that the factor in front of $\tau_\mu \tau_\nu$ is a sum of squares and so equation (\ref{L-EOM}) implies that 
\begin{align}\label{eq:TTNC}
    h^{\alpha \beta} h^{\rho \sigma} \tau_{\beta \sigma}=0\,,
\end{align}
which in the Newton--Cartan literature is known as the twistless torsional Newton--Cartan (TTNC) condition which is equivalent to  $\tau \wedge d\tau = 0$ \cite{Christensen:2013lma}. This condition tells us that the spacetime admits a foliation since by Frobenius' theorem this is equivalent to $\tau=NdT$ where $N$ and $T$ are two scalar fields. The function $N$ is like a non-relativistic lapse function that describes time dilation. 

In this work we will always assume that we can make a weak field approximation which corresponds to absolute time in the NC setting but it is perhaps interesting that in principle NC geometry can also describe what is called strong NR gravity. This simply means that $dN\wedge d\tau\neq 0$ so that $N$ describes time dilation. In \cite{VandenBleeken:2017rij,VandenBleeken:2019gqa} it has been shown that the Schwarzschild geometry admits a strong NR approximation, and that this regime of NR gravity can describe perihelion of mercury, and effects due to gravitational time dilation (in agreement with GR) \cite{Hansen:2019vqf}. It would be interesting to study this regime as a potential starting point for an approximation scheme that does not start with flat space (and a Newtonian potential).

In order to arrive at absolute time we must turn to the conservation of the energy-momentum tensor, given by
\begin{align}
    \nabla_\nu \mathcal{T}^{\mu \nu} =  0\,. \label{Conservation2}
\end{align}
Using the $1/c$-expansions of the previous section the LO term of the expansion of this equation is given by \cite{HHO},
\begin{align}
    0= E_{(\sm 2)} h^{\mu \sigma} u^\nu \tau_{\sigma \nu}\,, \label{abs1}
\end{align}
where $u^\mu$ is defined in equation \eqref{eq:LOfluidu_1}. From equation \eqref{eq:LOfluidu_2} it follows that we can write the fluid velocity field as
\begin{align}
    u^\nu = - v^\nu + h^{\nu \rho} X_{\rho}\,,
\end{align}
for some field $X_\rho$. Using this along with the TTNC condition, equation (\ref{abs1}) reduces to
\begin{align}
     0= h^{\mu \sigma} v^\nu \tau_{\sigma \nu}\,. 
\end{align}
If we contract this with $h_{\mu \rho}$ we get
\begin{align}
    0= v^\nu \tau_{\rho \nu}\,. 
\end{align}
This along with the TTNC condition \eqref{eq:TTNC} means that $\tau_{\rho \nu} = 0$ which is the condition for absolute time and so we set $\tau = dT$ for some scalar field $T$. We can and will always choose coordinates such that $T=t$. 

We want to arrive at Newtonian gravity, which means that we need to compute the metric up to order $c^0$ (i.e. up to $\tau_\mu^{(2)}$ and $h_{\mu \nu}$). So, we continue to expand  Einstein's field equations until we have solved the metric up to order $c^0$. 

The next non-trivial part of Einstein's field equations \eqref{PNR Einstein} comes at order $c^2$ in which case we get
\begin{align}
    \overset{(-2)}{R_{\mu \nu}^{[-4]}} + \overset{(0)}{R_{\mu \nu}^{[-2]}} = 0 \,,
\end{align}
where $\overset{(-2)}{R_{\mu \nu}^{[-4]}}$ denotes the order $c^2$ term in the $1/c$-expansion of $R_{\mu \nu}^{[-4]}$. Likewise, $\overset{(0)}{R_{\mu \nu}^{[-2]}}$ denotes the order $c^0$ term in the expansion of $R_{\mu \nu}^{[-2]}$. Using that $d \tau = 0$ this becomes
\begin{align}
    \frac{1}{4} \tau_\mu \tau_\nu h^{\alpha \beta} h^{\rho \sigma} \tau_{\alpha \rho}^{(1)} \tau_{\beta \sigma}^{(1)} =0\,,
\end{align}
and so we conclude that $\tau^{(1)} \wedge d\tau^{(1)} = 0$.

We then turn to the NLO equation in the expansion of (\ref{Conservation2}). 
Using again that $d \tau = 0$ we end up with the following expression
\begin{align}
    0= E_{(\sm 2)} h^{\mu \sigma} u^\nu \tau_{\sigma \nu}^{(1)}\,.
\end{align}
Using a similar argument as was used at LO, we conclude that $d\tau^{(1)}=0$, so that $\tau^{(1)}_\mu=\partial_\mu T^{(1)}$. The gauge transformation acting on $\tau^{(1)}_\mu$ is given in equation \eqref{gaugetau1}. We can use $\xi_{(1)}^t$ to set $\tau^{(1)}_\mu=0$. We will always assume this gauge choice. 

The next nonzero order in the expansion of Einstein's field equations is at order $c^0$, which is the Newtonian order and is given by
\begin{align}
    \overset{(4)}{R_{\mu \nu}^{[-4]}} + \overset{(2)}{R_{\mu \nu}^{[-2]}} + \overset{(0)}{R_{\mu \nu}^{[0]}}   = 4 \pi G \tau_\mu \tau_\nu E_{(\sm 2)}\,. \label{Newtonian order}
\end{align}
Using equations \eqref{ER-4}--\eqref{check} we have that
\begin{align}
    \overset{(4)}{R_{\mu \nu}^{[-4]}} &= \frac{1}{4} \tau_\mu \tau_\nu h^{\alpha \beta} h^{\rho \sigma} \tau_{\alpha \rho}^{(2)} \tau_{\beta \sigma}^{(2)}\,,
    \\
    \overset{(2)}{R_{\mu \nu}^{[-2]}} &= \check\nabla_\sigma \overset{(2)}{W^\sigma_{\mu \nu}} = \partial_\sigma \overset{(2)}{W^\sigma_{\mu \nu}} +\check\Gamma^\sigma_{\sigma \alpha} \overset{(2)}{W^\alpha_{\mu \nu}} - \check\Gamma^\alpha_{\sigma \mu} \overset{(2)}{W^\sigma_{\alpha \nu}} -\check\Gamma^\alpha_{\sigma \nu} \overset{(2)}{W^\sigma_{\mu \alpha}}\,, \label{R2}
    \\
    \overset{(0)}{R_{\mu \nu}^{[0]}} &= \check R_{\mu \nu}\,,
\end{align}
where we used that $d\tau=0$ and 
where $\check\Gamma^\rho_{\mu\nu}$ is the Newton-Cartan connection that is obtained as the LO term in the $1/c$-expansion of the $C$-connection. Explicitly, it is given by
\begin{equation}
    C^\rho_{\mu\nu}\Big|_{\mathcal{O}(c^0)}=\check\Gamma^\rho_{\mu\nu}=-v^\rho\partial_\mu\tau_\nu+\frac{1}{2}h^{\rho\sigma}\left(\partial_\mu h_{\nu\sigma}+\partial_\nu h_{\mu\sigma}-\partial_\sigma h_{\mu\nu}\right)\,.
\end{equation}
Quantities such as $\check\nabla_\mu$ and $\check R_{\mu \nu}$ are computed using the $\check\Gamma^\rho_{\mu\nu}$ connection. We furthermore have that
\begin{align}
    \overset{(2)}{W^\rho_{\mu \nu}} =  \frac{1}{2} \tau_\mu h^{\rho \sigma} \tau_{\sigma \nu}^{(2)} + \frac{1}{2} \tau_\nu h^{\rho \sigma} \tau_{\sigma \mu}^{(2)}\,. \label{W2}
\end{align}

In order to solve \eqref{Newtonian order} we start with the $ij$ component. This simply becomes
\begin{align}
    \check R_{ij} = 0\,. \label{Ricciflat}
\end{align}
Since $\tau_\mu h^{\mu\nu}=0$ and $\tau=dt$ we have that $h^{t\mu}=0$. We also fixed the local Lorentz boosts by setting $\Pi_{ti}=0$ which implies at LO that $h_{t\mu}=0$. Hence the only nonzero components of $\check\Gamma^\rho_{\mu \nu}$ are the $\check\Gamma^k_{ij}$ components. These are given by
\begin{align}
    \check\Gamma^k_{ij} = \frac{1}{2} h^{kl} (\partial_{i} h_{jl} + \partial_{j} h_{il} - \partial_{l} h_{ij})\,,
\end{align}
where $h_{ij}$ is a Riemannian metric on a constant $t$ slice. Equation \eqref{Ricciflat} states that $h_{ij}$ is Ricci flat.

Since we are working in three spatial dimensions we know that the Weyl tensor of the Riemannian geometry on the constant $t$ slices is zero. This means that if $h_{ij}$ is Ricci flat it is also Riemann flat. Since the constant time slices are assumed to be non-compact, there exist coordinates such that
\begin{align}
    h_{\mu \nu} = \delta_{ij} \delta^i_{\mu} \delta^i_{\nu}\,.
\end{align}
The $\mu = \nu = t$ and $\mu= t $, $\nu= i$ components of (\ref{Newtonian order}) then become
\begin{align}
    \frac{1}{4} \tau_{ij}^{(2)} \tau_{ij}^{(2)} + \partial_i \tau_{it}^{(2)} &= 4 \pi G E_{(\sm 2)}\,, \label{N1}
    \\
    \partial_j \tau_{ji}^{(2)} &=0\,. \label{Difftau2}
\end{align}

We use the fact that we are working in 3 spatial dimension to write $\tau^{(2)}_{ij}$ in terms of a (pseudo-)vector field  $F_k$
\begin{align}
    \tau^{(2)}_{ij} = \epsilon_{ijk} F_k\,. \label{DefF}
\end{align}
Equation (\ref{Difftau2}) then becomes
\begin{align}
    \partial_{[i}F_{k]} = 0\,.
\end{align}
This means that we can write $F_k = \partial_k F$ for some unknown function $F$. 
Recall that $\tau^{(2)}_{ij} = 2\partial_{[i} \tau^{(2)}_{j]}$ so it must satisfy the Bianchi identity
\begin{align}
    \partial_{[i} \tau^{(2)}_{jk]}=0\,.
\end{align}
If we contract this equation with $\epsilon^{ijk}$ and use \eqref{DefF} we find that $\partial_k F^k=0$, and thus that $F$ is harmonic.

Since $F$ is a harmonic function, any non-trivial solution to $\partial^2 F=0$ will lead to a singularity somewhere in space (independent of the matter distribution) if we include infinity. As discussed in the introduction the $1/c$-expansion only has a finite region of validity. Within this region that contains the matter source we need $F$ to be regular (and thus in particular we will demand that $F$ is regular at origin). In order for this harmonic function to be nonzero we need to match this onto a solution in the exterior region that is an order $G$ solution to the source-free Einstein equations. It turns out that by matching onto such a solution the harmonic function $F$ has to be zero. Another viewpoint is that we insist that in the NR regime Newtonian gravity is a good approximation and so we should be able to demand that the metric to order $c^0$ is asymptotically flat. This means that $F$ cannot be a nontrivial harmonic function and the only allowed solution for $F$ is $F=F(t)$. This means that $\tau_{ij}^{(2)} = 0$ and therefore equation (\ref{N1}) reduces to the Poisson equation whose solution is found by use of a Green's function
\begin{align}\label{U}
    \tau_t^{(2)}(x,t) = -G \int \frac{E_{(\sm2)}(t,x')}{|x-x'|} d^3x' = -U\,.
\end{align}
This is the Newtonian gravitational potential, as expected. The integration is over the matter source.

\subsection{The Newtonian matter equations}
 Having computed the Newtonian metric, we turn to the matter EOMs to see if we get the correct fluid equations. The Newtonian term in the expansion of (\ref{Conservation2}) is at order $c^0$ which evaluates to be
\begin{align}
    u^\mu u^\nu \partial_\nu E_{(\sm 2)} + 2E_{(\sm 2)} u^{(\mu} \partial_\nu u^{\nu)} + E_{(\sm 2)} h^{\mu \sigma} \tau^{(2)}_{\sigma \nu} u^\nu  + \partial_\nu P_{(0)} h^{\mu \nu}= 0\,. \label{MatEOM1}
\end{align}
This equation describes both mass conservation as well as momentum conservation. To see this we consider the $\mu=t,i$ components separately. For $\mu=t$ we obtain
\begin{align}
    0 = \partial_\nu (E_{(\sm 2)} u^\nu)\,. \label{masscons}
\end{align}
This equation corresponds to conservation of mass since $E_{(\sm 2)}>0$ is the nonzero mass density. Then if we take $\mu = i$ we get the Euler equation in a Newtonian potential
\begin{align}
    \partial_t v^i + v^j \partial_j v^i = - \frac{1}{E_{(\sm 2)}} \partial_i P_{(0)} - \partial_i \tau^{(2)}_t\,.
\end{align}

The latter equation together with \eqref{masscons} form 4 equations for 5 unknowns. The unknowns are velocity $v^i$, mass density $E_{(\sm2)}$ and temperature which enters $P_{(0)}$. Normally, the 5th equation is the energy conservation equation. However, this comes from the NLO correction to \eqref{masscons} which also depends on subleading fields, such as $v^i_{(2)}$, that appear in the expansion of the fluid variables.
Hence, we do not get a closed system of equations for just the LO fluid variables (see \cite{HHO} for more details).

\subsection{Gauging away the 0.5PN metric}\label{subsec:0.5PNmetric}

In this section we want to solve for the 0.5PN metric, which requires knowing $\tau_\mu^{(3)}$ and $h_{\mu \nu}^{(1)}$. We begin by expanding Einstein's field equations to one order higher in $1/c$ than the Newtonian order. Einstein's field equations at order $c^{-1}$ are 
\begin{align}
    \overset{(5)}{R_{\mu \nu}^{[-4]}} + \overset{(3)}{R_{\mu \nu}^{[-2]}} + \overset{(1)}{R_{\mu \nu}^{[0]}} = 0\,. \label{halfPNEin}
\end{align}
We note that there is no source term at this order. Using that $\tau_{ij}^{(2)}=0$ as well as $h_{(1)}^{tt}=0$, which we get from $\tau^{(1)}_\mu=0$ and the orthogonality conditions, it can be shown that $\overset{(5)}{R_{\mu \nu}^{[-4]}}=0$.
The other two terms in \eqref{halfPNEin} can be shown to be equal to
\begin{eqnarray}
    \overset{(3)}{R_{\mu \nu}^{[-2]}} &  = & \partial_\sigma \overset{(3)}{W^\sigma_{\mu \nu}} + \overset{(1)}{C^\sigma_{\sigma \alpha}}  \overset{(2)}{W^\alpha_{\mu \nu}} - 2 \overset{(1)}{C^\alpha_{\sigma (\mu}} \overset{(2)}{W^\sigma_{\nu) \alpha}}\,,\\
    \overset{(1)}{R_{\mu \nu}^{[0]}} & = & \partial_\lambda \overset{(1)}{C^\lambda_{ \nu \mu}} - \partial_\nu \overset{(1)}{C^\lambda_{ \lambda \mu}}\,,
\end{eqnarray}
where we used $\overset{(1)}{S^\rho_{\mu \nu}} = 0$. Finally, the gauge choice $\Pi_{it}=0$ tells us that $h_{it}^{(1)}=0$.

Using what we have just learned, we find that the $\mu = i$ and $\nu = j$ components of (\ref{halfPNEin}) give us
\begin{align}\label{h1ij}
    2\partial_k \partial_{(i} h^{(1)}_{j)k} - \partial_k \partial_k h^{(1)}_{ij} - \partial_i \partial_j h^{(1)}_{kk}=0\,.
\end{align}
The $\mu = t$ and $\nu = j$ components give us
\begin{align}\label{tau3ij}
  \partial_k \partial_{t} h^{(1)}_{ik} - \partial_i \partial_t h^{(1)}_{kk}  - \partial_k \tau^{(3)}_{ik}=0\, ,
\end{align}
Finally, for $\mu = \nu = t$ we find that
\begin{align}\label{tau3ti}
     -\partial_j \big( h^{(1)}_{ij} \tau_{it}^{(2)}\big) + \partial_k \tau^{(3)}_{kt} + \tau^{(2)}_{kt} \partial_k h^{(1)}_{ii} - \partial_t \partial_t h^{(1)}_{kk}=0\,.
\end{align}

We can without loss of generality decompose $h_{ij}^{(1)}$ into a transverse traceless (TT) part, a longitudinal traceless part and a trace part, using
\begin{equation}
    h_{ij}^{(1)}=h_{ij}^{(1)}(\text{TT})+\partial_i L_j^{(1)}+\partial_j L_i^{(1)}+\frac{1}{3}\delta_{ij} H^{(1)}\,,
\end{equation}
where $H^{(1)}$ is given by
\begin{equation}
    H^{(1)}=h_{kk}^{(1)}-2\partial_k L_k^{(1)}\,.
\end{equation}
From equation \eqref{gaugeh1} we learn that the gauge transformation acting on $h^{(1)}_{ij}$ is given by
\begin{equation}
    \delta h_{ij}^{(1)}=\partial_i\xi_{(1)}^j+\partial_j\xi_{(1)}^i+\mathcal{L}_\xi h_{ij}^{(1)}\,,
\end{equation}
where we used that $\tau=dt$, $\tau^1=0$ and $h=dx^idx^i$. We can thus gauge away $L^{(1)}_i$ using $\xi_{(1)}^i$. The trace of equation \eqref{h1ij} tells us that $H^{(1)}$ is harmonic. We require that $h_{ij}^{(1)}$ is globally well-defined and since there are no matter sources it follows that $H^{(1)}$ must be a function of time only. However we also require that the solution is asymptotically flat so that $h_{ij}^{(1)}$ goes to zero at infinity. Hence we find that $H^{(1)}$ is zero. The LHS of equation \eqref{h1ij} then reduces to $\partial^2 h^{(1)}_{ij}(\text{TT})$ and by similar arguments we conclude that $h^{(1)}_{ij}(\text{TT})=0$, so that in fact $h_{ij}^{(1)}=0$. The remaining equations \eqref{tau3ij} and \eqref{tau3ti} then simplify to
\begin{equation}
    \partial_k\tau^{(3)}_{ik}=0\,,\qquad \partial_k\tau^{(3)}_{tk}=0\,.
\end{equation}
Using similar arguments as in the case of \eqref{Difftau2} we find that we can choose a gauge (by using $\xi^t_{(3)}$) to set $\tau^{(3)}_i=0$. Finally, this implies that $\tau_t^{(3)}$ is harmonic without a source so that asymptotic flatness tells us that $\tau_t^{(3)}=0$. Hence we conclude that we can always choose a gauge such that $\tau^{(3)}_\mu=0$ and $h^{(1)}_{\mu\nu}=0$.

We emphasise that the above arguments used the assumptions that the metric up to and including 0.5PN terms is globally well-defined, that the spacetime is asymptotically flat, 4-dimensional and that constant time slices are topologically $\mathbb{R}^3$. 

To summarise, we have found that we can always choose a gauge such that
\begin{equation}\label{eq:0.5PNmetric}
    \tau=dt\,,\qquad h=dx^idx^i\,,\qquad\tau^{(1)}=0\,,\qquad h^{(1)}=0\,,\qquad
    \tau^{(2)}=-Udt\,,\qquad\tau^{(3)}=0\,,
\end{equation}
where $U$ is given in \eqref{U}. The residual gauge transformations are obtained by setting 
equations \eqref{gaugetau}--\eqref{gaugetau3} as well as \eqref{gaugeh} and \eqref{gaugeh1}, in which we substitute \eqref{eq:0.5PNmetric}, equal to zero with the exception of $\delta\tau^{(2)}_t$ which is simply equal to $-\delta U$. This leads to 
\begin{eqnarray}
    &&\xi^t=\text{cst}\,,\qquad\xi^i=a^i(t)+\lambda^i{}_jx^j\,,\qquad\xi_{(1)}^t=\text{cst}\,,\qquad\xi^i_{(1)}=a_{(1)}^i(t)+\lambda_{(1)}^i{}_jx^j\,,\nonumber\\
    &&\xi^t_{(2)}=x^i\dot a^i+f_{(2)}(t)\,,\qquad\xi^t_{(3)}=x^i\dot a^i_{(1)}+f_{(3)}(t)\,,\qquad\lambda_t=0\,,\\
    &&\lambda_i=-\dot a^i\,,\qquad\lambda^{(1)}_t=0\,,\qquad\lambda^{(1)}_i=-\dot a_{(1)}^i\,,\nonumber
\end{eqnarray}
where $\lambda_{ij}=-\lambda_{ji}$ corresponds to a rotation, $a^i(t)$ is any vector that only depends on $t$, and $f_{(2)}$ is any function that only depends on $t$. The Newtonian potential $U$ transforms under the residual gauge transformations as
\begin{equation}
    \delta U=\xi^\mu\partial_\mu U-x^i\ddot a^i-\partial_t f_{(2)}\,,
\end{equation}
which agrees with the results of \cite{Bergshoeff:2014uea}. The $\xi^\mu\partial_\mu$ generate the acceleration extended Galilei symmetries. Finally, the $\lambda_{(1)ij}=-\lambda_{(1)ji}$ are constant, and $\xi_{(1)}^\mu$ and $f_{(3)}$ have to correspond to a symmetry of $U$, i.e. they have to obey $\delta U=0$ or what is the same, they should solve the equation
\begin{equation}\label{eq:lasteqsec3}
    \xi_{(1)}^\mu\partial_\mu U=x^i\ddot a^i_{(1)}+\partial_t f_{(3)}\,.
\end{equation}

The arguments above assumed asymptotic flatness which means that we assume the $1/c$-expansion to be a good approximation all the way up to infinity. In actual fact we need to perform matched asymptotic expansion by matching with an order $G$ solution\footnote{Higher orders in $G$ would be too subleading in $1/c$. For example order $G^2$ is actually $G^2/c^2$ compared to $G$.} to the source-free Einstein equations (in the overlap region). 
That perspective, as we will see, leads to the same conclusion, namely that there is nothing at order 0.5PN. More concretely, if we had left the 0.5PN harmonic functions as undetermined and had matched the metric up to 0.5PN with the linear in $G$ solution we would have found the same result as what we just obtained assuming asymptotic flatness. From the matching perspective the absence of a 0.5PN solution can be shown to be a consequence of mass conservation.

\section{General structure of the post-Newtonian expansion}\label{sec:genPN}

Now that we have discussed the general framework of non-relativistic gravity and reviewed how it recovers the Newtonian regime, we will embark on the $1/c$-expansion of Einstein's equation to post-Newtonian orders in earnest. The framework developed here is valid in any gauge for which the vacuum is described in inertial coordinates and for which there is a Newtonian regime, but apart from that, it is fully general. We will on occasion discuss what happens for the harmonic gauge choice as well as for the transverse gauge (about which we will report more in \cite{companionpaper}).

We assume weak fields, so we are expanding around flat spacetime for which we use inertial coordinates denoted by $(t,x^i)$. The $1/c$-expansion is a general expansion that works off shell. The assumption that there exist fields that admit a Taylor series in $1/c$ (which is dimensionful) means that in a specific on shell context the expansion will organise itself in terms of a dimensionless ratio $v/c$ where the interpretation of $v$ depends on the context. For us the velocity $v$ is either the characteristic velocity of a bound gravitational system, i.e. $\frac{GM}{c^2l_c}\sim \frac{v_c^2}{c^2}$ where $M$ is the total mass of the fluid (as follows from the virial theorem), or $v$ is $l_c/t_c$ where $l_c$ and $t_c$ are the system's characteristic length and time, respectively. The latter is small compared to $c$ when the characteristic wavelength of the gravitational radiation $\lambda_c\sim t_c c$ is much larger than $l_c$. The general form of say a metric components' $1/c$-expansion is schematically
\begin{equation}
\sum_{n=1}^\infty\left(\frac{G}{c^2}\right)^n a_n\left(c^{-1};t,\vec x\right)\,,
\end{equation}
where the $a_n$ are independent of $G$ and admit a Taylor expansion in $1/c$ including odd powers. The latter assumption can breakdown signaling the need for the introduction of $\log c$ terms. We will not need to consider these terms, that are generically related to gravitational tails \cite{PhysRevD.25.2038,Blanchet:1985sp}, as they only appear at higher PN orders. We will restrict our attention up to and including 2.5PN order. We see that any order in $1/c$ will have a finite number of powers of $G$.

On sufficiently large scales retardation effects will no longer be perturbative in $1/c$, so the $1/c$-expansion is only valid in a finite region of space. The standard assumption is that the matter source behaves non-relativistically so that it is fully contained within the region where the $1/c$-expansion applies\footnote{The characteristic velocity of the fluid $v_c$ will be much smaller than the speed of light, i.e. $v_c\ll c$. If we multiply this with the characteristic time scale of the source we find $l_c\ll\lambda_c$ where $l_c$ is the length scale of the matter source and $\lambda_c$ is of the order of $t_c c$ which is the characteristic wavelength of the gravitational radiation. There is thus a separation of scales which is why there is an overlap region which allows us to use the method of matched asymptotic expansions.}. The latter will be called the near zone. The exterior zone will be all of space minus the compact matter source. These two zones overlap which is the region where the matching of the $1/c$-expansion (this section) and the $G$-expansion (next section) takes place.

\subsection{Equations of motion}\label{subsec:NZEOM}

Using the notation of section \ref{ref:sec2} (see in particular equation \eqref{eq:KSmetric}) we 
will expand the metric around flat Galilean spacetime as follows
\begin{equation}
ds^2=-c^2 \left(T_\mu dx^\mu\right)^2+\Pi_{ij}dx^i dx^j\,,
\end{equation}
where we made the choice $\Pi_{ti}=0$ (implying $\Pi_{tt}=0$) which can be done without loss of generality. We have
\begin{equation}
T_\mu=\delta^t_\mu+\frac{1}{c^2}\tau^{(2)}_t\delta^t_\mu+\sum_{n=4}^\infty c^{-n}\tau_\mu^{(n)}\,,\qquad\Pi_{ij}=\delta_{ij}+\sum_{n=2}^\infty c^{-n}h_{ij}^{(n)}\,,
\end{equation}
where we used the results from the previous section regarding the 0PN and 0.5PN orders in the expansion of $T_\mu$ and $\Pi_{ij}$. 

Following standard terminology the $n/2$PN order is the order at which we determine $\tau_\mu^{(n+2)}$ and $h_{ij}^{(n)}$. For the metric we have
\begin{eqnarray}
    g_{tt} & = & \cdots+c^{-n}\left(-2\tau_t^{(n+2)}+\cdots\right)+\mathcal{O}(c^{-n-1})\,,\\
    g_{ti} & = & \cdots+c^{-n}\left(-\tau_i^{(n+2)}+\cdots\right)+\mathcal{O}(c^{-n-1})\,,\\
    g_{ij} & = & \cdots+c^{-n}\left(h_{ij}^{(n)}+\cdots\right)+\mathcal{O}(c^{-n-1})\,,
\end{eqnarray}
where the dots on the left of $c^{-n}$ denote terms of lower order of $1/c$ while the dots in parenthesis denote terms that are of order $c^{-n}$ but that depend on $\tau^{(k+2)}_\mu$ and $h^{(k)}_{ij}$ for $k<n$.

We can expand Einstein's equations and only make explicit the appearance of the $n/2$PN fields. If we do this we find that at $n/2$PN the Einstein equations for $n\ge 2$ can be written as 
\begin{eqnarray}
S^{(n)}_{ij} & = & \partial^2 h_{ij}^{(n)}+\partial_i\partial_j h_{kk}^{(n)}-\partial_i\partial_k h_{kj}^{(n)}-\partial_j\partial_k h_{ki}^{(n)}\,,\label{eq:PNPoissoneq1} \\
S^{(n)}_i & = & \partial^2 \tau_i^{(n+2)}-\partial_i\partial_k\tau_k^{(n+2)}+\partial_t\left(\partial_k h_{ki}^{(n)}-\partial_i h_{kk}^{(n)}\right)\,,\label{eq:PNPoissoneq2}\\
S^{(n)} & = & \partial^2 \tau_t^{(n+2)}-\partial_t\partial_k\tau_k^{(n+2)}-\frac{1}{2}\partial_t^2 h^{(n)}_{kk} -\partial_i\tau^{(2)}_t\left(\partial_j h^{(n)}_{ij}-\frac{1}{2}\partial_i h_{jj}^{(n)}\right)\nonumber\\
&&-h_{ij}^{(n)}\partial_i\partial_j \tau^{(2)}_t\,,\label{eq:PNPoissoneq3}
\end{eqnarray}
where the sources $S^{(n)}, S^{(n)}_i, S^{(n)}_{ij}$ on the left-hand side depend on the expansion of the matter fields as well as lower order fields $h_{ij}^{(k)}$ and $\tau_\mu^{(k+2)}$ with $k<n$. Below we will give explicit expressions for these sources to 2.5PN. There is a natural order in which to solve the above PDEs by starting with \eqref{eq:PNPoissoneq1}, which can be solved for $h_{ij}^{(n)}$, and then moving on to \eqref{eq:PNPoissoneq3} by solving it for $\tau^{(n+2)}_i$, and ending with \eqref{eq:PNPoissoneq3}, which can be solved for $\tau^{(n+2)}_t$. It also follows from these equations (upon differentiation and combining equations) that 
\begin{eqnarray}
\frac{1}{2}\partial_t S^{(n)}_{ii}+\partial_i S^{(n)}_i & = & 0\,,\label{fluidcons1}\\
\partial_i S^{(n)}_{ij}-\frac{1}{2}\partial_j S^{(n)}_{ii} & = & 0\,.\label{fluidcons2}
\end{eqnarray}

The source $S^{(n)}_{ij}$ contains terms that are linear in lower order fields. If we isolate these we can write 
\begin{equation}\label{eq:NLsource}
    S^{(n)}_{ij}=\partial_t\left(\partial_i\tau_j^{(n)}+\partial_j\tau_i^{(n)}\right)+\partial_t^2 h_{ij}^{(n-2)}-2\partial_i\partial_j\tau_t^{(n)}+\tilde S_{ij}^{(n)}\,,
\end{equation}
where now $\tilde S_{ij}^{(n)}$ contains both the compact source terms as well as non-linear terms of lower-order fields. The sources $S_{i}^{(n)}$ and $S^{(n)}$ do not contain any linear terms in lower order fields. If we use \eqref{eq:NLsource} together with \eqref{eq:PNPoissoneq2} and \eqref{eq:PNPoissoneq3}, then equations \eqref{fluidcons1} and \eqref{fluidcons2} become
\begin{eqnarray}
    0 & = & \partial_t \left[S^{(n-2)}+\partial_k\tau_t^{(2)}\left(\partial_l h_{kl}^{(n-2)}-\frac{1}{2}\partial_l h_{kk}^{(n-2)}\right)+h_{kl}^{(n-2)}\partial_k\partial_l\tau_t^{(2)}-\frac{1}{2}\tilde S_{kk}^{(n)}\right]\nonumber\\
    &&-\partial_i S^{(n)}_i \,,\\
    0 & = & \partial_t S^{(n-2)}_j-\partial_j\left[S^{(n-2)}+\partial_k\tau_t^{(2)}\left(\partial_l h_{kl}^{(n-2)}-\frac{1}{2}\partial_l h_{kk}^{(n-2)}\right)+h_{kl}^{(n-2)}\partial_k\partial_l\tau_t^{(2)}\right]\nonumber\\
    &&+\partial_i\tilde S_{ij}^{(n)}-\frac{1}{2}\partial_j\tilde S_{ii}^{(n)}\,,
\end{eqnarray}
where $n\ge 2$ and where $S^{(0)}=0=S^{(0)}_i$. These lead to the fluid conservation equations, i.e. the $1/c$-expansion of \eqref{Conservation2}. We see that the $n/2$PN Einstein equations determine the $(n/2-1)$PN fluid equations.

In order to solve the expanded Einstein equations it will prove useful to decompose the $n/2$PN fields as follows
\begin{eqnarray}
h_{ij}^{(n)} & = & h_{ij}^{(n)}(\text{TT})+\partial_i L^{(n)}_j+\partial_j L^{(n)}_i+\frac{1}{3}\delta_{ij}H^{(n)}\,,\label{eq:decom1/cmetric1}\\
\tau_i^{(n+2)} & = & M_i^{(n)}(\text{T})-\partial_t L_i^{(n)}-\partial_i N^{(n)}\,,\label{eq:decom1/cmetric2}\\
\tau_t^{(n+2)} & = & M_t^{(n)}-\partial_t N^{(n)}\,,\label{eq:decom1/cmetric3}
\end{eqnarray}
where 
\begin{equation}
    H^{(n)}=h^{(n)}_{kk}-2\partial_k L^{(n)}_k\,.
\end{equation}
We will show that one can rewrite equations \eqref{eq:PNPoissoneq1}--\eqref{eq:PNPoissoneq3} schematically as $\partial^2 (\text{field})=\text{(known source)}$, so that they can in principle be solved by integration (we comment on issues that can arise in the integration step further below). 
Once we have solved for the fields appearing in the decomposition \eqref{eq:decom1/cmetric1}--\eqref{eq:decom1/cmetric3} we reassemble them to form the $n/2$PN fields $h_{ij}^{(n)}$, $\tau_\mu^{(n+2)}$ which are then used to write the source terms in \eqref{eq:PNPoissoneq1}--\eqref{eq:PNPoissoneq3} at the next order. Put differently, the decomposition \eqref{eq:decom1/cmetric1}--\eqref{eq:decom1/cmetric3} is only used on the right-hand side of \eqref{eq:PNPoissoneq1}--\eqref{eq:PNPoissoneq3} and not on the left-hand side.

Using \eqref{eq:decom1/cmetric1}--\eqref{eq:decom1/cmetric3} the $n/2$PN Einstein equations become 
\begin{eqnarray}
\partial^2 H^{(n)} & = & \frac{3}{4}S^{(n)}_{ii}\,,\\
\partial^2 h_{ij}^{(n)}(\text{TT}) & = & S^{(n)}_{ij}-\frac{1}{4}\delta_{ij}S^{(n)}_{kk}-\frac{1}{3}\partial_i\partial_j H^{(n)}\,,\\
\partial^2 M_i^{(n)}(\text{T}) &  = & S^{(n)}_i+\frac{2}{3}\partial_t\partial_i H^{(n)}\,,\\
\partial^2 M_t^{(n)} & = & S^{(n)}+\frac{1}{2}\partial_t^2 H^{(n)}+\partial^2 L_i^{(n)}\partial_i\tau_t^{(2)}-\frac{1}{6}\partial_i  H^{(n)}\partial_i\tau_t^{(2)}+h_{ij}^{(n)}\partial_i\partial_j\tau_t^{(2)}\,.
\end{eqnarray}
These equations can be rewritten as
\begin{eqnarray}
\partial^2 H^{(n)} & = & \frac{3}{4}S^{(n)}_{ii}\,,\label{eq:EOMcexp1}\\
\partial^2 \left(M_i^{(n)}(\text{T})
-\frac{1}{3}x^i\partial_t H^{(n)}\right) &  = & S^{(n)}_i-\frac{1}{4}x^i\partial_t S^{(n)}_{jj}\,,\label{eq:EOMcexp2}
\end{eqnarray}
\vspace{-0.6cm}
\begin{eqnarray}
    \hspace{-1cm}&&\partial^2\left(h^{(n)}_{ij}(\text{TT})+\frac{1}{12}\left[x^i\partial_j H^{(n)}+x^j\partial_i H^{(n)}-\frac{2}{3}\delta_{ij}x^k\partial_k H^{(n)}\right]\right)=\nonumber\\
    \hspace{-1cm}&&S^{(n)}_{ij}-\frac{1}{3}\delta_{ij}S^{(n)}_{kk}+\frac{1}{16}\left(x^i\partial_j S^{(n)}_{ll}+x^i\partial_j S^{(n)}_{ll}-\frac{2}{3}\delta_{ij}x^k\partial_k S^{(n)}_{ll}\right)\,,\label{eq:EOMcexp3}\\
\hspace{-1cm}&&\partial^2\left( M_t^{(n)} -\frac{1}{12}r^2\partial_t^2 H^{(n)}+\frac{1}{2}x^i\partial_t M_i^{(n)}(\text{T})\right) = \nonumber\\
\hspace{-1cm}&& S^{(n)} -\frac{1}{16}r^2\partial_t^2 S_{ii}^{(n)}+\frac{1}{2}x^i\partial_tS_i^{(n)}-\frac{1}{6}\partial_i  H^{(n)}\partial_i\tau_t^{(2)}+\partial^2 L_i^{(n)}\partial_i\tau_t^{(2)}+h_{ij}^{(n)}\partial_i\partial_j\tau_t^{(2)}\,.\label{eq:EOMcexp4}
\end{eqnarray}
We note that $N^{(n)}$ and $L^{(n)}_i$ do not appear on the left-hand side and that only $L^{(n)}_i$ appears on the right-hand side and only in the equation for $M_t^{(n)}$. However, the lower order longitudinal fields $N^{(k)}$ and $L^{(k)}_i$ for $k<n$ do appear inside the source terms. Hence, to have a well-defined set of equations we need to supplement the above equations with a gauge fixing condition that provides (solvable) equations for the longitudinal fields $N^{(n)}$ and $L^{(n)}_i$ at every order.

The right-hand side of \eqref{eq:EOMcexp4} depends on the solution for $h_{ij}^{(n)}$, so that equation is the last one to be integrated as we need to know what $h_{ij}^{(n)}$ is first. This can be determined by solving the other equations including the ones that determine the longitudinal fields.

\subsection{The source terms to 2.5PN}\label{subsec:sourceterms}

In this paper and in \cite{companionpaper} we will be interested in the near zone metric to 2.5PN. The metric up to this order is given by
\begin{eqnarray}
    g_{tt} & = & -c^2-2\tau_t^{(2)}-\frac{2}{c^2}\left(\tau_t^{(4)}+\frac{1}{2}(\tau_t^{(2)})^2\right)-\frac{2}{c^3}\tau_t^{(5)}-\frac{2}{c^4}\left(\tau_t^{(6)}+\tau_t^{(2)}\tau_t^{(4)}\right)\nonumber\\
    &&-\frac{2}{c^5}\left(\tau_t^{(7)}+\tau_t^{(2)}\tau_t^{(5)}\right)+\mathcal{O}(c^{-6})\,,\label{eq:gttexp}\\
    g_{ti} & = & -\frac{1}{c^2}\tau_i^{(4)}-\frac{1}{c^3}\tau_i^{(5)}-\frac{1}{c^4}\left(\tau_i^{(6)}+\tau_t^{(2)}\tau_i^{(4)}\right)-\frac{1}{c^5}\left(\tau_i^{(7)}+\tau_t^{(2)}\tau_i^{(5)}\right)+\mathcal{O}(c^{-6})\,,\label{eq:gtiexp}\nonumber\\
    &&\\
    g_{ij} & = & \delta_{ij}+\frac{1}{c^2}h^{(2)}_{ij}+\frac{1}{c^3}h^{(3)}_{ij}+\frac{1}{c^4}h^{(4)}_{ij}+\frac{1}{c^5}h^{(5)}_{ij}+\mathcal{O}(c^{-6})\,.\label{eq:gijexp}
\end{eqnarray}
In appendix \ref{app:1/cexpGR} we derive the Einstein equations to 2.5PN. These take the form of equations \eqref{eq:PNPoissoneq1}--\eqref{eq:PNPoissoneq3}. Here we will list the explicit form the source terms take.

Starting with the $ij$ components the nonzero sources are given by
\begin{eqnarray}
    S^{(2)}_{ij} & = & -8\pi G \overset{(2)}{\mathcal{S}}_{ij}-2\partial_i\partial_j\tau^{(2)}_t\,,\\
    S^{(4)}_{ij} & = & -8\pi G\overset{(4)}{\mathcal{S}}_{ij}+\partial_t\left(\partial_i\tau_j^{(4)}+\partial_j\tau_i^{(4)}\right)+\partial_t^2 h^{(2)}_{ij}-2\partial_i\partial_j \tau^{(4)}_t\nonumber\\
    &&+2\tau_t^{(2)}\partial_i\partial_j\tau_t^{(2)}+\partial_k\tau_t^{(2)}C_{ijk}^{(2)}-h^{(2)}_{kl}\left(\partial_k C^{(2)}_{ijl}-\partial_i C^{(2)}_{jkl}\right)\nonumber\\
    &&-\frac{1}{2}C_{kkl}^{(2)}C_{ijl}^{(2)}+\frac{1}{2}C_{ikl}^{(2)}C_{jkl}^{(2)}\,,\\
    S^{(5)}_{ij} & = & -8\pi G\overset{(5)}{\mathcal{S}}_{ij}+\partial_t\left(\partial_i\tau_j^{(5)}+\partial_j\tau_i^{(5)}\right)+\partial_t^2 h^{(3)}_{ij}-2\partial_i\partial_j \tau^{(5)}_t\nonumber\\
    &&+\partial_k\tau_t^{(2)}C_{ijk}^{(3)}-h^{(2)}_{kl}\left(\partial_k C^{(3)}_{ijl}-\partial_i C^{(3)}_{jkl}\right)
    -h^{(3)}_{kl}\left(\partial_k C^{(2)}_{ijl}-\partial_i C^{(2)}_{jkl}\right)\nonumber\\
    &&-\frac{1}{2}C_{kkl}^{(2)}C_{ijl}^{(3)}-\frac{1}{2}C_{kkl}^{(3)}C_{ijl}^{(2)}+\frac{1}{2}C_{ikl}^{(2)}C_{jkl}^{(3)}+\frac{1}{2}C_{ikl}^{(3)}C_{jkl}^{(2)}\,,
\end{eqnarray}
where we defined
\begin{equation}
    C_{ijk}^{(n)}=\partial_i h_{jk}^{(n)}+\partial_j h^{(n)}_{ik}-\partial_k h^{(n)}_{ij}\,.
\end{equation}
In here $\overset{(n)}{\mathcal{S}}_{ij}$ is the $c^{-n}$ term in the expansion of the source that appears in the trace reversed Einstein equations \eqref{EE} and \eqref{EE2}. For a perfect fluid this becomes (see appendix \ref{app:1/cexpGR} for details)
\begin{eqnarray}
    \overset{(2)}{\mathcal{S}}_{ij} & = & E_{(\sm2)}\delta_{ij}\,,\\
    \overset{(4)}{\mathcal{S}}_{ij} & = & E_{(\sm2)}h^{(2)}_{ij}+2E_{(\sm2)}v^i v^j+\left(E_{(0)}-P_{(0)}\right)\delta_{ij}\,,\\
    \overset{(5)}{\mathcal{S}}_{ij} & = & E_{(\sm2)}h^{(3)}_{ij}\,.
\end{eqnarray}
By expanding the $ti$ components of the Einstein equations we obtain the sources
\begin{eqnarray}
   \hspace{-1cm} S_{i}^{(2)} & = & 8\pi G\overset{(2)}{\mathcal{S}}_{ti}\,,\\
   \hspace{-1cm} S_{i}^{(4)} & = & 8\pi G\overset{(4)}{\mathcal{S}}_{ti}+\left(\partial_k h_{kl}^{(2)}-\frac{1}{2}\partial_l h^{(2)}_{kk}\right)\partial_t h_{il}^{(2)}-h_{kl}^{(2)}\partial_t\left(\partial_i h_{kl}^{(2)}-\partial_k h_{il}^{(2)}\right)\nonumber\\
  \hspace{-1cm}  &&\left(\partial_k h_{kl}^{(2)}-\frac{1}{2}\partial_l h^{(2)}_{kk}\right)\tau^{(4)}_{li}-\frac{1}{2}\partial_i h_{kl}^{(2)}\partial_t h_{kl}^{(2)}+\partial_k h_{ij}^{(2)}\tau_{kj}^{(4)}-\partial_t h^{(2)}_{kk}\partial_i\tau_t^{(2)}\nonumber\\
  \hspace{-1cm}  &&-\partial_k\tau_k^{(4)}\partial_i\tau_t^{(2)}-\tau^{(4)}_k\partial_k\partial_i\tau^{(2)}_t+2\partial_k\tau_t^{(2)}\partial_i\tau_k^{(4)}-\partial_k\tau_t^{(2)}\partial_k\tau^{(4)}_{i}\nonumber\\
  \hspace{-1cm}  &&-\tau^{(4)}_i\partial^2\tau_t^{(2)}+h_{kl}^{(2)}\partial_k\tau_{li}^{(4)}-\tau_t^{(2)}\partial_k\tau_{ki}^{(4)}+\partial_k\tau_t^{(2)}\partial_t h_{ik}^{(2)}\,,\\
  \hspace{-1cm}  S_{i}^{(5)} & = & 8\pi G\overset{(5)}{\mathcal{S}}_{ti}+\text{terms that follow from the `odd order rule' below}\,.
\end{eqnarray}
For a perfect fluid the matter sources are
\begin{eqnarray}
 \hspace{-1cm}   \overset{(2)}{\mathcal{S}}_{ti} & = & -2E_{(\sm2)}v^i\,,\\
  \hspace{-1cm}  \overset{(4)}{\mathcal{S}}_{ti} & = & E_{(\sm2)}\tau^{(4)}_i-2\tau_t^{(2)}E_{(\sm2)}v^i-E_{(\sm2)}v^2 v^i-2\left(E_{(0)}+P_{(0)}\right)v^i\nonumber\\
   \hspace{-1cm} &&-2E_{(\sm2)}h^{(2)}_{ij}v^j-2E_{(\sm2)}v_{(2)}^i\,,\\
   \hspace{-1cm} \overset{(5)}{\mathcal{S}}_{ti} & = & -2E_{(\sm2)}h^{(3)}_{ij}v^j+E_{(\sm2)}\tau_i^{(5)}\,.
\end{eqnarray}
Finally, the expansion of the $tt$ component of the Einstein equations tells us that
\begin{eqnarray}
  \hspace{-1cm}  S^{(2)} & = & 4\pi G\overset{(2)}{\mathcal{S}}_{tt}-\tau_t^{(2)}\partial^2 \tau_t^{(2)}\,,\\
  \hspace{-1cm}  S^{(4)} & = & 4\pi G\overset{(4)}{\mathcal{S}}_{tt}-\frac{1}{4}\tau^{(4)}_{ij}\tau^{(4)}_{ij}+\partial_t\tau^{(4)}_i\partial_i\tau^{(2)}_t+\tau^{(4)}_i\partial_i\partial_t\tau_t^{(2)}-\frac{1}{4}\partial_t h^{(2)}_{ij}\partial_t h^{(2)}_{ij}\nonumber\\
  \hspace{-1cm}  &&-\frac{1}{2}\partial_t h_{kk}^{(2)}\partial_t\tau_t^{(2)}-h^{(2)}_{ij}\partial_t\partial_i\tau^{(4)}_j+\tau^{(2)}_t\partial_t\partial_i\tau^{(4)}_i-\frac{1}{2}C^{(2)}_{iij}\partial_t\tau_j^{(4)}\nonumber\\
  \hspace{-1cm}  &&-\frac{1}{2}h^{(2)}_{ij}\partial_t^2 h^{(2)}_{ij}+h^{(2)}_{ij}\partial_i\partial_j\tau_t^{(4)}-\tau_t^{(2)}\partial^2\tau_t^{(4)}-\tau_t^{(4)}\partial^2\tau_t^{(2)}+\frac{1}{2}C^{(2)}_{iij}\partial_j\tau_t^{(4)}\nonumber\\
  \hspace{-1cm}  &&-h^{(2)}_{ik}h^{(2)}_{jk}\partial_i\partial_j\tau_t^{(2)}+\tau_t^{(2)}h^{(2)}_{ij}\partial_i\partial_j\tau_t^{(2)}-\frac{1}{2}h^{(2)}_{ij}C^{(2)}_{kkj}\partial_i\tau_t^{(2)}\nonumber\\
  \hspace{-1cm}  &&-\frac{1}{2}h^{(2)}_{ij}C^{(2)}_{ijk}\partial_k\tau^{(2)}_t+\frac{1}{2}\tau_t^{(2)}C^{(2)}_{iij}\partial_j\tau_t^{(2)}\,,\label{eq:S4}\\
  \hspace{-1cm}  S^{(5)} & = & 4\pi G\overset{(5)}{\mathcal{S}}_{tt}+\text{terms that follow from the `odd order rule' below}\,,
\end{eqnarray}
where in the case of a perfect fluid we have
\begin{eqnarray}
    \overset{(2)}{\mathcal{S}}_{tt} & = & E_{(0)}+3P_{(0)}+2E_{(\sm2)}v^2+2E_{(\sm2)}\tau_t^{(2)}\,,\\
    \overset{(4)}{\mathcal{S}}_{tt} & = & E_{(\sm2)}\left(2\tau_t^{(4)}+\left(\tau_t^{(2)}\right)^2+4v^2\tau_t^{(2)}+4v^i v^i_{(2)}+2h^{(2)}_{ij}v^i v^j\right)\nonumber\\
    &&+2\left(E_{(0)}+P_{(0)}\right)v^2+2\left(E_{(0)}+3P_{(0)}\right)\tau_t^{(2)}+E_{(2)}+3P_{(2)}\,,\\
    \overset{(5)}{\mathcal{S}}_{tt} & = & 2E_{(\sm2)}\tau_t^{(5)}+2E_{(\sm2)}h^{(3)}_{ij}v^i v^j+E_{(3)}\,.
\end{eqnarray}

The `odd order rule' is the following prescription that allows one to write down the 2.5PN source terms $S^{(5)}_{ij}$, $S^{(5)}_{i}$ and $S^{(5)}$ once we know the corresponding terms at 2PN. It also works to obtain the 1.5PN source terms from the 1PN ones, but the result is always zero. Given the 1PN and 2PN source terms, the 1.5PN and 2.5PN source terms follows from the following three prescriptions:
\begin{itemize}
    \item If the source term is linear in some field, say $X^{(k)}$, then we take the same term with $X^{(k)}$ and replace it by $X^{(k+1)}$.
    \item If the source term is quadratic in two fields, say $X^{(k)}Y^{(l)}$, then we take the same term and write it twice, once as $X^{(k+1)}Y^{(l)}$ and once as $X^{(k)}Y^{(l+1)}$. 
    \item If the source term is cubic in three fields, say $X^{(k)}Y^{(l)}Z^{(m)}$, then we take the same term and write it three times, $X^{(k+1)}Y^{(l)}Z^{(m)}+X^{(k)}Y^{(l+1)}Z^{(m)}+X^{(k)}Y^{(l)}Z^{(m+1)}$.
\end{itemize}
Note that when applying this rule one sometimes runs into coefficients that are zero such as $\tau^{(3)}_t$, so these terms need to be discarded. To illustrate these rules we give an example. Let us consider the following two terms that appear on the first line of \eqref{eq:S4} which we repeat here for convenience
\begin{equation}
    -\frac{1}{4}\tau^{(4)}_{ij}\tau^{(4)}_{ij}+\partial_t\tau^{(4)}_i\partial_i\tau^{(2)}_t\,.
\end{equation}
Applying the second prescription we obtain
\begin{equation}
    -\frac{1}{4}\tau^{(5)}_{ij}\tau^{(4)}_{ij}-\frac{1}{4}\tau^{(5)}_{ij}\tau^{(4)}_{ij}+\partial_t\tau^{(5)}_i\partial_i\tau^{(2)}_t+\partial_t\tau^{(4)}_i\partial_i\tau^{(3)}_t\,.
\end{equation}
Since $\tau^{(3)}_t=0$ we drop the final term. It would be interesting to check if this rule can be generalised to apply at higher orders such as 3.5PN.

One of the reasons why we do not write out the 2.5PN source terms explicitly is because these expressions become rather unwieldy, but as we shall see many terms will vanish after matching.

\subsection{Gauge fixing}\label{subsec:gaugefix}

As we mentioned before in order to have a complete set of equations of motion we need to choose a gauge, i.e. some equation that determines/constrains the longitudinal fields $N^{(n)}$ and $L_i^{(n)}$. In this paper we aim to set up a formalism that works for any gauge choice (for which the metric has a Newtonian limit described using inertial coordinates). We call this class of gauge choices `post-Newtonian gauges'. This rules out a gauge choice such as synchronous gauge in which case we have $g_{tt}=-c^2$ and $g_{it}=0$ as this has no Newtonian regime. 

From the results of the previous section we have seen that the source terms contain many non-linear terms. These will in general have non-compact support which complicates the integration step (see below for more details). It would therefore seem natural to choose a gauge to try to minimise the number of noncompact source terms at every order. However, it is not possible to completely remove all of them at each order \cite{Rendall}.

There are other considerations that concern a judicious choice of gauge that relate to being able to solve the $G$-expanded vacuum Einstein equations in the exterior zone. This will be discussed in the next section.

In this paper we aim to formulate the general framework in any post-Newtonian gauge and illustrate our methods for two specific gauge choices. The first is the harmonic gauge, chosen because this is the most common choice made in the literature and so this helps comparison and to show that the methods developed here reproduce existing results. The second is a gauge choice that is sometimes made for linearised GR that we call transverse gauge. This gauge has some interesting properties and illustrates that our framework also works outside the harmonic gauge. We will discuss the basics of the transverse gauge here and defer further analysis to the companion paper \cite{companionpaper}.


We next discuss the $1/c$-expansion of the harmonic gauge condition $\partial_\mu\left(\sqrt{-g}g^{\mu\nu}\right)=0$. 
Up to 2.5PN this tells us that
\begin{eqnarray}
    \partial_i\tau_i^{(4)}+\frac{1}{2}\partial_t h^{(2)}_{ii} & = & \partial_t\tau^{(2)}_t\,,\label{eq:HG1}\\
    \partial_i h^{(2)}_{ij}-\frac{1}{2}\partial_j h^{(2)}_{ii} & = & \partial_j\tau^{(2)}_t\,,\label{eq:HG2}\\
    \partial_i\tau_i^{(5)}+\frac{1}{2}\partial_t h^{(3)}_{ii} & = & 0\,,\\
    \partial_i h^{(3)}_{ij}-\frac{1}{2}\partial_j h^{(3)}_{ii} & = & 0\,,\\
    \partial_i\tau_i^{(6)}+\frac{1}{2}\partial_t h^{(4)}_{ii} & = & \partial_t\tau^{(4)}_t-\tau^{(2)}_t\partial_t\tau_t^{(2)}+\tau_i^{(4)}\partial_i\tau_t^{(2)}\nonumber\\
    &&-\tau_t^{(2)}\partial_i\tau_i^{(4)}+\frac{1}{2}h^{(2)}_{ij}\partial_t h^{(2)}_{ij}+h^{(2)}_{ij}\partial_i\tau^{(4)}_j\,,\\
    \partial_i h^{(4)}_{ij}-\frac{1}{2}\partial_j h^{(4)}_{ii} & = & h^{(2)}_{ik}\partial_i h^{(2)}_{jk}-\frac{1}{2}h_{ik}^{(2)}\partial_j h^{(2)}_{ik}\nonumber\\
    &&+\partial_j\tau^{(4)}_t-\partial_t\tau^{(4)}_j-\tau^{(2)}_t\partial_j\tau^{(2)}_t\,,
\end{eqnarray}
\begin{eqnarray}
    \partial_i\tau_i^{(7)}+\frac{1}{2}\partial_t h^{(5)}_{ii} & = & \partial_t\tau^{(5)}_t+\tau_i^{(5)}\partial_i\tau_t^{(2)}-\tau_t^{(2)}\partial_i\tau_i^{(5)}\nonumber\\
    &&+\frac{1}{2}h^{(2)}_{ij}\partial_t h^{(3)}_{ij}+\frac{1}{2}h^{(3)}_{ij}\partial_t h^{(2)}_{ij}+h^{(2)}_{ij}\partial_i\tau^{(5)}_j+h^{(3)}_{ij}\partial_i\tau^{(4)}_j\,,\\
      \partial_i h^{(5)}_{ij}-\frac{1}{2}\partial_j h^{(5)}_{ii} & = & h^{(2)}_{ik}\partial_i h^{(3)}_{jk}+h^{(3)}_{ik}\partial_i h^{(2)}_{jk}-\frac{1}{2}h_{ik}^{(2)}\partial_j h^{(3)}_{ik}-\frac{1}{2}h_{ik}^{(3)}\partial_j h^{(2)}_{ik}\nonumber\\
    &&+\partial_j\tau^{(5)}_t-\partial_t\tau^{(5)}_j\,.
\end{eqnarray}
Schematically and in terms of the decompositions \eqref{eq:decom1/cmetric1} and \eqref{eq:decom1/cmetric2}, this leads to equations of the form
\begin{eqnarray}
    \partial^2 N^{(n)}-\frac{1}{2}\partial_t H^{(n)} & = & K^{(n)}\,,\\
    \partial^2 L^{(n)}_i-\frac{1}{2}\partial_i H^{(n)} & = & K^{(n)}_i\,,
\end{eqnarray}
where $K^{(n)}$ and $K^{(n)}_i$ depend on the lower order fields. This can be rewritten as
\begin{eqnarray}
    \partial^2\left( N^{(n)}-\frac{1}{12}r^2\partial_t H^{(n)}+\frac{1}{2}x^i M^{(n)}_i(\text{T})\right) & = & K^{(n)}+\frac{1}{2}x^i S^{(n)}_i\,,\\
    \partial^2\left( L^{(n)}_i-\frac{1}{4}x^i H^{(n)}\right) & = & K^{(n)}_i-\frac{3}{16}x^i S^{(n)}_{jj}\,,
\end{eqnarray}
which is again of the form $\partial^2(\text{field})=\text{(known source)}$.

We are not necessarily saying that the ideal variables are the ones used in the decomposition \eqref{eq:decom1/cmetric1}--\eqref{eq:decom1/cmetric3}. The purpose of these variables is to show that the problem can be tackled in a rather large class of gauge choices. For a particular gauge it is perfectly possible that another set of variables is more convenient. For example in the harmonic gauge we can just work with $\tau^{(n+2)}_\mu$ and $h_{\mu\nu}^{(n)}$ as in that case we have 
\begin{eqnarray}
    \partial_i\tau_i^{(n+2)}+\frac{1}{2}\partial_t h^{(n)}_{ii} & = & -K^{(n)}\,,\label{eq:HGlinear1}\\
    \partial_j h^{(n)}_{ji}-\frac{1}{2}\partial_i h^{(n)}_{jj} & = & K^{(n)}_i\,,\label{eq:HGlinear2}
\end{eqnarray}
which allows us to write equations \eqref{eq:PNPoissoneq1}--\eqref{eq:PNPoissoneq3} as
\begin{eqnarray}
\partial^2 h_{ij}^{(n)} & = & S^{(n)}_{ij}+\partial_i K^{(n)}_j+\partial_j K^{(n)}_i\,, \label{eq:PNinHG_1}\\
\partial^2 \tau_i^{(n+2)} & = & S^{(n)}_i-\partial_t K_i^{(n)}-\partial_iK^{(n)}\,,\label{eq:PNinHG_2}\\
\partial^2 \tau_t^{(n+2)} & = & S^{(n)}+\partial_i\tau^{(2)}_t K_i^{(n)}-\partial_t K^{(n)}+h_{ij}^{(n)}\partial_i\partial_j \tau^{(2)}_t\,,\label{eq:PNinHG_3}
\end{eqnarray}
which is also of the form\footnote{By the definitions of $K^{(n)}$ and $K^{(n)}_i$ these equations are equivalent to \eqref{eq:PNPoissoneq1}--\eqref{eq:PNPoissoneq3}. Using the form of the sources at 1PN, one can ask whether there exists a choice for the $K^{(n)}$ and $K^{(n)}_i$ such that all sources are compact at 1PN. This is however not possible. For example if we make the right hand side of \eqref{eq:PNinHG_1} and \eqref{eq:PNinHG_2} compact we need to pick equations \eqref{eq:HG1} and \eqref{eq:HG2} which is the 1PN harmonic gauge, but then the equation for $\tau_t^{(4)}+\frac{1}{2}\left(\tau_t^{(2)}\right)^2$, i.e. the $tt$ component of the metric at order $c^{-2}$ via \eqref{eq:PNinHG_3} has a non-compact source which is $\partial_t K^{(2)}$. A weaker requirement, investigated in \cite{Rendall}, is to demand that the sources in \eqref{eq:PNinHG_1}-\eqref{eq:PNinHG_2} are such that we can write down a particular solution on all of $\mathbb{R}^3$ (that is asymptotically flat) using the Green's function for the Laplacian. It was found that this is possible up to 2PN for a judicious choice of the $K^{(n)}$ and $K^{(n)}_i$ with $n=2,4$. Of course, non of these requirements are necessary since the $1/c$-expansion has a finite regime of validity and so the solutions do not, and in general will not, be asymptotically flat. They simply need to be matched onto a $G$-expanded exterior solution that is asymptotically flat.} $\partial^2(\text{field})=\text{(known source)}$.
This is the form in which we will solve the $1/c$ expanded Einstein equations in harmonic gauge in subsequent sections. The EOM for $\tau_t^{(n+2)}$ depends on $h_{ij}^{(n)}$ which itself is given by the particular solution to its EOM plus a homogeneous solution. The latter is fixed by the matching process and so it is convenient to first match the $ij$ part of the metric before integrating the EOM for $\tau_t^{(n+2)}$.

In order to formulate the aforementioned transverse gauge condition we assume a metric that is of the form $g_{\mu\nu}=\eta_{\mu\nu}+h_{\mu\nu}$ where $h_{\mu\nu}$ is perturbative and we have chosen inertial coordinates for the Minkowski metric $\eta_{\mu\nu}$. This means that we can write the various components as
\begin{equation}
    g_{tt}=-c^2+h_{tt}\,,\qquad g_{ti}=h_{ti}\,,\qquad g_{ij}=\delta_{ij}+h_{ij}\,,
\end{equation}
where $(t,x^i)$ are the inertial coordinates. The transverse gauge condition is then the statement that
\begin{equation}\label{eq:Cgauge}
    \partial_i h_{ti}=0\,,\qquad \partial_i \left(h_{ij}-\frac{1}{3}\delta_{ij}h_{kk}\right)=0\,.
\end{equation}
We stress that this is to all orders in $1/c$ and $G$. This gauge choice is commonly made at the linearised level, i.e. when $h_{\mu\nu}$ is first order in $G$. There are infinitely many ways to extend this to a non-linear gauge choice (see footnote \ref{ftn:transversegauge} for more details). We found it convenient to use \eqref{eq:Cgauge} as our definition of transverse gauge in this work, but we do not rule out the possibility that allowing for certain non-linear terms on the right-hand side of \eqref{eq:Cgauge} would not be more preferential\footnote{In the same spirit one could consider modifying the harmonic gauge choice order by order by making different choices for the $K^{(n)}$ and $K^{(n)}_i$.}. 

Up to 2.5PN equation \eqref{eq:Cgauge} leads to
\begin{eqnarray}
    \partial_i\tau_i^{(4)} & = & 0\,,\\
    \partial_i\tau_i^{(5)} & = & 0\,,\\
    \partial_i\tau_i^{(6)} & = & -\tau_i^{(4)}\partial_i\tau_t^{(2)}\,,\\
    \partial_i\tau_i^{(7)} & = & -\tau_i^{(5)}\partial_i\tau_t^{(2)}\,,\\
    \partial_i\left( h^{(n)}_{ij}-\frac{1}{3}\delta_{ij}h^{(n)}_{kk}\right) & = & 0\,,\qquad\text{for $n=2,3,4,5$}\,.
\end{eqnarray}
Schematically, and in terms of the decompositions \eqref{eq:decom1/cmetric1} and \eqref{eq:decom1/cmetric2}, this leads to equations of the form
\begin{eqnarray}
    \partial^2 N^{(n)}+\partial_t\partial_k L_k^{(n)} & = & \tilde K^{(n)}\,,\\
    \partial^2 L^{(n)}_i+\frac{1}{3}\partial_i\partial_k L_k^{(n)} & = & \tilde K^{(n)}_i\,,
\end{eqnarray}
where $\tilde K^{(n)}$ and $\tilde K^{(n)}_i$ depend on lower order fields. These equations can be rewritten as 
\begin{eqnarray}
    \partial^2\left(L^{(n)}_i+\frac{1}{6}x^i\partial_k L_k^{(n)}\right) & = & \tilde K_i^{(n)}+\frac{1}{8}x^i\partial_k \tilde K_k^{(n)}\,,\\
    \partial^2\left(N^{(n)}+\frac{1}{30}r^2\partial_t\partial_k  L_k^{(n)}+\frac{2}{5}x^i\partial_t L_i^{(n)}\right) & = & \tilde K^{(n)}+\frac{2}{5}x^i\partial_t\tilde K^{(n)}_i+\frac{1}{40}r^2\partial_t\partial_k \tilde K_k^{(n)}\,,\nonumber\\
    &&
\end{eqnarray}
which are again of the form that allows for integration.

\subsection{Comments on integrating the equations of motion}\label{subsec:intEOM}

Both in the harmonic and in the transverse gauge we now have a complete set of equations that are all schematically of the form $\partial^2(\text{field})=\text{(source)}$. The generic way to solve the equations at order $n$ is as follows. We start with equation \eqref{eq:EOMcexp1} which can be formally integrated using a Green's function for the Laplacian. The general solution is thus a harmonic function plus a Poisson integral over the source. We then continue solving \eqref{eq:EOMcexp2} and \eqref{eq:EOMcexp3} in the same way. When writing down the solutions for $M_i(\text{T})$ and $h_{ij}(\text{TT})$ in terms of homogeneous solutions and Poisson integrals we still need to ensure that the solutions are transverse. We then use the gauge condition to solve for the longitudinal fields. We then finally use all of the above solutions to determine the source for \eqref{eq:EOMcexp4}, so that we can integrate that equation as well. We subsequently impose the boundary condition that the solutions are all regular for small $r$. Once we have found the most general solution we re-assemble the fields into the fields $\tau^{(n)}_{\mu}$ and $h^{(n)}_{\mu\nu}$ at order $c^{-n}$. This is then used to compute the sources $S^{(n+1)}_{\mu\nu}$ at the next order in the $1/c$-expansion.

However, when performing the above recipe for constructing solutions a few issues can arise that we now address in general terms. To aid the discussion, let us consider equations \eqref{eq:EOMcexp1} and \eqref{eq:EOMcexp2}. These can be formally solved by using the Green's function of the Laplacian leading to
\begin{eqnarray}
  \hspace{-1cm}  H^{(n)} & = & F^{(n)}-\frac{1}{4\pi}\int_{\Omega_{R_\star}} d^3 x' \frac{3}{4}\frac{S^{(n)}_{ii}(t,x')}{\vert x-x'\vert}\,,\\
  \hspace{-1cm}  M^{(n)}_i(\text{T}) & = & H_i^{(n)}+\frac{1}{3}x^i\partial_t H^{(n)}-\frac{1}{4\pi}\int_{\Omega_{R_\star}} d^3 x' \frac{S^{(n)}_{i}(t,x')-\frac{1}{4}x'^i\partial_t S^{(n)}_{jj}(t,x')}{\vert x-x'\vert}\,,\label{eq:solMi}
\end{eqnarray}
where $F^{(n)}$ and $H_i^{(n)}$ are harmonic functions (solutions to the homogeneous equation) that are regular at $r=0$. The domain of integration has been chosen to be $\Omega_{R_\star}$ which is a ball of radius $R_\star$ centered around the origin $r=0$. We are solving the equations within the region of validity of the PN expansion, i.e. the near zone, so we assume that $R_\star$ is large enough that $x\in\Omega_{R_\star}$. We will introduce the notation
\begin{equation}\label{eq:Pintegral}
    P_{\Omega_{R_\star}}[S]=\frac{1}{4\pi}\int_{\Omega_{R_\star}} d^3 x'\frac{S^{(n)}(t,x')}{\vert x-x'\vert}\,,
\end{equation}
for a Poisson integral over a source $S$ with integration region $\Omega_{R_\star}$. The source is in general non-compact. This is due to the non-linearities of GR. The actual matter source is assumed to be compact. As a result the Poisson integrals over the source when the integration range is $\mathbb{R}^3$ become indefinite integrals and these can and will eventually lead to divergences. To regulate these integrals we introduce a cut-off radius $R_\star$.

There are now four possible scenarios concerning these Poisson integrals:
\begin{enumerate}
    \item The Poisson integral over the source converges for large $R_\star$. In this case we can extend the integration range to $\mathbb{R}^3$. 
    \item A power-counting argument applied to the integrand (when the integration measure is $d^3x'$) suggests that the Poisson integral diverges but the divergent terms in $R_\star$ have zero coefficients (the naive divergence goes away after performing the angular integrations when expressing $d^3 x'$ in spherical coordinates). Again in this case we can extend the range of integration to $\mathbb{R}^3$.
    \item The integral is divergent for large $R_\star$. However there exists a particular harmonic function (depending on $R_\star$) such that when added to the integral the sum does have a large $R_\star$ limit. In other words the divergence can be removed/absorbed by an appropriate harmonic function. \label{type3}
    \item The integral is divergent for large $R_\star$ and the divergence cannot be removed by adding a harmonic function.
\end{enumerate}
In the latter case the $1/c$-expansion has broken down and we need to add $\log c$ terms. However, this does not happen at the orders that we are interested in, which is up to and including 2.5PN (at least not in the harmonic and transverse gauges). Such $\log c$ terms are associated with the appearance of tail terms \cite{Blanchet:1985sp,PhysRevD.25.2038}. We will not have to consider option 4 here.

Let us consider again the integral in \eqref{eq:Pintegral} where $x$ is a point in the near zone. The integration region $\Omega_{R_{\star}}$ is a ball of radius $R_{\star}$ (which is large enough to contain the near zone) with origin $x=0$. This integral will diverge for large $R_{\star}$ if $\int d\Omega S(t,x)=O(r^{-n})$ for $n\le 2$ where $\int d\Omega$ are all the angular integrations when we express the integral in spherical coordinates with centre at $x=0$. A simple diagnostic is to check how the source behaves for large $r$. If $S$ goes to zero strictly faster than $r^{-2}$ the limit $R_{\star}\to\infty$ exists. If $S$ goes to zero as $r^{-2}$ or slower then we need to check what happens to $\int d\Omega S$. If the latter also goes to zero as $r^{-2}$ or slower then the integral is divergent. We then need to check whether or not we can add a harmonic function that is regular close to $x=0$ to make the result finite again.

Finally, we point out that a solution such as \eqref{eq:solMi} still has to obey the transversality condition. For simplicity we will assume that the Poisson integrals are of type 1 or 2. Taking the divergence of  \eqref{eq:solMi} we obtain after some rewriting\footnote{We used that $\partial_i\vert x-x'\vert^{-1}=-\partial'_i\vert x-x'\vert^{-1}$ as well as the identity
\begin{equation}
    \int d^3x' x^i\partial_i \frac{S(t,x')}{\vert x-x'\vert}=\int d^3x' x'^i\partial_i \frac{S(t,x')}{\vert x-x'\vert}-\int d^3x' \frac{S(t,x')}{\vert x-x'\vert}\,,
\end{equation}
where we replaced by $x^i$ in the first integral by $x^i-x'^i+x'^i$ and used that $(x^i-x'^i)\partial_i\vert x-x'\vert^{-1}=-\vert x-x'\vert^{-1}$.}
\begin{equation}
    \partial_i H^{(n)}_i+\partial_t F^{(n)}+\frac{1}{3}x^i\partial_i\partial_t F^{(n)}+\frac{1}{4\pi}\int_{\mathbb{R}^3} d^3 x' \partial'_i\frac{S^{(n)}_i(t,x')}{\vert x-x'\vert}=0\,.
\end{equation}
Since, by assumption the integral $\int_{\mathbb{R}^3} d^3 x'\frac{S_i(t,x')}{\vert x-x'\vert}$ converges, the fall off of $S_i(t,x')$ is such that the boundary term at infinity, that results from applying Stokes' theorem to the last term in the above equation, vanishes. We thus end up with the condition on the homogeneous part of the solution
\begin{equation}
    \partial_i H^{(n)}_i+\partial_t F^{(n)}+\frac{1}{3}x^i\partial_i\partial_t F^{(n)}=0\,.
\end{equation}
We can solve this for $F^{(n)}$ in terms of $\partial_i H^{(n)}_i$ and substitute the result into \eqref{eq:solMi} via $H^{(n)}$.
Similar comments apply to equation \eqref{eq:EOMcexp3} where we need to ensure that the solution for $h_{ij}^{(n)}(\text{TT})$ is transverse.

\section{The covariant $G$-expansion}\label{sec:covGexp}

So far we have focused on the near zone of the spacetime. In this section we will consider the exterior zone where we have vaccuum Einstein's equations. In this part of spacetime we will use an expansion in $G$. Just like before we will be general concerning the gauge choice. We start by expanding the metric around Minkowski spacetime (in inertial coordinates) in powers of $G$
\begin{align}
    g_{\mu \nu} = \eta_{\mu \nu} + G h_{\mu \nu}^{[1]} +  G^2 h_{\mu \nu}^{[2]} + \cdots\,.
\end{align}
We want to approach this in a similar fashion as to what we did in the near zone. This means that, at each order, we want to first expand the equations, apply the gauge conditions, and finally solve the PDEs subject to appropriate boundary conditions.

\subsection{Equations of motion}

We will solve the vacuum Einstein equations in an expansion in $G$, outside the source. Hence the equation of interest is $R_{\mu\nu}=0$. From our knowledge of linearised gravity (expanding $R_{\mu\nu}=0$) we know the form of the equation at every order is going to be
\begin{align}\label{eq:expEinsteq}
    - \square h_{\mu \nu}^{[n]} + \eta^{\rho\sigma}\left(2 \partial_\rho \partial_{(\mu} h^{[n]}_{\nu)\sigma}- \partial_\mu \partial_\nu h^{[n]}_{\rho\sigma}\right) = \tau^{[n]}_{\mu \nu}\,,
\end{align}
where $\eta_{\mu\nu}=\text{diag}(-1,1,1,1)$ ($\mu=0,i$ with $x^0=ct$) and where $\tau^{[n]}_{\mu \nu}$ is a non-linear object that will only depend on products of lower order fields $h^{[n-1]}_{\mu \nu},\ldots,h^{[1]}_{\mu \nu}$ and their derivatives, and thus can be thought of as a source term. We stress that in this section we will find it convenient to use $x^0=ct$. To second order in $G$ we have
\begin{eqnarray}
    \tau^{[1]}_{\mu\nu} & = & 0\,,\\
    \tau^{[2]}_{\mu\nu} & = & 2\overset{[1]}{\Gamma}{}_{\nu\lambda}^{\sigma}\overset{[1]}{\Gamma}{}^\lambda_{\sigma\mu}-2\overset{[1]}{\Gamma}{}^\sigma_{\sigma\lambda}\overset{[1]}{\Gamma}{}^\lambda_{\mu\nu}+2\partial_\sigma\left(h_{[1]}^\sigma{}_\kappa\overset{[1]}{\Gamma}{}^\kappa_{\mu\nu}\right)-\partial_\nu\left(h_{[1]}^{\rho\sigma}\partial_\mu h^{[1]}_{\rho\sigma}\right)\nonumber\\
    & = & h_{[1]}^{\rho\sigma}\left(\partial_\sigma \overset{[1]}{C}_{\nu\mu\rho}-\partial_\nu \overset{[1]}{C}_{\sigma\mu\rho}\right)+\frac{1}{2}\overset{[1]}{C}_{\sigma}{}^\sigma{}_\rho \overset{[1]}{C}_{\mu\nu}{}^\rho-\frac{1}{2}\overset{[1]}{C}_\mu{}^{\rho\sigma}\overset{[1]}{C}_{\nu\rho\sigma}\,,\label{eq:tau2}
\end{eqnarray}
where $\overset{[1]}{\Gamma}{}^\rho_{\mu\nu}$ is the order $G$ term in the expansion of the Levi-Civita connection, i.e.
\begin{equation}
    \overset{[1]}{\Gamma}{}^\rho_{\mu\nu}=\frac{1}{2}\eta^{\rho\sigma}\overset{[1]}{C}_{\mu\nu\sigma}\,,\qquad \overset{[1]}{C}_{\mu\nu\sigma}=\partial_\mu h^{[1]}_{\nu\sigma}+\partial_\nu h^{[1]}_{\mu\sigma}-\partial_\sigma h^{[1]}_{\mu\nu}\,.
\end{equation}

\subsection{Gauge transformations}

The gauge transformation of $g_{\mu\nu}$ is $\delta g_{\mu\nu}=\mathcal{L}_\Xi g_{\mu\nu}$ where we expand $\Xi^\mu$ in powers of $G$ as
\begin{equation}\label{expXi}
    \Xi^\mu=\xi^\mu_{[0]}+G\xi^\mu_{[1]}+G^2\xi^\mu_{[2]}+\mathcal{O}(G^3)\,,
\end{equation}
and where $\xi^\mu_{[0]}$ must be an isometry of $\eta_{\mu\nu}$ in order to preserve the form of the expansion of $g_{\mu\nu}$, i.e. $\xi^\mu_{[0]}$ is given by
\begin{equation}
\xi^\mu_{[0]}=A^\mu+L^\mu{}_\nu x^\nu    \,,
\end{equation}
where $A^\mu$ and $L_{\mu\nu}=-L_{\nu\mu}$ are constant (corresponding to spacetime translations, Lorentz boosts and spatial rotations). Indices are raised and lowered with the Minkowski metric. The gauge transformations acting on $h_{\mu \nu}^{[1]}$, $h_{\mu \nu}^{[2]}$ and $h_{\mu \nu}^{[n]}$ are 
\begin{eqnarray}
    \delta h_{\mu \nu}^{[1]} & = & \mathcal{L}_{\xi_{[0]}}h^{[1]}_{\mu\nu}+\partial_\mu\xi^{[1]}_\nu+\partial_\nu\xi^{[1]}_\mu\,,\\
    \delta h_{\mu \nu}^{[2]} & = & \mathcal{L}_{\xi_{[0]}}h^{[2]}_{\mu\nu}+\mathcal{L}_{\xi_{[1]}}h^{[1]}_{\mu\nu}+\partial_\mu\xi^{[2]}_\nu+\partial_\nu\xi^{[2]}_\mu\,,\\
    \delta h_{\mu \nu}^{[n]} & = & \sum_{k=0}^{n-1}\mathcal{L}_{\xi_{[k]}}h_{\mu\nu}^{[n-k]}+\partial_\mu\xi^{[n]}_\nu+\partial_\nu\xi^{[n]}_\mu\,.\label{gaugetrafoh[n]}
\end{eqnarray}

We next split the index $\mu$ into $(0,i)$ where $x^0=ct$. Furthermore, we introduce the following decomposition
\begin{eqnarray}
    h_{ij}^{[n]} & = & h_{ij}^{[n]}(\text{TT})+\partial_i L_j^{[n]}+\partial_j L_i^{[n]}+\frac{1}{3}\delta_{ij} H^{[n]}\,,\label{eq:decomp1}\\
    h_{0i}^{[n]} & = & -M_i^{[n]}(\text{T})+\partial_0 L_i^{[n]}+\partial_i N^{[n]}\,,\label{eq:decomp2}\\
    h_{00}^{[n]} & = & -2 M^{[n]}_0+2\partial_0 N^{[n]}\,,\label{eq:decomp3}
\end{eqnarray}
where $h_{ij}^{[n]}(\text{TT})$ is transverse traceless, $M_i^{[n]}(\text{T})$ is transverse and where $H^{[n]}$ is given by
\begin{equation}
    H^{[n]}=h_{ii}^{[n]}-2\partial_i L_i^{[n]}\,.
\end{equation}
In terms of these variables the $G$-expanded vacuum Einstein equations \eqref{eq:expEinsteq} can be written as
\begin{eqnarray}
    \partial^2 H^{[n]} & = & -\frac{3}{4}\left(\tau_{00}^{[n]}+\tau_{kk}^{[n]}\right)\,,\label{eq:einst1}\\
    \partial^2 M_0^{[n]} & = & \frac{1}{2}\partial_0^2 H^{[n]}+\frac{1}{2}\tau_{00}^{[n]}\,,\\
    \partial^2 M_i^{[n]}(\text{T}) & = & \frac{2}{3}\partial_0\partial_i H^{[n]}+\tau_{0i}^{[n]}\,,\label{eq:einst3}\\
    -\square h_{ij}^{[n]}(\text{TT}) & = & -2\partial_0\partial_{(i}M_{j)}^{[n]}(\text{T})+2\partial_{\langle i}\partial_{j\rangle}M_0^{[n]}+\frac{1}{3}\partial_{ i}\partial_{j} H^{[n]}+\tau^{[n]}_{\langle ij\rangle}\,,\label{eq:einst4}
\end{eqnarray}
where $\partial^2=\partial_i\partial_i$ and where we have split the $ij$ part of \eqref{eq:expEinsteq} into a trace part (first equation) and a traceless part (last equation). The notation $\langle ij\rangle$ denotes the symmetric tracefree part of $ij$. The equations are presented in the order in which they should be solved in. 

The longitudinal fields $L_i^{[n]}$ and $N^{[n]}$ do not appear at all on the left-hand side of these equations. These fields are fixed by an appropriate gauge fixing condition. The physical propagating degrees of freedom are described by $h_{ij}^{[n]}(\text{TT})$. The right-hand side, through $\tau^{[n]}_{\mu\nu}$, does depend on $L_i^{[k]}$ and $N^{[k]}$ for $k<n$. For $\tau^{[2]}_{\mu\nu}$ this can be seen by using the second equality in \eqref{eq:tau2} and the fact that
\begin{equation}
    \overset{[1]}{C}_{\mu\nu\sigma}=2\partial_\mu\partial_\nu L^{[1]}_\sigma+\overset{[1]}{\tilde C}_{\mu\nu\sigma}\,,
\end{equation}
where $\overset{[1]}{\tilde C}_{\mu\nu\sigma}$ does not depend on the longitudinal fields and where we defined $L^{[1]}_0=N^{[1]}$.

When we are solving \eqref{eq:einst1}--\eqref{eq:einst4} at order $G^n$ (in a particular gauge) the object $\tau^{[n]}_{\mu\nu}$ is known from solving lower orders and matching the result to the near zone. We use the variables $h_{ij}^{[n]}(\text{TT})$, $M_i^{[n]}(\text{T})$ etc only on the left-hand side, i.e. only at order $G^n$. For the source we use $h^{[k]}_{\mu\nu}$ with $k<n$ which are known functions obtained after integration and matching.

Equations \eqref{eq:einst1}--\eqref{eq:einst4} imply
\begin{eqnarray}
    -\frac{1}{2}\partial_0\left(\tau_{00}^{[n]}+\tau_{kk}^{[n]}\right)+\partial_i\tau_{0i}^{[n]} & = & 0\,,\label{eq:conservation1}\\
    -\partial_0\tau_{0j}^{[n]}+\partial_i\left(\frac{1}{2}\left(\tau_{00}^{[n]}-\tau_{kk}^{[n]}\right)\delta_{ij}+\tau_{ij}^{[n]}\right) & = & 0\,.\label{eq:conservation2}
\end{eqnarray}
This is obtained by taking the divergence of equations \eqref{eq:einst3} and \eqref{eq:einst4} and using the other equations to eliminate all but $\tau_{\mu\nu}^{[n]}$. This can also be written as
\begin{equation}
    \partial_\mu \left(\tau^{[n]}{}^\mu{}_\nu-\frac{1}{2}\delta^\mu_\nu\tau^{[n]}{}^\rho{}_\rho\right)=0\,,
\end{equation}
which follows from the divergence of \eqref{eq:expEinsteq}.

The decomposition \eqref{eq:decomp1}--\eqref{eq:decomp3} suffers from the following ambiguity
\begin{eqnarray}
    {h'}^{[n]}_{ij}(\text{TT}) & = & h_{ij}^{[n]}(\text{TT})+\partial_i\chi^{[n]}_j+\partial_j\chi^{[n]}_i-\frac{2}{3}\delta_{ij}\partial_k\chi_k^{[n]}\,,\label{ambi1}\\
    {L'}_i^{[n]} & = & L_i^{[n]}-\chi_i^{[n]}\,,\label{ambi2}\\
    {M'}_i^{[n]}(\text{T}) & = & M_i^{[n]}(\text{T})-\partial_0\chi_i^{[n]}-\partial_i\chi^{[n]}\,,\label{ambi3}\\
    {N'}^{[n]} & = & N^{[n]}-\chi^{[n]}\,,\label{ambi4}\\
    {M'}^{[n]}_0 & = & M_0^{[n]}-\partial_0\chi^{[n]}\,,\label{ambi5}\\
    {H'}^{[n]} & = & H^{[n]}+2\partial_i\chi_i^{[n]}\,,\label{ambi6}
\end{eqnarray}
where $\chi_i^{[n]}$ and $\chi^{[n]}$ satisfy the equations
\begin{eqnarray}
    0 & = & \partial^2\chi_i^{[n]}+\frac{1}{3}\partial_i\partial_j\chi_j^{[n]}\,,\label{eq:chii}\\
    0 & = & \partial_0\partial_i\chi_i^{[n]}+\partial^2\chi^{[n]}\,.\label{eq:chi}
\end{eqnarray}
The latter two equations follow from the transversality of ${h'}^{[n]}_{ij}(\text{TT})$ and ${M'}_i^{[n]}(\text{T})$. These ambiguities are St\"uckelberg-like transformations in the sense that they do not act on the metric $h_{\mu\nu}^{[n]}$, but only on the terms in the decomposition \eqref{eq:decomp1}--\eqref{eq:decomp3}. Equations \eqref{eq:chii} and \eqref{eq:chi} can be written as
\begin{eqnarray}
    0 & = & \partial^2\left(\chi^{[n]}_i+\frac{1}{6}x^i\partial_j\chi^{[n]}_j\right)\,,\\
    0 & = & \partial^2\left(\chi^{[n]}+\frac{2}{5}x^i\partial_0\chi^{[n]}_i+\frac{1}{30}r^2\partial_0\partial_i\chi^{[n]}_i\right)\,,
\end{eqnarray}
where we used that $\partial^2\partial_i\chi_i^{[n]}=0$ which follows from the divergence of equation \eqref{eq:chii}. Hence the solution for $\chi^{[n]}_i$ and $\chi^{[n]}$ is 
\begin{eqnarray}
    \chi^{[n]}_i & = & \Sigma_i^{[n]}-\frac{1}{6}x^i\partial_j\chi_j^{[n]}\,,\label{eq:newchii}\\
    \chi^{[n]} & = & \Sigma^{[n]}-\frac{2}{5}x^i\partial_0 \Sigma^{[n]}_i+\frac{1}{30}r^2\partial_0\partial_i\chi_i^{[n]}\,,\label{eq:newchi}
\end{eqnarray}
where $\Sigma_i^{[n]}$ and $\Sigma^{[n]}$ are harmonic and where we still need to express $\partial_i\chi_i^{[n]}$ in terms of $\partial_i\Sigma_i^{[n]}$, which can be achieved by taking the divergence of \eqref{eq:newchii} and solving the subsequent equation\footnote{If we denote $\partial_i\chi_i^{[n]}$ and $\partial_i\Sigma_i^{[n]}$ as $f_\chi$ and $f_\Sigma$, respectively, then equation \eqref{eq:divchi} reads $\frac{3}{2}f_\chi+\frac{r}{6}\frac{\partial f_\chi}{\partial r}=g_\Sigma$. This equation can be integrated to give $f_\chi=\frac{6}{r^9}\int_c^r dr' r'^8 g_\Sigma$ where $c$ is some constant giving the homogeneous solution.}
\begin{equation}\label{eq:divchi}
    \frac{3}{2}\partial_i\chi_i^{[n]}+\frac{1}{6}x^j\partial_j \partial_i\chi_i^{[n]}=\partial_i\Sigma_i^{[n]}\,,
\end{equation}
for $\partial_i\chi_i^{[n]}$.

The gauge transformation with parameter $\xi^\mu_{[n]}$ acting on $h^{[n]}_{\mu\nu}$, see equation \eqref{gaugetrafoh[n]}, can be realised entirely on the longitudinal fields $L^{[n]}_i$ and $N^{[n]}$ via
\begin{equation}\label{eq:gaugetrafosatGn}
    \delta_{\xi_{[n]}}L_i^{[n]}=\xi_i^{[n]}\,,\qquad\delta_{\xi_{[n]}} N^{[n]}=\xi^{[n]}_0\,.
\end{equation}
Together with equations \eqref{ambi1}--\eqref{ambi6}, these are all the gauge transformations acting on the fields appearing in the decomposition \eqref{eq:decomp1}--\eqref{eq:decomp3} with the exception of lower order gauge transformations with parameters $\xi_{[k]}^\mu$ ($k<n$) that appear in \eqref{gaugetrafoh[n]}. However, once we get to order $G^n$ these lower order transformations will not concerns us because the matching of the solution at the previous orders will have fixed these lower order gauge transformations sufficiently for them to no longer be of interest once we get to the next order in the $G$-expansion. 

Finally, we mention that what is a gauge transformation at the level of the vacuum exterior solution is not necessarily a gauge transformation of the whole solution obtained after matching. This is because in a given gauge the PDEs that the residual gauge transformation parameters have to obey need to satisfy different boundary conditions in the near zone and the exterior zone.

\subsection{Gauge fixing}

Our formalism assumes that the full metric $g_{\mu\nu}$ can be written as $\eta_{\mu\nu}+h_{\mu\nu}$ where $h_{\mu\nu}$ represents either the $1/c$ or $G$-expansion of the metric and where $\eta_{\mu\nu}$ is the Minkowski metric in inertial coordinates. The class of allowed gauge choices to which our formalism applies involves conditions imposed on $h_{tt}, h_{ti}, h_{ij}$, and requires there to be a Newtonian regime. As mentioned previously, we will refer to this class as post-Newtonian gauge choices. This restriction rules out for example a gauge choice such as Bondi gauge (because it does not describe flat spacetime in inertial coordinates) or synchronous gauge (because it does not allow for a Newtonian regime). It would be interesting to develop similar methods that are more covariant with regards to the coordinates used to describe Minkowski spacetime.

In order to solve \eqref{eq:einst1}--\eqref{eq:einst4} we need to impose a gauge fixing condition that tells us what $L_i^{[n]}$ and $N^{[n]}$ are for otherwise the sources $\tau^{[n+1]}_{\mu\nu}$ at the next order depend on the undetermined fields $L_i^{[n]}$ and $N^{[n]}$. Furthermore, at order $G^n$ the choice for $L_i^{[n]}$ and $N^{[n]}$ influences the matching process.

A common gauge choice is the harmonic gauge. In order to show that our methods reproduce existing results, we will employ the harmonic gauge in this paper. An alternative gauge choice is what we refer to as a transverse gauge\footnote{At the linearised level this can be thought as the GR analogue of the Coulomb gauge used in electromagnetism and is also known as the Poisson gauge \cite{Fragkos:2022tbm}. There are of course infinitely many nonlinear gauge choices that reduce to the transverse gauge at the linearised level. One common nonlinear gauge choice is to set $\partial_i N^i=0$ and $\partial_i\left(\gamma^{1/3}\gamma^{ij}\right)=0$ where we used ADM variables to write the metric as
\begin{equation}
    ds^2=-N^2 dt^2+\gamma_{ij}\left(dx^i+N^idt\right)\left(dx^j+N^jdt\right)\,,
\end{equation}
with $\gamma$ is the determinant and $\gamma^{ij}$ the inverse of $\gamma_{ij}$. The derivative $\partial_i$ is with respect to inertial coordinates of a flat background metric. The condition $\partial_i\left(\gamma^{1/3}\gamma^{ij}\right)=0$ is due to Dirac \cite{Smarr:1978dia,Dirac:1958jc}.\label{ftn:transversegauge}} in which case we set
\begin{equation}
 L_i^{[n]}=0\,,\qquad N^{[n]}=0\,,   
\end{equation}
at every order in the $G$-expansion. We will study this gauge choice in the companion paper \cite{companionpaper}.

The harmonic gauge choice is the choice
\begin{equation}\label{harmonicgauge}
    g^{\mu\nu}\Gamma^\rho_{\mu\nu}=0\quad\longleftrightarrow\quad\partial_\mu\left(\sqrt{-g}g^{\mu\nu}\right)=0\,.
\end{equation}
If we expand this in powers of $G$ we find
\begin{eqnarray}
     \eta^{\mu\rho}\left(\partial_\mu h^{[1]}_{\rho\nu}-\frac{1}{2}\partial_\nu h^{[1]}_{\mu\rho}\right) & = & 0\,,\\
     \eta^{\mu\rho}\left(\partial_\mu h^{[2]}_{\rho\nu}-\frac{1}{2}\partial_\nu h^{[2]}_{\mu\rho}\right)  & = & h_{[1]}^{\mu\rho}\partial_\mu h^{[1]}_{\rho\nu}-\frac{1}{2}h_{[1]}^{\mu\rho}\partial_\nu h^{[1]}_{\mu\rho}\,,
\end{eqnarray}
to first and second order in $G$, respectively. To order $G^n$ it takes the form
\begin{equation}\label{eq:HGchoiceordern}
    \eta^{\mu\rho}\left(\partial_\mu h^{[n]}_{\rho\nu}-\frac{1}{2}\partial_\nu h^{[n]}_{\mu\rho}\right) =  K^{[n]}_\nu\,,
\end{equation}
where $K^{[n]}_\nu$ depends on lower order fields. If we use the decomposition \eqref{eq:decomp1}--\eqref{eq:decomp3} then we find
\begin{eqnarray}
    \square L_i^{[n]} & = & \frac{1}{6}\partial_i H^{[n]}+\partial_i M_0^{[n]}-\partial_0 M^{[n]}_i(\text{T})+K^{[n]}_i\,,\label{Ln}\\
    \square N^{[n]} & = & \frac{1}{2}\partial_0 H^{[n]}-\partial_0 M^{[n]}_0+K^{[n]}_0\,.\label{Nn}
\end{eqnarray}
These equations should be added to the list \eqref{eq:einst1}--\eqref{eq:einst4}.

In the formulation \eqref{eq:expEinsteq} the Einstein equations become
\begin{equation}\label{eq:HGeinsteq}
    \square h^{[n]}_{\mu\nu} = -\tau^{[n]}_{\mu\nu}+\partial_\mu K^{[n]}_\nu+\partial_\nu K^{[n]}_\mu\,,
\end{equation}
where the right-hand side now only depends on the lower order fields, and where at order $G^2$ we have
\begin{equation}
    \partial_\mu K^{[2]}_\nu+\partial_\nu K^{[2]}_\mu=\frac{1}{2}h_{[1]}^{\rho\sigma}\left(\partial_\mu \overset{[1]}{C}_{\rho\sigma\nu}+\partial_\nu \overset{[1]}{C}_{\rho\sigma\mu}\right)+\frac{1}{2}\overset{[1]}{C}_{\mu\rho\sigma}\overset{[1]}{C}{}^{\rho\sigma}{}_\nu+\frac{1}{2}\overset{[1]}{C}_{\nu\rho\sigma}\overset{[1]}{C}{}^{\rho\sigma}{}_\mu\,.
\end{equation}
We will show that in any post-Newtonian gauge, for as much as the fields $h^{[n]}_{ij}(\text{TT})$, $M_i^{[n]}(\text{T})$, etc. are concerned, we can reduce the problem of solving the Einstein equations to inverting Laplacian and d'Alembertian operators.

The residual gauge transformations of the choice \eqref{harmonicgauge} are those diffeomorphisms generated by $\Xi^\mu$ for which we have
\begin{equation}
    g^{\mu\nu}\partial_\mu\partial_\nu\Xi^\rho=0\,.
\end{equation}
The diffeomorphism generator is expanded as in \eqref{expXi}. We will ignore the LO term $\xi^\mu_{[0]}$ as this is constrained to be an isometry of Minkowski spacetime. Hence, setting $\xi^\mu_{[0]}=0$ we find the well-known result that the residual gauge transformations are
\begin{eqnarray}
    \square \xi_{[1]}^\rho & = & 0\,,\label{eq:residualharmonic1}\\
    \square \xi_{[2]}^\rho & = & h_{[1]}^{\mu\nu}\partial_\mu\partial_\nu\xi_{[1]}^\rho\,,\label{eq:residualharmonic2}
\end{eqnarray}
to first and second order in $G$.

\subsection{Asymptotic boundary conditions} \label{sec:Asymp_bdy_cond}

The equations of motion that we need to solve are \eqref{eq:einst1}--\eqref{eq:einst4} supplemented with a gauge fixing condition and an appropriate set of asymptotic boundary conditions. First of all, we will demand that the spacetime is asymptotically flat so $h_{\mu\nu}$ will go to zero for large $r$. We will formulate all boundary conditions for a coordinate system that is asymptotically inertial, i.e. the metric approaches flat spacetime described in inertial coordinates $(t,x^i)$. We will demand that $H^{[n]}$, $M_0^{[n]}$, $M_i^{[n]}(\text{T})$ and $h_{ij}^{[n]}(\text{TT})$ are all $\mathcal{O}(r^{-1})$ for large $r=\sqrt{x^i x^i}$.

\subsubsection{The non-propagating sector}

We start with the fields 
$H^{[n]}$, $M_0^{[n]}$, $M_i^{[n]}(\text{T})$ which obey Poisson-type PDEs \eqref{eq:einst1}--\eqref{eq:einst3}, and so do not correspond to propagating fields. For these fields a Dirichlet boundary condition will suffice. Equations \eqref{eq:einst1}--\eqref{eq:einst3} can be rewritten as follows
\begin{eqnarray}
\partial^2 H^{[n]} & = & -\frac{3}{4}\left(\tau_{00}^{[n]}+\tau_{kk}^{[n]}\right)\,,\label{eq:neweinst1}\\
    \partial^2\left(M_0^{[n]}-\frac{r^2}{12}\partial_0^2 H^{[n]}+\frac{x^i}{2}\partial_0 M^{[n]}_i(\text{T})\right) & = & \frac{1}{2}\tau^{[n]}_{00}+\frac{r^2}{16}\partial_0^2\left(\tau^{[n]}_{00}+\tau^{[n]}_{kk}\right)+\frac{x^i}{2}\partial_0\tau^{[n]}_{0i}\,,\label{eq:neweinst2}\nonumber\\
    &&\\
    \partial^2\left(M^{[n]}_i(\text{T})-\frac{1}{3}x^i\partial_0 H^{[n]}\right) & = & \tau_{0i}^{[n]}+\frac{x^i}{4}\partial_0\left(\tau^{[n]}_{00}+\tau^{[n]}_{kk}\right)\,,\label{eq:neweinst3}
\end{eqnarray}
turning them into genuine Poisson equations. The right-hand side can be rewritten using \eqref{eq:conservation1} and \eqref{eq:conservation2} but we will not attempt this as the focus will be on the left-hand side. The solutions thus take the general form
\begin{eqnarray}
    H^{[n]} & = & K^{[n]}+\cdots\,,\label{eq:H}\\
    M_0^{[n]} & = & F^{[n]}-\frac{r^2}{12}\partial_0^2 H^{[n]}-\frac{x^i}{2}\partial_0 H^{[n]}_i+\cdots\,,\label{eq:M0}\\
    M^{[n]}_i(\text{T}) & = & H^{[n]}_i+\frac{x^i}{3}\partial_0 H^{[n]}+\cdots\,,\label{eq:Mi}
\end{eqnarray}
where the dots denote terms resulting from the non-linear sources in $\tau_{\mu\nu}^{[n]}$ and where $K^{[n]}$, $F^{[n]}$ and $H^{[n]}_{i}$ (for every $i$) are all harmonic functions and where $H^{[n]}_{i}$ obeys
\begin{equation}\label{eq:divHi}
    \partial_i H^{[n]}_{i}=-\partial_0 H^{[n]}-\frac{1}{3}x^i\partial_i\partial_0 H^{[n]}\,,
\end{equation}
resulting from the fact that $M^{[n]}_i(\text{T})$ is transverse. The boundary condition that the fields $H^{[n]}$, $M_0^{[n]}$, $M_i^{[n]}(\text{T})$ are $\mathcal{O}(r^{-1})$ can now be seen to have a number of consequences. From the fact that $K^{[n]}$, $F^{[n]}$ and $H^{[n]}_{i}$ (for every $i$) are all harmonic functions it follows that all the terms on the right-hand side of \eqref{eq:H}, \eqref{eq:M0} and \eqref{eq:Mi} have to separately be $\mathcal{O}(r^{-1})$. Hence, we conclude that
\begin{equation}\label{eq:boundaryconditions}
    \partial_0 H^{[n]}=\mathcal{O}(r^{-2})\,,\qquad\partial^2_0 H^{[n]}=\mathcal{O}(r^{-3})\,,\qquad     \partial_0 H_i^{[n]}=\mathcal{O}(r^{-2})\,.
\end{equation}

The homogeneous solution to equations \eqref{eq:neweinst1}--\eqref{eq:neweinst3} can be solved asymptotically (for large $r$) as follows.
We start with $K^{[n]}$. Since it is harmonic and decaying for large $r$ we know that\footnote{In appendix \ref{app:PDEstuff} we collect some results about multipole expansions of solutions to the Laplace and the free wave equation. For the problem at hand see equation \eqref{eq:decayharm}.}
\begin{equation}
    K^{[n]}=\frac{A^{[n]}}{r}+\partial_i\left(\frac{A^{[n]}_i}{r}\right)+\frac{1}{2}\partial_i\partial_j\left(\frac{A^{[n]}_{ij}}{r}\right)+\mathcal{O}(r^{-4})\,,
\end{equation}
where the coefficients $A^{[n]}$, $A^{[n]}_i$ and $A^{[n]}_{ij}$ (symmetric tracefree) are in general functions of $t$. However, the conditions \eqref{eq:boundaryconditions} tell us that 
\begin{equation}
    \dot A^{[n]}=0\,,\qquad \ddot A^{[n]}_i=0\,,
\end{equation}
where the dots denote $x^0$ derivatives.
There are no conditions on $F^{[n]}$ other than it being a decaying harmonic, so we have 
\begin{equation}
    F^{[n]}=\frac{B^{[n]}}{r}+\mathcal{O}(r^{-2})\,,
\end{equation}
where $B^{[n]}$ is a function of $t$. Finally, since $H_i^{[n]}$ obeys $\partial^2 H_i^{[n]}=0$ we know that we must have the following large $r$ expansion
\begin{equation}
    H_i^{[n]}=\frac{C^{[n]}_i}{r}+\partial_j\left(\frac{C^{[n]}_{i,j}}{r}\right)+\mathcal{O}(r^{-3})\,,
\end{equation}
where a priori $C^{[n]}_i$ and $C^{[n]}_{i,j}$ are functions of $t$ and where the comma between the indices in $C^{[n]}_{i,j}$ is to indicate that, a priori, there is no symmetry between them. Equation \eqref{eq:divHi} then tells us that we must have
\begin{equation}
    C^{[n]}_i=-\frac{1}{3}\dot A^{[n]}_i\,,\qquad C^{[n]}_{i,j}=\frac{1}{3}\delta_{ij}C^{[n]}_{l,l}+C^{[n]}_{[i,j]}\,,
\end{equation}
i.e. the traceless symmetric part of $C^{[n]}_{i,j}$ is zero. This leads to the following asymptotic homogeneous solution for $H^{[n]}$, $M_0^{[n]}$, $M_i^{[n]}(\text{T})$, 
\begin{eqnarray}
    H^{[n]} & = & \frac{A^{[n]}}{r}+\partial_i\left(\frac{A^{[n]}_i}{r}\right)+\frac{1}{2}\partial_i\partial_j\left(\frac{A^{[n]}_{ij}}{r}\right)+\mathcal{O}(r^{-4})\,,\label{eq:Hsol}\\
    M^{[n]}_0 & = & \frac{B^{[n]}}{r}-\frac{1}{6}\frac{\dot C^{[n]}_{l,l}}{r}-\frac{1}{8}\frac{x^i x^j}{r^3}\ddot A^{[n]}_{ij}+\mathcal{O}(r^{-2})\,,\label{eq:M0sol}\\
    M^{[n]}_i(\text{T}) & = & -\frac{1}{3}\frac{\dot A^{[n]}_i}{r}-\frac{1}{3}\frac{x^i x^k}{r^3}\dot A^{[n]}_k+\frac{1}{2}\frac{x^i}{r^5} x^k x^l\dot A^{[n]}_{kl}-\frac{1}{3}\frac{x^i}{r^3}C^{[n]}_{l,l}-\frac{x^k}{r^3}C^{[n]}_{[i,k]}+\mathcal{O}(r^{-3})\,,\nonumber\\
    &&\label{eq:Misol}
\end{eqnarray}
where $A^{[n]}$ and $\dot A^{[n]}_i$ are constants (as is $A^{[n]}_i-x^0\dot A^{[n]}_i$ since $A^{[n]}_i$ is linear in $x^0$). The other coefficients $B^{[n]}$, $A_{ij}^{[n]}$, $C^{[n]}_{l,l}$, and $C^{[n]}_{[i,j]}$ are at this stage arbitrary functions of time. 

Further below we will see that part of the above asymptotic solution for $H^{[n]}$, $M_0^{[n]}$, $M_i^{[n]}(\text{T})$ takes the form of an ambiguity transformation. In other words parts of the solution can be shown to correspond to coefficients in the asymptotic expansion of the parameters $\chi^{[n]}$ and $\chi^{[n]}_i$ that describe the ambiguities \eqref{ambi1}--\eqref{ambi6}. These ambiguities get intertwined with the gauge transformations \eqref{eq:gaugetrafosatGn} when specifying the gauge choice\footnote{For example if we choose the gauge $L_i^{[n]}=0$ and $N^{[n]}=0$ then we can perform the transformation \eqref{ambi1}--\eqref{ambi5} provided we also perform a compensating gauge transformation \eqref{eq:gaugetrafosatGn} with $\xi_0^{[n]}=\chi^{[n]}$ and $\xi_i^{[n]}=\chi_i^{[n]}$ to ensure that the transformed $L_i^{[n]}$ and $N^{[n]}$ remain zero. More precisely, under the combination of the ambiguity and an order $G^n$ gauge transformation the longitudinal fields transform as ${L'}^{[n]}_i=L_i^{[n]}-\xi_i^{[n]}+\chi_i^{[n]}$ and ${N'}^{[n]}=N^{[n]}-\xi_0^{[n]}+\chi^{[n]}$. Setting this to zero gives $\xi_0^{[n]}=\chi^{[n]}$ and $\xi_i^{[n]}=\chi_i^{[n]}$.}. We stress that even though these may appear as gauge artefacts we cannot set these ambiguity parameters equal to zero as this would amount to setting the residual gauge transformations equal to zero and these are not actual residual gauge transformations of the whole matched solution. The process of matching tells us to find the most general solution to the PDEs on both sides of the matching and this most general solution includes what appear to be residual gauge transformations\footnote{For example if a residual gauge parameter has to be a harmonic function in the exterior region then it must be a decaying harmonic to respect the boundary conditions, but in the near zone the same equation would have to be solved by a harmonic function that is regular at the origin. There is no harmonic function that obeys both these properties at the same time. Hence, what appears to be a residual gauge transformation is not a gauge transformation of the whole matched solution. In fact for well-chosen gauge conditions and boundary conditions there are no globally well-defined gauge transformations left to perform.}. 

To be more explicit about the nature of the effect of the ambiguities we solve equation \eqref{eq:divchi} asymptotically so that we can apply the transformations \eqref{ambi1}--\eqref{ambi6} with $\chi_i^{[n]}$ and $\chi^{[n]}$ as given in \eqref{eq:newchii} and \eqref{eq:newchi}. To respect the boundary conditions both $\chi_i^{[n]}$ and $\chi^{[n]}$ need to be $\mathcal{O}(r^{-1})$. Equations \eqref{eq:newchii} and \eqref{eq:newchi} then tell us that the harmonic functions $\Sigma^{[n]}_i$ and $\Sigma^{[n]}$ need to decay for large $r$ and $\partial_0\Sigma^{[n]}_i=\mathcal{O}(r^{-2})$ as well as $\partial_0\partial_i\chi^{[n]}_i=\mathcal{O}(r^{-3})$. Using that $\partial_i\chi^{[n]}_i$ and $\Sigma^{[n]}_i$ are harmonic we can write
\begin{eqnarray}
    \partial_i\chi^{[n]}_i & = & \frac{D^{[n]}}{r}+\partial_i\left(\frac{D^{[n]}_i}{r}\right)+\frac{1}{2}\partial_i\partial_j\left(\frac{D^{[n]}_{ij}}{r}\right)+\mathcal{O}(r^{-4})\,,\\
    \Sigma^{[n]}_i & = & \frac{E^{[n]}_i}{r}+\partial_j\left(\frac{E^{[n]}_{i,j}}{r}\right)+\mathcal{O}(r^{-3})\,.
\end{eqnarray}
From the above boundary conditions we learn that $D^{[n]}=0$ and that $D^{[n]}_i$ and $E^{[n]}_i$ are time-independent. Solving \eqref{eq:divchi} we find that 
\begin{equation}
    D^{[n]}_i=\frac{6}{7}E^{[n]}_i\,,\qquad D^{[n]}_{ij}=2E^{[n]}_{\langle i,j\rangle}\,.
\end{equation}
We can use this to determine $\chi^{[n]}_i$ at the orders $r^{-1}$ and $r^{-2}$ and $\chi^{[n]}$ at the leading $r^{-1}$ order. We will denote the leading order part in the expansion of $\Sigma^{[n]}$ by $E^{[n]}r^{-1}$. Using \eqref{ambi3}, \eqref{ambi5} and \eqref{ambi6} we then find the following asymptotic ambiguities in $H^{[n]}$, $M^{[n]}_0$ and $M^{[n]}_i(\text{T})$,
\begin{eqnarray}
    {H^{[n]}}' & = & H-\frac{12}{7}\partial_i\left(\frac{E_i}{r}\right)-2\partial_i\partial_j\left(\frac{E_{ij}}{r}\right)+\mathcal{O}(r^{-4})\,,\label{eq:asympambiforH}\\
    {M^{[n]}_0}' & = & M^{[n]}_0-\frac{\dot E^{[n]}}{r}-\frac{2}{15}\frac{\ddot E^{[n]}_{i,i}}{r}-\frac{1}{2}\frac{x^i x^j}{r^3}\ddot E^{[n]}_{\langle i,j\rangle}+\mathcal{O}(r^{-2})\,,\\
    {M^{[n]}_i}'(\text{T}) & = & M^{[n]}_i(\text{T})+\frac{x^j}{r^3}\dot E^{[n]}_{[i,j]}+2\frac{x^ix^jx^k}{r^5}\dot E^{[n]}_{\langle j,k\rangle}+\frac{7}{15} \frac{x^i}{r^3}\dot E^{[n]}_{j,j}+\frac{x^i}{r^3}E^{[n]}+\mathcal{O}(r^{-3})\,.\label{eq:ambiNLO}\nonumber\\
    &&
\end{eqnarray}
This can be matched with the appearance of the functions $B^{[n]}$, $A^{[n]}_{ij}$, $C^{[n]}_{l,l}$ and $C^{[n]}_{[i,j]}$ (as well as the constant $A^{[n]}_i-x^0\dot A^{[n]}_i$ via $E^{[n]}_i$) in the solution \eqref{eq:Hsol}, \eqref{eq:M0sol} and \eqref{eq:Misol}. The ambiguity transformation does not affect $A^{[n]}$ and $\dot A^{[n]}_i$.

\subsubsection{ADM charges}

Before we continue our discussion of the boundary conditions for the remaining fields, we show that our boundary conditions are such that the homogeneous solutions lead to well-defined ADM charges. The Landau--Lifshitz (LL) energy-momentum pseudotensor is defined as
\begin{equation}
    T^{\mu\nu}_{\text{LL}}=-\frac{c^4}{8\pi G}G^{\mu\nu}+\frac{c^4}{16\pi G (-g)}\partial_\rho\partial_\sigma\left((-g)\left(g^{\mu\nu}g^{\rho\sigma}-g^{\mu\rho}g^{\nu\sigma}\right)\right)\,.
\end{equation}
Hence, upon using the Einstein equations we see that $\mathcal{T}^{\mu\nu}:=(-g)(T^{\mu\nu}+T^{\mu\nu}_{\text{LL}})$ is conserved, i.e. $\partial_\mu \mathcal{T}^{\mu\nu}=0$. 
We can thus define conserved charges (energy-momentum) as follows
\begin{equation}\label{eq:ADM_EM}
    P^\nu=\int_{t=\text{cst}}d^3x (-g)(T^{0\nu}+T^{0\nu}_{\text{LL}})\,.
\end{equation}
The integrand can be written as 
\begin{equation}
     (-g)(T^{0\nu}+T^{0\nu}_{\text{LL}})=\frac{c^4}{16\pi G}\partial_j J^{j\nu}\,,
\end{equation}
where we defined
\begin{eqnarray}
    J^{j\nu} & = & \partial_0I^{\nu 0 j}+\partial_0 I^{\nu j0}+\partial_k I^{\nu jk}\,,\\
    I^{\nu\rho\sigma} & = & (-g)\left(g^{0\nu}g^{\rho\sigma}-g^{0\rho}g^{\nu\sigma}\right)\,.
\end{eqnarray}
The energy-momentum vector $P^\nu$ can thus be expressed as a surface integral over the boundary of the constant $t$ slices, i.e. at spatial infinity as
\begin{equation}
    P^\nu=\frac{c^4}{16\pi G}\int_{S^2_\infty} d\Omega r^2 n^j J^{j\nu}\,,
\end{equation}
where $n^j=x^j/r$ and the integral is over the 2-sphere at spatial infinity.

Due to the symmetry of $\mathcal{T}^{\mu\nu}:=(-g)(T^{\mu\nu}+T^{\mu\nu}_{\text{LL}})$ we can build another conserved current $J^{\mu\nu\rho}$ given by
\begin{equation}
    J^{\mu\nu\rho}=\mathcal{T}^{\mu\nu}x^\rho-\mathcal{T}^{\mu\rho}x^\nu\,,
\end{equation}
and hence we can define the angular momentum and Lorentz boost charges
\begin{equation}
    M^{\nu\rho}=\int_{t=\text{cst}}d^3x J^{0\nu\rho}\,.
\end{equation}

If we work to first order in $G$ it can be readily shown that
\begin{eqnarray}
    J^{i0} & = & -\frac{2}{3}\partial_i H^{[1]}+\mathcal{O}(G^2)\,,\label{eq:Ji0}\\
    J^{ij} & = & \frac{2}{3}\delta_{ij}\partial_0 H^{[1]}-\partial_i M_j^{[1]}(\text{T})-\partial_j M_i^{[1]}(\text{T})+\mathcal{O}(G^2)\,.\label{eq:Jij}
\end{eqnarray}
At higher orders in $G$ we get the above terms but with the superscript $[1]$ replaced by $[n]$ as well as new non-linear terms. The boundary conditions that $H^{[n]}$, $M_0^{[n]}$, $M_i^{[n]}(\text{T})$ are $\mathcal{O}(r^{-1})$ as well as \eqref{eq:boundaryconditions} ensures that the contribution at order $G^n$ coming from the linear terms in $J^{i\nu}$, i.e. \eqref{eq:Ji0} and \eqref{eq:Jij}, is finite. It can be shown that $A^{[n]}$ contributes to the ADM energy $P^0$ while $\dot A^{[n]}_i$ contributes to the ADM momentum. For example, at order $G$ we have that $P^0$ is proportional to $A^{[1]}$ and $P^i$ is proportional to $\dot A^{[1]}_i$. Furthermore, the angular momentum $M^{ij}$ at order $G$ is proportional{}\footnote{This can be shown by using that $J^{0ij}=\frac{c^4}{16\pi G}\partial_k\left(x^j J^{ki}-x^i J^{kj}\right)$.} to $C^{[1]}_{[i,j]}$. Finally, the Lorentz boost $M^{0i}$ is proportional\footnote{This follows from $J^{00i}=\mathcal{T}^{00}x^i-x^0 \mathcal{T}^{0i}=\frac{c^4}{16\pi G}\partial_k\left(x^i J^{k0}-x^0J^{ki}+\frac{2}{3}\delta_{ik}H^{[1]}\right)$.} to $A_i-x^0\dot A_i$, i.e. the $t$-independent part of $A_i$.

Earlier we said that the coefficients $C^{[n]}_{[i,j]}$ suffer from the ambiguity described by the transformations \eqref{ambi1}--\eqref{ambi6} (due to the appearance of $E^{[n]}_{[i,j]}$ in \eqref{eq:ambiNLO}). Now we see that the angular momentum at leading order in $G$ is proportional to $C^{[1]}_{[i,j]}$ which therefore suffers from the ambiguity as well\footnote{One might wonder where the ambiguity in the angular momentum comes from since the Landau--Lifshitz energy-momentum pseudo tensor depends on $g_{\mu\nu}$ which is free from these ambiguities. The step where this happens is when we write $J^{ij}$ in equation \eqref{eq:Jij}. The integrand of $M^{ij}$ when written as an integral over 3-space is the divergence $\partial_i J^{ij}$ that can be written in terms of $H^{[n]}$ which does not suffer from the ambiguities, but when we apply Stokes' theorem, the object $J^{ij}$ contains $M_i(\text{T})$ which does suffer from it at order $r^{-2}$.}. We expect this to be related to the known ambiguities in defining angular momentum for asymptotically flat spacetimes. Relatedly, we point out that the appearance of the constant vector $E_i$ in the ambiguity of $H^{[n]}$ (see equation \eqref{eq:asympambiforH}) implies that there is an ambiguity in the Lorentz boost charge as well.

\subsubsection{The propagating sector}

We next turn to the field $h^{[n]}_{ij}(\text{TT})$ which solves equation \eqref{eq:einst4} and hence describes propagating degrees of freedom. If we differentiate \eqref{eq:einst4} twice with respect to $x^0$ and use equations \eqref{eq:neweinst2} and \eqref{eq:neweinst3} we obtain
\begin{eqnarray}
    &&\square\left(\partial_0^2 h^{[n]}_{ij}(\text{TT})+\partial_i\partial_0 M^{[n]}_j(\text{T})+\partial_j\partial_0 M^{[n]}_i(\text{T})-2\partial_i\partial_j M^{[n]}_0+\frac{1}{3}\delta_{ij}\partial_0^2 H^{[n]}\right)=\nonumber\\
    &&-\partial_0^2\tau^{[n]}_{ij}+\partial_0\partial_i\tau^{[n]}_{0j}+\partial_0\partial_j\tau^{[n]}_{0i}-\partial_i\partial_j\tau^{[n]}_{00}\,.
\end{eqnarray}
Let us first consider the homogeneous part. We note that we can write we can write the homogeneous equation as
\begin{eqnarray}
    0&=&\square\partial_0^2\left(h^{[n]}_{ij}(\text{TT}) +\partial_i U^{[n]}_j+\partial_j U^{[n]}_i+\partial_i\partial_j\left(x^k U^{[n]}_k\right)-2\partial_i\partial_j U^{[n]}+\frac{r^2}{6}\partial_i\partial_j H^{[n]}\right.\nonumber\\
    &&\left.+\frac{2}{3}\partial_i\left(x^j H^{[n]}\right)+\frac{2}{3}\partial_j\left(x^i H^{[n]}\right)\right)\,,\label{eq:ddothij}
\end{eqnarray}
where we defined the functions $U^{[n]}$ and $U^{[n]}_i$ which satisfy
\begin{equation}
    \partial_0^2 U^{[n]}=F^{[n]}\,,\qquad \partial_0 U^{[n]}_i=H^{[n]}_i\,.
\end{equation}
The functions $F$ and $H_i$ are harmonic and appeared for the first time in \eqref{eq:M0} and \eqref{eq:Mi}. At this point we go through a long rewriting of \eqref{eq:ddothij} and its solution in order to be able to arrive at the homogeneous solution as expressed in equation \eqref{eq:intermediatesol_hTT2} which will prove very convenient later on\footnote{If the following calculation is not of interest to the reader they can skip ahead to the summary at the top page 56.}. 
Recall that $H_i$ obeys the condition \eqref{eq:divHi} which can now be written as 
\begin{equation}
    \partial_0\left(\partial_i U^{[n]}_i+H^{[n]}+\frac{1}{3}x^k\partial_k H^{[n]}\right)=0\,.
\end{equation}
Integrating this we find that 
\begin{equation}
\partial_i U^{[n]}_i+H^{[n]}+\frac{1}{3}x^k\partial_k H^{[n]}=\tilde U^{[n]}\,,    
\end{equation}
where $\tilde U^{[n]}$ is some time-independent function that is not a new function as it is entirely determined by the left-hand side. It is merely a useful shorthand notation.

If we define $\hat\chi^{[n]}_i$ and $\hat\chi^{[n]}$ as follows
\begin{eqnarray}
    \hat\chi^{[n]}_i & = & -U^{[n]}_i+\partial_i U^{[n]}-\frac{1}{2}\partial_i\left(x^k U^{[n]}_k\right)-\frac{1}{2}x^i H^{[n]}-\frac{1}{12}r^2\partial_i H^{[n]}\,,\label{eq:hatchii}\\
    \hat\chi^{[n]} & = & \frac{1}{12}r^2\partial_0 H^{[n]}+\frac{1}{2}x^i H^{[n]}_i-\partial_0 U^{[n]}\,,\label{eq:hatchi}
\end{eqnarray}
then we can write
\begin{eqnarray}
    H^{[n]} & = & 2\partial_k\hat\chi^{[n]}_k+4\tilde U^{[n]}-2\partial^2 U^{[n]}+x^i\partial^2 U^{[n]}_i\,,\label{eq:finalhomogsol2}\\
    M^{[n]}_0 & = & -\partial_0\hat\chi^{[n]}\,,\label{eq:finalhomogsol3}\\
    M^{[n]}_i(\text{T}) & = & -\partial_0\hat\chi^{[n]}_i-\partial_i\hat\chi^{[n]}\,.\label{eq:finalhomogsol4}
\end{eqnarray}
This is almost of the form of an ambiguity transformation \eqref{ambi1}--\eqref{ambi6}. Referring to equations \eqref{eq:chii} and \eqref{eq:chi} we see that we have
\begin{eqnarray}
    \partial^2\hat\chi^{[n]}_i+\frac{1}{3}\partial_i\partial_j\hat\chi^{[n]}_j & = & -\frac{5}{3}\partial_i\tilde U^{[n]}-\partial^2 U^{[n]}_i+\frac{4}{3}\partial_i\partial^2 U^{[n]}-\frac{2}{3}\partial_i\left(x^j\partial^2 U^{[n]}_j\right)\,,\\
    \partial^2\hat\chi^{[n]}+\partial_0\partial_i\hat\chi^{[n]}_i & = & 0\,.
\end{eqnarray}
We thus see that the failure for this to be an ambiguity transformation is measured by $\tilde U^{[n]}$, $\partial^2 U^{[n]}$ and $\partial^2 U^{[n]}_i$. 

Furthermore, we have
\begin{eqnarray}
    &&\partial_i U^{[n]}_j+\partial_j U^{[n]}_i+\partial_i\partial_j\left(x^k U^{[n]}_k\right)-2\partial_i\partial_j U^{[n]}+\frac{r^2}{6}\partial_i\partial_j H^{[n]}+\frac{2}{3}\partial_i\left(x^j H^{[n]}\right)\nonumber\\
    &&+\frac{2}{3}\partial_j\left(x^i H^{[n]}\right)=-\partial_i\hat\chi^{[n]}_j-\partial_j\hat\chi^{[n]}_i+\frac{1}{3}\delta_{ij}H^{[n]}\,,
\end{eqnarray}
so that we can write \eqref{eq:ddothij} as 
\begin{equation}\label{eq:intermediatewveeq1}
    0=\square\partial_0^2\left(h^{[n]}_{ij}(\text{TT})-\partial_i\hat\chi^{[n]}_j-\partial_j\hat\chi^{[n]}_i+\frac{2}{3}\delta_{ij}\partial_k\hat\chi^{[n]}_k\right)\,,
\end{equation}
where we used that 
\begin{equation}
    \partial_0^2 H^{[n]}=2\partial_0^2\partial_k\hat\chi^{[n]}_k\,.
\end{equation}
The second time derivative of the term in parenthesis in \eqref{eq:intermediatewveeq1} is transverse traceless. The most general solution to equation \eqref{eq:intermediatewveeq1} necessarily must be of the form\footnote{We use here that the solution to an equation of the form $\square\partial_0^2f=0$ is a sum $W+T$ where $\square W=0$ and $\partial_0^2 T=0$. We checked this for the class of solutions that can be obtained by the method of separation of variables.}
\begin{equation}\label{eq:intermediatesol_hTT}
    h^{[n]}_{ij}(\text{TT})=W^{[n]}_{ij}+\partial_i\hat\chi^{[n]}_j+\partial_j\hat\chi^{[n]}_i-\frac{2}{3}\delta_{ij}\partial_k\hat\chi^{[n]}_k+A^{[n]}_{ij}+x^0 B^{[n]}_{ij}\,,
\end{equation}
where $A^{[n]}_{ij}$ and $B^{[n]}_{ij}$ are time-independent and traceless and where $W^{[n]}_{ij}$ is traceless and obeys the free wave equation. We can decompose $W^{[n]}_{ij}$ into a TT and longitudinal traceless part as 
\begin{equation}\label{eq:decomW}
    W^{[n]}_{ij}=W^{[n]}_{ij}(\text{TT})+\partial_i C^{[n]}_j+\partial_j C^{[n]}_i-\frac{2}{3}\delta_{ij}\partial_k C^{[n]}_k\,.
\end{equation}

Since $\partial_0^2 \partial_{\langle i}\hat\chi^{[n]}_{j\rangle}$ is TT it follows from \eqref{eq:intermediatesol_hTT} that $\partial_0^2 \partial_{\langle i}C^{[n]}_{j\rangle}$ is also TT. This means that we have
\begin{equation}\label{eq:intermediateC}
    \partial_0^2\left(\partial^2 C^{[n]}_j+\frac{1}{3}\partial_j\partial_i C^{[n]}_i\right)=0\,.
\end{equation}
Since $\square W_{ij}^{[n]}=0$, if we act with $\square\partial_j$ on the decomposition \eqref{eq:decomW} the result \eqref{eq:intermediateC} also tells us that $\partial^2 C^{[n]}_j+\frac{1}{3}\partial_j\partial_i C^{[n]}_i$ is harmonic. 

The decomposition \eqref{eq:decomW} suffers from the following ambiguity transformation
\begin{eqnarray}
    {W'}^{[n]}_{ij}(\text{TT}) & = & W^{[n]}_{ij}(\text{TT})+2\partial_{\langle i}\psi^{[n]}_{j\rangle}\,,\\
    {C'}^{[n]}_i & = & C^{[n]}_i+\psi^{[n]}_i\,,\label{eq:ambiC}
\end{eqnarray}
where in order for ${W'}^{[n]}_{ij}(\text{TT})$ to be TT we need that 
\begin{equation}\label{eq:ambpsi}
    \partial^2\psi^{[n]}_j+\frac{1}{3}\partial_j\partial_k\psi_k^{[n]}=0\,.
\end{equation}
We have just proven that both $\partial_0^2 C_i^{[n]}$ and $\partial^2 C_i^{[n]}$ are solutions to \eqref{eq:ambpsi}. Let us define these solutions as 
\begin{equation}\label{eq:Cprime}
    \check\psi_i^{[n]}=\partial_0^2 C_i^{[n]}\,,\qquad\hat\psi_i^{[n]}=\partial^2 C_i^{[n]}\,.
\end{equation}
These are not independent since we have $\partial^2\check\psi_i^{[n]}=\partial_0^2\hat\psi^{[n]}_i$. Under an ambiguity transformation \eqref{eq:ambiC} we have
\begin{equation}
    \square {C'}^{[n]}_i=\square C^{[n]}_i+\square \psi^{[n]}_i=-\check\psi_i^{[n]}+\hat\psi_i^{[n]}+\square \psi^{[n]}_i\,.
\end{equation}
Hence, if we can write $-\check\psi_i^{[n]}+\hat\psi_i^{[n]}=\square X^{[n]}_i$ where $X_i$ solves \eqref{eq:ambpsi} (up to a solution to the free wave equation), then we can without of loss of generality set $\square {C'}^{[n]}_i=0$ by taking $\psi^{[n]}_i$ such that $\square(\psi^{[n]}_i+X^{[n]}_i)=0$. The equation $-\check\psi_i^{[n]}+\hat\psi_i^{[n]}=\square X^{[n]}_i$ implies that $O_{ij}\square X_j^{[n]}=0$ where we defined the operator $O_{ij}=\delta_{ij}\partial^2+\frac{1}{3}\partial_i\partial_j$. Since $O_{ij}$ and $\square$ commute and are different operators it follows that $X_i^{[n]}$ is a sum $W_i^{[n]}+\bar\psi_i^{[n]}$ where $W_i^{[n]}$ obeys $\square W_i^{[n]}=0$ and $\bar\psi_i^{[n]}$ solves \eqref{eq:ambpsi}. We thus conclude that without loss of generality we can set $\square C_i^{[n]}=0$ and thus $\square W_{ij}^{[n]}(\text{TT})=0$ as follows from \eqref{eq:decomW}.

As an intermediate result we now know that $h^{[n]}_{ij}(\text{TT})$ must take the form 
\begin{equation}\label{eq:intermediatesol_hTT_2}
    h^{[n]}_{ij}(\text{TT})=W^{[n]}_{ij}(\text{TT})+2\partial_{\langle i}\left(\hat\chi^{[n]}_{j\rangle}+C^{[n]}_{j\rangle}\right)+A^{[n]}_{ij}+x^0 B^{[n]}_{ij}\,,
\end{equation}
where $\square W_{ij}^{[n]}(\text{TT})=0$. On this result we still have to enforce that the right-hand side is transverse and that the original equation for $h^{[n]}_{ij}(\text{TT})$, i.e. equation \eqref{eq:einst4}, is satisfied (where as usual we ignore the nonlinear sources described by $\tau^{[n]}_{\mu\nu}$). We start with the latter. Equation \eqref{eq:einst4} can alternatively be written as
\begin{equation}\label{eq:finalhomogsol5}
    \square h_{ij}^{[n]}(\text{TT})=-2\partial_0^2\partial_{\langle i}\hat\chi_{j\rangle}^{[n]}-\frac{1}{3}\partial_i\partial_j H^{[n]}\,.
\end{equation}
We substitute \eqref{eq:intermediatesol_hTT_2} into \eqref{eq:finalhomogsol5} which leads to 
\begin{equation}\label{eq:intermediatewveeq2}
    \partial^2\left(\partial_i\hat\chi^{[n]}_j+\partial_j\hat\chi^{[n]}_i-\frac{2}{3}\delta_{ij}\partial_k\hat\chi^{[n]}_k\right)+\partial^2 A^{[n]}_{ij}+x^0\partial^2 B^{[n]}_{ij}+\frac{1}{3}\partial_i\partial_j H^{[n]}=0\,.
\end{equation}
If we differentiate this equation with respect to $x^0$ we obtain an equation for $B^{[n]}_{ij}$ that is solved by
\begin{equation}\label{eq:intermediatewveeq3}
    B^{[n]}_{ij}=H^{[n]}_{ij}-2\partial_0\partial_i\partial_j U^{[n]}+\frac{2}{3}\delta_{ij}\partial_0\partial^2 U^{[n]}\,,
\end{equation}
where $H^{[n]}_{ij}$ is harmonic and traceless. Substituting this into \eqref{eq:intermediatewveeq2} we obtain an equation for $A^{[n]}_{ij}$. Rather than working with $A^{[n]}_{ij}$ it will prove convenient to write the right-hand side of \eqref{eq:intermediatesol_hTT_2} as follows
\begin{eqnarray}\label{eq:intermediatesol_hTT2}
    h^{[n]}_{ij}(\text{TT}) & = & W^{[n]}_{ij}(\text{TT})+2\partial_{\langle i}C_{j\rangle}^{[n]}+\hat A^{[n]}_{ij}+x^0 H^{[n]}_{ij}-\frac{1}{6}r^2\partial_i\partial_j H^{[n]}\nonumber\\
    &&-\frac{2}{3}\left[\partial_i\left(x^j H^{[n]}\right)+\partial_j\left(x^i H^{[n]}\right)-\frac{2}{3}\delta_{ij}\partial_k\left(x^k H^{[n]}\right)\right]\,,
\end{eqnarray}
where we defined $\hat A^{[n]}_{ij}$ which is traceless as
\begin{eqnarray}
    \hat A^{[n]}_{ij} & = & A^{[n]}_{ij}-\partial_i U^{[n]}_j-\partial_j U^{[n]}_i+\frac{2}{3}\delta_{ij}\partial_k U^{[n]}_k-\partial_i\partial_j\left(x^k U^{[n]}_k\right)+\frac{1}{3}\delta_{ij}\partial^2\left(x^k U^{[n]}_k\right)\nonumber\\
    &&+2\partial_i\partial_j\left(U^{[n]}-x^0\partial_0 U^{[n]}\right)-\frac{2}{3}\delta_{ij}\partial^2\left(U^{[n]}-x^0\partial_0 U^{[n]}\right)\,.\label{eq:hat F0}
\end{eqnarray}
Equations \eqref{eq:intermediatewveeq2} and \eqref{eq:intermediatewveeq3} then lead to
\begin{equation}\label{eq:prophatF1_1}
    \partial^2\hat A^{[n]}_{ij}=2\partial_i\partial_j\left(H^{[n]}+\frac{1}{3}x^k\partial_k H^{[n]}\right)\,.
\end{equation}
From the fact that $A^{[n]}_{ij}$ and $B^{[n]}_{ij}$ are time-independent and the redefinitions \eqref{eq:intermediatewveeq3} and \eqref{eq:hat F0} we see that
\begin{eqnarray}
    \partial_0 H^{[n]}_{ij} & = & 2\partial_i\partial_j F^{[n]}\,,\label{eq:timederH1}\\
    \partial_0 \hat A^{[n]}_{ij} & = & -2x^0\partial_i\partial_j F^{[n]}-\partial_i\partial_j\left(x^k H^{[n]}_k\right)-\partial_i H^{[n]}_j-\partial_j H^{[n]}_i+\frac{4}{3}\delta_{ij}\partial_k H^{[n]}_k\,.\label{eq:prophatF1_2}
\end{eqnarray}
We still need to require that the right-hand side of \eqref{eq:intermediatesol_hTT_2} is transverse. By taking the divergence of \eqref{eq:intermediatesol_hTT2} and differentiating with respect to $x^0$ we find that $H^{[n]}_{ij}$ obeys
\begin{equation}\label{eq:divH1}
    \partial_i H_{ij}^{[n]}+\left(\delta_{ij}\partial^2 +\frac{1}{3}\partial_j\partial_i\right)\partial_0C^{[n]}_i=0\,.
\end{equation}
Substituting this into the divergence of \eqref{eq:intermediatesol_hTT2} we learn that
\begin{equation}\label{eq:prophatF1_3}
    \partial_i\hat A^{[n]}_{ij}+\left(\delta_{ij}\partial^2+\frac{1}{3}\partial_j\partial_i\right) \left( C^{[n]}_i-x^0\partial_0 C^{[n]}_i\right)=\frac{5}{3}\partial_j\left(H^{[n]}+\frac{1}{3}x^k\partial_k H^{[n]}\right)\,.
\end{equation}
We recall that $C^{[n]}_i$ obeys \eqref{eq:intermediateC} and so the $C$-dependent terms in the above two equations are time-independent.

To summarise, the solution for $h_{ij}^{[n]}(\text{TT})$ is given by \eqref{eq:intermediatesol_hTT2}. In here $H^{[n]}_{ij}$ is traceless and harmonic and obeys equations \eqref{eq:timederH1} and \eqref{eq:divH1}. Furthermore, $\hat A^{[n]}_{ij}$ is traceless and obeys equations \eqref{eq:prophatF1_1}, \eqref{eq:prophatF1_2} and \eqref{eq:prophatF1_3}.

The field $h_{ij}^{[n]}(\text{TT})$ obeys a wave equation and we will demand that $h_{ij}^{[n]}(\text{TT})$ obeys the Sommerfeld no-incoming radiation condition at past null infinity $\mathcal{I}^-$. If we write the Minkowski line element in spherical coordinates and define retarded and advanced time as $u=t-r/c$ and $v=t+r/c$, respectively, then this means that we will require that
\begin{equation}
    \lim_{\overset{r\rightarrow\infty}{v=\text{cst}}}\partial_v\left(rh_{ij}^{[n]}(\text{TT})\right)=0\,.
\end{equation}
Apart from $W^{[n]}_{ij}(\text{TT})$ and $C^{[n]}_i$ the only terms on the right-hand side of the solution for $h^{[n]}_{ij}(\text{TT})$ in equation \eqref{eq:intermediatesol_hTT2} that are $\mathcal{O}(r^{-1})$ come from the terms with $H^{[n]}$, $\hat A^{[n]}_{ij}$ and $H^{[n]}_{ij}$. Using their asymptotic solutions, which for
$\hat A^{[n]}_{ij}$ follows from solving \eqref{eq:prophatF1_1}, tells us that the Sommerfeld condition on $h^{[n]}_{ij}(\text{TT})$ translates into a Sommerfeld condition on $W^{[n]}_{ij}(\text{TT})$ and the symmetric tracefree derivative of $C_i^{[n]}$. Hence we need to require that
\begin{equation}
    \lim_{\overset{r\rightarrow\infty}{v=\text{cst}}}\partial_v\left(rW_{ij}^{[n]}(\text{TT})\right)=0\,,
\end{equation}
and similarly for $\partial_{\langle i}C_{j\rangle}^{[n]}$.

We are however not done yet. In order to determine the homogeneous part of the metric at order $G^n$ we need to include the longitudinal fields $L_i^{[n]}$ and $N^{[n]}$. The equations \eqref{eq:einst1}--\eqref{eq:einst4} for $n\ge 2$ do not form a closed set of equations. The reason being that the source terms $\tau_{\mu\nu}^{[n]}$ depend on the longitudinal fields $L_i^{[k]}$ and $N^{[k]}$ for $k<n$. 

The fields $L_i^{[n]}$ and $N^{[n]}$ are fixed by imposing a gauge fixing condition and if the latter take the form of a PDE then we need boundary conditions for the $L_i^{[n]}$ and $N^{[n]}$ fields as well. This is furthermore relevant since these fields will be part of the matching process.
In order that the metric satisfies $g_{\mu\nu}=\eta_{\mu\nu}+\mathcal{O}(r^{-1})$ we will need to impose that $L_i^{[n]}$ and $N^{[n]}$ are at most $\mathcal{O}(1)$ and such that both their $\partial_0$ and $\partial_i$ derivatives are $\mathcal{O}(r^{-1})$. This is because the metric only depends on the $\partial_0$ and $\partial_i$ derivatives of $L_i^{[n]}$ and $N^{[n]}$.

\subsection{Parametrising the harmonic gauge metric}\label{subsec:paramHG}

In harmonic gauge the homogeneous part of $h_{\mu\nu}^{[n]}$ obeys the free wave equation. However, the gauge condition relates the various components of $h_{\mu\nu}^{[n]}$. In this section we show that we can parametrise $h_{\mu\nu}^{[n]}$ into a number of independent solutions to the free wave equation. The final result is given in equations \eqref{eq:HGparam1}--\eqref{eq:HGparam3}.

In the harmonic gauge the longitudinal fields obey the wave equations \eqref{Ln} and \eqref{Nn}. The time-derivative of the homogeneous part of these latter two equations is equivalent to $\square h^{[n]}_{0i}=0=\square h^{[n]}_{00}$. Using equations \eqref{eq:M0} and \eqref{eq:Mi} (as well as the properties of $\hat\chi^{[n]}$ and $\hat\chi^{[n]}_i$) we can rewrite (the homogeneous part of) equations \eqref{Ln} and \eqref{Nn} as follows
\begin{eqnarray}
\square\left(N^{[n]}+\hat\chi^{[n]}\right)  & = & -\partial_0\partial^2 U^{[n]}   \,,\label{eq:N}\\
\square\left(L^{[n]}_i+\hat\chi^{[n]}_i\right) & = & -\partial_i \tilde U^{[n]}-\partial^2 U^{[n]}_i+\partial_i\partial^2 U^{[n]}-\frac{1}{2}\partial_i\left(x^j\partial^2 U_j^{[n]}\right)\,.\label{eq:Li}
\end{eqnarray}

Using that $\partial_0^2 U^{[n]}=F^{[n]}$ is harmonic we can differentiate the first of these two equations to get 
$\partial_0\square\left(N^{[n]}+\hat\chi^{[n]}\right)=0$ whose solution is of the form $N^{[n]}+\hat\chi^{[n]}=W^{[n]}+A^{[n]}$ where $W^{[n]}$ obeys $\square W^{[n]}=0$ and where $A^{[n]}$ is independent of $x^0$. Substituting this into \eqref{eq:N} we find that $\partial^2\left(A^{[n]}+\partial_0 U^{[n]}\right)=0$ so that $A^{[n]}=\tilde H^{[n]}-\partial_0 U^{[n]}$ with $\tilde H^{[n]}$ harmonic. Since $A^{[n]}$ is time-independent we learn that $\partial_0 \tilde H^{[n]}=F^{[n]}$. We thus conclude that
\begin{equation}\label{eq:finalN}
    N^{[n]}=W^{[n]}-\hat\chi^{[n]}+\tilde H^{[n]}-\partial_0 U^{[n]}=W^{[n]}-\frac{1}{12}r^2\partial_0 H^{[n]}-\frac{1}{2}x^i H_i^{[n]}+\tilde H^{[n]}\,.
\end{equation}

Next we consider equation \eqref{eq:Li}. We start by observing that $\partial_0^2\square\left(L^{[n]}_i+\hat\chi^{[n]}_i\right)=0$, so that 
\begin{equation}
    L^{[n]}_i+\hat\chi^{[n]}_i=W^{[n]}_i+A^{[n]}_i+x^0 B^{[n]}_i\,,
\end{equation}
where $A^{[n]}_i$ and $B^{[n]}_i$ are time-independent. Substituting this into \eqref{eq:Li} we obtain
\begin{equation}\label{eq:intermediateLi_1}
    \partial^2 A^{[n]}_i+x^0\partial^2 B^{[n]}_i=-\partial_i \tilde U^{[n]}-\partial^2 U^{[n]}_i+\partial_i\partial^2 U^{[n]}-\frac{1}{2}\partial_i\left(x^j\partial^2 U_j^{[n]}\right)\,.
\end{equation}
If we differentiate this with respect to $x^0$ we find $\partial^2\left(B^{[n]}_i-\partial_i\partial_0 U^{[n]}\right)=0$ so that we obtain
\begin{equation}
   B^{[n]}_i=\tilde H^{[n]}_i+\partial_i\partial_0 U^{[n]} \,,
\end{equation}
where $\tilde H^{[n]}_i$ is harmonic and for which $\partial_0 \tilde H_i^{[n]}=-\partial_i F^{[n]}$. Equation \eqref{eq:intermediateLi_1} now reduces to an equation for $A^{[n]}_i$ which can be simplified by defining $\hat A^{[n]}_i$ as
\begin{equation}
    \hat A^{[n]}_i=A^{[n]}_i+U^{[n]}_i+\frac{1}{2}\partial_i\left( x^k U^{[n]}_k\right)-\partial_i\left(U^{[n]}-x^0\partial_0 U^{[n]}\right)\,.
\end{equation}
This object then obeys the following two equations
\begin{eqnarray}
    \partial^2\hat A^{[n]}_i & = & -\partial_i\left(H^{[n]}+\frac{1}{3}x^k\partial_k H^{[n]}\right)\,,\\
    \partial_0\hat A^{[n]}_i & = & x^0\partial_i F^{[n]}+H^{[n]}_i+\frac{1}{2}\partial_i\left(x^k H^{[n]}_k\right)\,.
\end{eqnarray}
We conclude that the solution for $L^{[n]}_i$ is
\begin{equation}\label{eq:finalLi}
    L^{[n]}_i=W^{[n]}_i-\hat\chi^{[n]}_i+A^{[n]}_i+x^0 B^{[n]}_i=W^{[n]}_i+\frac{1}{2}x^i H^{[n]}+\frac{1}{12}r^2\partial_i H^{[n]}+\hat A^{[n]}_i+x^0 \tilde H^{[n]}_i\,.
\end{equation}

The terms to the right of $W^{[n]}$ and $W^{[n]}_i$ in equations \eqref{eq:finalN} and \eqref{eq:finalLi}, respectively, are at most $\mathcal{O}(1)$, but such that if we differentiate these terms with respect to $x^0$ or $x^i$ they are $\mathcal{O}(r^{-1})$. Since only derivatives of $N^{[n]}$ and $L^{[n]}_i$ appear in the metric we will demand that they obey the boundary condition that their $x^0$ and $x^i$ derivatives are $\mathcal{O}(r^{-1})$.

The solutions for $N^{[n]}$ and $L^{[n]}_i$ can now be used to write $h^{[n]}_{\mu\nu}$ in harmonic gauge. We find
\begin{eqnarray}
    h^{[n]}_{00} & = & -2M^{[n]}_0+2\partial_0 N^{[n]}=2\partial_0 W^{[n]}\,,\\
    h^{[n]}_{0i} & = & -M^{[n]}_i(\text{T})+\partial_0 L^{[n]}_i+\partial_i N^{[n]}=\partial_0 W^{[n]}_i+\partial_i W^{[n]}+\tilde H^{[n]}_i+\partial_i \tilde H^{[n]}\,,\\
    h^{[n]}_{ij} & = & h^{[n]}_{ij}(\text{TT})+\partial_i L^{[n]}_j+\partial_j L^{[n]}_i+\frac{1}{3}\delta_{ij}H^{[n]}\nonumber\\
    & = & W^{[n]}_{ij}(\text{TT})+\partial_i W^{[n]}_j+\partial_j W^{[n]}_i+2\partial_{\langle i}C_{j\rangle}^{[n]}+\hat H^{[n]}_{ij}+x^0 \tilde H^{[n]}_{ij}\,,
\end{eqnarray}
where we defined
\begin{eqnarray}
    \hat H^{[n]}_{ij} & = & \hat A^{[n]}_{ij}+\partial_i \hat A^{[n]}_j+\partial_j \hat A^{[n]}_i+\frac{4}{3}\delta_{ij}\left(H^{[n]}+\frac{1}{3}x^k\partial_k H^{[n]}\right)\,,\\
    \tilde H^{[n]}_{ij} & = & H^{[n]}_{ij}+\partial_i \tilde H^{[n]}_j+\partial_j \tilde H^{[n]}_i\,,
\end{eqnarray}
where we used the solutions for $M^{[n]}_i(\text{T})$ and $h^{[n]}_{ij}(\text{TT})$ obtained previously. It can be shown using properties derived previously that $\tilde H^{[n]}_i+\partial_i \tilde H^{[n]}$, $\hat H^{[n]}_{ij}$ and $\tilde H^{[n]}_{ij}$ are all time-independent and harmonic and furthermore that
\begin{eqnarray}
    \partial_i \hat H^{[n]}_{ij} & = & \partial_j\partial_i\hat A^{[n]}_i+2\partial_j\left(H^{[n]}+\frac{1}{3}x^k\partial_k H^{[n]}\right)-O_{ij}\left( C^{[n]}_i-x^0\partial_0 C^{[n]}_i\right)\,,\label{eq:propsG01}\\
    \hat H^{[n]}_{ii} & = & 2\partial_i\hat A^{[n]}_i+4\left(H^{[n]}+\frac{1}{3}x^k\partial_k H^{[n]}\right)\,,\label{eq:propsG02}\\
    \partial_i \tilde H^{[n]}_{ij} & = & \partial_j\partial_i \tilde H^{[n]}_i-O_{ij} \partial_0 C^{[n]}_i\,,\label{eq:G1props1}\\
    \tilde H^{[n]}_{ii} & = & 2\partial_i \tilde H^{[n]}_i\,,\label{eq:G1props2}
\end{eqnarray}
where we defined (as before)
\begin{equation}
    O_{ij}=\delta_{ij}\partial^2+\frac{1}{3}\partial_i\partial_j\,.
\end{equation}

Since $\tilde H^{[n]}_i+\partial_i \tilde H^{[n]}$ is time-independent and harmonic we can absorb it into $W^{[n]}_i$ in the expression for $h^{[n]}_{it}$ by defining 
\begin{equation}
    \tilde W^{[n]}_i = W^{[n]}_i+x^0\left(\tilde H^{[n]}_i+\partial_i \tilde H^{[n]}\right)\,.
\end{equation}
This gives $h^{[n]}_{it}=\partial_0\tilde W^{[n]}_i+\partial_i W^{[n]}$ where both $\tilde W^{[n]}_i$ and $W^{[n]}$ satisfy the free wave equation. In terms of $\tilde W^{[n]}_i$ the expression for $h^{[n]}_{ij}$ becomes
\begin{equation}
    h^{[n]}_{ij} =W^{[n]}_{ij}(\text{TT})+\partial_i\tilde W^{[n]}_j+\partial_j\tilde W^{[n]}_i+2\partial_{\langle i}C_{j\rangle}^{[n]}+\hat H^{[n]}_{ij}+x^0 \check H^{[n]}_{ij}\,,
\end{equation}
where 
\begin{equation}
    \check H^{[n]}_{ij}=\tilde H^{[n]}_{ij}-\partial_i \tilde H^{[n]}_j-\partial_j \tilde H^{[n]}_i-2\partial_i\partial_j \tilde H^{[n]}\,,
\end{equation}
which is harmonic, traceless, time-independent and whose divergence is given by
\begin{equation}
    \partial_i\check H^{[n]}_{ij}=-O_{ij}\partial_0 C_i^{[n]}\,.
\end{equation}

Finally, we define $G^{[n]}_{ij}$ as
\begin{equation}\label{eq:defG}
    G^{[n]}_{ij}=2\partial_{\langle i}C^{[n]}_{j\rangle}+\hat H^{[n]}_{ij}+x^0 \check H^{[n]}_{ij}\,,
\end{equation}
so that 
\begin{equation}\label{eq:paramhij}
    h^{[n]}_{ij} =W^{[n]}_{ij}(\text{TT})+ G^{[n]}_{ij}+\partial_i\tilde W^{[n]}_j+\partial_j\tilde W^{[n]}_i\,.
\end{equation}
The object $G^{[n]}_{ij}$ has the following properties (that follow from its definition)
\begin{eqnarray}
    \square  G^{[n]}_{ij} & = & 0\,,\label{eq:propG1}\\
    \partial_0^2 G^{[n]}_{ij} & = & \text{transverse traceless}\,,\label{eq:propG2}\\
    \partial_i G^{[n]}_{ij}-\frac{1}{2}\partial_j G_{ii}^{[n]} & = & 0\,.\label{eq:propG3}
\end{eqnarray}
Furthermore, these properties are equivalent to its definition\footnote{We will suppress the $[n]$ superscript here. To show this we first use $\partial_0^2\square G_{ij}=0$ so that $G_{ij}=Z_{ij}+A_{ij}+x^0 B_{ij}$ where $\square Z_{ij}=0$ and where $A_{ij}$ and $B_{ij}$ are time-independent. Using next that $\square G_{ij}=0$ it follows that $A_{ij}$ and $B_{ij}$ are harmonic. Next we decompose $Z_{ij}=Z_{ij}(\text{TT})+2\partial_{\langle i}Y_{j\rangle}+\frac{2}{3}\delta_{ij}Y$. Using that $\partial_0^2 G_{ij}$ is traceless we see that $\partial_0^2 Y=0$, but we also know that $Z_{ij}$ and hence its trace obeys the free wave equation so that $Y$ must harmonic. We can therefore absorb $Y$ into $A_{ij}$ and $B_{ij}$. We have thus arrived at $G_{ij}=Z_{ij}(\text{TT})+2\partial_{\langle i}Y_{j\rangle}+A_{ij}+x^0 B_{ij}$. Using that $\partial_0^2 G_{ij}$ is transverse we conclude that $O_{ij}\partial_0^2 C_j=0$. By acting with $\partial^2$ on \eqref{eq:propG3} we also find that $O_{ij}\partial^2 C_j=0$. The decomposition of $Z_{ij}$ suffers from the usual ambiguity and we can now repeat the argument around equation \eqref{eq:Cprime} which leads to the conclusion that without loss of generality we can take $Y_i$ to obey $\square Y_i=0$. Finally, by simply writing out $\partial_i G_{ij}$ and $G_{ii}$ we see that we have recovered \eqref{eq:defG} (up to a TT solution to the free wave equation).} \eqref{eq:defG} (up to a TT solution to the free wave equation), so that we can take \eqref{eq:paramhij} as the final form of the harmonic gauge parametrisation of $h_{ij}^{[n]}$.

We thus conclude that in harmonic gauge we can parametrise (the homogeneous part of) $h^{[n]}_{\mu\nu}$ as follows\footnote{There is another parametrisation of the homogeneous part of the harmonic gauge metric that is commonly used in the literature on post-Newtonian expansions (see introduction). We will use this parametrisation in section \ref{sec:NZmetric2.5}.}
\begin{eqnarray}
    h^{[n]}_{tt} & = & 2\partial_t W^{[n]}\,,\label{eq:HGparam1}\\
    h^{[n]}_{ti} & = & \partial_t W^{[n]}_i+\partial_i W^{[n]}\,,\label{eq:HGparam2}\\
    h^{[n]}_{ij} & = & W^{[n]}_{ij}(\text{TT})+\partial_i W^{[n]}_j+\partial_j W^{[n]}_i+G^{[n]}_{ij}\,,\label{eq:HGparam3}
\end{eqnarray}
where we dropped tildes on $W^{[n]}_i$ and $W^{[n]}_{ij}$ and where we used $x^0=ct$. We absorbed a factor of $c$ into $W^{[n]}$ (i.e. we defined $cW^{[n]}=\tilde W^{[n]}$ and subsequently dropped the tilde on $W^{[n]}$). In here $W^{[n]}$, $W^{[n]}_i$, $W^{[n]}_{ij}(\text{TT})$ and $G^{[n]}_{ij}$ all obey the free wave equation and $G^{[n]}_{ij}$ furthermore satisfies \eqref{eq:propG2} and \eqref{eq:propG3}. In the expression for $G^{[n]}_{ij}$ we assume that there is no TT part that separately solves the free wave equation (for if that existed we could absorb it into $W^{[n]}_{ij}$).

The functions $W^{[n]}$ and $W^{[n]}_i$ can be viewed as corresponding to the residual gauge transformations of the harmonic gauge conditions. The function $W^{[n]}_{ij}(\text{TT})$ describes the physical degrees of freedom. Finally, the object $G^{[n]}_{ij}$ is needed to ensure that the spacetime has the appropriate ADM energy as neither $W^{[n]}_{ij}(\text{TT})$ nor $W^{[n]}_i$ contribute to the ADM energy $P^0$ defined in \eqref{eq:ADM_EM} (for $\nu=0$).

There is a slight freedom in the choice of functions $W^{[n]}$, $W_j^{[n]}$, $W_{ij}^{[n]}(\text{TT})$. This freedom is parametrised by time-independent harmonic functions $\Lambda^{[n]}$ and $\Lambda^{[n]}_i$ and are given by the following transformations
\begin{eqnarray}
    W'^{[n]} & = & W^{[n]}+\Lambda^{[n]}\,,\\
    W_i'^{[n]} & = & W^{[n]}_i-t\partial_i\Lambda^{[n]}+\Lambda^{[n]}_i\,,\\
    G'^{[n]}_{ij} & = & G^{[n]}_{ij}+2t\partial_i\partial_j\Lambda^{[n]}-\partial_i\Lambda^{[n]}_j-\partial_j\Lambda^{[n]}_i\,.
\end{eqnarray}
The properties of $\Lambda^{[n]}$ and $\Lambda^{[n]}_i$ follow from writing $h^{[n]}_{tt}= 2\partial_t W'^{[n]}$ and $h^{[n]}_{ti} = \partial_t W'^{[n]}_i+\partial_i W'^{[n]}$ and demanding that $W'^{[n]}$ and $W'^{[n]}_i$ obey the free wave equation.

A natural choice of boundary conditions in the harmonic gauge is to demand that $h^{[n]}_{\mu\nu}$ obeys the Sommerfeld no-radiation condition at $\mathcal{I}^{-}$. We will abbreviate this boundary condition simply by ``$\mathcal{S}$''. This means in particular that $\partial_t W^{[n]}$ obeys $\mathcal{S}$. This does not imply that $W^{[n]}$ itself obeys $\mathcal{S}$ but we can choose $\Lambda^{[n]}$ such that it does. This means that $\partial_i W^{[n]}$ obeys $\mathcal{S}$ and hence so does $\partial_t W^{[n]}_i$ (since $h^{[n]}_{it}$ is required to obey $\mathcal{S}$). Again we can choose $\Lambda^{[n]}_i$ such that $W^{[n]}_i$ obeys $\mathcal{S}$. Turning to the $ij$ component of the metric we already know that $h^{[n]}_{ij}(\text{TT})$ and $W^{[n]}_{ij}(\text{TT})$ obey $\mathcal{S}$ so we conclude that $G^{[n]}_{ij}$ must obey $\mathcal{S}$. Finally, we also want that $h^{[n]}_{\mu\nu}=\mathcal{O}(r^{-1})$. This implies that we can allow $W^{[n]}$ and $W^{[n]}_i$ to be $\mathcal{O}(1)$ as long as their $\partial_t$ and $\partial_i$ derivatives are $\mathcal{O}(r^{-1})$. Furthermore, we need that both $W^{[n]}_{ij}(\text{TT})$ and $G^{[n]}_{ij}$ are each $\mathcal{O}(r^{-1})$.

The boundary conditions as formulated above in the harmonic gauge are compatible with the boundary conditions as formulated by Trautman in \cite{Trautman:1958zdi}. For the transverse-like gauge the boundary conditions used here are almost but not quite in agreement with \cite{Trautman:1958zdi}. However, we have previously shown that the boundary conditions in that gauge result in finite expressions for the ADM energy and momentum.

As an illustration of the harmonic gauge parametrisation we consider linearised Schwarschild in isotropic coordinates.
The Schwarzschild line element in isotropic coordinates is given by
\begin{equation}
    ds^2=-\frac{\left(1-\frac{GM}{2c^2 r}\right)^2}{\left(1+\frac{GM}{2c^2 r}\right)^2}c^2 dt^2+\left(1+\frac{GM}{2c^2 r}\right)^4dx^i dx^i\,.
\end{equation}
To first order in $G$ this is
\begin{equation}
    ds^2=\left(-c^2+2\frac{GM}{r}\right)dt^2+\left(1+2\frac{GM}{c^2 r}\right)dx^i dx^i+\mathcal{O}(G^2)\,,
\end{equation}
so that 
\begin{equation}
    h^{[1]}_{tt}=\frac{2M}{r}\,,\qquad h^{[1]}_{ti}=0\,,\qquad h^{[1]}_{ij}=\frac{2M}{c^2 r}\,.
\end{equation}
This can be written in the form \eqref{eq:HGparam1}--\eqref{eq:HGparam3} if we choose 
\begin{eqnarray}
    W^{[1]} &= & \frac{Mu}{r}\,,\\
    W^{[1]}_i & = & -\frac{M}{2}\partial_i\left(\frac{u^2}{r}\right)\,,\label{eq:WiSchwarz}\\
    W^{[1]}_{ij}(\text{TT}) & = & 0\,,\\
    G^{[1]}_{ij} & = & \frac{2M}{c^2 r}\delta_{ij}+M\partial_i\partial_j\left(\frac{u^2}{r}\right)\,.\label{eq:WijSchwarz}
\end{eqnarray}
It can be readily verified that these all obey the free wave equation (with Sommerfeld boundary conditions), as well as \eqref{eq:propG2} and \eqref{eq:propG3}.

\subsection{Summary}\label{subsec:summary}

We briefly summarise the main findings of this section. In transverse gauge the $G$-expanded vacuum Einstein equations are 
\begin{eqnarray}
\partial^2 H^{[n]} & = & -\frac{3}{4}\left(\tau_{00}^{[n]}+\tau_{kk}^{[n]}\right)\,,\label{eq:summary1}\\
    \partial^2\left(M_0^{[n]}-\frac{r^2}{12}\partial_0^2 H^{[n]}+\frac{x^i}{2}\partial_0 M^{[n]}_i(\text{T})\right) & = & \frac{1}{2}\tau^{[n]}_{00}+\frac{r^2}{16}\partial_0^2\left(\tau^{[n]}_{00}+\tau^{[n]}_{kk}\right)+\frac{x^i}{2}\partial_0\tau^{[n]}_{0i}\,,\label{eq:summary2}\nonumber\\
    &&\\
    \partial^2\left(M^{[n]}_i(\text{T})-\frac{1}{3}x^i\partial_0 H^{[n]}\right) & = & \tau_{0i}^{[n]}+\frac{x^i}{4}\partial_0\left(\tau^{[n]}_{00}+\tau^{[n]}_{kk}\right)\,,\label{eq:summary3}
\end{eqnarray}
as well as
\begin{equation}\label{eq:summary4}
     -\square h_{ij}^{[n]}(\text{TT}) = -2\partial_0\partial_{(i}M_{j)}^{[n]}(\text{T})+2\partial_{\langle i}\partial_{j\rangle }M_0^{[n]}+\frac{1}{3}\partial_{\langle i}\partial_{j\rangle} H^{[n]}+\tau^{[n]}_{\langle ij\rangle}\,.
\end{equation}
The latter implies
\begin{eqnarray}
    &&\square\left(\partial_0^2 h^{[n]}_{ij}(\text{TT})+\partial_i\partial_0 M^{[n]}_j(\text{T})+\partial_j\partial_0 M^{[n]}_i(\text{T})-2\partial_i\partial_j M^{[n]}_0+\frac{1}{3}\delta_{ij}\partial_0^2 H^{[n]}\right)=\nonumber\\
    &&-\partial_0^2\tau^{[n]}_{ij}+\partial_0\partial_i\tau^{[n]}_{0j}+\partial_0\partial_j\tau^{[n]}_{0i}-\partial_i\partial_j\tau^{[n]}_{00}\,.\label{eq:summary5}
\end{eqnarray}
The homogeneous solution is given by \eqref{eq:finalhomogsol2}--\eqref{eq:finalhomogsol4} and \eqref{eq:intermediatesol_hTT2}. For the particular solution to the sourced equations we need to invert the Laplacian and the d'Alembertian. The boundary conditions are such that $H^{[n]}$, $M_0^{[n]}$ and $M^{[n]}_i(\text{T})$ are $\mathcal{O}(r^{-1})$ for large $r$ and $h^{[n]}_{ij}(\text{TT})$ obeys the Sommerfeld no-incoming radiation condition at past null infinity.

In harmonic gauge the equations are 
\begin{equation}\label{eq:HGGexp}
    \square h^{[n]}_{\mu\nu} = -\tau^{[n]}_{\mu\nu}+\partial_\mu K^{[n]}_\nu+\partial_\nu K^{[n]}_\mu\,.
\end{equation}
The homogeneous solutions are just the most general solutions to $\square h^{[n]}_{\mu\nu}=0$. The boundary conditions are such that $h^{[n]}_{\mu\nu}$ obey the Sommerfeld no-incoming radiation condition at past null infinity. Not all the components of $h_{\mu\nu}^{[n]}$ are independent in the harmonic gauge which is why it is convenient to use the parametrisation given in \eqref{eq:HGparam1}--\eqref{eq:HGparam3}.

In a general gauge the equation \eqref{eq:summary1}--\eqref{eq:summary5} together with the above boundary conditions are also valid but then we still need to specify what the longitudinal fields $L^{[n]}_i$, $N^{[n]}$ are by making a gauge choice, and an appropriate boundary condition. 
The problem of solving the $G$-expansion can thus always be reduced to that of inverting the operators $\partial^2$ and $\square$ as well as solving the equations that result from the gauge choice.

If we compare the transverse gauge with the harmonic gauge then the former has the advantage of a smaller set of residual gauge transformations. In harmonic gauge the residual gauge transformations are \eqref{eq:residualharmonic1} and \eqref{eq:residualharmonic2} which involves homogeneous solutions to the free wave equation whereas for the transverse gauge the residual gauge transformations are given by the ambiguities \eqref{eq:chii} and \eqref{eq:chi} which involves harmonic functions. The latter are much easier to deal with. Another feature of the transverse gauge is that the traceless part of $h_{ij}$ is automatically transverse so we do not need to resort to transverse traceless projectors which are used in harmonic gauge to find an expression for the waveform.

\subsection{Non-linear sources}

In this section we focused on the homogeneous solution and the consequences the boundary conditions have for these solutions. We end this section with a few remarks about the non-linear sources described by the $\tau_{\mu\nu}^{[n]}$. We already gave an expression for $\tau^{[2]}_{\mu\nu}$ in \eqref{eq:tau2}. Here we give an explicit expression in terms of the metric and its derivatives for $\tau^{[2]}_{\mu\nu}$ and $\tau^{[3]}_{\mu\nu}$. These are 
\begin{align}
    \tau^{[2]}_{\mu \nu} 
    =& \partial_\alpha \left(h^{\alpha \beta}_{[1]} - \frac{1}{2} \eta^{\alpha \beta} h^\gamma_{{[1]} \gamma} \right)\left( 2 \partial_{(\mu} h_{\nu) \beta}^{[1]} - \partial_\beta h_{\mu \nu}^{[1]} \right) - \frac{1}{2}\partial_\mu h^{\alpha \beta}_{[1]} \partial_\nu h_{\alpha \beta}^{[1]}  \nonumber
    \\
    & -  \partial^\beta h^{\alpha}_{[1] \mu} \partial_\beta h_{\alpha \nu}^{[1]} +  \partial^\beta h^{\alpha}_{[1] \mu} \partial_\alpha h_{\beta \nu}^{[1]} + h^{\alpha \beta}_{[1]} \left( 2 \partial_{\beta (\mu}h_{\nu) \alpha}^{[1]} - \partial_{\mu \nu}h_{ \alpha \beta}^{[1]}- \partial_{\alpha \beta} h_{ \mu \nu}^{[1]} \right) \,, \label{eq:tau^2}
\end{align}
\begin{align}
    \tau^{[3]}_{\mu \nu}
    = &-\tfrac{1}{2} h^{\alpha \beta }_{[1]} \partial_{\alpha }h_{\mu \nu }^{[1]} \partial_{\beta }h^{\gamma }_{[1]\gamma }+  h^{\alpha \beta }_{[1]} \partial_{\alpha }h_{\mu }^{[1]\gamma } \partial_{\beta }h_{\nu \gamma }^{[1]} + h^{\alpha \beta }_{[1]} \partial_{\alpha }h_{\mu \nu }^{[1]} \partial_{\gamma }h_{\beta }^{{[1]}\gamma } - h^{\alpha \beta }_{[1]} \partial_{\alpha }h_{\mu }^{{[1]}\gamma } \partial_{\gamma }h_{\nu \beta }^{[1]} \nonumber
    \\
    & + h_{\alpha }^{{[1]}\gamma } h^{\alpha \beta }_{[1]} \partial_{\gamma }\partial_{\beta }h_{\mu \nu }^{[1]}- h^{\alpha \beta }_{[1]} \partial_{\beta }h_{\nu \gamma }^{[1]} \partial^{\gamma }h_{\mu \alpha }^{[1]} + h^{\alpha \beta }_{[1]} \partial_{\gamma }h_{\nu \beta }^{[1]} \partial^{\gamma }h_{\mu \alpha }^{[1]}  + h^{\alpha \beta }_{[1]} \partial_{\beta }h_{\alpha \gamma }^{[1]} \partial^{\gamma }h_{\mu \nu }^{[1]} \nonumber
    \\
    & - \tfrac{1}{2} h^{\alpha \beta }_{[1]} \partial_{\gamma }h_{\alpha \beta }^{[1]} \partial^{\gamma }h_{\mu \nu }^{[1]} +  \tfrac{1}{2} h^{\alpha \beta }_{[1]} \partial_{\beta }h^{\gamma }_{{[1]}\gamma } \partial_{\mu }h_{\nu \alpha }^{[1]} - h^{\alpha \beta }_{[1]} \partial_{\gamma }h_{\beta }^{{[1]}\gamma } \partial_{\mu }h_{\nu \alpha }^{[1]} - h^{\alpha \beta }_{[1]} \partial_{\beta }h_{\alpha \gamma }^{[1]} \partial_{\mu }h_{\nu }^{{[1]} \gamma }\nonumber
    \\
    &  +  \tfrac{1}{2} h^{\alpha \beta }_{[1]} \partial_{\gamma }h_{\alpha \beta }^{[1]} \partial_{\mu }h_{\nu }^{[1]\gamma } - h_{\alpha }^{[1]\gamma } h^{\alpha \beta }_{[1]} \partial_{\mu }\partial_{\gamma }h_{\nu \beta }^{[1]} + h^{\alpha \beta }_{[1]} \partial_{\mu }h_{\alpha }^{{[1]} \gamma } \partial_{\nu }h_{\beta \gamma }^{[1]} \nonumber 
    \\
    &+  \tfrac{1}{2} h^{\alpha \beta }_{[1]} \partial_{\beta }h^{\gamma }_{[1] \gamma } \partial_{\nu }h_{\mu \alpha }^{[1]}  - h^{\alpha \beta }_{[1]} \partial_{\gamma }h_{\beta }^{[1] \gamma } \partial_{\nu }h_{\mu \alpha }^{[1]} - h^{\alpha \beta }_{[1]} \partial_{\beta }h_{\alpha \gamma }^{[1]} \partial_{\nu }h_{\mu }^{[1] \gamma }\nonumber
    \\
    & +  \tfrac{1}{2} h^{\alpha \beta }_{[1]} \partial_{\gamma }h_{\alpha \beta }^{[1]} \partial_{\nu }h_{\mu }^{[1] \gamma } - h_{\alpha }^{{[1]}\gamma } h^{\alpha \beta }_{[1]} \partial_{\nu }\partial_{\gamma }h_{\mu \beta }^{[1]} + h_{\alpha }^{[1]\gamma } h^{\alpha \beta }_{[1]} \partial_{\nu }\partial_{\mu }h_{\beta \gamma }^{[1]} \nonumber
    \\
    &+\partial_\alpha \left(h^{\alpha \beta}_{[2]} - \frac{1}{2} \eta^{\alpha \beta} h^\gamma_{{[2]} \gamma} \right)\left( 2 \partial_{(\mu} h_{\nu) \beta}^{[1]} - \partial_\beta h_{\mu \nu}^{[1]} \right) - \frac{1}{2}\partial_\mu h^{\alpha \beta}_{[2]} \partial_\nu h_{\alpha \beta}^{[1]}  \nonumber
    \\
    & -  \partial^\beta h^{\alpha}_{[2] \mu} \partial_\beta h_{\alpha \nu}^{[1]} +  \partial^\beta h^{\alpha}_{[2] \mu} \partial_\alpha h_{\beta \nu}^{[1]} + h^{\alpha \beta}_{[2]} \left( 2 \partial_{\beta (\mu}h_{\nu) \alpha}^{[1]} - \partial_{\mu \nu}h_{ \alpha \beta}^{[1]}- \partial_{\alpha \beta} h_{ \mu \nu}^{[1]} \right) \nonumber
    \\
    &+\partial_\alpha \left(h^{\alpha \beta}_{[1]} - \frac{1}{2} \eta^{\alpha \beta} h^\gamma_{{[1]} \gamma} \right)\left( 2 \partial_{(\mu} h_{\nu) \beta}^{[2]} - \partial_\beta h_{\mu \nu}^{[2]} \right) - \frac{1}{2}\partial_\mu h^{\alpha \beta}_{[1]} \partial_\nu h_{\alpha \beta}^{[2]}  \nonumber
    \\
    & -  \partial^\beta h^{\alpha}_{[1] \mu} \partial_\beta h_{\alpha \nu}^{[2]} +  \partial^\beta h^{\alpha}_{[1] \mu} \partial_\alpha h_{\beta \nu}^{[2]} + h^{\alpha \beta}_{[1]} \left( 2 \partial_{\beta (\mu}h_{\nu) \alpha}^{[2]} - \partial_{\mu \nu}h_{ \alpha \beta}^{[2]}- \partial_{\alpha \beta} h_{ \mu \nu}^{[2]} \right) \,.
\end{align}
In harmonic gauge we also need 
\begin{align}
    K_\mu^{[2]} =& h^{\alpha \beta}_{[1]} \partial_{\alpha} h_{\beta \mu}^{[1]} - \frac{1}{2} h^{\alpha \beta}_{[1]} \partial_\mu h_{\alpha \beta}^{[1]}\,,
    \\
    K_\mu^{[3]} =&  h_{[1]}^{\sigma \rho}\left( \partial_\sigma h^{[2]}_{\rho\mu}-\frac{1}{2}\partial_\mu h^{[2]}_{\sigma \rho} \right) +\left(h_{[2]}^{\sigma \rho} -\eta_{\alpha \beta} h_{[1]}^{\sigma \alpha} h_{[1]}^{\beta \rho}\right)\left( \partial_\sigma h^{[1]}_{\rho \mu}-\frac{1}{2}\partial_\mu h^{[1]}_{\sigma\rho} \right)\,,
\end{align}
as these feature in the source in \eqref{eq:HGGexp}.

For the case of the transverse gauge we will give the sources to order $G^2$. In this case the equations are given in the summary section \ref{subsec:summary}. Using 
equation \eqref{eq:tau2} together with the transverse gauge condition and the order $G$ equations of motion, we find 
\begin{eqnarray}
 \hspace{-1cm}   \tau_{00}^{[2]} & = & -\frac{1}{2}\overset{[1]}{C}_{ll0}\overset{[1]}{C}_{000}+\frac{1}{2}\overset{[1]}{C}_{llk}\overset{[1]}{C}_{00k}+\frac{1}{2}\overset{[1]}{C}_{k00}\overset{[1]}{C}_{k00}-\frac{1}{2}\overset{[1]}{C}_{0kl}\overset{[1]}{C}_{0kl}\nonumber\\
 \hspace{-1cm}   &&+h^{[1]}_{kl}(\text{TT})\left(\partial_l \overset{[1]}{C}_{00k}-\partial_0 \overset{[1]}{C}_{0kl}\right)\,,\\
 \hspace{-1cm}   \tau_{00}^{[2]}+\tau_{ii}^{[2]} & = & -\frac{1}{2}\overset{[1]}{C}_{kk0}\overset{[1]}{C}_{ll0}+\frac{1}{2}\overset{[1]}{C}_{kki}\overset{[1]}{C}_{lli}+\frac{1}{2}\overset{[1]}{C}_{kl0}\overset{[1]}{C}_{kl0}-\frac{1}{2}\overset{[1]}{C}_{ijk}\overset{[1]}{C}_{ijk}\nonumber\\
\hspace{-1cm}    &&+h^{[1]}_{kl}(\text{TT})\left(\partial_l \overset{[1]}{C}_{00k}-\partial_0 \overset{[1]}{C}_{l0k}\right)+h^{[1]}_{kl}(\text{TT})\left(\partial_l \overset{[1]}{C}_{iik}-\partial_i \overset{[1]}{C}_{lik}\right)\,,\\
 \hspace{-1cm}   \tau_{0i}^{[2]} & = & -\frac{1}{2}\overset{[1]}{C}_{ll0}\overset{[1]}{C}_{i00}+\frac{1}{2}\overset{[1]}{C}_{llk}\overset{[1]}{C}_{0ik}+\frac{1}{2}\overset{[1]}{C}_{k00}\overset{[1]}{C}_{ik0}-\frac{1}{2}\overset{[1]}{C}_{0kl}\overset{[1]}{C}_{ikl}\nonumber\\
 \hspace{-1cm}   &&+M^{[1]}_k(\text{T})\left(\partial_0\overset{[1]}{C}_{0ik}-\partial_i\overset{[1]}{C}_{00k}\right)+h^{[1]}_{kl}(\text{TT})\left(\partial_l\overset{[1]}{C}_{0ik}-\partial_i\overset{[1]}{C}_{0lk}\right)\,,\\
  \hspace{-1cm}   \tau^{[2]}_{ij} & = &  
    \frac{1}{2}\left(\overset{[1]}{C}_{000}-\overset{[1]}{C}_{kk0}\right)\overset{[1]}{C}_{ij0}-\frac{1}{2}\left(\overset{[1]}{C}_{00k} -\overset{[1]}{C}_{llk}\right) \overset{[1]}{C}_{ijk}\nonumber\\
\hspace{-1cm}    &&-\frac{1}{2}\overset{[1]}{C}_{i00}\overset{[1]}{C}_{j00}+\frac{1}{2}\overset{[1]}{C}_{0ik}\overset{[1]}{C}_{0jk}+\frac{1}{2}\overset{[1]}{C}_{ik0}\overset{[1]}{C}_{jk0}-\frac{1}{2}\overset{[1]}{C}_{ikl}\overset{[1]}{C}_{jkl}\nonumber\\
 \hspace{-1cm}   &&+\left(2M^{[1]}_0-\frac{1}{3}H^{[1]}\right)\left(\partial^2 h^{[1]}_{ij}(\text{TT})+\frac{1}{3}\partial_i\partial_j H^{[1]}\right)\nonumber\\
\hspace{-1cm}    &&+h^{[1]}_{kl}(\text{TT})\left(\partial_l \overset{[1]}{C}_{ijk}-\partial_j \overset{[1]}{C}_{ilk}\right)\nonumber\\
\hspace{-1cm}    &&+M^{[1]}_k(\text{T})\left[\partial_k \overset{[1]}{C}_{ij0}+\partial_0 \overset{[1]}{C}_{ijk}+2\partial_i\partial_j M_k(\text{T})\right]\,,
\end{eqnarray}
where in transverse gauge we have
\begin{eqnarray}
    \overset{[1]}{C}_{000} & = & -2\partial_0 M_0^{[1]}\,,\\
    \overset{[1]}{C}_{00k} & = & 2\partial_k M_0^{[1]}-2\partial_0 M_k^{[1]}(\text{T})\,,\\
    \overset{[1]}{C}_{k00} & = & -2\partial_k M_0^{[1]}\,,\\
    \overset{[1]}{C}_{ij0} & = & -\partial_0 h^{[1]}_{ij}(\text{TT})-\partial_i M^{[1]}_j(\text{T})-\partial_j M^{[1]}_i(\text{T})-\frac{1}{3}\delta_{ij}\partial_0 H^{[1]}\,,\\
    \overset{[1]}{C}_{0ij} & = & -\overset{[1]}{C}_{ij0}-2\partial_i M_j^{[1]}(\text{T})\,,\\
    \overset{[1]}{C}_{ijk} & = & \partial_i h_{jk}^{[1]}(\text{TT})+\partial_j h_{ik}^{[1]}(\text{TT})-\partial_k h_{ij}^{[1]}(\text{TT})\nonumber\\
    &&+\frac{1}{3}\left(\delta_{jk}\partial_i H^{[1]}+\delta_{ik}\partial_j H^{[1]}-\delta_{ij}\partial_k H^{[1]}\right)\,.
\end{eqnarray}

When solving the inhomogeneous PDEs at order $G^2$ and higher we will have to use Green's functions to write down the particular solution and these will involve integration over the exterior zone which has a boundary (or a lower cut-off). Therefore, just like we encountered in section \ref{subsec:intEOM} when we discussed the integration of the near zone PDEs, we will have to worry about dependence of the particular solution on said boundary. We will come back to this in the next section.

\section{Near zone metric to 1.5PN}\label{sec:NZto1.5}

The purpose of this section is to determine the near zone metric to 1.5PN order by solving the $1/c$ expanded Einstein equations. The latter are of the form $\partial^2(\text{field})=\text{(source)}$ and so the most general solution will involve near zone regular harmonic functions. In order to determine these harmonic functions we will use the matching with the exterior zone metric. We will from now on exclusively work in the harmonic gauge. For a similar analysis in the transverse gauge we refer the reader to \cite{companionpaper}. For the homogeneous part of the harmonic gauge metric in the exterior zone we will use the parametrisation \eqref{eq:HGparam1}--\eqref{eq:HGparam3}. The purpose of this section and the next is to show that our methods work. The results that will be derived are well-known (see \cite{Blanchet2014} and references therein). Nevertheless, seeing them emerge in this way will help when using a very different gauge.

Before we can start the matching process we first need some general results about expanding the exterior zone metric in $1/c$ which is valid only in the part of the spacetime where the exterior zone overlaps with the near zone.

\subsection{$1/c$-expansion of the exterior zone metric}\label{subsec:expextzone}

Here we will collect some general results about $1/c$-expansions of the solutions $W^{[n]}$, $W^{[n]}_i$, etc. Since the free indices will play no role in this section we will suppress them. We will also suppress the superscript $[n]$. We refer to appendix \ref{app:PDEstuff} for some standard results about multipole expansions of solutions to the free wave equation using inertial coordinates.

Using equation \eqref{eq:gensol_S} we know that if $W$ is a solution to the free wave equation (obeying Sommerfeld) it can be expanded as
\begin{equation}\label{eq:freewave}
    W=\frac{U(u)}{r}+\partial_i\left(\frac{U_i(u)}{r}\right)+\frac{1}{2}\partial_i\partial_j\left(\frac{U_{ij}(u)}{r}\right)+\ldots\,,
\end{equation}
where the $U_{ij}$ are symmetric tracefree (STF) and the dots denote higher multipole moments. If we Taylor expand this around $u=t$ we obtain
\begin{eqnarray}
    W & = & \sum_{n=0}^\infty\frac{1}{n!}\left(\frac{-1}{c}\right)^n\left[r^{n-1}\partial^n_tU(t)+\partial_i r^{n-1}\partial^n_tU_i(t)+\frac{1}{2}\partial_i\partial_j r^{n-1}\partial^n_tU_{ij}(t)+\ldots\right]\nonumber\\
    & = & \sum_{l=0}^\infty\sum_{n=0}^\infty\frac{1}{n!}\left(\frac{-1}{c}\right)^n\frac{1}{l!}\partial_{L} r^{n-1}\partial^n_tU_{L}(t)\nonumber\\
    & = & \sum_{l=0}^\infty\sum_{n=0}^\infty\frac{1}{n!}\left(\frac{-1}{c}\right)^n\frac{1}{l!}(n-1)(n-3)\cdots(n-2l+1)x^{L} r^{n-2l-1}\partial^n_tU_{L}(t)\nonumber\\
    & = & \sum_{l=0}^\infty\frac{1}{l!}\partial_{L} r^{-1}U_{L}(t)-\frac{1}{c}\partial_t U(t)+\frac{1}{2c^2}\sum_{l=0}^\infty\frac{1}{l!}\partial_{L} r\partial_t^2 U_{L}(t)\nonumber\\
    &&-\frac{1}{6c^3}\left(r^2\partial_t^3 U(t)+2x^i\partial_t^3 U_i(t)\right)+\mathcal{O}(c^{-4})\,,\label{eq:expW}
\end{eqnarray}
where we use the multi-index notation $L = i_1 \cdots i_l$. We see that the even powers of $1/c$ lead to all order multipole expansions whereas the odd powers lead to truncated expansions with only a finite number of multipole moments contributing. We also see from this that we have a harmonic function that is regular in the interior
whenever $n=2l+1$. For example for $n=1$ and $l=0$ we have the term $-\frac{1}{c}\partial_t U(t)$. For $n=3$ and $l=1$ the harmonic function is $-\frac{x^i}{3c^3}\partial_t^3 U_i(t)$, and for $n=5$ and $l=2$ we get $-\frac{x^ix^j}{30c^5}\partial_t^5 U_{ij}(t)$.
The harmonic part (regular at $r=0$) of $W$ in the overlap region is given by
\begin{equation}\label{eq:harmonicpartW}
    -\sum_{l=0}^\infty\frac{2^l}{(2l+1)!}c^{-2l-1}x^{L}\partial_t^{2l+1}U_{L}(t)\,,
\end{equation}
where we used that $U_{L}$ is STF.

If we $1/c$ expand the multipole coefficients $U_{i_1\cdots i_l}$, which we will assume is an expansion in even powers (which is related to the even power expansion of the fluid variables discussed in section \ref{subsec:PFvariables}), as
\begin{equation}
    U_{L}=U^{(0)}_{L}+\frac{1}{c^2}U^{(2)}_{L}+\mathcal{O}(c^{-4})\,,
\end{equation}
then we get for $W$ the expansion\footnote{We let the expansion of $W$ start at order $c^0$ in order to recover the Newtonian limit from the $1/c$-expansion of the exterior solution.}
\begin{eqnarray}
    W & = & \sum_{l=0}^\infty\frac{1}{l!}\partial_{L} r^{-1}U^{(0)}_{L}(t)-\frac{1}{c}\partial_t U^{(0)}(t)+\frac{1}{2c^2}\sum_{l=0}^\infty\frac{1}{l!}\partial_{L} r\partial_t^2 U^{(0)}_{L}(t)\nonumber\\
    &&+\frac{1}{c^2}\sum_{l=0}^\infty\frac{1}{l!}\partial_{L} r^{-1}U^{(2)}_{L}(t)-\frac{1}{6c^3}\left(r^2\partial_t^3 U^{(0)}(t)+2x^i\partial_t^3 U^{(0)}_i(t)\right)\nonumber\\
    &&-\frac{1}{c^3}\partial_t U^{(2)}+\mathcal{O}(c^{-4})\,.\label{eq:1/cexpext}
\end{eqnarray}

We can write similar expressions for $W_i$ and $W_{ij}+G_{ij}$. This leads to multipole coefficients of the form\footnote{We are suppressing the $[n]$ index. In general we will have multipole coefficients of the form $V^{[n](m)}_{i, i_1\cdots i_l}(t)$ at order $G^n c^{-m}$.} $V^{(0)}_{i, i_1\cdots i_l}(t)$ and $Z^{(0)}_{ij, i_1\cdots i_l}(t)$, etc, where the comma between the $i$ index and the remaining indices indicates that there is no symmetry assumed between interchanging $i$ with any of the other indices. The indices after the comma are assumed to be STF\footnote{We remind the reader that our conventions regarding the manipulation of indices can be found in appendix \ref{app:conventions}.}. Objects such as $V_{i,i_1\cdots i_l}$ can be decomposed into irreducible representations of $SO(3)$, but we will refrain from implementing this decomposition until we are forced to do so (by the matching process) as this will lead to a proliferation of terms.

Using the above results together with the parametrisation of the harmonic gauge metric given in \eqref{eq:HGparam1}--\eqref{eq:HGparam3}, which we repeat here for convenience
\begin{eqnarray}
    g_{tt} & = & -c^2+2G\partial_t W^{[1]}+\mathcal{O}(G^2)\,,\\
    g_{ti} & = & G\partial_t W^{[1]}_i+G\partial_i W^{[1]}+\mathcal{O}(G^2)\,,\\
    g_{ij} & = & \delta_{ij}+G\left(W^{[1]}_{ij}(\text{TT})+G^{[1]}_{ij}\right)+G\left(\partial_i W^{[1]}_j+\partial_j W^{[1]}_i\right)+\mathcal{O}(G^2)\,,
\end{eqnarray}
we can match the exterior and near zone metrics to first order in $G$. The above results only concern the homogeneous solution in the exterior zone so if we want to match at order $G^2$ or higher we need to include a discussion of the particular solution to the inhomogeneous PDE in the exterior zone.

Let us write the metric in the exterior region as we did at the start of the previous section,
\begin{equation}
    g_{\mu\nu}=\eta_{\mu\nu}+G h^{[1]}_{\mu\nu}+G^2 h^{[2]}_{\mu\nu}+\cdots\,.
\end{equation}
Then at order $G^n$ the object $h^{[n]}_{\mu\nu}$ solves the following PDE
\begin{equation}\label{eq:extPDE}
    \square h^{[n]}_{\mu\nu} = S^{[n]}_{\mu\nu}=-\tau^{[n]}_{\mu\nu}+\partial_\mu K^{[n]}_\nu+\partial_\nu K^{[n]}_\mu\,,
\end{equation}
where we used harmonic gauge and notation introduced in the previous section. The full solution that obeys Sommerfeld's no-incoming radiation boundary condition at $\mathcal{I}^-$ is 
\begin{equation}\label{eq:solhn}
    h^{[n]}_{\mu\nu}=W^{[n]}_{\mu\nu}-\frac{1}{4\pi}\int_{\mathcal{E}} d^3 x'\frac{S_{\mu\nu}^{[n]}(t-\vert x-x'\vert/c,x')}{\vert x-x'\vert}+B_{\mu\nu}^{[n]}\,,
\end{equation}
where $W^{[n]}_{\mu\nu}$ obeys the free wave equation with Sommerfeld boundary conditions. The last two terms represent the retarded Green's function on the exterior zone $\mathcal{E}$, constituting the particular solution to \eqref{eq:extPDE}. Here $x$ is a point in the exterior zone and so does not lie on its boundary. The term $B_{\mu\nu}^{[n]}$ also obeys the free wave equation but it only has support on the inner boundary of $\mathcal{E}$ and depends on the source $S_{\mu\nu}^{[n]}$ in a specific way. It is in general of the following form
\begin{equation}\label{eq:bdryterm}
    B_{\mu\nu}^{[n]}=\frac{1}{4\pi}\int_{\mathcal{E}} d^3x'\partial'_i\left(\frac{J_{\mu\nu}^{[n]i}(t-|x-x'|/c,x')}{|x-x'|}\right)\,,
\end{equation}
where $J^{[n]i}_{\mu\nu}$ depends on the source. 

The reason that we need this term can be understood as follows. We want that the particular solution obeys the harmonic gauge condition which can be written as (see equation \eqref{eq:HGchoiceordern})
\begin{equation}\label{eq:HGchoice_n}
    H^{\rho\sigma}{}_\nu h^{[n]}_{\rho\sigma}=K^{[n]}_\nu\,,
\end{equation}
where we defined
\begin{equation}
    H^{\rho\sigma}{}_\nu=\left(\eta^{\lambda\rho}\delta^\sigma_\nu-\frac{1}{2}\eta^{\rho\sigma}\delta^\lambda_\nu\right)\partial_\lambda\,.
\end{equation}
Let us formally denote the particular solution to $\square h_{\mu\nu}=S_{\mu\nu}$ (satisfying Sommerfeld) by 
\begin{equation}\label{eq:partsol_n}
    h^{[n]}_{\mu\nu}=\square^{-1}_{\text{ret}} S^{[n]}_{\mu\nu}\,.
\end{equation}
By taking the d'Alembertian of \eqref{eq:HGchoice_n} we see that the harmonic gauge operator $H^{\rho\sigma}{}_\nu$ acting on $S_{\mu\nu}$ gives
\begin{equation}
    H^{\rho\sigma}{}_\nu S^{[n]}_{\rho\sigma}=\square K^{[n]}_\nu\,.
\end{equation}
In order that the particular solution \eqref{eq:partsol_n} obeys the harmonic gauge condition we need the following set of formal manipulations to be valid
\begin{equation}\label{eq:formal}
    H^{\rho\sigma}{}_\nu h^{[n]}_{\rho\sigma}=H^{\rho\sigma}{}_\nu \square^{-1}_{\text{ret}} S^{[n]}_{\rho\sigma}= \square^{-1}_{\text{ret}} H^{\rho\sigma}{}_\nu S^{[n]}_{\rho\sigma}=\square^{-1}_{\text{ret}}\square K^{[n]}_\nu=K^{[n]}_\nu\,.
\end{equation}
The non-trivial steps are the second and fourth equalities. For example, if we take for 
$\square^{-1}_{\text{ret}} S^{[n]}_{\mu\nu}$ just the middle term in \eqref{eq:solhn} without the $B_{\mu\nu}$ term, then the second and fourth equalities in \eqref{eq:formal} would only be true up to boundary terms of the form \eqref{eq:bdryterm}. This is the rationale for adding $B_{\mu\nu}^{[n]}$ to the particular solution. Rather than explicitly constructing $B_{\mu\nu}^{[n]}$ we will simply drop boundary terms in the exterior metric that can be absorbed into $B_{\mu\nu}^{[n]}$. With this in mind we will not explicitly write this term.

Let us introduce the following notation. Let $R[S_{\mu\nu}^{[n]}]$ and $A[S_{\mu\nu}^{[n]}]$ denote the retarded and advanced Green's functions given by
\begin{eqnarray}
    R[S_{\mu\nu}^{[n]}] & = & \frac{1}{4\pi}\int_{\mathcal{E}} d^3 x'\frac{S_{\mu\nu}^{[n]}(t-\vert x-x'\vert/c,x')}{\vert x-x'\vert}\,,\label{eq:defretGreen}\\
    A[S_{\mu\nu}^{[n]}] & = & \frac{1}{4\pi}\int_{\mathcal{E}} d^3 x'\frac{S_{\mu\nu}^{[n]}(t+\vert x-x'\vert/c,x')}{\vert x-x'\vert}\,,
\end{eqnarray}
where the integrations are over the exterior zone $\mathcal{E}$ and $x$ is a point in the exterior zone (not on its boundary). Using the retarded and advanced Green's functions we can write the solution \eqref{eq:solhn} as
\begin{equation}\label{eq:fullextsol}
    h^{[n]}_{\mu\nu}=W^{[n]}_{\mu\nu}-\frac{1}{2}\left(R[S_{\mu\nu}^{[n]}]+A[S_{\mu\nu}^{[n]}]\right)-\frac{1}{2}\left(R[S_{\mu\nu}^{[n]}]-A[S_{\mu\nu}^{[n]}]\right) + B_{\mu \nu}^{[n]}\,.
\end{equation}
The sum of the retarded and advanced Green's functions is even in $1/c$ and is a particular  solution to \eqref{eq:extPDE}. The difference of the retarded and advanced Green's functions is odd in $1/c$ and is a homogeneous solution to \eqref{eq:extPDE}. By $1/c$ expanding the particular solution we obtain
\begin{eqnarray}
    -\frac{1}{2}\left(R[S_{\mu\nu}^{[n]}]+A[S_{\mu\nu}^{[n]}]\right) & = & -\frac{1}{4\pi}\int_{\mathcal{E}} d^3x'\frac{S_{\mu\nu}^{[n]}(t,x')}{\vert x-x'\vert}\nonumber\\
    &&-\frac{1}{8\pi}\frac{1}{c^2}\int_{\mathcal{E}} d^3x'\partial_t^2S_{\mu\nu}^{[n]}(t,x')\vert x-x'\vert+\mathcal{O}(c^{-4})\,.
\end{eqnarray}
When we discussed the homogeneous solution we concluded that the harmonic part only appears at odd powers of $1/c$ (see equation \eqref{eq:harmonicpartW}. Here we see that also for the particular solution the even powers of $1/c$ will never give rise to harmonic functions that are regular at the origin. We thus arrive at the important conclusion that the near zone harmonic functions obtained in solving the $1/c$ expanded Einstein equations at even powers of $1/c$ must be set to zero\footnote{We assume here that the near zone integrals have already been made well-defined (which sometimes requires the use of a specific harmonic function as discussed in section \ref{subsec:intEOM}) so that the particular solution does not depend on any cut-off. Furthermore, we assume that the integrals in the exterior zone are also well-defined and (lower) cut-off independent (by a judicious choice of $B_{\mu\nu}^{[n]}$).}. Furthermore, we learn that the odd powers of $1/c$ in the exterior region obey the free wave equation.

All of the above is based on the assumption that the dependence on $1/c$ is real analytic so that we can perform a Taylor series in $1/c$. As soon as this assumption breaks down these comments need to be revisited. It is known that the breakdown of the Taylor expansion in $1/c$ is associated with the presence of tail terms \cite{Blanchet:1985sp,PhysRevD.25.2038}. To the order we are working such terms do not arise in the near zone. For more details we refer the reader to the review paper \cite{Blanchet2014}.

\subsection{Fixing the inertial coordinates}

Before we start the matching process it will be useful to fix our choice of inertial coordinates by choosing an appropriate origin. So far we have been using inertial coordinates that describe our vacuum Minkowski spacetime, but we have not chosen any particular origin yet. At this stage we are still free to perform Poincar\'e transformations on our inertial coordinates (if we are using the $G$-expanded Einstein equations) or the $1/c$ expanded Poincar\'e transformations (see e.g. \cite{HHO} for the construction of the $1/c$ expanded Poincar\'e algebra) if we are using the $1/c$ expanded Einstein equations. 

We will choose inertial coordinates such that the origin is at the centre of mass of the matter distribution. To define this we need to use the fluid conservation equations which as discussed in appendix \ref{app:conservation} can be written as 
\begin{equation}
    \partial_t\mathcal{T}^{t\nu}+\partial_i\mathcal{T}^{i\nu}=0\,,
\end{equation}
where $\mathcal{T}^{\mu\nu}$ is defined with the help of the Landau--Lifshitz energy-momentum pseudo tensor (see equation \eqref{eq:EMTplusLL}). The ADM charges 
\begin{equation}
    \int_{t=\text{cst}}d^3x\mathcal{T}^{t\nu}\,,
\end{equation}
form a Lorentz vector with respect to the Lorentz symmetries of the vacuum. We can always perform a Lorentz boost to set the total momentum equal to zero, i.e.
\begin{equation}
    \int d^3x \mathcal{T}^{ti}=0\,. 
\end{equation}
Having made this choice we can show that the dipole moment $\int d^3x x^i\mathcal{T}^{tt}$ is constant. We can thus perform a translation to set this to zero, i.e.
\begin{equation}
    \int d^3x x^i\mathcal{T}^{tt}=0\,.
\end{equation}

If we expand the latter two equations in $1/c$ then at leading order we get 
\begin{eqnarray}
    \int d^3x E_{(\sm2)}v^i & = & 0\,,\label{eq:restframe}\\
    \int d^3x E_{(\sm2)}x^i & = & 0\,,
\end{eqnarray}
which simply state that the centre of mass momentum is zero and that the origin of our coordinate system coincides with the centre of mass and so the dipole moment of the mass distribution is zero. We can always use Galilei boosts and translations to achieve this. At higher orders in $1/c$ we get subleading boosts and translations (as unfixed diffeomorphisms) that can be used to set 
\begin{eqnarray}
    \int d^3x \mathcal{T}^{0i}_{(n)} & = & 0\,,\\
    \int d^3x x^i\mathcal{T}^{tt}_{(n)} & = & 0\,,
\end{eqnarray}
where $\mathcal{T}^{tt}_{(n)}$ is the coefficient of $c^{-n}$ in the $1/c$-expansion of $\mathcal{T}^{\mu\nu}$.

\subsection{Matching to 0.5PN}

We start with the Newtonian order. From the near zone metric we know that we have for the $tt$ component,
\begin{equation}
    g_{tt}=-c^2+2U+\mathcal{O}(c^{-2})\,,\qquad U=G\int d^3x' \frac{E_{(\sm2)}(t,x')}{|x-x'|}\,.
\end{equation}
From the $1/c$-expansion of the exterior metric at order $G$ we know that 
\begin{equation}
    g_{tt}=-c^2+2G \sum_{l=0}^\infty\frac{1}{l!}\partial_{L} r^{-1}\partial_t U^{[1](0)}_{L}(t)-\frac{2G}{c}\partial^2_t U^{(0)}(t)+\mathcal{O}(c^{-2})\,,
\end{equation}
where we used \eqref{eq:1/cexpext}. Comparing the two results leads to 
\begin{equation}\label{eq:NewtonianMatching}
    \sum_{l=0}^\infty\frac{1}{l!}\partial_{L} r^{-1}\partial_t U^{[1](0)}_{L}(t)=\int d^3x' \frac{E_{(\sm2)}(t,x')}{|x-x'|}\,.
\end{equation}
For the integral on the right-hand side, the point $x$ is in the overlap region and the point $x'$ is inside the matter distribution. We can thus expand 
\begin{equation}
    \frac{1}{\vert x-x'\vert}=\frac{1}{r}-x'^i\partial_i\frac{1}{r}+\frac{1}{2}x'^i x'^j\partial_i\partial_j\frac{1}{r}+\cdots=\sum_{l=0}^\infty\frac{(-1)^l}{l!}\partial_{L}\frac{1}{r}\,.
\end{equation}
Hence, equation \eqref{eq:NewtonianMatching} tells us that
\begin{equation}\label{eq:0PNmatch_tt}
    \partial_t U^{[1](0)}_{L}=(-1)^l\int d^3x' x'^{\langle L\rangle}E_{(\sm2)}(t,x')\,,
\end{equation}
where the $\langle \rangle$ denotes the symmetric tracefree combination of the indices inside.

We thus see that the $\partial_t U^{[1](0)}_{L}(t)$ are related to the multipole moments of the mass distribution. Furthermore, since the PN near zone metric has no term at order $c^{-1}$ in the expansion of $g_{tt}$ we conclude that $\partial^2_t U^{[1](0)}=0$, which means that the total mass as measured by $\int d^3 x' E_{(\sm2)}(t,x')$ is constant. This also follows from the leading order fluid conservation equation given by the 4-divergence of \eqref{eq:fluidconLO_tt} and \eqref{eq:fluidconLO_ti}. We see that this has the effect of removing the term proportional to $r^2$ at order $c^{-3}$ making the entire $c^{-3}$ term in \eqref{eq:1/cexpext} harmonic. 

Further below we will often denote the constant total mass by $M$, so we have
\begin{equation}\label{eq:totalmass}
    M=\partial_t U^{[1](0)}=\int d^3x' E_{(\sm2)}(t,x')\,.
\end{equation}
Furthermore, as we discussed in the previous subsection, we will choose inertial coordinates for which the mass dipole moment vanishes, so that
\begin{equation}\label{eq:nomassdipole}
    \partial_t U^{[1](0)}_i=-\int d^3x' x'^i E_{(\sm2)}(t,x')=0\,.
\end{equation}

We next consider the $ti$ component of the metric. From the exterior solution we know that this is 
given by $G\left(\partial_t W^{[1]}_i+\partial_i W^{[1]}\right)$ at order $G$. We know from the $1/c$-expansion that at order $c^0$ the metric $g_{ti}$ is simply zero. However, from the matching of the $tt$ component we know that $W^{[1]}$ starts at order $c^0$. This means that $W^{[1]}_i$ must also start at order $c^0$ in order that we can have a cancellation at order $c^0$ between the $\partial_t W^{[1]}_i$ and $\partial_i W^{[1]}$ terms. The $1/c$-expansion of $W^{[1]}_i$ follows from \eqref{eq:1/cexpext} and we will denote the multipole moments by $V^{[1]}_{i,L}$. Explicitly, we have
\begin{eqnarray}
    W^{[1]}_i & = & \sum_{l=0}^\infty\frac{1}{l!}\partial_{L} r^{-1}V^{[1](0)}_{i,L}(t)-\frac{1}{c}\partial_t V^{[1](0)}_i(t)+\frac{1}{2c^2}\sum_{l=0}^\infty\frac{1}{l!}\partial_{L} r\partial_t^2 V^{[1](0)}_{i,L}(t)\nonumber\\
    &&+\frac{1}{c^2}\sum_{l=0}^\infty\frac{1}{l!}\partial_{L} r^{-1}V^{[1](2)}_{i,L}(t)-\frac{1}{6c^3}\left(r^2\partial_t^3 V^{[1](0)}_i(t)+2x^j\partial_t^3 V^{[1](0)}_{i,j}(t)\right)\nonumber\\
    &&-\frac{1}{c^3}\partial_t V^{[1](2)}_i(t)+\mathcal{O}(c^{-4})\,,\label{eq:1/cexpWi}
\end{eqnarray}
where we also $1/c$ expanded $V^{[1]}_{i,L}$ in even powers of $1/c$.

In order that the $c^0$ contribution from $\partial_t W_i^{[1]}$ cancels the one from $\partial_i W^{[1]}$ we need that
\begin{equation}
    \sum_{l=0}^\infty\frac{1}{l!}\partial_{L} r^{-1}\partial_t V^{[1](0)}_{i,L}(t)=-\partial_i\sum_{l=0}^\infty\frac{1}{l!}\partial_{L} r^{-1}U^{[1](0)}_{L}(t)\,.
\end{equation}
At low multipole moments this implies that we have
\begin{eqnarray}
    \partial_t V_i^{[1](0)} & = & 0\,,\label{eq:dtVi0}\\
    \partial_t V_{i,j}^{[1](0)} & = & -U^{[1](0)}\delta_{ij}\,,\label{eq:dtVij0}\\
    \partial_t V_{i,jk}^{[1](0)} & = & -\left(\delta_{ij}U_k^{[1](0)}+\delta_{ik}U_j^{[1](0)}-\frac{2}{3}\delta_{jk}U_i^{[1](0)}\right)\,.\label{eq:dtVijk0}
\end{eqnarray}
At a general order this is solved by
\begin{equation}\label{eq:0PNmatch_it}
    \partial_t V^{[1](0)}_{i,i_1\cdots i_{l+1}}=-(l+1)\delta_{i\langle i_{l+1}}U^{[1](0)}_{i_1\cdots i_l\rangle}\,.
\end{equation}
From equation \eqref{eq:dtVi0} we learn that $W^{[1]}_i$ is zero at order $c^{-1}$. Since the same is true for $W^{[1]}$ we immediately see that there cannot be anything at 0.5PN. In other words we have $\partial_t W_i^{[1]}+\partial_i W^{[1]}=\mathcal{O}(c^{-2})$.

Finally, we turn to the $ij$ components. We know from the $1/c$-expansion that at order $c^0$ this is just $\delta_{ij}$. At the same time $W^{[1]}_i$ has terms at order $c^0$ so we need to ensure that $W^{[1]}_{ij}(\text{TT})+G^{[1]}_{ij}$ has an order $c^0$ term that cancels the one from $\partial_i W^{[1]}_j+\partial_j W^{[1]}_i$. For the time being we will consider the sum $W^{[1]}_{ij}(\text{TT})+G^{[1]}_{ij}$. We know that this solves the free wave equation so we have the following $1/c$-expansion
\begin{eqnarray}
    W^{[1]}_{ij}(\text{TT})+G^{[1]}_{ij}     & = & \sum_{l=0}^\infty\frac{1}{l!}\partial_{L} r^{-1}Z^{[1](0)}_{ij,L}(t)-\frac{1}{c}\partial_t Z^{[1](0)}_{ij}(t)+\frac{1}{2c^2}\sum_{l=0}^\infty\frac{1}{l!}\partial_{L} r\partial_t^2 Z^{[1](0)}_{ij,L}(t)\nonumber\\
    &&+\frac{1}{c^2}\sum_{l=0}^\infty\frac{1}{l!}\partial_{L} r^{-1}Z^{[1](2)}_{ij,L}(t)-\frac{1}{6c^3}\left(r^2\partial_t^3 Z^{[1](0)}_{ij}(t)+2x^k\partial_t^3 Z^{[1](0)}_{ij,k}(t)\right) \nonumber\\
    &&-\frac{1}{c^3}\partial_t Z^{[1](2)}_{ij}(t)+\mathcal{O}(c^{-4})\,,\label{eq:1/cexpij}
\end{eqnarray}
where we followed the same steps as with the $1/c$-expansions of $W^{[1]}$ and $W_i^{[1]}$.
To get the right cancellation between $W^{[1]}_{ij}(\text{TT})+G^{[1]}_{ij}$ and  $\partial_i W^{[1]}_j+\partial_j W^{[1]}_i$, we require that
\begin{equation}
    \sum_{l=0}^\infty\frac{1}{l!}\partial_{L} r^{-1}Z^{[1](0)}_{ij,L}(t)=-\partial_i \sum_{l=0}^\infty\frac{1}{l!}\partial_{L} r^{-1}V^{[1](0)}_{j,L}(t)+(i\leftrightarrow j)\,.
\end{equation}
For low multipole moments this equation leads to
\begin{eqnarray}
    Z^{[1](0)}_{ij} & = & 0\,,\label{eq:Zij}\\
    Z^{[1](0)}_{ij,k} & = & -\delta_{ik}V_j^{[1](0)}-\delta_{jk}V_i^{[1](0)}\,,\label{eq:Zij,k}\\
    Z^{[1](0)}_{ij,kl} & = & -2\left(V_{j,\langle k}^{[1](0)}\delta_{l\rangle i}+V_{i,\langle k}^{[1](0)}\delta_{l\rangle j}\right)\,.\label{eq:Zij,kl}
\end{eqnarray}
For general $l$ we have
\begin{equation}
    Z^{[1](0)}_{ij,i_1\cdots i_l}=-(l+1)\left(\delta_{i\langle i_{l+1}}V^{[1](0)}_{|j|,i_1\cdots i_l\rangle}+\delta_{j\langle i_{l+1}}V^{[1](0)}_{|i|,i_1\cdots i_l\rangle}\right)\,.
\end{equation}

We still need to ensure that $W^{[1]}_{ij}(\text{TT})+G^{[1]}_{ij}$ satisfies the properties that we have derived earlier, i.e. that $W^{[1]}_{ij}(\text{TT})$ is a TT solution to the free wave equation and $G^{[1]}_{ij}$ obeys \eqref{eq:propG2} and \eqref{eq:propG3}. Since $W^{[1]}_{ij}(\text{TT})$ is TT the sum $W^{[1]}_{ij}(\text{TT})+G^{[1]}_{ij}$ also obeys \eqref{eq:propG2} and \eqref{eq:propG3}. Furthermore, since at order $c^0$ we have 
\begin{eqnarray}
    \left(W^{[1]}_{ij}(\text{TT})+G^{[1]}_{ij}\right)\Big|_{\mathcal{O}(c^0)} & = &  \sum_{l=0}^\infty\frac{1}{l!}\partial_{L} r^{-1}Z^{[1](0)}_{ij,L}(t)\nonumber\\
    & = & -\left(\partial_i W^{[1]}_j+\partial_j W^{[1]}_i\right)\Big|_{\mathcal{O}(c^0)}\,,
\end{eqnarray}
property \eqref{eq:propG3} becomes $\partial^2 W^{[1]}_j\Big|_{\mathcal{O}(c^0)}=0$ which is automatically fulfilled. Next we consider property \eqref{eq:propG2}. The second time derivative of $Z^{[1](0)}_{ij,L}$ can be evaluated using \eqref{eq:0PNmatch_it} and \eqref{eq:0PNmatch_tt} and in order for this to be TT we need
\begin{equation}
    \partial_t^2 Z^{[1](0)}_{i\langle j, i_1\cdots i_l\rangle}=0\,.
\end{equation}
Using equation \eqref{eq:0PNmatch_it} this can be shown to be satisfied. The part of $Z^{[1](0)}_{ij,i_1\cdots i_l}$ that satisfies $Z^{[1](0)}_{i\langle j,i_1\cdots i_l\rangle}=0$ and that is furthermore tracefree with respect to $ij$ can be attributed to $W^{[1]}_{ij}(\text{TT})\Big|_{\mathcal{O}(c^0)}$. The trace of $Z^{[1](0)}_{ij,i_1\cdots i_l}$ with respect to $ij$ is zero if and only if $V^{[1](0)}_{\langle i,i_1\cdots i_l\rangle}=0$. An example of such a term is given by
\begin{eqnarray}
    Z^{[1](0)}_{ij,kl} & = & \delta_{ik}V^{[1](0)}_{[l,j]}+\delta_{jk}V^{[1](0)}_{[l,i]}+\delta_{il}V^{[1](0)}_{[k,j]}+\delta_{jl}V^{[1](0)}_{[k,i]}\nonumber\\
    &&-\frac{2}{3}\left(\delta_{ik}\delta_{jl}+\delta_{jk}\delta_{il}-\frac{2}{3}\delta_{ij}\delta_{kl}\right)V^{[1](0)}_{n,n}\,,
\end{eqnarray}
which is traceless with respect to $ij$ and satisfies $Z^{[1](0)}_{i\langle j,kl\rangle}=0$.

From equation \eqref{eq:Zij} it follows that the order $c^{-1}$ term in \eqref{eq:1/cexpij} vanishes so that $g_{ij}=\delta_{ij}+\mathcal{O}(c^{-2})$.

The results of this subsection are in agreement with the comments made in section \ref{subsec:0.5PNmetric} (see below equation \eqref{eq:lasteqsec3}) regarding asymptotic flatness in Newtonian gravity and at 0.5PN order. Sufficiently close to the matter source we can ignore retardation effects. Far away from it we cannot but they do not invalidate the assumption of asymptotic flatness at 0PN and 0.5PN order from the point of view of a near zone observer.

\subsection{Matching to 1.5PN}

We now move on to the 1PN and 1.5PN metric. Einstein's field equations at 1PN order are given by equations \eqref{eq:PNinHG_1}--\eqref{eq:PNinHG_3} for $n=2$, which using the results of subsections \ref{subsec:sourceterms} and \ref{subsec:gaugefix} can be shown to be
\begin{eqnarray}
\partial^2 h_{ij}^{(2)} & = & -8\pi G E_{(\sm2)}\delta_{ij}\,,\label{eq:1PN_1}\\
\partial^2 \tau_i^{(4)} & = & -16\pi G E_{(\sm2)}v^i\,,\label{eq:1PN_2}\\
\partial^2 \tau_t^{(4)} & = & 4\pi G\left(E_{(0)}+3P_{(0)}+2E_{(\sm2)}v^2\right)+\frac{1}{2}\partial^2\left(\tau^{(2)}_t\right)^2\nonumber\\
&&+\partial^2_t \tau_t^{(2)}+h_{ij}^{(2)}\partial_i\partial_j \tau^{(2)}_t\,.\label{eq:1PN_3}
\end{eqnarray}
The 0PN solution is $\tau_t^{(2)}=-U$ where $U$ is defined in equation \eqref{U}. Using the $1/c$-expansion of the fluid equations given in appendix \ref{app:conservation} we can rewrite this as 
\begin{eqnarray}
\partial^2 h_{ij}^{(2)} & = & -8\pi G \mathcal{T}_{(0)}^{tt}\delta_{ij}\,,\\
\partial^2 \tau_i^{(4)} & = & -16\pi G \mathcal{T}_{(0)}^{ti}\,,\\
\partial^2 \tau_t^{(4)} & = & 4\pi G\left(\mathcal{T}_{(2)}^{tt}+\mathcal{T}_{(0)}^{ii}\right)+\frac{5}{2}\partial^2 U^2\nonumber\\
&&-\partial^2_t U-\left(h_{ij}^{(2)}-2U\delta_{ij}\right)\partial_i\partial_j U\,,\label{eq:1PN_3mod}
\end{eqnarray}
where \eqref{eq:1PN_1} tells us that $h_{ij}^{(2)}-2U\delta_{ij}$ is harmonic.

At 1.5PN order the Einstein equations are \eqref{eq:PNinHG_1}--\eqref{eq:PNinHG_3} for $n=3$. Using the results of subsections \eqref{subsec:sourceterms} and \eqref{subsec:gaugefix} we have in harmonic gauge
\begin{eqnarray}
    \partial^2 h^{(3)}_{ij} & = & 0\,,\\
    \partial^2 \tau_i^{(5)} & = & 0\,,\\
    \partial^2 \tau_t^{(5)} & = & 0\,.
\end{eqnarray}

If we solve \eqref{eq:1PN_1} and \eqref{eq:1PN_2} the most general solution is given by
\begin{eqnarray}
    h^{(2)}_{ij} & = & 2 U \delta_{ij}+\mathcal{H}^{(2)}_{ij}\,, \label{h2ij_mostgen}\\
    \tau^{(4)}_i & = & 4 G \int d^3x' \frac{\big(E_{(\sm2)} v^i\big)(t,x')}{|x-x'|} +\mathcal{H}^{(4)}_i = 4 U^i+\mathcal{H}^{(4)}_i\,,\label{Soltau4i_mostgen}
\end{eqnarray}
where $\mathcal{H}^{(2)}_{ij}$ and $\mathcal{H}^{(4)}_i$ are near zone harmonics, and where the second equality defines $U^i$. The solution is first order in $G$ and must therefore be matched by a homogeneous solution in the exterior zone. From the results of subsection \ref{subsec:expextzone} we know that the harmonic functions that come from the $1/c$-expansion of the homogeneous part of the exterior metric only show up at odd orders in $1/c$. So using that 1PN is an even order in $1/c$ we conclude that 
\begin{equation}\label{eq:noharm1PN}
    \mathcal{H}^{(2)}_{ij}=0\,,\qquad \mathcal{H}^{(4)}_{i}=0\,.
\end{equation}
Using this we see that the last term in \eqref{eq:1PN_3mod} vanishes. With this extra information the most general solution for $\tau_t^{(4)}$ is given by
\begin{equation} \label{eq:soltau^4}
    \tau_t^{(4)}=- G \int d^3x'\frac{\left(E_{(0)}+3P_{(0)}+2E_{(\sm2)}(v^2+U)\right)(t,x')}{|x-x'|} - \frac{1}{2} \partial_t^2 X + \frac{1}{2} U^2+\mathcal{H}^{(4)}\,,
\end{equation}
where $\mathcal{H}^{(4)}$ is a near zone harmonic, and where the integral is over the matter source. 

Furthermore, $X$ is the superpotential given by
\begin{align}
    X(t,x) =  G \int {E_{(\sm2)}(t,x')}{|x-x'|} d^3x'\,. \label{SuperPot}
\end{align}
The superpotential obeys the defining equation 
\begin{align}
    \partial^2 X = 2U\,. \label{eq:SupPot}
\end{align}
To see how we arrive at this, consider first the most general solution to \eqref{eq:SupPot} given by
\begin{align}
    X(x,t) = - \frac{1}{2 \pi} \int_{\Omega_{R_\star}} d^3 x'\frac{U(t,x')}{|x-x'|}  + X_0(x,t)\,,
\end{align}
where $X_0(x,t)$ is a harmonic function and $\Omega_{R_\star}$ is a ball of radius $R_\star$ centered around the origin and containing $x$. The integrand is non-compact and the integral diverges as we send $R_\star$ to infinity. However, this divergence can be removed by a judicious choice of $X_0$ as is well known\footnote{In fact the divergent terms get annihilated by $\partial_t^2$ in the expression for $\tau_t^{(4)}$.}. This is an example of a type 3 integral (see section \ref{type3}).

To find $X_0$ first consider the identity
\begin{align}
    \frac{1}{2}\int_{\Omega_{R_\star}}d^3x'\partial'_i\left(\frac{\partial'_i\vert x-x'\vert}{\vert y-x'\vert}-\vert x-x'\vert\partial'_i\frac{1}{\vert y-x'\vert}\right)=\int_{\Omega_{R_\star}}\frac{d^3x'}{\vert x-x'\vert \vert y-x'\vert}+2\pi\vert x-y\vert\,,\label{eq:identitysuperpot}
\end{align}
where we used 
\begin{equation}
    \partial^2\vert x-x'\vert=\frac{2}{\vert x-x'\vert}\,.
\end{equation}
Using this we obtain
\begin{align}
    X(x,t) = & - \frac{G}{2 \pi} \int d^3y\int_{\Omega_{R_\star}}  d^3 x'\frac{E_{(\sm2)}(t,y)}{|y-x'||x-x'|} + X_0(x,t)\nonumber\\
    = &  G \int E_{(\sm2)}(t,y)|y-x|d^3y+ X_0(x,t)\nonumber\\
    &- \frac{G}{4 \pi} \int d^3y \int_{\Omega_{R_\star}} d^3x' E_{(\sm2)}(t,y)\partial'_i\left(\frac{\partial'_i\vert x-x'\vert}{\vert y-x'\vert}-\vert x-x'\vert\partial'_i\frac{1}{\vert y-x'\vert}\right) \,.
\end{align}
All the $y$-integrals are over the compact source. The second line is a harmonic function of $x$ so we can choose $X_0$ to cancel this function which leads to the result \eqref{SuperPot}. It can be checked that the harmonic function diverges linearly with $R_\star$ for large $R_\star$.

From the $1/c$-expansion we know that to 1PN $g_{tt}$ is given by
\begin{equation}\label{eq:expgttto1PN}
    g_{tt}=-c^2+2U-\frac{2}{c^2}\left(\tau_t^{(4)}+\frac{1}{2}U^2\right)+\mathcal{O}(c^{-3})\,.
\end{equation}
This expression contains terms that are order $G^2$ so in order to match this onto the exterior solution we need to know the latter to order $G^2$ (at least for as much as
the $tt$ component is concerned). To order $G^2$ the exterior solution is
\begin{eqnarray}
    \hspace{-.4cm}g_{tt}=-c^2+2G\partial_t W^{[1]}+\frac{2G^2}{c^2}\partial_t W^{[2]}-\frac{G^2}{4\pi}\int_{\mathcal{E}} d^3 x'\frac{S_{tt}^{[2]}\left(t-|x-x'|/c,x'\right)}{\vert x-x'\vert}+\mathcal{O}(G^3)\,,
\end{eqnarray}
where we rescaled $W^{[2]}$ with a factor of $c^{-2}$ since $W^{[1]}$ already matches onto the 0PN metric, and where $S^{[2]}_{tt}$ is defined by $\square h^{[2]}_{tt}=S^{[2]}_{tt}$ where according to equation \eqref{eq:tau^2} we have
\begin{align}
    S^{[2]}_{tt} =& - \frac{1}{c^2} \partial_k h^{[1]}_{tt} \partial_k h^{[1]}_{tt} + h^{[1]}_{kl} \partial_{\langle kl \rangle} h^{[1]}_{tt} +\frac{2}{c^4} h^{[1]}_{tt} \partial_t^2 h^{[1]}_{tt} + \frac{4}{c^4} \partial_t h^{[1]}_{tt} \partial_t h^{[1]}_{tt} \nonumber 
    \\
    &- \frac{2}{c^2}  h^{[1]}_{kt} \partial_k \partial_t h^{[1]}_{tt} + 2\partial_{[k}h^{[1]}_{l]t} \partial_k h^{[1]}_{lt}\,.\label{eq:S2tt}
\end{align}
We now wish to $1/c$ expand the right-hand side. 

We know from the matching at 0PN and 0.5PN that the near zone metric is such that $g_{it}=\mathcal{O}(c^{-2})$ and $g_{ij}=\delta_{ij}+\mathcal{O}(c^{-2})$. Using \eqref{h2ij_mostgen} and \eqref{eq:noharm1PN} we also know that the $ij$ components of the near zone metric at order $c^{-2}$ is pure trace. We thus conclude that $h_{it}^{[1]}=\mathcal{O}(c^{-2})$ and $h_{\langle ij\rangle}^{[1]}=\mathcal{O}(c^{-4})$. Furthermore, from the matching of the $tt$ component at 0PN and 0.5PN we derive that $h^{[1]}_{tt}=2G^{-1}U+\mathcal{O}(c^{-2})$. Thus, expanding the right-hand side of \eqref{eq:S2tt} in $1/c$ we see that 
\begin{equation}
    S^{[2]}_{tt} = - \frac{1}{G^2 c^2} \partial_k U \partial_k U+\mathcal{O}(c^{-4})\,.
\end{equation}

The $tt$-component of the 1PN matching equation becomes
\begin{eqnarray}
-\frac{2}{c^2}\left(\tau_t^{(4)}+\frac{1}{2}U^2\right)&=&\frac{2G}{c^2}\partial_t W^{[1]}\Big|_{\mathcal{O}(c^{-2})}+\frac{2G^2}{c^2}\partial_t W^{[2]}\Big|_{\mathcal{O}(c^{0})}\nonumber\\
&&+\frac{1}{\pi c^2}\int_{\mathcal{E}} d^3x'\frac{\left(\partial'_k U\partial'_k U\right)(t,x')}{\vert x-x'\vert}\,, \label{eq:1PNttMatching}
\end{eqnarray}
where $W^{[1]}\Big|_{\mathcal{O}(c^{-2})}$ denotes the coefficient of $1/c^2$ in the $1/c$-expansion of $W^{[1]}$ as given in \eqref{eq:1/cexpext}. Likewise, $W^{[2]}\Big|_{\mathcal{O}(c^{0})}$ denotes the coefficient of $c^0$ in the $1/c$-expansion of $W^{[2]}$. 

We can rewrite the last integral in \eqref{eq:1PNttMatching} as
\begin{eqnarray}
   && \frac{1}{\pi c^2}\int_{\mathcal{E}} d^3x'\frac{\left(\partial'_k U\partial'_k U\right)(t,x')}{\vert x-x'\vert} = \frac{1}{2\pi c^2}\int_{\mathcal{E}} d^3x'\frac{\partial'^2 U^2(t,x')}{\vert x-x'\vert}=\nonumber\\
   &&\frac{1}{2\pi c^2}\int_{\mathbb{R}^3} d^3x'\frac{\partial'^2 U^2(t,x')}{\vert x-x'\vert}- \frac{1}{2\pi c^2}\int_{\mathcal{I}} d^3x'\frac{\partial'^2 U^2(t,x')}{\vert x-x'\vert}=\nonumber\\
   &&-\frac{2}{c^2}U^2+ \frac{1}{2\pi c^2}\int_{\mathcal{E}} d^3x'\partial'_i\left(\frac{\partial'_i U^2(t,x')-U^2(t,x')\frac{x^i-x'^i}{|x-x'|^2}}{\vert x-x'\vert}\right)\,,\label{eq:rewriting}
\end{eqnarray}
where in the first equality we used that $\partial^2 U=0$ for $x \in \mathcal{E}$ and in the second equality we simply used that $\mathbb{R}^3$ is the disjoint union of $\mathcal{E}$ and $\mathcal{I}$. In the last equality\footnote{We also used that 
\begin{equation}
    - \frac{1}{2\pi c^2}\int_{\mathcal{I}} d^3x'\frac{\partial'^2 U^2(t,x')}{\vert x-x'\vert}=- \frac{1}{2\pi c^2}\int_{\mathcal{I}} d^3x'\partial'_i\left(\frac{\partial'_i U^2(t,x')}{\vert x-x'\vert}-U^2(t,x')\partial'_i\frac{1}{|x-x'|}\right)\,.
\end{equation}} we used that $x$ is an interior point of $\mathcal{E}$. 
The last term in \eqref{eq:rewriting} can be absorbed into the $1/c$-expansion of the boundary term in the particular solution \eqref{eq:bdryterm} and so will be dropped.

With this result we see that the matching equation in \eqref{eq:1PNttMatching} becomes
\begin{eqnarray}
&&    \frac{2G}{c^2} \int d^3x'\frac{\left(E_{(0)} + 3P_{(0)}+2E_{(\sm2)}v^2 +2E_{(\sm2)}U\right)(x')}{|x-x'|}  + \frac{1}{c^2} \partial_t^2 X-\frac{2}{c^2}\mathcal{H}^{(4)}=\nonumber\\
&&\frac{2G}{c^2}\partial_t W^{[1]}\Big|_{\mathcal{O}(c^{-2})}+\frac{2G^2}{c^2}\partial_t W^{[2]}\Big|_{\mathcal{O}(c^{0})}\,.\label{eq:matching1PN}
\end{eqnarray}
Since the right-hand side cannot give rise to near zone regular harmonic functions (as we are at even orders in $1/c$) we conclude that 
\begin{equation}
\mathcal{H}^{(4)}=0\,.  
\end{equation}

From the general result \eqref{eq:1/cexpext} we know that
\begin{eqnarray}
    W^{[1]}\Big|_{\mathcal{O}(c^{-2})} & = & \frac{1}{2}\sum_{l=0}^\infty\frac{1}{l!}\partial_{L} r\partial_t^2 U^{[1](0)}_{L}(t)+\sum_{l=0}^\infty\frac{1}{l!}\partial_{L} r^{-1}U^{[1](2)}_{L}(t)\,,\\
    W^{[2]}\Big|_{\mathcal{O}(c^{0})} & = & \sum_{l=0}^{\infty}\frac{1}{l!}\partial_{L}r^{-1}U^{[2](0)}_{L}\,.
\end{eqnarray}
 Using equation \eqref{eq:1/cexpext} and \eqref{eq:0PNmatch_tt} from the matching at the Newtonian order we can write\footnote{In deriving this we used
\begin{equation}
    \vert x-x'\vert=\sum_{l=0}^{\infty}\frac{(-1)^l}{l!}x^{L}\partial_{L}r=\sum_{l=0}^{\infty}\frac{(-1)^l}{l!}x^{\langle L\rangle}\partial_{L}r+\frac{1}{3}x'^2r^{-1}-\frac{1}{5}x'^2 x'^i\partial_i r^{-1}+\cdots\,,
\end{equation}
where the dots denote higher multipole terms.}
\begin{eqnarray}\label{eq:intermW2}
    \partial_t W^{[1]}\Big|_{\mathcal{O}(c^{-2})} & = & \frac{1}{2G}\partial_t^2 X+\frac{1}{2}\left[-\frac{1}{3}r^{-1}\partial_t^2\int d^3 x' x'^2 E_{(\sm2)}(t,x')\right.\\
    &&\left. +\frac{1}{5}\partial_i r^{-1}\partial_t^2\int d^3 x' x'^2 x'^i E_{(\sm2)}(t,x')+\cdots\right]+\sum_{l=0}^\infty\frac{1}{l!}\partial_{L}r^{-1}\partial_t U^{[1](2)}_{L}\nonumber\,,
\end{eqnarray}
where the dots denote higher multipole moments. If we consider the monopole term in the multipole expansion of the matching equation \eqref{eq:matching1PN} we find
\begin{eqnarray}
&&  \partial_t U^{[1](2)}+G\partial_t U^{[2](0)}= \nonumber\\
&&   +\int d^3x\left(E_{(0)} + 3P_{(0)}+2E_{(\sm2)}v^2 +2E_{(\sm2)}U\right) +\frac{1}{6}\partial_t^2\int d^3 x x^2 E_{(\sm2)}\,.\label{eq:matching1PN_l=0}
\end{eqnarray}
We are particularly interested in this term since this is what is needed to fix the 1.5PN term as we will show now.

At 1.5PN order the $tt$-component of the exterior metric reads
\begin{equation}
    2G\partial_t W^{[1]}\Big|_{\mathcal{O}(c^{-3})}+\frac{2G^2}{c^2}\partial_t W^{[2]}\Big|_{\mathcal{O}(c^{-1})}\,.
\end{equation}
Using equation \eqref{eq:1/cexpext} this is equal to 
\begin{equation}
   -\frac{2G}{3c^3}x^i\partial_t^4 U^{[1](0)}_i(t)-\frac{2G}{c^3}\partial^2_t U^{[1](2)} -\frac{2G^2}{c^3}\partial_t^2 U^{[2](0)}\,,
\end{equation}
where we used that $\partial_t^2 U^{[1](0)}(t)=0$.
From the PN expansion we know that the 1.5PN metric is given by $-2c^{-3}\tau_t^{(5)}$.
Since we use coordinates for which the mass dipole moment is zero, i.e. $\partial_t U^{[1](0)}_i(t)=0$ we conclude that
\begin{equation}
    \tau_t^{(5)}=G\partial^2_t U^{[1](2)}+G^2\partial_t^2 U^{[2](0)}\,.
\end{equation}
Equation \eqref{eq:matching1PN_l=0} then tells us that 
\begin{eqnarray}\label{eq:tau5t}
    \tau_t^{(5)} & = & \frac{G}{6}\partial_t^3\int d^3x x^2 E_{(\sm2)}\nonumber\\
    &&+G\partial_t\int_{\mathcal{I}} d^3 x\left(E_{(0)}+3P_{(0)}+2 E_{(\sm2)}v^2+2E_{(\sm2)}U\right)\,.
\end{eqnarray}
We thus see that $\tau_t^{(5)}$ only depends on time and is thus harmonic (as it should be). Using the fluid conservation equations from appendix \ref{app:conservation} we can simplify the expression for $\tau_t^{(5)}$ to
\begin{equation} \label{eq:soltau^5}
    \tau_t^{(5)} = \frac{2G}{3}\partial_t^3\int d^3x x^2 E_{(\sm2)}\,.
\end{equation}

To 1.5PN we thus have for $g_{tt}$
\begin{eqnarray}
    g_{tt} & = & -c^2+2G\int d^3x'\frac{E_{(\sm2)}(t,x')}{\vert x-x'\vert}+\frac{2G}{c^2} \int d^3x'\frac{\left(E_{(0)} + 3P_{(0)}+E_{(\sm2)} (2 v^2 + 2U) \right)(t,x')}{|x-x'|} \nonumber\\
    &&+\frac{1}{c^2}\partial^2 X-\frac{2}{c^2}U^2-\frac{2}{c^3}\tau_t^{(5)}+\mathcal{O}(c^{-4})\\
    & = & -c^2+2G\int d^3x'\frac{E_{(\sm2)}(t-|x-x'|/c,x')}{\vert x-x'\vert}\nonumber\\
    &&+\frac{2G}{c^2} \int d^3x'\frac{\left(E_{(0)} + 3P_{(0)}+E_{(\sm2)} (2 v^2 + 2U)\right)(t-|x-x'|/c,x')}{|x-x'|} \nonumber\\
    &&-\frac{2}{c^2}U^2+\mathcal{O}(c^{-4})\,,
\end{eqnarray}
where in the second way of writing $g_{tt}$ we have used retarded potentials\footnote{We used the following $1/c$-expansion of the retarded Newtonian potential 
\begin{eqnarray}
    && G \int d^3x' \frac{E_{\sm2}(t-|x-x'|/c,x')}{|x-x'|} =U  - \frac{G}{c} \partial_t \int d^3x'{E_{-2}(t,x')} + \frac{1}{2}\frac{G}{c^2} \partial_t^2 \int d^3x' {E_{-2}(t,x')} |x-x'| \nonumber\\
    &&- \frac{1}{6} \frac{G}{c^3} \partial_t^3 \int d^3x' {E_{-2}(t,x')} |x-x'|^2 + \mathcal{O}(c^{-4})= U   + \frac{1}{2}\frac{1}{c^2} \partial_t^2 X - \frac{1}{6} \frac{G}{c^3} \partial_t^3 \int d^3x' {E_{-2}(t,x')} x'^2  + \mathcal{O}(c^{-4})\,,\nonumber\\
\end{eqnarray}
where we used that $\partial_t \int d^3 x' E_{-2}(t,x') =0$ and $\int d^3 x' x'^i E_{-2}(t,x') =0$.}. We thus see that the superpotential $X$ and $\tau_t^{(5)}$ can be viewed as originating from retardation effects. The first term in \eqref{eq:tau5t} can be shown to be a 1.5PN retardation effect of the 0PN term and the second is a 0.5PN retardation effect of the 1PN term. The potential $U$ does not give rise to a 0.5PN retardation term due to the total mass being constant and thus the $U^2$ term does not have a retardation effect at 1.5PN order.

Before continuing the matching process for the other components we note that there is a certain asymmetry in the $1/c$ and $G$-expansions. When we expand in $1/c$ we expand all variables (both metric and fluid). However, when we expand in $G$ we only expand the metric. This is because the exterior zone metric solves the vacuum Einstein equations. However, when we perform the matching one might wonder whether we should have expanded the fluid variables in $G$ as well (and we know that they must depend on $G$ because the fluid conservation equations\footnote{For example, the 0PN fluid equations are given by $\partial_t\mathcal{T}^{t\nu}_{(0)}+\partial_i\mathcal{T}^{i\nu}_{(0)}=0$ where $\mathcal{T}^{\mu\nu}_{(0)}$ is given in appendix \ref{app:conservation}. Explicitly, these equations are
\begin{equation}
    \partial_t E_{(\sm2)}+\partial_i\left(E_{(\sm2)}v^i\right)=0\,,\qquad \partial_t v^i+v^j\partial_j v^i+\frac{1}{E_{(\sm2)}}\partial_i P_{(0)}=-\partial_i U\,.
\end{equation}
The right-hand side of the second equation is order $G$. Hence, the solution for the fluid variables featuring in these equations must contain terms that are at least order $G^0$ and order $G$.}  contain terms proportional to $G$). This asymmetry comes about because we are treating the $G$ dependence of the fluid variables as implicit and in the matching process we only match explicit $G$ dependent terms. So when we expand the exterior zone metric we simply say that the coefficients $h^{[n]}_{\mu\nu}$ should not have any explicit $G$ dependence.

Next, we consider the $ti$ components of the metric. From the near zone and exterior metric we know that at order $c^{-2}$ we must have
\begin{equation}
    -\frac{1}{c^2}\tau_i^{(4)}=\frac{G}{c^2}\left(\partial_t W^{[1]}_i+\partial_i W^{[1]}\right)\Big|_{\mathcal{O}(c^{-2})}\,.
\end{equation}
Hence using \eqref{eq:1/cexpext} we derive the condition
\begin{eqnarray}
    -4\int d^3x' \frac{\big(E_{(\sm2)} v^i\big)(x')}{|x-x'|}  & = & \frac{1}{2}\sum_{l=0}^\infty\frac{1}{l!}\partial_{L}r\partial_t^3 V^{[1](0)}_{i,L}+\sum_{l=0}^\infty\frac{1}{l!}\partial_{L}r^{-1}\partial_t V^{[1](2)}_{i,L}\\
    &&+\frac{1}{2}\partial_i\sum_{l=0}^\infty\frac{1}{l!}\partial_{L}r\partial_t^2 U^{[1](0)}_{L}+\partial_i\sum_{l=0}^\infty\frac{1}{l!}\partial_{L}r^{-1} U^{[1](2)}_{L}\,, \nonumber
\end{eqnarray}
where we used the solution for $\tau_i^{(4)}$ given in \eqref{Soltau4i_mostgen} and \eqref{eq:noharm1PN}. If we multipole expand the left-hand side then the monopole term from this equation tells us that
\begin{equation}
    \partial_t V_i^{[1](2)}=-4\int d^3x E_{(\sm2)}v^i=0\,,
\end{equation}
where we used equations \eqref{eq:dtVi0}--\eqref{eq:dtVijk0} as well as the fact that $\partial_t^2 U^{[1](0)}=0=\partial_t U^{[1](0)}_i$ (see equations \eqref{eq:totalmass} and \eqref{eq:nomassdipole}), and \eqref{eq:restframe}.

At 1.5PN order the $ti$ component of the matching equation is
\begin{equation}
    -\frac{1}{c^3}\tau_i^{(5)}=\frac{G}{c^3}\left(\partial_t W^{[1]}_i+\partial_i W^{[1]}\right)\Big|_{\mathcal{O}(c^{-3})}\,,
\end{equation}
which can be seen to simplify to
\begin{equation}
    \tau_i^{(5)}=G\partial_t^2 V_i^{[1](2)}=0\,.
\end{equation}

Finally, we consider the $ij$ component of the metric. At order $c^{-2}$ the matching equation reads
\begin{equation}
    \frac{1}{c^2}h^{(2)}_{ij}=\frac{G}{c^2}\left(W^{[1]}_{ij}(\text{TT})+ G^{[1]}_{ij}+\partial_i W^{[1]}_j+\partial_j W^{[1]}_i\right)\Big|_{\mathcal{O}(c^{-2})}\,.
\end{equation}
Using the solution for $h_{ij}^{(2)}$ given in equations \eqref{Soltau4i_mostgen} and \eqref{eq:noharm1PN}, and using furthermore \eqref{eq:1/cexpext} and \eqref{eq:1/cexpij} this becomes
\begin{equation}\label{eq:1PNmatchingij}
    2U\delta_{ij}=G\left(W^{[1]}_{ij}(\text{TT})+ G^{[1]}_{ij}+\partial_i W^{[1]}_j+\partial_j W^{[1]}_i\right)\Big|_{\mathcal{O}(c^{-2})}\,,
\end{equation}
where
\begin{eqnarray}
    W^{[1]}_i\Big|_{\mathcal{O}(c^{-2})} & = & \frac{1}{2}\sum_{l=0}^\infty\frac{1}{l!}\partial_{L}r\partial_t^2 V^{[1](0)}_{i,L}+\sum_{l=0}^\infty\frac{1}{l!}\partial_{L}r^{-1} V^{[1](2)}_{i,L}\,,\\
    \left(W^{[1]}_{ij}(\text{TT})+ G^{[1]}_{ij}\right)\Big|_{\mathcal{O}(c^{-2})} & = & \frac{1}{2}\sum_{l=0}^\infty\frac{1}{l!}\partial_{L} r\partial_t^2 Z^{[1](0)}_{ij,L}+\sum_{l=0}^\infty\frac{1}{l!}\partial_{L} r^{-1}Z^{[1](2)}_{ij,L}\,.
\end{eqnarray}

The monopole term in \eqref{eq:1PNmatchingij} can be evaluated using equations \eqref{eq:Zij}--\eqref{eq:Zij,kl} as well as \eqref{eq:dtVi0} and \eqref{eq:dtVij0} leading to 
\begin{equation}
    \frac{1}{4}\partial_k\partial_l r\partial_t^2 Z^{[1](0)}_{ij,kl}+r^{-1}Z^{[1](2)}_{ij}+\frac{1}{2}\partial_i\partial_k r\partial_t^2 V^{[1](0)}_{j,k}+\frac{1}{2}\partial_j\partial_k r\partial_t^2 V^{[1](0)}_{i,k}=2\partial_t U^{[1](0)}r^{-1}\delta_{ij}\,.
\end{equation}
Using that (see equation \eqref{eq:Zij,kl} and \eqref{eq:dtVij0})
\begin{equation}
    \partial_t^2 Z^{[1](0)}_{ij,kl}=2\partial_t U^{[1](0)}\left(\delta_{ik}\delta_{jl}+\delta_{jk}\delta_{il}-\frac{2}{3}\delta_{ij}\delta_{kl}\right)\,,
\end{equation}
we obtain
\begin{equation}
    Z^{[1](2)}_{ij}=\frac{8}{3}\delta_{ij}\partial_t U^{[1](0)}\,,
\end{equation}
where $\partial_t U^{[1](0)}$ is the total mass of the source. We then find
\begin{equation}
    \left(W^{[1]}_{ij}(\text{TT})+ G^{[1]}_{ij}\right)\Big|_{\mathcal{O}(c^{-2})}=\partial_t U^{[1](0)}\left[\partial_i\partial_j r+2\delta_{ij}r^{-1}\right]+\cdots\,,
\end{equation}
where the dots denote higher multipole moments. This agrees with the $c^{-2}$ terms in the expression we found for the linearised Schwarzschild solution in isotropic coordinates \eqref{eq:WijSchwarz}. It obeys the properties \eqref{eq:propG2} and \eqref{eq:propG3}.

At 1.5PN the matching equation is 
\begin{equation}
    \frac{1}{c^3}h^{(3)}_{ij}=\frac{G}{c^3}\left(W^{[1]}_{ij}(\text{TT})+ G^{[1]}_{ij}+\partial_i W^{[1]}_j+\partial_j W^{[1]}_i\right)\Big|_{\mathcal{O}(c^{-3})}\,.
\end{equation}
Using equations \eqref{eq:1/cexpWi} and \eqref{eq:1/cexpij} and the results obtained above we find that 
\begin{equation}
    h_{ij}^{(3)}=0\,.
\end{equation}

We point out that the vanishing of $\tau_i^{(5)}$ and $h^{(3)}_{ij}$ is consistent with the interpretation that the low odd orders in $1/c$ (at least to 2.5PN) are entirely due to retardation effects. Since $\tau_i^{(4)}$ and $h_{ij}^{(2)}$ involve moments of conserved quantities (momentum and mass, respectively) it follows from the Taylor expansion of the corresponding retarded potentials (where in the integrands we replace $t$ by $t-|x-x'|/c$) that $\tau_i^{(5)}$ and $h^{(3)}_{ij}$ must vanish.

\section{Near zone metric to 2.5PN}\label{sec:NZmetric2.5}

In this section we go one and a half PN order higher in the determination of the near zone metric.

\subsection{Solving the near zone equations of motion}

To determine the near zone metric at 2PN and 2.5PN order we consider the $1/c$ expanded Einstein equations \eqref{eq:PNinHG_1}--\eqref{eq:PNinHG_3} for $n=4,5$. The source terms are given in subsections \ref{subsec:sourceterms} and \ref{subsec:gaugefix}. Using the results from the previous section, in particular that $\tau_i^{(5)}=0=h_{ij}^{(3)}$ as well as $h^{(2)}_{ij}=2U\delta_{ij}$, the near zone PDEs in harmonic gauge are
\begin{eqnarray}
    \partial^2 h^{(4)}_{ij}&=&- 16\pi G E_{(\sm2)} (v^i v^j + \delta_{ij} U) - 8 \pi G(E_{(0)} - P_{(0)}) \delta_{ij} - 4 \partial_i U \partial_j U  \nonumber    \\
    && +2 \delta_{ij} \partial^2 U^2 + 2 \delta_{ij} \partial_t^2 U\,,\label{2PNspatialfield}\\
    \partial^2 \tau^{(6)}_i & = & -16 \pi G  \left[-E_{(\sm2)} U_i +E_{(\sm2)}v^i_{(2)} + \left(\frac{1}{2}E_{(\sm2)}v^2+4E_{(\sm2)}U+E_{(0)} + P_{(0)}\right)v^i  \right] \nonumber \\
    &&+8 \partial_k U \partial_k  U_i - 16 \partial_k U \partial_i  U_k - 12 \partial_i U \partial_t U  + 4 \partial^2_t U_i\,, \label{eq:2PNit}\\
    \partial^2 \tau^{(6)}_t & = &  4 \pi G \Big[ E_{(\sm 2)} \left(\tau^{(4)}_t + 4 v_{(2)}^k v^k + 2U(3v^2+2U) \right)  + E_{(0)}(U + 2v^2) \nonumber\\
    &&+ P_{(0)} (3U + 2v^2)  + E_{(2)}+ 3P_{(2)} \Big] - 8(\partial_j U_k \partial_j U_k - \partial_j U_k \partial_k U_j) \nonumber\\
    && - 8 U_k \partial_t \partial_k U -\frac{11}{2} \partial_k U \partial_k U^2 + 2 \partial^2 U^3  - 7 \partial_t U \partial_{t} U- 4U \partial_t \partial_t U\nonumber\\
    &&  
      - h_{kl}^{(4)} \partial_k \partial_l U- 2 \partial_k U \partial_k \tau^{(4)}_{t}+\partial^2_t \tau^{(4)}_t \,. \label{eq:tau_t^6}
\end{eqnarray} 
Similarly, we find that the 2.5PN equations of motion are
\begin{eqnarray}
    \partial^2 h^{(5)}_{ij} & = &  0\,,\label{eq:NZ2.5PNij}\\
    \partial^2 \tau^{(7)}_i & = &  0 \,,\label{eq:NZ2.5PNti}\\
    \partial^2 \tau^{(7)}_t & = & \partial_t^2 \tau^{(5)}_t - h^{(5)}_{kl} \partial_k\partial_l U  +4 \pi G \left(E_{(\sm 2)} \tau^{(5)}_t+E_{(3)}\right)\,.\label{eq:NZ2.5PNtt}
\end{eqnarray}
All these field equation are consistent with the 2.5PN metric given in \cite{Pati:2000vt}.

In order to solve for $h^{(4)}_{ij}$ in equation (\ref{2PNspatialfield}) we start by applying the diagnostics of section \ref{subsec:intEOM} to the non-compact source terms. These are $\partial_i U \partial_j U$, $\partial^2 U^2$ and $\partial_t^2 U$. The first two of these sources goes like $r^{-4}$ for large $r$ and so there are no issues with extending the range of the Poisson integral over all of $\mathbb{R}^3$. The non-compact source $\partial_t^2 U$ is one we already encountered when solving for $\tau_t^{(4)}$ and this leads to a superpotential $X$ contribution to the solution for $h^{(4)}_{ij}$. The solution for $h^{(4)}_{ij}$ is then given by
\begin{eqnarray}
    h^{(4)}_{ij} & = & 2U^2\delta_{ij}+ \partial_t^2X\delta_{ij}+4P[\partial_i U\partial_j U]\nonumber\\
    &&+8\pi G P[(E_{(0)}-P_{(0)}+2E_{(\sm2)} U)\delta_{ij}+2E_{(\sm2)}v^i v^j]\,,\label{eq:Solh3ij}
\end{eqnarray}
where we have introduced the following notation
\begin{equation}
    P[S]=\frac{1}{4\pi}\int_{\mathbb{R}^3} d^3x'\frac{S(t,x')}{\vert x-x'\vert}\,.
\end{equation}
In principle we could add a harmonic function to equation (\ref{eq:Solh3ij}) that is regular in the interior. However we know from the argument given at the start of section 6 (below equation \eqref{eq:fullextsol}) that such a function cannot occur at this order in the exterior solution and so we must set it equal to zero.

We continue with the discussion of the non-compact source terms by looking at the equation for $\tau_i^{(6)}$. All the non-compact source terms are written on the second line in \eqref{eq:2PNit}. Using that $U_i$, which is defined in \eqref{Soltau4i_mostgen} and \eqref{eq:noharm1PN}, goes as $r^{-1}$ for large $r$ we see that the first two terms, i.e. $\partial_k U \partial_k  U_i$ and $\partial_k U \partial_i U_k$ go as $r^{-4}$ and so are well-behaved. The third term $\partial_i U \partial_t U $ is naively $\mathcal{O}(r^{-3})$ but since $U$ is the Poisson integral of a conserved quantity (the mass), the monopole term is constant in time. Furthermore, we chose coordinates such that the mass dipole moment is zero. This means that $\partial_t U$ actually goes as $r^{-3}$ and so $\partial_i U \partial_t U $ is also well-behaved. This leaves us with $\partial_t^2 U_i$ which naively goes as $r^{-1}$. However, just like $U$, the quantity $U_i$ involves the Poisson integral of a conserved quantity, namely the momentum.
So, analogously to the introduction of the superpotential $X$ (see below \eqref{SuperPot}), we define
\begin{equation}\label{eq:Xi}
    X_i=G\int d^3 x' \vert x-x'\vert (E_{(\sm2)}v^i)(t,x')\,,
\end{equation}
which satisfies
\begin{equation}\label{eq:diffXi}
    \partial^2 X_i=2U_i\,.
\end{equation}
The argument leading to the existence of $X_i$ is identical to the case of $X$. The most general solution to \eqref{eq:diffXi} is given by
\begin{equation}
    X_i=-\frac{1}{2\pi}\int_{\Omega_{R_\star}}d^3x'\frac{U_i(t,x')}{\vert x-x'\vert}+X^0_i\,,
\end{equation}
where $X^0_i$ is harmonic and where $\Omega_{R_\star}$ is a ball of radius $R_\star$ with centre at $0$. Using the identity \eqref{eq:identitysuperpot} we can show that there exists an $X^0_i$ such that we get \eqref{eq:Xi}. We thus find the following solution for $\tau_i^{(6)}$
\begin{eqnarray}
    \tau_i^{(6)} & = & 16\pi GP\left[-E_{(\sm2)} U_i +E_{(\sm2)}v^i_{(2)} + \left(\frac{1}{2}E_{(\sm2)}v^2+4E_{(\sm2)}U+E_{(0)} + P_{(0)}\right)v^i\right]\nonumber\\
    &&+2\partial_t^2 X_i-4P\left[2 \partial_k U \partial_k  U_i - 4 \partial_k U \partial_i  U_k - 3 \partial_i U \partial_t U\right]\,,
\end{eqnarray}
where again we do not add a near zone harmonic function as we already know that these will be set to zero by the matching.

We now turn to the equation for $\tau^{(6)}_t$. If we consider the right-hand side of \eqref{eq:tau_t^6} we see that the first three lines consist of either compact source terms or non-compact sources that decay fast enough for the Poisson integrals to exist.
That leaves us with the last line of \eqref{eq:tau_t^6}. The trace part of $h^{(4)}_{ij}$ in $h^{(4)}_{kl} \partial_{k}\partial_l U$ gives rise to a compact source. Meanwhile, the traceless symmetric part of $h^{(4)}_{ij}$ falls off like $r^{-1}$, so combined with the fact that $\partial_k\partial_l U$ goes like $r^{-3}$, we can conclude that the Poisson integral over $h^{(4)}_{kl} \partial_{k}\partial_l U$ is well-behaved. The next term is $\partial_k U \partial_k \tau^{(4)}_{t}$ but this also goes like $r^{-4}$ for large $r$ where we used that $\tau_t^{(4)}$ goes like\footnote{This uses the fact that $\partial_t^2 X$ goes like $r^{-1}$ which follows from mass conservation and the vanishing of the mass dipole moment.} $r^{-1}$. The last term to consider is $\partial_t^2\tau_t^{(4)}$. The solution for $\tau_t^{(4)}$ is given in \eqref{eq:soltau^4}. This means that
\begin{equation}
    \partial_t^2\tau_t^{(4)}=\int d^3x'\frac{C_1(t,x')}{\vert x-x'\vert}+\int d^3x' \vert x-x'\vert C_2(t,x')+\frac{1}{2}\partial_t^2 U^2\,,
\end{equation}
where $C_1$ and $C_2$ denote terms with compact support given by
\begin{eqnarray}
    C_1 & = & -G\partial_t^2\left(E_{(0)}+3P_{(0)}+2E_{(\sm2)}(v^2+U)\right)\,,\\
    C_2 & = & -\frac{G}{2}\partial_t^4 E_{(\sm2)}\,.
\end{eqnarray}
We know that $\frac{1}{2}\partial_t^2 U^2$ goes like $r^{-4}$ for $r \rightarrow \infty$ and so the Poisson integral for this term is well-behaved. For the remaining two term we use the following identity
\begin{equation}
    \partial^2 \vert x-x'\vert^n=n(n+1)\vert x-x'\vert^{n-2}\,,
\end{equation}
to see that
\begin{equation}
    \partial_t^2\left(\tau_t^{(4)}-\frac{1}{2}U^2\right)=\partial^2\left[\int d^3x'\frac{1}{2}\vert x-x'\vert C_1(t,x')+\int d^3x' \frac{1}{12}\vert x-x'\vert^3 C_2(t,x')\right]\,. \label{eq:C_1C_2}
\end{equation}
Thus, up to a harmonic function the Poisson integral of $\partial_t^2\left(\tau_t^{(4)}-\frac{1}{2}U^2\right)$ is equal to the term in square brackets on the right-hand side of the above equation. So, even though the Poisson integral 
\begin{equation}
    \int_{\Omega_{R^\star}}  d^3x'\frac{\partial_t^2\left(\tau_t^{(4)}-\frac{1}{2}U^2\right)}{\vert x-x'\vert}\,,
\end{equation} 
is divergent in the limit $R_\star \rightarrow \infty$, equation (\ref{eq:C_1C_2}) shows that there exists a harmonic function such that when its added to the latter Poisson integral the limit $R_\star \rightarrow \infty$ becomes finite. Using this we can rewrite \eqref{eq:tau_t^6} as
\begin{eqnarray}
    &&\partial^2\left( \tau^{(6)}_t-2U^3-\int d^3x'\frac{1}{2}\vert x-x'\vert C_1(t,x')-\int d^3x' \frac{1}{12}\vert x-x'\vert^3 C_2(t,x')\right)=\nonumber\\
    &&  4 \pi G \Big[ E_{(\sm 2)} \left(\tau^{(4)}_t + 4 v_{(2)}^k v^k + 2U(3v^2+2U) \right)  + E_{(0)}(U + 2v^2) \nonumber\\
    &&+ P_{(0)} (3U + 2v^2) + 3P_{(2)} + E_{(2)} \Big]- h_{kl}^{(4)} \partial_k \partial_l U- 2 \partial_k U \partial_k \tau^{(4)}_{t}+\frac{1}{2}\partial_t^2 U^2\nonumber\\
    && - 8 U_k \partial_t \partial_k U -\frac{11}{2} \partial_k U \partial_k U^2  - 7 \partial_t U \partial_{t} U- 4U \partial_t \partial_t U\nonumber\\
    &&  
      - 8(\partial_j U_k \partial_j U_k - \partial_j U_k \partial_k U_j)\,,
\end{eqnarray}
where now the Poisson integral of the right-hand side is convergent and so we find 
\begin{align}
    \tau^{(6)}_t =&2U^3 -\frac{G}{2} \partial_t^2\int d^3x'\vert x-x'\vert\left(E_{(0)}+3P_{(0)}+2E_{(\sm2)}(v^2+U)\right)(t,x') \nonumber
    \\
    &-\frac{G}{24}\partial_t^4 \int d^3x' \vert x-x'\vert^3 E_{(\sm2)}(t,x') +P[ h_{kl}^{(4)} \partial_k \partial_l U] + 2 P[\partial_k U \partial_k \tau^{(4)}_{t}] \nonumber
    \\
    &-\frac{1}{2}P[\partial_t^2 U^2]  + 8 P[U_k \partial_t \partial_k U] +\frac{11}{2} P[\partial_k U \partial_k U^2]  +7 P[\partial_t U \partial_{t} U]\nonumber
    \\
    &+ 4P[U \partial_t \partial_t U] + 8P[\partial_j U_k \partial_j U_k - \partial_j U_k \partial_k U_j] \nonumber
    \\
     &-4 \pi G P\Big[ E_{(\sm 2)} \left(\tau^{(4)}_t + 4 v_{(2)}^k v^k + 2U(3v^2+2U) \right)  + E_{(0)}(U + 2v^2) \nonumber\\
    &\qquad \qquad+ P_{(0)} (3U + 2v^2) + 3P_{(2)} + E_{(2)} \Big]\,.
\end{align}

Lastly, we want to solve the equations for the 2.5PN metric in (\ref{eq:NZ2.5PNij})-(\ref{eq:NZ2.5PNtt}). The first two equation are simply solved by a harmonic function. For equation (\ref{eq:NZ2.5PNtt}) we see that the first two terms are non-compact. In equation \eqref{eq:soltau^5} we found that $\tau^{(5)}_t$ is just a function of time and so the source term $\partial_t^2 \tau^{(5)}_t$ gives rise to a biharmonic function, more specifically it is solved by $\frac{1}{6}r^2\partial_t^2 \tau^{(5)}_t$. Finally, for the term $h^{(5)}_{kl} \partial_k\partial_l U$ we will assume that $h^{(5)}_{kl}$ is only a function of time which later in this section is shown to be the case. Given this we can write $h^{(5)}_{kl}(t) \partial_k\partial_l U=\frac{1}{2}\partial^2\left(h^{(5)}_{kl}(t) \partial_k\partial_l X\right)$ where $X$ is the superpotential. So, we end up with the following solution to the 2.5PN metric
\begin{eqnarray}
    h^{(5)}_{ij} &=& \mathcal{H}_{ij}^{(5)}\,,\label{eq:NZH5ij}\\
    \tau^{(7)}_i &=& \mathcal{H}_i^{(7)}\,,\label{eq:NZH5i}
        \\
    \tau^{(7)}_t &=& \frac{1}{6}r^2 \partial_t \tau^{(5)}_t - 4 \pi G P[ E_{(3)}] - U\tau^{(5)}_t - \frac{1}{2}h^{(5)}_{kl}(t) \partial_k\partial_l X + \mathcal{H}^{(7)} \, ,\label{eq:NZH5}
\end{eqnarray}
where $\mathcal{H}_{ij}^{(5)}$, $\mathcal{H}_{i}^{(7)}$ and $\mathcal{H}^{(7)}$ are the undetermined near zone harmonics. The purpose for the rest of this section is to determine these.

\subsection{Exterior zone metric and matching to 2.5PN order} \label{sec:Exteriorzone7}

In section \ref{sec:NZto1.5} we worked with the parametrisation of the homogeneous part of the exterior zone metric given in \eqref{eq:HGparam1}--\eqref{eq:HGparam3}. In this section we will (for the sake of contrast) use the more conventional parametrisation given in equations (\ref{hK1})-(\ref{eq:MPMgauge2}). We will use this to determine the near zone harmonic functions to 2.5PN order.

Equations \eqref{eq:HGparam1}--\eqref{eq:HGparam3} imply that the homogeneous part of $h_{\mu \nu}$ can be written as 
\begin{subequations}
    \begin{align}
        h^{\text{hom}}_{tt} =& 2 \sum_{l=0}^\infty \frac{(-)^l}{l!} \partial_L \left( \frac{I_L(u)}{r} \right)+\frac{8}{c^2} \sum_{l=0}^\infty \frac{(-)^l}{l!} \partial_L \left( \frac{ \dot W_L(u)}{r} \right)\,, \label{eq:Homh_tt}
        \\
        h^{\text{hom}}_{it} =& \frac{4}{c^2} \sum_{l=1}^\infty \frac{(-)^l}{l!} \left[  \partial_{L-1} \left( \frac{ \dot I_{i L-1}(u)}{r} \right) + \frac{l}{l+1} \epsilon_{iab} \partial_{a L-1} \left( \frac{J_{bL-1}(u)}{r}\right) \right] \nonumber\\
        &+ \frac{4}{c^2} \sum_{l=0}^\infty \frac{(-)^l}{l!} \partial_{iL} \left( \frac{W_L(u)}{r} \right) - \frac{4}{c^4} \sum_{l=0}^\infty \frac{(-)^l}{l!} \partial_{iL} \left( \frac{\dot X_L(u)}{r} \right)\nonumber
        \\
        & - \frac{4}{c^4} \sum_{l=1}^\infty \frac{(-)^l}{l!} \left[\partial_{L-1} \left( \frac{\dot Y_{iL-1}(u)}{r}\right) + \frac{l}{l+1} \partial_{aL-1} \left( \frac{\epsilon_{iab} \dot 
        Z_{bL-1}(u)}{r} \right) \right]\,,
        \\
        h^{\text{hom}}_{ij} =& \frac{4}{c^4} \sum_{l=2}^\infty \frac{(-)^l}{l!} \left[  \partial_{L-2} \left( \frac{ \ddot I_{ij L-2}(u)}{r} \right) + \frac{2l}{l+1}  \partial_{a L-2} \left( \frac{\epsilon_{ab(i} \dot J_{|b|j)L-2}(u)}{r}\right) \right] \nonumber
        \\
        &+ \delta_{ij} \frac{2}{c^2} \sum_{l=0}^\infty \frac{(-)^l}{l!} \partial_L \left( \frac{I_L(u)}{r} \right) - \frac{8}{c^4} \sum_{l=0}^\infty \frac{(-)^l}{l!} \partial_{ijL} \left( \frac{X_L(u)}{r} \right)\nonumber
        \\
        & - \frac{8}{c^4} \sum_{l=1}^\infty \frac{(-)^l}{l!} \left[\partial_{L-1(i} \left( Y_{j)L-1}(u)/r\right) + \frac{l}{l+1} \epsilon_{ab(i}\partial_{j)aL-1} \left( \frac{ Z_{bL-1}(u)}{r} \right) \right] \,,  \label{eq:Homh_ij}
    \end{align}
\end{subequations}
where $I_L, J_L, W_L, X_L, Y_L, Z_L$ are all STF tensors that can be thought of as having an expansion in $G$ themselves. In other words, at each order in $G$ these multipole moments get corrected. It also follows from the harmonic gauge condition that 
\begin{equation}
\dot I =0\,,\qquad \dot J_a =0\,,\qquad \ddot I_{k} =0\,.
\end{equation}
Matching this result with the Newtonian metric we find that 
\begin{align}
    I_L = \int d^3x E_{(\sm 2)} x^{\langle L \rangle} + \mathcal{O}(c^{-2})\,. \label{eq:NewtonianI}
\end{align}
This means that the mass $I=M$ is constant. As discussed in the previous section our choice of coordinates is such that $I_k =0$. 

In order to catch up with what we did in the previous section (using the parameterisation given in \eqref{eq:HGparam1}--\eqref{eq:HGparam3}), we will match the exterior zone metric with the 1.5PN near zone metric. For this we use the multipole expansion of the particular solution we found in \eqref{eq:rewriting}, as well equation (\ref{eq:Homh_tt})-(\ref{eq:Homh_ij}). We find that that the $1/c$-expansion of the exterior zone metric to 1.5PN order is given by
\begin{subequations} \label{eq:HomExtMet}
\begin{align}
    \mathcal{C} \big(g_{tt}^{\mathcal{E}} \big) =& -c^2 + \frac{2GM}{r} + G \partial_{kl} \Big( \frac{I_{kl}^{(0)}(t)}{r}\Big) - \frac{G}{3} \partial_{klm} \Big( \frac{I_{klm}^{(0)}(t)}{r}\Big) + \frac{G}{4!}  \partial_{klmn} \Big( \frac{I_{klmn}^{(0)}(t)}{r}\Big) \nonumber
    \\
    & + \frac{2GI^{(2)}}{c^2r} + \frac{G}{c^2} \partial_{kl} \Big( \frac{I_{kl}^{(2)}(t)}{r}\Big) + \frac{G}{2c^2} \partial_{kl} \Big( r \ddot I_{kl}^{(0)}(t)\Big)  - \frac{1}{6} \frac{G}{c^2} \partial_{klm} \Big( r \ddot I_{klm}^{(0)}(t)\Big)    \nonumber
    \\
    & + \frac{1}{24} \frac{G}{c^2} \partial_{klmn} \Big( r \ddot I_{klmn}^{(0)}(t) \Big) +\frac{8G}{c^2} \bigg[ \frac{\dot W^{(0)}(t)}{r}- \frac{\ddot W^{(0)}(t)}{c}  - \partial_k \Big( \frac{\dot W_k^{(0)}(t)}{r} \Big)   \nonumber
    \\
    &+ \frac{1}{2} \partial_{kl} \left( \frac{\dot W_{kl}^{(0)}(t)}{r}\right) \bigg] - 2 \frac{G^2M^2}{c^2 r^2} + \cdots 
    +\mathcal{O} (c^{-4}) \,, \label{eq:MPMmet00}
    \\
    \mathcal{C} \big(g_{ti}^{\mathcal{E}} \big) =& \frac{4G}{c^2} \Bigg[ \frac{1}{2} \partial_k \Big( \frac{\dot I_{ik}^{(0)}(t)}{r} \Big) - \frac{1}{6} \partial_{kl} \Big( \frac{\dot I_{ikl}^{(0)}(t)}{r} \Big) + \frac{1}{4!} \partial_{klm} \Big( \frac{\dot I_{iklm}^{(0)}(t)}{r} \Big)  + \frac{1}{2} \epsilon_{iab} \frac{n_a J_b^{(0)}}{r^2}\nonumber
    \\
    &\qquad   +\frac{1}{3} \epsilon_{iab} \partial_{ak} \Big( \frac{J_{bk}^{(0)}(t)}{r} \Big) - \frac{1}{8} \epsilon_{iab} \partial_{akl} \Big( \frac{J_{bkl}^{(0)}(t)}{r} \Big) + \partial_i   \bigg(\frac{W^{(0)}(t)}{r} \bigg)\nonumber
    \\
     &\qquad  - \partial_{ik} \bigg( \frac{W_k^{(0)}(t)}{r} \bigg) + \frac{1}{2} \partial_{ikl} \bigg( \frac{W_{kl}^{(0)}(t)}{r} \bigg)   \Bigg] + \cdots + \mathcal{O} (c^{-4})\,, \label{eq:MPMmet0i}
    \\
    \mathcal{C} \big( g_{ij}^{\mathcal{E}} \big)=& \delta_{ij} \left[1 + \frac{2GM}{r}   + G \partial_{kl} \Big( \frac{I^{(0)}_{kl}(t)}{r}\Big)  \right]
    + \cdots + \mathcal{O} (c^{-4}) 
        \label{eq:MPMmetij}
\end{align}
\end{subequations}
where $\mathcal{C}$ denotes the operation of $1/c$-expanding and where $Q_L^{(n)}$ denotes the $n$th order coefficient in the $1/c$-expansion of the multipole moments $Q_L = I_L,J_L,W_L,X_L,Y_l,Z_L$. Furthermore, the dots denote higher order terms in the multipole expansion.

For the other side of the matching condition we need the multipole expansion of the 1.5PN near zone metric which we work out in section \ref{sec:MultNearZ}. Matching the multipole expanded near zone metric components in (\ref{eq:PNmet00}), (\ref{eq:PNmet0i}) and \eqref{eq:2PNMNZmetric} with equations (\ref{eq:MPMmet00})--(\ref{eq:MPMmetij}) we find that 
\begin{align}
    &J_k^{(0)} = \mathcal{J}_k^{(0)}\,,
    \qquad
    J_{kl}^{(0)}(t) = \mathcal{J}_{(kl)}(t)\,, \qquad J_{klm}^{(0)}(t) = \mathcal{J}_{(klm)}^{(0)}(t)\,,
    \qquad
   W^{(0)}(t) = \frac{1}{6} \dot{\mathcal{I}}_{nn}^{(0)}(t)\,, \nonumber
    \\
    & W_k^{(0)}(t) = \frac{1}{15} \dot{\mathcal{I}}_{knn}^{(0)} - \frac{1}{6} \epsilon_{klm} \mathcal{J}_{lm}^{(0)}\,, \qquad W_{kl}^{(0)}(t) = \frac{1}{28} \dot{\mathcal{I}}_{klnn}^{(0)} -  \frac{1}{6} \epsilon_{(k|mp} \mathcal{J}_{mp|l)}^{(0)}\,, \nonumber
    \\
    &I^{(2)} = M^{(2)}\,, \qquad I_{kl}^{(2)} =  \mathcal{I}_{\langle kl \rangle}^{(2)} +  \frac{1}{84} \ddot{\mathcal{I}}_{\langle kl \rangle nn}^{(0)} + \frac{8}{3} \mathcal{P}_{nn \langle kl \rangle}^{(0)} +\frac{1}{6} \epsilon_{(k|mp} \mathcal{J}_{mp|l)}^{(0)}\,, \label{eq:1.5match}
\end{align}
where $\mathcal{I}_L^{(n)}$, $\mathcal{J}_{iL}^{(n)}$ and $\mathcal{P}_{ijklL}^{(n)}$ are given in terms of multipole moments associated with the fluid. See equation (\ref{eq:DefI&J&P1/c}) for their definitions. We also use the notation $M^{(n)}:=\mathcal{I}^{(n)}$ (and $M = M^{(0)}$).  Additionally, we see that at 1.5PN order in (\ref{eq:MPMmet00}) we just have $-8\ddot W^{(0)}$, so through the matching condition we also find that $\tau^5_t = \frac{2}{3} \partial_t^3 \mathcal{I}_{kk}$, as we had already learned in the previous section.

At this point we have just repeated what we did in the previous section, so now we are going to move on to determine the 2.5PN near zone harmonics and for that we need to be able to work out the particular solution to higher orders in the multipole expansion and in the $G$-expansion.

\subsubsection{Solving the inhomogeneous equation}
Our formal solution to the sourced wave equation \eqref{eq:extPDE} was given in equation (\ref{eq:solhn}) which we restate here for convenience
\begin{align}
    h^{[n]}_{\mu\nu}&=W^{[n]}_{\mu\nu}-R[S_{\mu \nu}]+B_{\mu\nu}^{[n]}\,, \qquad R[S_{\mu \nu}]= \frac{1}{4\pi}\int_{\mathcal{E}} d^3 x'\frac{S_{\mu\nu}^{[n]}(t-\vert x-x'\vert/c,x')}{\vert x-x'\vert}\,,
    \\
    B_{\mu\nu}^{[n]}&=\frac{1}{4\pi}\int_{\mathcal{E}} d^3x'\partial'_i\left(\frac{J^{[n]i}(t-|x-x'|/c,x')}{|x-x'|}\right)\,,
\end{align}
where $W^{[n]}_{\mu \nu}$ is the coefficient of $G^n$ in the expansion of $h_{\mu \nu}^{\text{hom}}$. We recall that the role of $B_{\mu\nu}^{[n]}$ is to cancel any boundary term that comes from the regularised retarded Green's function, $R[S]$.

To the order we are interested in, the source for the exterior zone wave equation consists of terms taking the following form (this breaks down when tail terms show up in the source term), 
\begin{align}
    S = \frac{f_L(u) n^{\langle L \rangle}}{r^m}\,, \label{eq:MulPSource}
\end{align}
where we have suppressed any free indices. Using that the source term takes the form in (\ref{eq:MulPSource}) we get
\begin{equation}
    R[S](t,x) = \frac{n^{\langle L \rangle}}{r} \bigg[\int_0^{l_c} f(u-2s/c) A(s,r) ds + \int_{l_c}^\infty f(u-2s/c) B(s,r) ds \bigg]\,, \label{eq:W&PInt}
\end{equation}
with
\begin{align}
    A(s,r) := \int_{l_c}^{r+s} dr' \frac{P_l(\xi)}{r'^{(m-1)}}, \qquad B(s,r) := \int_s^{r+s} dr' \frac{P_l(\xi)}{r'^{(m-1)}}.
\end{align}
where $P_l$ is the Legendre polynomial of degree $l$ and where $\xi = (r+2s)/r-2s(r+s)/(rr')$. 
This is the same integral as is used in the DIRE approach (see equation  (\ref{eq:WWext})). It can be derived by performing a  change of variables and integrating over the azimuthal angle by making use of the connection between the set of STF unit vectors, $n^{\langle L \rangle}$, and spherical harmonics. For a full derivation and an accompanying geometrical interpretation, see section 6.3 of \cite{poisson2014gravity}. 

Once we have a specific source term, we can compute $A(s,r)$ and $B(s,r)$. From there one can use integration by parts, resulting in higher and higher derivatives of $f(u-2s/c)$, while throwing away boundary terms that depend explicitly on $l_c$ as they are expected to be cancelled by $B^{[n]}_{\mu \nu}$. This process eventually truncates, usually with a tail term or with the remaining integral being associated with a boundary term that is again expected to be cancelled by $B_{\mu \nu}^{[n]}$. 

At this point we have the tools to work out whether the particular solution contributes to the near zone harmonics at 2.5PN order. Source terms for 2PM/3PM equations are worked out in appendix \ref{sec:Match_inhom_sol}. We will make use of equations (\ref{eq:S_ij^2}), (\ref{eq:S_it^2}), (\ref{eq:S_tt^2}) and (\ref{eq:S_tt^3}) in this section. Let us start by taking a look at the the $(it)$-components as an example
\begin{align}
    \square h^{[2]}_{it} = - \frac{8 \epsilon_{ikb} J^b M n^k}{r^5 c^4} + \frac{36}{5} \frac{M \dot{\mathcal{I}}_{kk}(u) n^i }{c^4 r^5}  - \frac{28}{5} \frac{M \dot{\mathcal{I}}_{ik}(u) n^k}{c^4 r^5} + \frac{6 M \dot{\mathcal{I}}_{kl}(u)  n^{\langle ilk \rangle}}{c^4 r^5}  +\cdots\,,  \label{eq:h2it}
\end{align}
where the dots denote terms that are higher order in the multipole expansion of the source term or $\mathcal{O}(c^{-5})$. An $\mathcal{O}(c^{-5})$-term in the particular solution cannot give rise to near zone harmonics at 2.5PN. This is because it only contributes with the leading order term of its $1/c$-expansion and since the leading order term corresponds to just replacing $u$ by $t$, that term must go to zero as $r\rightarrow \infty$ per the exterior zone boundary conditions, thus excluding it from producing a near zone harmonic term.

Considering the terms in equation (\ref{eq:h2it}), we see that the first term is constant in $u$ and can therefore not produce any near zone harmonics. The next two terms can be written as 
\begin{align}
    \frac{ F_{il}(u) n^l}{c^4r^5} \qquad \text{for} \qquad F_{il}(u) = \frac{36}{5}M \dot{\mathcal{I}}_{kk}^{(0)}(u) \delta_{il}- \frac{28}{5}M \dot{\mathcal{I}}_{il}^{(0)}(u)\,.
\end{align}
This takes the form of (\ref{eq:MulPSource}) with $m=5$ and $l=1$. We plug this into equation (\ref{eq:W&PInt}) and use integration by parts to find
\begin{align}
\square^{-1}_{\text{ret}} \left( \frac{ F_{il}(u) n^l}{c^4r^5} \right) =& -\frac{ n^{l}}{c^4r} \bigg( \left[F_{il}(u)(u-2s/c) \bar A(s,r) \right]_0^{l_c} + \left[F_{il}(u)(u-2s/c) \bar B(s,r) \right]_{l_c}^\infty \nonumber
\\
&\qquad + \frac{2}{c}\int_0^{l_c} \dot{F}_{il}(u-2s/c) \bar 
 A(s,r) ds + \frac{2}{c} \int_{l_c}^\infty \dot{F}_{il}(u-2s/c) \bar 
 B(s,r) ds \bigg), \label{eq:PartSol}
\end{align}
where $\partial_s \bar A = A$ and $\partial_s \bar B = B$. Specifically, we use
\begin{align}
     \bar A(s,r) = \frac{4 l s (r + s)^3 + (r - 2 s) (r + s)^4 + 
 l^4 (3 r + 2 s)}{12 l^4 r (r + s)^2}\,,\qquad \bar B(s,r) = \frac{r^2}{12 s^2 (r+s)^2}\,.
\end{align}
The last two terms in (\ref{eq:PartSol}) can be ignored as they are $\mathcal{O}(c^{-5})$ and as 
explained earlier they cannot contribute to the near zone harmonics. For the first two terms we find (after dropping boundary terms),
\begin{align}
    \square^{-1}_{\text{ret}} \left( \frac{ F_{il}(u) n^l}{c^4r^5} \right) =&  \frac{ F_{il}(u) n^l }{4r^3c^4} + \mathcal{O}(c^{-5})\,.
\end{align}
Now, just from the power of $1/r$ in the equation above we can conclude that this term will not produce near zone harmonics until order $c^{-7}$.

Similarly, we find for the last term in (\ref{eq:h2it}) that
\begin{align}
    \square^{-1}_{\text{ret}}\left( \frac{6 M \dot{\mathcal{I}}^{(0)}_{kl}(u)  n^{\langle ilk \rangle}}{c^4 r^5} \right)   =& - \frac{7}{5} \frac{M\dot{\mathcal{I}}^{(0)}_{ik}(u) }{c^4 r^3} n^{k} +\mathcal{O}(c^{-5})\,,
\end{align}
where again, because of the power of $1/r$, it is obvious that the $1/c$-expansion of this term in the overlap will not lead to any near zone harmonics at 2.5PN. In (\ref{eq:h2it}) we have of course also ignored terms that are higher order in the multipole expansion but these only come with higher powers in $1/r$, so they will not contribute to the near zone harmonics either. Thus, we conclude that the particular solution for $h_{it}^{[2]}$ does not give rise to near zone harmonic terms at 2.5PN order.

Similar analysis can be carried out for the $ij$ and $tt$ components in which case we find
\begin{align}
    h_{ij}^{[2]} =& W^{[2]}_{ij} + \delta_{ij} \frac{M^2}{c^4 r^2} +\frac{ M^2}{c^4 r^2} n^i n^j + \cdots + \mathcal{O}(c^{-5})\,,
    \\
    h_{tt}^{[2]} +h_{tt}^{[3]} =& W^{[2]}_{tt}+W^{[3]}_{tt} + \frac{2M^3}{c^4r^3} - \frac{4M^2}{c^2r^4}  - \frac{4Mn^{\langle kl \rangle}}{c^2} \left[ \frac{9\mathcal{I}^{(0)}_{kl}(u)}{r^6} + \frac{9 \dot{\mathcal{I}}^{(0)}_{kl}(u)}{c r^5} - \frac{2 \ddot{\mathcal{I}}_{kl}^{(0)}(u)}{c^2 r^4} \right]  \nonumber 
 \\
 &- \frac{16M}{3c^2} \frac{ \ddot{\mathcal{I}}^{(0)}_{kk}(u)}{c^2 r^4} + \frac{16Mn^l \epsilon_{lmn} \dot{\mathcal{J}}^{(0)}_{mn}(u)}{c^4 r^5}   +\cdots + \mathcal{O}(c^{-5})\,.
\end{align}
Following similar arguments as for the $it$-components we see that the particular solution for the $tt$- and $ij$-components will not produce near zone harmonic terms at 2.5PN. Hence, those can only come from the homogeneous solution.

\subsubsection{Matching with the 2PN metric}
With what we have learned in the previous subsection we are ready to determine the 2.5PN near zone harmonics. If we $1/c$ expand the homogeneous solution in (\ref{eq:Homh_tt})--(\ref{eq:Homh_tt}) in the overlap region we get 
\begin{subequations}
\begin{align}
    \mathcal{C} \big(g_{tt}^{\mathcal{E}} \big) =& \cdots- \frac{8 \ddot W^{(0)}(t)}{c^3} + \frac{1}{c^5} \bigg[\frac{8}{3} x^k \partial_t^4 W_k^{(0)}(t) -8 \ddot W^{(2)}(t) - \frac{1}{15}x^{\langle kl \rangle} \partial_t^5 I_{kl}^{(0)}(t)  \bigg] + \mathcal{O}(c^{-6})\,,\label{eq:HarmonicFctInExt_tt}
    \\
    \mathcal{C} \big(g_{ti}^{\mathcal{E}} \big) =& \cdots + \frac{1}{c^5} \bigg[ - \frac{2}{3} x^k \partial_t^4 I_{ik}^{(0)} - \frac{4}{3} x^i \partial_t^3 W^{(0)}(t) + \frac{4}{3} \partial_t^3 W_i^{(0)}(t) - 4 \ddot Y_i^{(0)}(t) \bigg] + \mathcal{O}(c^{-6})\,,
    \\
    \mathcal{C} \big( g_{ij}^{\mathcal{E}} \big)=& \cdots- \frac{1}{c^5} 2 \partial_t^3 I_{ij}^{(0)}(t) + \mathcal{O}(c^{-6})\,,
\end{align}
\end{subequations}
where the dots here denote any term that is not a near zone harmonic. 

Using that the particular solutions will not contribute to the near zone harmonics up to 2.5PN order we find from the matching condition that
\begin{subequations} \label{eq:NearZonHarmfull}
\begin{align}
    &\mathcal{H}^{(7)}(t) =  4 \ddot{W}^{(2)} +\frac{2}{9} x^k\epsilon_{klm} \dot J_{lm}^{(0)} -\frac{4}{3} x^k\partial_t^4 W_{k}^{(0)}(t) + \frac{1}{30} x^{\langle kl \rangle} \partial_t^5 I_{kl}^{(0)} (t)\,,
    \\
    &\mathcal{H}^{(7)}_i (t) =4 \ddot Y_i^{(0)}(t)- \frac{4}{3} \partial_t^3 W_i^{(0)}(t) +\frac{2}{3} x^k \partial_t^4 I_{ik}^{(0)} + \frac{4}{3} x^i \partial_t^3 W^{(0)}(t)\,,
    \\
    &\mathcal{H}^{(5)}_{ij}(t) = 2 \partial_t^5 I_{ij}^{(0)}(t)\,, \label{eq:NearZonHarm}
\end{align}
\end{subequations}
where $\mathcal{H}_{ij}^{(5)}$, $\mathcal{H}_{i}^{(7)}$ and $\mathcal{H}^{(7)}$ are the near zone harmonics defined in equations \eqref{eq:NZH5ij}, \eqref{eq:NZH5i} and \eqref{eq:NZH5}.

Using equation \eqref{eq:1.5match} we see that all of the multipole moments in the above expressions have already been determined with the exception of $W^{(2)}(t)$ and $Y_i^{(0)}(t)$. So the goal now is to determine $W^{(2)}(t)$ and $Y_i^{(0)}(t)$. These both appear at order $c^{-4} r^{-1}$ in the $1/c$-expansion of $g_{tt}^{\mathcal{E}}$ and $g_{it}^{\mathcal{E}}$, respectively. Therefore, we need to match with the 2PN metric up to the monopole order, $r^{-1}$, in the multipole expansion. The multipole expanded 2PN metric is derived in (\ref{eq:Mult2PNtt}) and (\ref{eq:Mult2PNit}) and given here at the monopole order,
\begin{align}
     \mathcal{M} \left( g^{2PN}_{tt} \right) 
    =&\frac{2M^{(4)}}{r} + \frac{4\ddot{\mathcal{I}}_{kk}^{(2)}}{3r} + \frac{1}{2} \partial_{kl} \left( r\ddot{\mathcal{I}}_{\langle kl \rangle}^{(2)} \right) +\frac{1}{24} \partial_{ kl } \left( r^3 \partial_t^4\mathcal{I}_{\langle kl \rangle}^{(0)} \right) - \frac{1}{72} \partial_{klm} \left( r^3 \partial_t^4 \mathcal{I}_{\langle klm \rangle}^{(0)} \right)\nonumber
    \\
    &  + \frac{1}{12\cdot 4!} \partial_{klmn} \left( r^3 \partial_t^4 \mathcal{I}_{\langle klmn \rangle}^{(0)} \right) +\frac{2}{3} r \partial_t^4 \mathcal{I}_{kk}^{(0)} - \frac{4}{15} \partial_k \left( r \partial_t^4 \mathcal{I}_{knn}^{(0)} \right)  \nonumber
    \\
    &+ \frac{13}{168} \partial_{kl} \left( r \partial_t^4 \mathcal{I}_{\langle kl \rangle nn}^{(0)} \right)  + \frac{2}{45} \frac{\partial_t^4 \mathcal{I}_{llnn}^{(0)}}{r}  + \frac{2}{3} \partial_k \Big( r \epsilon_{kab} \dddot{\mathcal{J}}^{(0)}_{ab} \Big) - \frac{1}{4} \partial_{kl} \left( r \epsilon_{kab} \dddot{\mathcal{J}}^{(0)}_{abl} \right) \nonumber
    \\
    &+ \frac{4}{3} \partial_{kl} \left( r \ddot{\mathcal{P}}_{mm\langle kl \rangle}^{(0)} \right)  + \frac{8}{9} \frac{\ddot{\mathcal{P}}_{kkll}^{(0)}}{r} + \mathcal{O}(r^{-2})  \,,
    \\
    \mathcal{M} \Big(g_{it}^{\text{2PN}} \Big) =&\frac{4G}{c^4}\Bigg[ \frac{1}{4}\partial_k (r \dddot{\mathcal{I}}^{(0)}_{\langle ik \rangle}) + \frac{1}{12} \partial_i (r \dddot{\mathcal{I}}_{kk}^{(0)}) - \frac{1}{12} \partial_{kl} \big( r \dddot{\mathcal{I}}^{(0)}_{\langle ikl \rangle}\big)- \frac{1}{30} \partial_{ik} \big( r \dddot{\mathcal{I}}^{(0)}_{knn}\big) - \frac{1}{30} \frac{\dddot{\mathcal{I}}^{(0)}_{inn}}{r} \nonumber
    \\
    &\qquad + \frac{1}{6} \partial_{kl} \big( r \epsilon_{ikm} \ddot{\mathcal{J}}_{ml}^{(0)}\big) \bigg] + \mathcal{O}(r^{-2}) \,.
\end{align} 
Next, we collect the $c^{-4}$ terms in the $1/c$-expansion of the exterior zone metric
\begin{align}
    \mathcal{C}\big( g^{\mathcal{E}}_{\mu \nu} \big) =& \eta_{\mu \nu} +\overset{(0)}{g^{\mathcal{E}}_{\mu \nu}} + \frac{1}{c^2} \overset{(2)}{g^{\mathcal{E}}_{\mu \nu}}+ \frac{1}{c^3} \overset{(3)}{g^{\mathcal{E}}_{\mu \nu}} + \frac{1}{c^4} \overset{(4)}{g^{\mathcal{E}}_{\mu \nu}}+\cdots\,,
\end{align}
where the $it$- and $tt$-components of the 2PN term are given by
\begin{align}
    \overset{(4)}{g^{\mathcal{E}}_{tt}} =& \frac{2GI^{(4)}}{r} + \frac{1}{2c^2} \partial_{kl} \left( Gr \ddot I_{kl}^{(2)}(t) \right)  + \frac{1}{4!c^4} \partial_{kl} \left( Gr^3 \partial_t^4 I_{kl}^{(0)}(t) \right)   \nonumber
    \\
    & - \frac{1}{3\cdot 4!c^4}\partial_{klm} \left(Gr^3 \partial_t^4 I_{klm}^{(0)}(t) \right) + \frac{1}{12\cdot 4!c^4}\partial_{klmn} \left(Gr^3 \partial_t^4 I_{klmn}^{(0)}(t) \right) \nonumber
    \\
    & + 8 \frac{ \dot W^{(2)}(t)}{c^2r} + \frac{4r}{c^4} \dddot W^{(0)}(t) -  \frac{4}{c^4} \partial_k (r \dddot W_k^{(0)}(t)) + \frac{2}{c^4} \partial_{kl} (r \dddot W_{kl}^{(0)}(t)) + \mathcal{O}(r^{-2})\,,
    \\
    \overset{(4)}{g^{\mathcal{E}}_{it}} =&  \partial_l \Big( r \partial_t^3 {I}^{(0)}_{il}(t) \Big) - \frac{1}{3} \partial_{kl} \Big( r\partial_t^3 I^{(0)}_{ikl}(t) \Big)   + 2 \partial_i \left(r \partial_t^2 W^{(0)}(t) \right) \nonumber
    \\
    &-2 \partial_{ik} \left(r \partial_t^2 W^{(0)}_k(t) \right) +\frac{4\dot Y_i(t)}{r} + \mathcal{O}(r^{-2})\,,
\end{align}
where we used that the particular solution given in the previous subsection are all $\mathcal{O}(r^{-4})$ in the multipole expansion and thus do not contribute to the equations above.

The matching condition tells us that
\begin{align}
    \overset{(4)}{g^{\mathcal{E}}_{tt}} = \mathcal{M} \left( g^{2PN}_{tt} \right), \qquad \overset{(4)}{g^{\mathcal{E}}_{it}} = \mathcal{M} \left( g^{2PN}_{it} \right).
\end{align}
If we use this along with what we learned in (\ref{eq:1.5match}) we can conclude that 
\begin{align}
    \dot Y_i =& - \frac{1}{30} \partial_t^3 \mathcal{I}_{ikk}^{(0)} - \frac{1}{6} \epsilon_{ipq} \partial_t^2 \mathcal{J}_{pq}^{(0)}\,, \\ 
    \dot W^{(2)}(t) =& \frac{1}{6} \ddot{\mathcal{I}}^{(2)}_{kk} + \frac{1}{180} \partial_t^4 \mathcal{I}_{llnn}^{(0)} + \frac{1}{9} \ddot{\mathcal{P}}_{kkll}^{(0)} + W_0^{(2)}\,,
\end{align}
where $W_0^{(2)}$ is a constant, which can be shown to be zero by matching at higher order in the multipole expansion. However, this is not necessary as we are only interested in $\ddot W^{(2)}(t)$. Thus, at this point we can use equation (\ref{eq:NearZonHarmfull}) to fix the undetermined function in the 2.5PN metric. Finally, we find that the 2.5PN metric variables are given by
\begin{align}
    h^{(5)}_{ij} =& -2 \partial_t^3 \mathcal{I}_{\langle ij \rangle}^{(0)}\,,
    \\
    \tau^{(7)}_i =& - \frac{2}{9} \partial_t^4 \mathcal{I}_{ikk}^{(0)} -\frac{4}{9}\epsilon_{ikl} \partial_t^3 \mathcal{J}_{kl}^{(0)}+\frac{2}{3} x^k \partial_t^4 \mathcal{I}_{ik}^{(0)} \,,
      \\
    \tau^{(7)}_t =& \frac{1}{9}r^2 \partial_t^5 \mathcal{I}_{kk}^{(0)} - 4 \pi G P[E_{(3)}] -\frac{2}{3}U \partial_t^3\mathcal{I}_{kk}^{(0)}- \frac{1}{2}h^{(5)}_{kl}(t) \partial_k\partial_l X +\frac{2}{9} x^k \epsilon_{klm} \partial_t^4 \mathcal{J}_{lm}^{(0)} \nonumber
    \\
    &- \frac{4}{45} x^k \partial_t^5 \mathcal{I}_{kll}^{(0)} + \frac{1}{30} x^{\langle kl \rangle} \partial_t^5 \mathcal{I}_{kl}^{(0)} + \frac{2}{3} \partial_t^3 \mathcal{I}^{(2)}_{kk} + \frac{1}{45} \partial_t^5 \mathcal{I}_{kkll}^{(0)} + \frac{4}{9} \partial_t^3 \mathcal{P}_{kkll}^{(0)}\, .
\end{align}
This along with the lower order fields has been checked against and is in agreement with the result from \cite{Pati:2000vt}.

\newpage
\section{Outlook}

This work leads to a number of natural follow-up questions which we discuss here in turn.

The first concerns the use of new gauge choices. In \cite{companionpaper} we will work out the details of the matching process in transverse gauge. Are there other useful gauge choices with particular computational advantages? Is there a systematic set of conditions at every order in $1/c$ and $G$ that singles out a preferred gauge choice? 

Going beyond the scope of this work a natural question is to what extend it is possible to further covariantise the approach taken here. As is well-known the Newtonian description is a gauge-fixed version of Newton--Cartan gravity and so it would be natural to extend this work in the direction of a fully covariant post-Newton--Cartan gravity theory. In the near zone something like that is certainly possible at the level of the expansion of Einstein's equations. The question is whether something similar can be done for the $G$-expansion and the matching process. At which point does one have to choose a gauge to make progress?

Then there is the issue of tail terms and the associated breakdown of the $1/c$ Taylor expansion. Is there a systematic way to incorporate these radiation reaction effects into the $1/c$-expansion framework?

Is it possible to reorganise the $1/c$-expansion, by expanding around a non-vacuum configuration? We know that non-relativistic gravity is not necessarily a weak field approximation and so it might be interesting to explore this option further.

It would also be interesting to change the vacuum to say an FLRW spacetime which can be incorporated into the framework for Newtonian gravity and to develop similar techniques in such a setting.

Finally, post-Newtonian theory is also used in the study of quantum theory in curved gravitational backgrounds (see e.g. \cite{dimopoulos2007testing, zych2011quantum, zych2012general, Pikovski:2013qwa, zych2016general} and \cite{Hartong:2023yxo} for the use of post-Newtonian methods in that context). These applications require a different class of sources and so it would be interesting to see if we can extend our methods to include more general sources such as scalar fields and electromagnetic fields.

\addcontentsline{toc}{section}{Acknowledgements}
\section*{Acknowledgements}

We are grateful for discussions with Gerben Oling. 

JH was supported by the Royal Society University Research Fellowship
Renewal “Non-Lorentzian String Theory” (grant number URF\textbackslash
R\textbackslash 221038), and in part by the Leverhulme Trust Research
Project Grant (RPG-2019-218) ``What is Non-Relativistic Quantum
Gravity and is it Holographic?''.

\newpage

\appendix

\section{Notation, abbreviations and conventions}\label{app:conventions}

For indices we use the following:

\begin{itemize}
\item Lower case Greek indices are coordinate indices, $\mu=0,\ldots,d$ or $\mu=t,\ldots,d$ depending on whether $x^0=ct$ or $x^0=t$.
\item $i,j,k$, etc. are spatial indices in Cartesian coordinates.
\end{itemize}

A superscript of the type $\overset{(n)}{X}$ corresponds to the coefficient of $c^{-n}$ in a Taylor series expansion in $1/c$ of $X$. Likewise, unless explicitly states otherwise, a superscript of the type $\overset{[n]}{X}$  denotes the coefficient of a Taylor expansion in $G$ of $X$ at order $G^n$.

We denote a totally symmetrised collection of indices with round brackets, $(ijkl\cdots)$, and a totally antisymmetrised collection with square brackets, $[ijkl\cdots]$. The symmetrisation and anti-symmetrisation of indices is done with the following normalisation
\begin{eqnarray}
    T_{(i_1\cdots i_l)} & = & \frac{1}{l!}\sum T_{i_{\sigma(1)}\cdots i_{\sigma(1)}}\,,\\
    T_{[i_1\cdots i_l]} & = & \frac{1}{l!}\sum \text{sgn}(\sigma)T_{i_{\sigma(1)}\cdots i_{\sigma(1)}}\,,
\end{eqnarray}
where $\sigma$ is a permutation of $1\cdots l$. We use angle brackets, $\langle ijkl\cdots\rangle$, to denote the traceless part of the totally symmetrised pair of indices $(ijkl\cdots)$. Finally, we use vertical bars to indicate that the (anti-)symmetrisation does not affect the enclosed indices. Sometimes we use a multi-index $L$ to denote a collection of $l$ indices $i_1\cdots i_l$, so instead of writing $T_{i_1\cdots i_l}$ we simply write $T_L$.

We will use mostly plus signature for $g_{\mu\nu}$. We define the Riemann tensor as
\begin{eqnarray}
\left[\nabla_\mu,\nabla_\nu\right]X_\sigma & = & R_{\mu\nu\sigma}{}^\rho X_\rho-T^\rho{}_{\mu\nu}\nabla_\rho X_\sigma\,,\\
\left[\nabla_\mu,\nabla_\nu\right]X^\rho & = & -R_{\mu\nu\sigma}{}^\rho X^\sigma-T^\sigma{}_{\mu\nu}\nabla_\sigma X^\rho\,,
\end{eqnarray}
where $\nabla_\mu$ is any affine connection with connection coefficients $\Gamma^\rho_{\mu\nu}$. Explicitly, this means that
\begin{eqnarray}
{R}_{\mu\nu\sigma}{}^{\rho}&\equiv&-\partial_{\mu}{\Gamma}_{\nu\sigma}^{\rho}+\partial_{\nu}{\Gamma}_{\mu\sigma}^{\rho}-{\Gamma}_{\mu\lambda}^{\rho}{\Gamma}_{\nu\sigma}^{\lambda}+{\Gamma}_{\nu\lambda}^{\rho}{\Gamma}_{\mu\sigma}^{\lambda}\,,\\
T^\rho{}_{\mu\nu} &\equiv& 2\Gamma^\rho_{[\mu\nu]}\,.
\end{eqnarray}
The Ricci tensor is defined as
\begin{equation}
R_{\mu\nu} = {R}_{\mu\rho\nu}{}^{\rho}\,.
\end{equation}

We frequently use the following two abbreviations:
\begin{itemize}
    \item STF: symmetric trace free
    \item TT: transverse traceless
\end{itemize}

\section{The $1/c$-expansion of the Einstein equations}\label{app:1/cexpGR}

In this appendix we will provide some details regarding the $1/c$-expansion of the Einstein equations using the PNR variables \eqref{PNR Einstein} in the KS gauge which means $\Pi_{ti}=0$ (see below equation \eqref{eq:KSmetric}). In this paper we will perform this expansion to 2.5PN order. This appendix provides some background to the derivation of the results presented in section \ref{subsec:sourceterms}.

\subsection{KS gauge}

We start with the left-hand side of \eqref{PNR Einstein} which is given by \eqref{eq:PNRdecomRicci}--\eqref{ER2}. The main objects are $W^\rho_{\mu\nu}$, $C^\rho_{\mu\nu}$, $S^\rho_{\mu\nu}$ and $V^\rho_{\mu\nu}$. We will first study these in KS gauge and then consider how they behave in the $1/c$-expansion in the next subsection. 

We will first consider the objects $V^\rho_{\mu\nu}$ and $S^\rho_{\mu\nu}$ defined in \eqref{C(2)} and \eqref{C(0)}. The nonzero components in KS gauge are
\begin{eqnarray}
    V^t_{ij} & = & \frac{1}{2 T_t^2}\partial_t\Pi_{ij}\,,\\
    S^t_{ti} & = & \frac{1}{T_t}T_{it}\,,\\
    S^t_{ij} & = & -\frac{1}{2T_t}T_{ij}+\frac{1}{2T_t^2}\left(T_{it}T_j+T_{jt}T_i\right)\,,
\end{eqnarray}
where we remind the reader that $T_{\mu\nu}=\partial_\mu T_\nu-\partial_\nu T_\mu$.
The nonzero components of the $C$-connection are
\begin{eqnarray}
    C^t_{tt} & = & \frac{1}{T_t}\partial_t T_t\,,\\
    C^t_{ti} & = & \frac{1}{T_t}\partial_t T_i-\frac{1}{2T_t}\Pi^{kl}T_k\partial_t\Pi_{il}\,,\\
    C^t_{it} & = & \frac{1}{T_t}\partial_i T_t-\frac{1}{2T_t}\Pi^{kl}T_k\partial_t\Pi_{il}\,,\\
    C^t_{ij} & = & \frac{1}{T_t}\partial_i T_j-\frac{1}{T_t}\tilde C^k_{ij}T_k-\frac{1}{2T_t^2}\Pi^{kl}T_k T_l\partial_t \Pi_{ij}\,,\\
    C^k_{ti}=C^k_{it} & = & \frac{1}{2}\Pi^{kl}\partial_t\Pi_{li}\,,\\
    C^k_{ij} & = & \tilde C^k_{ij}+\frac{1}{2T_t}\Pi^{kl}T_l\partial_t\Pi_{ij}\,,
\end{eqnarray}
where we defined
\begin{equation}
    \tilde C^k_{ij}=\frac{1}{2}\Pi^{kl}\left(\partial_i\Pi_{jl}+\partial_j\Pi_{il}-\partial_l\Pi_{ij}\right)\,,
\end{equation}
which is the Levi-Civita connection for a Riemannian manifold with metric $\Pi_{ij}$.
The components of $W^\rho_{\mu\nu}$ are all generically nonzero without any obvious simplification but it is useful to note that
\begin{eqnarray}
    W^k_{tt} & = & T_t^2\Pi^{kl}S^t_{tl}\,,\\
    W^k_{ti} & = & T_t^2\Pi^{kl}S_{il}^t\,.
\end{eqnarray}

From this it follows that in KS gauge we have
\begin{eqnarray}
    \RD_{tt} & = & 0\,,\\
    \RD_{ti} & = & 0\,,\\
    \RC_{tt} & = & \overset{(C)}{R}_{tt}=-\partial_t C^k_{kt}+C^k_{kt}C^t_{tt}-C^l_{kt}C^k_{lt}\,,\\
    \RC_{ti} & = & \overset{(C)}{R}_{ti}+C^k_{kt}S^t_{ti}-W^k_{tt}V^t_{ki}=\overset{(C)}{R}_{it}+C^k_{it}S^t_{tk}-W^k_{tt}V^t_{ki}\,,\\
    \RB_{tt} & = & Y_{tt}+\Pi^{kl}T_{kt}T_{lt}\,,\\
    \RB_{ti} & = & Y_{ti}-W^t_{tt}S^t_{ti}-\Pi^{kl}T_{lt}T_{ik}\,,
\end{eqnarray}
where we defined 
\begin{equation}
    Y_{\mu\nu}=\overset{(C)}{\nabla}_\sigma W^\sigma_{\mu\nu}\,,
\end{equation}
and where 
\begin{equation}
    \overset{(C)}{R}_{ti}=\partial_k C^k_{ti}+C^l_{lk}C^k_{ti}-C^k_{li}C^l_{tk}+\partial_t C^k_{ki}+C^k_{kt}C^t_{ti}\,.
\end{equation}
We left out the spatial components $\RD_{ij}$, $\RC_{ij}$ and $\RB_{ij}$ as the main simplification for those objects comes only once we start $1/c$ expanding.

\subsection{The equations of motion up to 2.5PN}

The Einstein equations \eqref{PNR Einstein} are repeated here for convenience
\begin{align}\label{PNR Einstein2}
    R_{\mu \nu} = c^4 \RA_{\mu \nu} + c^2 \RB_{\mu \nu}+  \RC_{\mu \nu}+ c^{-2} \RD_{\mu \nu}=4\pi G \mathcal{S}_{\mu\nu}\,,
\end{align}
where $\mathcal{S}_{\mu\nu}$ is a compact perfect fluid matter source.
The goal is to expand these to 2.5PN, i.e. to $c^{-5}$. This requires knowing 
\begin{equation}
    \RA_{\mu \nu}=\mathcal{O}(c^{-10})\,,\qquad \RB_{\mu \nu}=\mathcal{O}(c^{-8})\,,\qquad  \RC_{\mu \nu}=\mathcal{O}(c^{-6})\,,\qquad \RD_{\mu \nu}=\mathcal{O}(c^{-4})\,.
\end{equation}
Based on results from the previous subsection we have in general for the $tt$, $ti$ and $ij$ components of \eqref{PNR Einstein2},
\begin{align}
    c^4 \RA_{tt} + c^2 \left(Y_{tt}+\Pi^{kl}T_{kt}T_{lt}\right)-\partial_t C^k_{kt}+C^k_{kt}C^t_{tt}-C^l_{kt}C^k_{lt}  = &\ 4\pi G\mathcal{S}_{tt}\,,\label{eq:tt}\\
    c^4 \RA_{ti} + c^2 \left(Y_{ti}-W^t_{tt}S^t_{ti}-\Pi^{kl}T_{lt}T_{ik}\right)+\overset{(C)}{R}_{ti}+C^k_{kt}S^t_{ti}-W^k_{tt}V^t_{ki}  = &\ 4\pi G\mathcal{S}_{ti}\,,\label{eq:ti}\\
    c^4 \RA_{ij} + c^2 \RB_{ij}+  \RC_{ij}+ c^{-2} \RD_{ij}  = &\ 4\pi G\mathcal{S}_{ij}\,.\label{eq:ij}
\end{align}

Explicitly, the expansion of the metric variables to 2.5PN in KS gauge is
\begin{eqnarray}
    T_t & = & 1+c^{-2}\tau_t^{(2)}+c^{-4}\tau_t^{(4)}+c^{-5}\tau_t^{(5)}+c^{-6}\tau_t^{(6)}+c^{-7}\tau_t^{(7)}+\mathcal{O}(c^{-8})\,,\label{eq:Tt_2.5}\\
    T_i & = & c^{-4}\tau_i^{(4)}+c^{-5}\tau_i^{(5)}+c^{-6}\tau_i^{(6)}+c^{-7}\tau_i^{(7)}+\mathcal{O}(c^{-8})\,,\label{eq:Ti_2.5}\\
    \Pi_{ij} & = & \delta_{ij}+c^{-2}h^{(2)}_{ij}+c^{-3}h^{(3)}_{ij}+c^{-4}h^{(4)}_{ij}+c^{-5}h^{(5)}_{ij}+\mathcal{O}(c^{-6})\,.\label{eq:Piij_2.5}
\end{eqnarray}
Using these expansions we see that 
\begin{equation}
    \Pi^{\alpha \beta} \Pi^{\rho \sigma} T_{\alpha \rho} T_{\beta \sigma}=\mathcal{O}(c^{-8})
\end{equation}
which appears in $\RA_{\mu\nu}$. Since we only need to know the Einstein equations up to terms that are $\mathcal{O}(c^{-6})$ we can discard $\RA_{ij}$ and $\RA_{ti}$ but not $\RA_{tt}$. Furthermore we can determine
\begin{eqnarray}
    W^{t}_{tt} &= & \mathcal{O}(c^{-6})\,,\\
    W^{t}_{ti} &= & \mathcal{O}(c^{-8})\,,\\
    W^{t}_{ij} &= & \mathcal{O}(c^{-10})\,,\\
    W^{k}_{tt} &= & \mathcal{O}(c^{-2})\,,\\
    W^{k}_{ti} &= & \mathcal{O}(c^{-4})\,,\\
    W^{k}_{ij} &= & \mathcal{O}(c^{-8})\,.
\end{eqnarray}
This allows us to write a version of \eqref{eq:tt}--\eqref{eq:ij} that is only correct to 2.5PN, which is
\begin{align}
    c^4 \RA_{tt} + c^2 \left(Y_{tt}+\Pi^{kl}T_{kt}T_{lt}\right)-\partial_t C^k_{kt}+C^k_{kt}C^t_{tt}-C^l_{kt}C^k_{lt} = &\, 4\pi G\mathcal{S}_{tt}+\mathcal{O}(c^{-6})\,,\label{eq:tt_exp}\\
    c^2 \left(Y_{ti}-\Pi^{kl}T_{lt}T_{ik}\right)+\overset{(C)}{R}_{ti}+C^k_{kt}S^t_{ti}-W^k_{tt}V^t_{ki} = &\, 4\pi G\mathcal{S}_{ti}+\mathcal{O}(c^{-6})\,,\label{eq:ti_exp}\\
    c^2 \RB_{ij}+  \RC_{ij}+ c^{-2} \RD_{ij} = &\, 4\pi G\mathcal{S}_{ij}+\mathcal{O}(c^{-6})\,,\label{eq:ij_exp}
\end{align}
where
\begin{eqnarray}
    Y_{tt} & = & \partial_t W^t_{tt}+\partial_k W^k_{tt}+\left(C^t_{tk}-2C^t_{kt}+\tilde C^l_{lk}\right)W^k_{tt}-2C^l_{kt}W^k_{lt}+\mathcal{O}(c^{-8})\,,\\
    Y_{ti} & = & \tilde D_k W^k_{ti}+\left(C^t_{tk}-C^t_{kt}\right)W^k_{ti}-C^t_{ki}W^k_{tt}+\mathcal{O}(c^{-8})\nonumber\\
    & = & \tilde D_k W^k_{ti}+\frac{1}{2}\Pi^{kl}T_{lt}T_{ik}-\Pi^{kl}T_{lt}\tilde D_k T_i+\mathcal{O}(c^{-8})\,,\\
    \RA_{tt} & = & \frac{1}{4}T_t^2\Pi^{ij}\Pi^{kl}T_{ik}T_{jl}+\mathcal{O}(c^{-10})\,.
\end{eqnarray}
In here $\tilde D_k$ is a 3-dimensional covariant derivative with connection $\tilde C^k_{ij}$ and $W^k_{ti}$ is viewed as a 3-dimensional (1,1) tensor.

By inserting the expressions for the relevant components of $W$ and $C$, the $tt$ component, equation \eqref{eq:tt_exp}, can be further rewritten as
\begin{align}
    & c^4\frac{1}{4}T_t^2\Pi^{ij}\Pi^{kl}T_{ik}T_{jl}+c^2\Pi^{kl}\left(-\partial_t T_l \partial_k T_t-T_l\partial_t\partial_k T_t+T_t\tilde D_k\partial_l T_t-T_t\tilde D_k\partial_t T_l\right)\nonumber\\
    &-\frac{1}{2}\Pi^{kl}\partial_t^2\Pi_{kl}+\frac{1}{4}\Pi^{ki}\Pi^{lj}\partial_t\Pi_{ij}\partial_t\Pi_{kl}+\frac{1}{2T_t}\Pi^{kl}\partial_t\Pi_{kl}\partial_t T_t=4\pi G\mathcal{S}_{tt}+\mathcal{O}(c^{-6})\,.
\end{align}
Performing a similar rewriting for the $ti$ component leads to
\begin{eqnarray}
    &&c^2 \left(\tilde D_k \tilde W^k_{ti}+\frac{1}{2}\Pi^{kl}\left(\partial_k T_l\partial_i T_t+T_l\partial_k\partial_i T_t+T_i\partial_k\partial_l T_t-\partial_l T_t\partial_i T_k\right)\right)-\frac{1}{2T_t}\Pi^{kl}T_{lt}\partial_t\Pi_{ik}\nonumber\\
    &&+\tilde D_k C^k_{ti}-\frac{1}{2}\partial_t\left(\Pi^{kl}\partial_i\Pi_{kl}\right) +\frac{1}{2T_t}\Pi^{kl}\partial_t\Pi_{kl}\partial_i T_t = 4\pi G\mathcal{S}_{ti}+\mathcal{O}(c^{-6})\,,
\end{eqnarray}
where we defined
\begin{equation}
    \tilde W^k_{ti}=\frac{1}{2}T_t\Pi^{kl}T_{li}\,.
\end{equation}

To simplify the $ij$ components of the Einsteins equation we need to use
\begin{eqnarray}
\RD_{ij} & = &  \frac{1}{2 T_t^2}\partial^2_t\Pi_{ij}+\mathcal{O}(c^{-4})\,,\\
    \RB_{ij} &=& \mathcal{O}(c^{-8}) \,,\\
\RC_{ij} &=& \overset{(\tilde C)}{R}_{ij}+\frac{1}{2T_t}\partial_{t}\left(\partial_i T_j+\partial_j T_i\right) - \frac{1}{T_t}\partial_{i}\partial_j T_t  + \frac{1}{T_t}\partial_k T_t \tilde C^{k}_{ij}   +\mathcal{O}(c^{-6}) \,,
\end{eqnarray}
where 
\begin{equation}
    \overset{(\tilde C)}{R}_{ij}=\partial_{k} \tilde C^{k}_{ij}- \partial_{i} \tilde C^{k}_{k j}+ \tilde C^{l}_{l k} \tilde C^{k}_{ij}- \tilde C^{l}_{i k}\tilde C^{k}_{l j}\,.
\end{equation}
This allows us to write
\begin{align}
      \overset{(\tilde C)}{R}_{ij}+\frac{1}{2T_t}\partial_{t}\left(\partial_i T_j+\partial_j T_i\right) - \frac{1}{T_t}\partial_{i}\partial_j T_t  + \frac{1}{T_t}\partial_k T_t \tilde C^{k}_{ij} + c^{-2} \frac{1}{2 T_t^2}\partial^2_t\Pi_{ij}=4\pi G \mathcal{S}_{ij}+\mathcal{O}(c^{-6})\,.
\end{align}

We have so far focused on the left-hand side of the Einstein equation \eqref{PNR Einstein2}. The source in \eqref{PNR Einstein2} is given by \eqref{EE2} with a perfect fluid energy-momentum tensor given in \eqref{PNREMT}. In the PNR variables the right-hand side is as in \eqref{PNR Einstein}. For a perfect fluid in KS gauge using the leading order $1/c$ behavior of all the fields involved we can write for the various components of the source $\mathcal{S}_{\mu\nu}$ the following,
\begin{eqnarray}
    \mathcal{S}_{tt} & = & \frac{2}{c^4}(E+P)T_t^2\Pi_{ij}U^i U^j+\frac{1}{c^2}(E+3P)T_t^2\,,\\
    \mathcal{S}_{ti} & = & -\frac{2}{c^4}(E+P)T_t\Pi_{ij}U^j-\frac{1}{c^6}ET_t\Pi_{ij}U^j\Pi_{kl}U^k U^l\nonumber\\
    &&+\frac{1}{c^2}T_t T_i(E+3P)+\mathcal{O}(c^{-6})\,,\\
    \mathcal{S}_{ij} & = & \frac{1}{c^4}(E-P)\Pi_{ij}+\frac{2}{c^6}E\Pi_{ik}U^k\Pi_{jl}U^l+\mathcal{O}(c^{-6})\,,
\end{eqnarray}
where we used that 
\begin{equation}
    \left(T_\mu U^\mu\right)^2=1+\frac{1}{c^2}\Pi_{ij}U^i U^j\,.
\end{equation}

We are now ready to insert the explicit $1/c$-expansions \eqref{eq:Tt_2.5}--\eqref{eq:Piij_2.5} for the PNR variables as well as \eqref{eq:E_2.5}--\eqref{eq:Ui_2.5} for the fluid variables leading to \eqref{eq:PNPoissoneq1}--\eqref{eq:PNPoissoneq3} with the source terms given in section \ref{subsec:sourceterms}.

\section{Multipole expansions}\label{app:PDEstuff}

In this appendix we collect some standard results regarding the multipole expansion of the solution to the free wave equation $\square f=0$. We suppress any potential free indices $f$ might have.

Using 3D spherical coordinates the wave equation reads
\begin{equation}
    \left(-\frac{1}{c^2}\partial_t^2+\partial_r^2+\frac{2}{r}\partial_r+\frac{1}{r^2}\nabla_{S^2}\right)\phi=0\,,
\end{equation}
where $\nabla_{S^2}$ is the Laplacian on the round 2-sphere. Going to Fourier space by writing $\psi=e^{-i\omega t}\psi(x)$ we obtain the Helmholtz equation for $\psi$,
\begin{equation}
    \left(k^2+\partial^2\right)\psi=0\,,
\end{equation}
where $k^2=\omega^2/c^2$. This equation can be solved by the method of separation of variables and the well-known solution is given by
\begin{equation}
    \psi(x)=\sum_{l=0}^\infty\sum_{m=-l}^l\left(A_{lm}Y_{lm}(\theta,\varphi)h^{(1)}_{l}(kr)+B_{lm}Y_{lm}(\theta,\varphi)h^{(2)}_{l}(kr)\right)\,,
\end{equation}
where $A_{lm}$ and $B_{lm}$ are constants and $Y_{lm}(\theta,\varphi)$ are the usual spherical harmonics with respect to spherical coordinates $(\theta,\varphi)$ that are such that the round sphere metric is $d\theta^2+\sin^2\theta d\varphi^2$. Finally, the functions 
$h^{(1)}_{l}(kr)$ and $h^{(2)}_{l}(kr)$ are the spherical Hankel functions of the first and second kind, i.e.
\begin{equation}
    h^{(1)}(x)=-i(-x)^l\left(\frac{1}{x}\frac{d}{dx}\right)^l\left(\frac{e^{ix}}{x}\right)\,,
\end{equation}
and with $h^{(2)}_{l}(x)$ the complex conjugate of $h^{(1)}_{l}(x)$.

Since we use inertial coordinates it will be useful to write this in terms of Cartesian coordinates. This can be achieved by the following useful map\footnote{This map is a consequence of the fact that both the left- and the right-hand side of \eqref{eq:map} represent the most general solution to the Laplacian on $\mathbb{R}^3$ for solutions that are homogeneous of degree $l$. Alternatively, STF polynomials (on the unit sphere) form a finite dimensional irreducible representation for the group $SO(3)$, but so do the spherical harmonics and since these irreps are unique (for a given finite dimension) there must exist a map relating them.}
\begin{equation}\label{eq:map}
    r^l\sum_{m=-l}^l A_{lm}Y_{lm}=d_{i_1\cdots i_l}x^{i_1}\cdots x^{i_l}\,,
\end{equation}
where the constants $d_{i_1\cdots i_l}$ are STF. Using a similar expression for the $B_{lm}$ coefficients and by absorbing some $k$-dependent constants into these STF coefficients we can write
\begin{equation}
    \psi(x)=\sum_{l=0}^\infty d^{(1)}_{i_1\cdots i_l}x^{i_1}\cdots x^{i_l}\left(\frac{1}{r}\frac{d}{dr}\right)^l\left(\frac{e^{ir}}{r}\right)+\sum_{l=0}^\infty d^{(2)}_{i_1\cdots i_l}x^{i_1}\cdots x^{i_l}\left(\frac{1}{r}\frac{d}{dr}\right)^l\left(\frac{e^{-ir}}{r}\right)\,.
\end{equation}
Using 
\begin{equation}
    x^{i_1}\cdots x^{i_l}\left(\frac{1}{r}\frac{d}{dr}\right)^l\left(\frac{e^{ir}}{r}\right)=\partial_{\langle i_1}\cdots\partial_{i_l\rangle}\left(\frac{e^{ir}}{r}\right)\,,
\end{equation}
we can also write the solution to the free wave equation with a single frequency as 
\begin{equation}
    \psi(x)e^{-i\omega t}=\sum_{l=0}^\infty \partial_{i_1}\cdots\partial_{i_l} \left(d^{(1)}_{i_1\cdots i_l}\frac{e^{-i\omega (t-r/c)}}{r}\right)+\sum_{l=0}^\infty \partial_{i_1}\cdots\partial_{i_l}\left( d^{(2)}_{i_1\cdots i_l}\frac{e^{-i\omega (t+r/c)}}{r}\right)\,.
\end{equation}
Integrating over $\omega$ we then obtain the most general solution to the free wave equation as a multipole expansion (in Cartesian coordinates) that is given by
\begin{equation}\label{eq:gensol}
    \phi(x)=\sum_{l=0}^\infty \partial_{i_1}\cdots\partial_{i_l} \left(\frac{U_{i_1\cdots i_l}(u)}{r}\right)+\sum_{l=0}^\infty \partial_{i_1}\cdots\partial_{i_l} \left(\frac{V_{i_1\cdots i_l}(v)}{r}\right)\,,
\end{equation}
where we used retarded $u=t-r/c$ and advanced time $v=t+r/c$ and where the functions $U_{i_1\cdots i_l}$ and $V_{i_1\cdots i_l}$ are STF.

Asymptotically, at leading order in $1/r$, the solution behaves as 
\begin{equation}
    \sum_{l=0}^\infty r^{-l}x^{i_1}\cdots x^{i_l}\frac{1}{r}\left(\frac{-1}{c}\right)^lU^{(l)}_{i_1\cdots i_l}(u)+\sum_{l=0}^\infty r^{-l}x^{i_1}\cdots x^{i_l}\frac{1}{r}\left(\frac{1}{c}\right)^lV^{(l)}_{i_1\cdots i_l}(v)\,,
\end{equation}
where $U^{(l)}_{i_1\cdots i_l}(u)$ denotes the $l$th derivative of $U_{i_1\cdots i_l}(u)$ and similarly for $V^{(l)}_{i_1\cdots i_l}(v)$.

Hence if we impose the Sommerfeld boundary condition of no-incoming radiation at $\mathcal{I}^-$, i.e.
\begin{equation}
    \lim_{\overset{r\rightarrow\infty}{v=\text{cst}}}\partial_v\left(r\phi\right)=0\,,
\end{equation}
then this leads to
\begin{equation}
    V^{(l+1)}_{i_1\cdots i_l}(v)=0\,,
\end{equation}
so that
\begin{equation}
    V_{i_1\cdots i_l}(v)=\sum_{n=0}^l A^{(n)}_{i_1\cdots i_l}v^n\,,
\end{equation}
which is a polynomial in $v$ of degree $l$. By using the following observation
\begin{equation}
    \partial_{i_1}\cdots\partial_{i_l} \left(\frac{V_{i_1\cdots i_l}(v)-V_{i_1\cdots i_l}(u)}{r}\right)=\partial_{i_1}\cdots\partial_{i_l}\sum_{n=0}^l A^{(n)}_{i_1\cdots i_l}\left(\frac{v^n-u^n}{r}\right)=0\,.
\end{equation}
This follows from the fact that $u^n-v^n$ is an odd function of $r$ so that $(u^n-v^n)/r$ only contains even powers of $r$. The function $(u^n-v^n)/r$ is a polynomial in $x^i$ of degree $n-2$ for $n=\text{even}$ and $n-1$ for $n=\text{odd}$. Thus, for the solution in \eqref{eq:gensol} we can replace the $v$ by a $u$ in the second term when we impose Sommerfeld, and then subsequently absorb this term into the first one. We thus conclude that the most general solution obeying Sommerfeld is given by 
\begin{equation}\label{eq:gensol_S}
    \phi(x)=\sum_{l=0}^\infty \partial_{i_1}\cdots\partial_{i_l} \left(\frac{U_{i_1\cdots i_l}(u)}{r}\right)\,.
\end{equation}

At leading order in $1/r$ this solution is given by 
\begin{equation}
    \sum_{l=0}^\infty r^{-l}x^{i_1}\cdots x^{i_l}\frac{1}{r}\left(\frac{-1}{c}\right)^lU^{(l)}_{i_1\cdots i_l}(u)\,.
\end{equation}
Another boundary condition that we will impose is that $\phi=\mathcal{O}(r^{-1})$ for large $r$. We can send $r$ to infinity in different ways depending on what we do with $t$. We can keep $t$ fixed in which case we approach spatial infinity, we can keep $v$ fixed in which case we approach past null infinity or we can keep $u$ fixed in which case we approach future null infinity. We want that $\phi$ is order $r^{-1}$ in all these cases. This means that $U_{i_1\cdots i_l}(u)$ and all its derivatives must be bounded for large negative values of its argument.

Regarding the Laplace equation $\partial^2 f=0$ we can use very similar arguments to show that the most general solution that decays to zero for large $r$ is given by
\begin{equation}\label{eq:decayharm}
    f=\sum_{l=0}^\infty \partial_{i_1}\cdots\partial_{i_l} \left(\frac{f_{i_1\cdots i_l}(t)}{r}\right)\,,
\end{equation}
where now the coefficients are STF functions of $t$. If we want the function to go to zero close to $r=0$ the solution is given by
\begin{equation}
    f=\sum_{l=0}^\infty g_{i_1\cdots i_l}(t)x^{i_1}\cdots x^{i_l} \,,
\end{equation}
where the $g_{i_1\cdots i_l}(t)$ are STF functions of $t$.

\section{Fluid conservation equations and identities}\label{app:conservation}

In this appendix we will consider the matter equations of motion and how to extract useful identities that play a crucial role in the multipole expansion and the subsequent matching procedure for the near zone metric. 

There is more than one way to write down the equations of motion for the matter source in general relativity. 
For the multipole expansion of the post-Newtonian metric it is beneficial to express the fluid equations of motion in the form of conserved currents, i.e. $\partial_\mu \mathcal{T}^{\mu \nu} =0 $ where the derivatives are with respect to inertial coordinates. This can be achieved with the help of the Landau--Lifshitz energy-momentum pseudo-tensor as follows
\begin{align}
    \partial_\mu \mathcal{T}^{\mu \nu} &= 0\,, \label{eq:Conservation}
    \\
    \mathcal{T}^{\mu \nu}&:= (-g) (T^{\mu \nu} + T_{LL}^{\mu \nu})\,,\label{eq:EMTplusLL} 
    \\
    T_{LL}^{\mu \nu} &:=-\frac{c^4}{8\pi G}G^{\mu\nu}+\frac{c^4}{16\pi G (-g)}\partial_\rho\partial_\sigma\left((-g)\left(g^{\mu\nu}g^{\rho\sigma}-g^{\mu\rho}g^{\nu\sigma}\right)\right)\,.
\end{align}

The reason this is a useful way of expressing the fluid equations is because the multipole moments of the near zone metric are time-derivatives of expressions of the form
\begin{align}
     \int d^3 x \mathcal{T}^{\mu \nu} x^L\,.
\end{align}
Due to the conservation equation (\ref{eq:Conservation}) these are not all independent. 

We can use equation (\ref{eq:Conservation}) to derive the following set of identities that will relate time-derivatives of different multipole moments upon integration,
\begin{subequations} \label{eq:FluidIden}
\begin{align} 
    \mathcal{T}^{ti} x^L =& \frac{1}{l+1} \partial_t \left( \mathcal{T}^{tt} x^{iL}\right)  + \frac{l}{l+1} \epsilon^{im(k_1} \mathcal{A}^{|m|k_2\cdots k_l)}+ \frac{1}{l+1} \partial_m \left( \mathcal{T}^{tm} x^{iL} \right)\,,
    \\
    \mathcal{T}^{ij} =& \frac{1}{2} \partial_t^2 \left( \mathcal{T}^{tt} x^{ij}\right) + \frac{1}{2} \partial_m \left( \mathcal{T}^{m p} \partial_p \big( x^{ij}\big) + \partial_t \mathcal{T}^{tm} x^{ij}\right)\,, 
    \\
    \mathcal{T}^{ij} x^k =&  \frac{1}{6} \partial_t^2 \left( \mathcal{T}^{tt} x^{ijk}\right) + \frac{2}{3} \epsilon^{mk(i} \partial_t \mathcal{A}^{|m|j)} + \frac{1}{6}  \partial_m \left( \mathcal{T}^{m p} \partial_p \big( x^{ijk}\big) + \partial_t \mathcal{T}^{tm} x^{ijk}\right)  \nonumber
    \\
    &- \frac{2}{3} \partial_m \left( \mathcal{T}^{mk} x^{ij} - \mathcal{T}^{m(i} x^{j)k} \right),
    \\
    \mathcal{T}^{ij} x^L =& \frac{1}{(l+1)(l+2)} \partial_t^2 \left( \mathcal{T}^{tt} x^{ijL} \right) + \frac{1}{l+2} \partial_t\left( \epsilon^{im(k_1} \mathcal{A}^{|m|k_2\cdots k_l)j} + \epsilon^{jm(k_1} \mathcal{A}^{|m|k_2...k_l)i} \right) \nonumber
    \\
    &+ \frac{8(l-1)}{(l+1)} \mathcal{B}^{ij(k_1...k_l)} + \frac{1}{(l+1)(l+2)} \partial_m \left( \mathcal{T}^{m p} \partial_p \big( x^{ijL}\big) + \partial_t \mathcal{T}^{tm} x^{ijL}\right)\nonumber
    \\
    &+ \frac{2}{l+2} \partial_m \left( \mathcal{T}^{m(i} x^{j)L} - \mathcal{T}^{m (k_1} x^{k_2\cdots k_l)ij} \right)\,,\label{eq:Iden4}
\end{align}
\end{subequations}
where the last identity, (\ref{eq:Iden4}), only holds for $l\geq 2$ and where we defined
\begin{align}
    \mathcal{A}^{iL} = \epsilon^{ijk} ~x^j\mathcal{T}^{tk}  x^L\,, \qquad \mathcal{B}^{ij kl L}= x^{[k} \mathcal{T}^{i][j} x^{l]L}\,,
\end{align}
where $\epsilon^{ijk}$ is the Levi-Civita symbol with $\epsilon^{123}=1$.

For the purposes of this paper all multipole moments that we will work with can be expressed as time-derivatives of the following set of multipole moments: $\mathcal{I}_L, \mathcal{J}_{iL}, \mathcal{P}_{ijklL}$ which are defined as
\begin{align}
    \mathcal{I}_L := \int d^3x \mathcal{T}^{tt}\,,
    \qquad
    \mathcal{J}_{iL} := \int d^3x \mathcal{A}^{iL}\,,
    \qquad
    \mathcal{P}_{ijklL} 
:= \int d^3x ~ \mathcal{B}^{ijklL}\,. \label{eq:DefI&J&P}
\end{align}

Finally, we the $1/c$-expand  $\mathcal{T}^{\mu \nu}$ as well as the multipole moments 
\begin{align}
    \mathcal{T}^{\mu \nu} =& \mathcal{T}^{\mu \nu}_{(0)} + \frac{1}{c^2} \mathcal{T}^{\mu \nu}_{(2)} + \frac{1}{c^4}\mathcal{T}^{\mu \nu}_{(4)} + \frac{1}{c^5} \mathcal{T}^{\mu \nu}_{(5)} + \mathcal{O}(c^{-6})\,,
    \\
    \mathcal{Q}_L =& \mathcal{Q}_L^{(0)} + \frac{1}{c^2} \mathcal{Q}_L^{(2)} + \frac{1}{c^4}\mathcal{Q}_L^{(4)} + \frac{1}{c^5} \mathcal{Q}_L^{(4)} + \mathcal{O}(c^{-6})\,, \label{eq:DefI&J&P1/c}
\end{align}
for $\mathcal{Q}_L = \mathcal{I}_L,\mathcal{J}_L, \mathcal{P}_L$. Additionally, we define $M^{(n)}:=\mathcal{I}^{(n)}$.
In harmonic gauge the coefficients are given by (see for example \cite{Pati:2000vt})
\begin{subequations} \label{eq:FluidConsCurnt}
\begin{align} 
    \mathcal{T}^{tt}_{(0)} =& E_{(\sm 2)}\,, 
    \label{eq:fluidconLO_tt}\\
    \mathcal{T}^{ti}_{(0)} =& E_{(\sm 2)}v^i\,,\label{eq:fluidconLO_ti}
    \\
    \mathcal{T}^{ij}_{(0)} =& E_{(\sm 2)} v^i v^j + P_{(0)} \delta_{ij} + \frac{1}{4\pi G} \left( \partial_i U \partial_j U - \frac{1}{2} \delta_{ij} \partial_k U \partial_k U\right)\,,
    \label{eq:calTij0}\\
    \mathcal{T}^{tt}_{(2)} =&E_{(0)} + E_{(\sm 2)} (v^2+6U)  - \frac{7}{8\pi G} \partial_l U \partial_l U\,,
    \\
    \mathcal{T}^{it}_{(2)} =&(E_{(0)}+P_{(0)}) v^i + E_{(\sm 2)} v^i_2 + \left(\frac{1}{2}v^2+5U\right) v^i E_{(\sm 2)} \nonumber\\
    &+ \frac{1}{4\pi G} \left[ 3\partial_t U \partial_i U + 4\left(\partial_i U_k -\partial_k U_i\right) \partial_k U)\right] \,, 
    \\
    \mathcal{T}^{ij}_{(2)} =& 2 E_{(\sm 2)} v^{(i} v^{j)}_{(2)} + \left(E_{(0)} + 4U E_{(\sm 2)} + P_{(0)}\right) v^i v^j + \delta_{ij}\left(P_{(2)} + 2U P_{(0)}\right) \nonumber
    \\
    & +\frac{1}{4\pi G} \Bigg[ 2 \partial_{(i} U \partial_{j)} \Psi + \partial_{(i} U \partial_{j)} \partial_t^2 X -  16\partial_{[i} U_{k]} \partial_{[j} U_{k]} + 8 \partial_{(i} U \partial_t U_{j)} \nonumber 
    \\
    & - \delta_{ij} \bigg( \partial_{k} U \partial_{k} \Psi + \frac{1}{2} \partial_{k} U \partial_{k} \partial_t^2 X  -  4\partial_{k} U_{l} \partial_{[k} U_{l]} + 4 \partial_t U_k \partial_k U + \frac{3}{2} \partial_t U\partial_t U\bigg)\Bigg]\,,
    \\
    \mathcal{T}^{tt}_{(4)} =& E_{(2)} + E_{(0)} v^2 + 6E_{(0)} U + P_{(0)} v^2 \nonumber\\
    &+ E_{(\sm 2)} \left[ 3 \partial_t^2 X - 8 U_k v^k + 2 v_{(2)}^k v^k + 17 U^2 + 8U v^2\right] \nonumber
    \\
    &+4\pi G E_{(\sm 2)} P \left[6 E_{(0)} - 2P_{(0)} +14 E_{(\sm 2)} U + 4 E_{(\sm 2)} v^2 \right]\nonumber\\
    &+ \frac{1}{4 \pi G} \bigg\{\frac{5}{2} \partial_t U \partial_t U\nonumber - 4 U \partial_t^2 U + 4 \partial_t U_k \partial_k U- 7 \partial_k U\partial_k \left(\Psi+ \frac{1}{2} \partial_t^2 X\right)
    \\
    & - 8 U_k \partial_k \partial_t U + 2 \partial_l U_k \left( 3 \partial_k U_l + \partial_l U_k \right)-10 U \partial_k U \partial_k U \nonumber\\
    &- 4 \left( P[E_{(\sm 2)} v^k v^l] + P[\partial_k U \partial_l U] \right) \partial_k\partial_l U\bigg\} \nonumber
    \\
    & + 8\partial_k U \partial_k\left( P\Big[3P_{(0)} + E_{(\sm 2)} v^2 - \frac{1}{2}E_{(\sm 2)}U\Big]\right)\,,
    \\
     \mathcal{T}^{tt}_{(5)} =& E_{(3)} + \frac{1}{2\pi} \mathcal{I}_{kl}^{(0)} \partial_{kl}U\,,
\end{align}
\end{subequations}
where $X$ is the superpotential given in equation \eqref{SuperPot} and where $\Psi$ is 
\begin{equation}
    \Psi = -4\pi G P[E_{(0)} + 3P_{(0)} + 2 E_{(\sm 2)}v^2 + 2 E_{(\sm 2)}U ]\,.
\end{equation}
In writing down the expressions for $\mathcal{T}^{\mu \nu}_{(n)}$ (for a given n) we used the matched near zone solution to the metric at lower orders. For example, in computing $\mathcal{T}^{ij}_{(0)}$ and $\mathcal{T}^{tt}_{(2)}$ (which appear for the first time as source terms at 1PN) we used the 0PN near zone metric. Likewise, when computing $\mathcal{T}^{it}_{(2)}$ and $\mathcal{T}^{ij}_{(2)}$ we used the 1PN near zone metric (after matching). It would be interesting to compute both the $1/c$ and $G$-expansions of \eqref{eq:EMTplusLL} for the general class of gauges used in sections \ref{sec:genPN} and \ref{sec:covGexp}.

\section{Solving for the exterior zone metric} \label{sec:Match_inhom_sol}

In the first part of this appendix we will focus on the exterior zone metric. We will go through some generalities and then make consistency checks of our treatment of the exterior zone metric. In doing so we need to also perform the multipole expansion of the near zone metric which will be done in the second half of this appendix.

However, we will begin by discussing how to get a consistent treatment of error terms in the double expansion that we perform on the exterior zone metric. The relativistic multipole expansion schematically takes the following form
\begin{align}
    \frac{\mathcal{F}(u)}{r} + \partial_k \left( \frac{\mathcal{F}_k(u)}{r}\right) + \partial_{kl} \left( \frac{\mathcal{F}_{kl}(u)}{r}\right) + \partial_{klm} \left( \frac{\mathcal{F}_{klm}(u)}{r}\right) +\cdots
\end{align}
where the $\mathcal{F}_L(u)$ are associated with near zone multipole moments through the matching procedure, thus we can assume that $F_L \sim (l_c)^L F$ and $\dot F_L \sim \tfrac{1}{t_c} F_L$. Using this we see that for the $l$th order term in the expansion we get 
\begin{align}
    \partial_{i_1...i_l} \left( \frac{\mathcal{F}_{i_1...i_l}(u)}{r} \right)  \sim \left( \frac{l_c}{r} \right)^l \left[ 1 + \frac{r}{\lambda_c} + \left(\frac{r}{\lambda_c}\right)^2 +... + \left(\frac{r}{\lambda_c}\right)^l \right] \frac{F}{r}. \label{eq:multp_par}
\end{align}
Now, in general $l_c/r$ is going to be completely unrelated to the post-Minkowskian expansion parameter, $\epsilon = \frac{GM}{c^2l_c}$. Thus, a priori there is no good answer to the question of how many orders in the multipole expansion one needs to keep if we truncate the $G$-expansion at $n$th order. At the very least the answer will be $r$-dependent. In other words there is no consistent analogy to the $n$PN metric for the exterior zone, it simply depends on what one is interested in calculating.

However, if we restrict ourselves to the wave zone, $\lambda_c \leq r$, then we know that  $ \left(\frac{l_c}{r} \right)^2 \leq \epsilon$. This allows us to put an upper limit on the order in the multipole expansion that we need to keep for any finite order in the $G$-expansion. For example, if the highest order correction we are interested in is the monopole correction to the $n$PM metric. Then we know that we at most need to keep up to the $2(n-k)$th order correction in the multipole expansion of the $k$PM metric ($k<n$). Anything higher in the multipole expansion of the $k$PM correction is guaranteed to be subleading. However, this is often much more than what is actually needed.

For example, in gravitational wave physics the goal is usually just to compute the waveform for which we only need the $1/r$-piece of the metric\footnote{The subleading correction to this is completely negligible for any physically relevant sources.}. More precisely, the waveform is constructed by taking the transverse traceless projection of the $1/r$-part of $g_{ij}$. At leading order the waveform is given by the famous quadrupole formula
\begin{align}
    h_{ij}^{TT} = \frac{2G \ddot{\mathcal{I}}^{TT}_{ij}}{c^4 r}   + \mathcal{O}(c^{-5})\,.
\end{align}
Using this along with (\ref{eq:multp_par}) we see that in order to compute the full $n$th order $1/c$-corrections to the quadruple moment, one needs to keep up to $n+2-2m$ orders in the multipole expansion of the $G^m$ correction to $h_{ij}$. 

Now, returning to the main aim of this appendix, in section 7.2 we have laid out how to compute the particular solution to the exterior zone metric, but we have only made very limited use of it since it does not contribute to the determination of the near zone harmonic functions. Therefore, we want give more examples of the matching process. One way to do this is to compute the relevant part of wave zone metric (following the counting argument laid out above) to a given order and match it against the 2.5PN near zone metric. We have chosen to do this to up to the order where we get the leading order (in the multipole expansion) contribution to the 3PM particular solution for the $tt$-component and the leading order 2PM particular solution for the $it$- and $ij$-components.

\subsection{Solving the inhomogeneous wave equation}
We know from equation (\ref{eq:HGeinsteq}) that the 2PM equations of motion are given by
\begin{align}
    \square h^{[2]}_{\mu \nu} =&  S^{[2]}_{\mu \nu}\,,
\\
    S^{[2]}_{\mu \nu} =&- \tau^{[2]}_{\mu\nu} +\partial_\mu \left( h_{[1]}^{\alpha \beta} \partial_\alpha h^{[1]}_{\beta \nu} \right) + \partial_\nu \left( h_{[1]}^{\alpha \beta} \partial_\alpha h^{[1]}_{\beta \mu}\right) - \partial_\mu \left( h_{[1]}^{\alpha \beta} \partial_{\nu} h_{\alpha \beta}^{[1]} \right)\,. 
\end{align}
We also already matched the exterior zone metric to 1.5PN order (see section \ref{sec:Exteriorzone7}), and from this we know that 
\begin{subequations} \label{eq:1.5PNWaveMet}
\begin{align} 
    g_{tt}^{\mathcal{E}} =& -c^2 + \frac{2G(M+c^{\sm 2}M^{(2)})}{r} - 2 \frac{G^2M^2}{c^2 r^2} +\frac{4G}{3c^2} \partial_k \Big( \frac{\epsilon_{klm} \dot{\mathcal{J}}^{(0)}_{lm}(u)}{r}\Big) + G\partial_{kl} \Big( \frac{\mathcal{I}^{(0)}_{kl}(u)}{r}\Big)\nonumber
       \\
    & - \frac{G}{3} \partial_{klm} \Big( \frac{\mathcal{I}^{(0)}_{klm}(u)}{r}\Big) + \frac{G \ddot{\mathcal{I}}^{(0)}_{ll}(u)}{r} - \frac{1}{3}\partial_k \left(\frac{G \ddot{\mathcal{I}}^{(0)}_{kll}(u)}{r} \right) +\cdots\,,
       \\
    g_{ti}^{\mathcal{E}} =& \frac{4G}{c^2} \bigg[ \frac{1}{2} \epsilon_{iab} \frac{n_a \mathcal{J}^{(0)}_b}{r^2} + \frac{1}{2} \partial_l \Big( \frac{\dot{\mathcal{I}}^{(0)}_{il}(u)}{r} \Big) \nonumber
    \\
    &\qquad+ \frac{1}{3} \epsilon_{iab} \partial_l \partial_a \Big( \frac{\mathcal{J}^{(0)}_{bl}(u)}{r} \Big) - \frac{1}{6} \partial_{kl} \Big( \frac{\dot{\mathcal{I}}^{(0)}_{ikl}(u)}{r} \Big)\bigg] + \cdots\,,
    \\
    g_{ij}^{\mathcal{E}} =& \delta_{ij} \left(1+ \frac{2GM}{c^2r} \right) + \cdots\,,
\end{align}
\end{subequations}
where the dots denote terms that are subleading according to the wave zone counting. This is not all we know, for example the $I^{(0)}_L$ have been fixed for all $l$ in the Newtonian expansion, see equation (\ref{eq:NewtonianI}). Equation (\ref{eq:1.5PNWaveMet}) is simply stated here for convenience when computing the source terms for the 2PM equations.

\subsubsection{2PM spatial components}
Starting with the spatial components, 
we find that the leading order correction to $S^{[2]}_{ij}$ is given by
\begin{align}
     S^{[2]}_{ij} = - \frac{4M^2}{c^4 r^4} \left( n^{\langle ij \rangle} - \frac{2}{3} \delta_{ij}\right) +\cdots\,, \label{eq:S_ij^2}
\end{align}
where the dots denote terms that are either higher order in the multipole expansion or $\mathcal{O}(c^{-6})$ as in this section we are not interested in the part of the wave zone metric that we cannot match with the 2.5PN metric. Using the integral equation in (\ref{eq:W&PInt}) we find
\begin{align}
    -\frac{1}{4\pi}\int_{\mathcal{E}} d^3 x'\frac{S_{ij}^{[2]}(t-\vert x-x'\vert/c,x')}{\vert x-x'\vert}= \frac{M^2}{c^4 r^2}\left( (n^i n^j + \delta_{ij})  - \frac{8}{3}\delta_{ij} \frac{r}{ l_c} + \frac{4}{5} \frac{ l_c}{r} n^{\langle ij \rangle} \right)+ \cdots\,.
\end{align}
The first two terms make up the particular solution while the last two terms above are boundary terms that are assumed to be cancelled by $B^{[2]}_{ij}$.

Adding the homogeneous solution to the 2PM particular solution we find that the exterior zone metric is given by
\begin{align}
    g_{ij}^{\mathcal{E}} =& \delta_{ij} \bigg[1 + \frac{2(M+c^{\sm 2}M^{(2)})}{c^2r} +   \frac{1}{c^2} \partial_{kl} \Big( \frac{\mathcal{I}^{(0)}_{kl}(u)}{r} \Big) + \frac{M^2}{c^4 r^2} - \frac{2}{3c^4} \frac{\ddot{\mathcal{I}}^{(0)}_{kk}(u)}{r} \bigg] \nonumber
    \\
   & + \frac{2}{c^4} \frac{\ddot{\mathcal{I}}^{(0)}_{ij}(u)}{r}  +\frac{ M^2}{c^4 r^2} n^i n^j + \frac{1}{c^4} \frac{8}{3} \partial_a \bigg( \frac{\epsilon_{ab(i} \dot{J}_{|b|j)}(u)}{r}\bigg) - \frac{1}{c^4} \frac{2}{3} \partial_k \bigg( \frac{\ddot{\mathcal{I}}^{(0)}_{ijk}(u)}{r}\bigg) \nonumber
   \\
   &
   -\delta_{ij} \bigg[   \frac{1}{3c^2}\partial_{klm} \Big( \frac{\mathcal{I}^{(0)}_{klm}(u)}{r} \Big) \bigg] +  \frac{4}{15 c^4}  \partial_{(i} \Big(\ddot{\mathcal{I}}^{(0)}_{j)ll}(u) /r\Big)\nonumber
   \\
   &  + \delta_{ij} \left[ \frac{1}{5c^2} \partial_{mml} \bigg( \frac{\mathcal{I}^{(0)}_{lkk}(u)}{r} \bigg)- \frac{1}{3c^2} \partial_{ll} \bigg( \frac{\mathcal{I}^{(0)}_{kk}(u)}{r} \bigg) + \frac{2}{15c^4} \partial_{k} \bigg( \frac{\ddot{\mathcal{I}}^{(0)}_{kll}(u)}{r} \bigg) \right] \nonumber 
   \\
   &-\frac{8}{c^4}  
    + \partial_{(i} \Big(Y_{j)}(u)/r \Big)
     + \cdots\,. \label{eq:2PNWZmetric1}
\end{align}
In order to match with the near zone we go to the overlap region and $1/c$-expand
\begin{align}
    \mathcal{C} \big(g_{ij}^{\mathcal{E}} \big) =& \delta_{ij} \bigg(1 + \frac{2(M+c^{\sm 2}M^{(2)})}{c^2r} +   \frac{1}{c^2} \partial_{kl} \Big( \frac{\mathcal{I}^{(0)}_{kl}(t)}{r} \Big) +  \frac{1}{2c^4} \partial_{kl} \Big( r \ddot{\mathcal{I}}^{(0)}_{kl}(t) \Big) + \frac{M^2}{c^4 r^2} - \frac{1}{c^4} \frac{\ddot{\mathcal{I}}^{(0)}_{kk}(t)}{r} \bigg) \nonumber
    \\
   & + \frac{2}{c^4} \frac{\ddot{\mathcal{I}}^{(0)}_{ij}(t)}{r}  +\frac{ M^2}{c^4 r^2} n^i n^j + \frac{1}{c^4} \frac{8}{3} \partial_a \bigg( \frac{\epsilon_{ab(i} \dot{J}_{|b|j)}(t)}{r}\bigg) - \frac{1}{c^4} \frac{2}{3} \partial_k \bigg( \frac{\ddot{\mathcal{I}}^{(0)}_{ijk}(t)}{r}\bigg) \nonumber
   \\
   &
   +\delta_{ij} \bigg[ \frac{1}{3c^4} \partial_k \bigg( \frac{\ddot{\mathcal{I}}^{(0)}_{kll}(t)}{r}\bigg) - \frac{1}{3c^2}\partial_{klm} \Big( \frac{\mathcal{I}^{(0)}_{klm}(t)}{r} \Big)\bigg) - \frac{1}{6c^4} \partial_{klm} \Big( r\ddot{\mathcal{I}}^{(0)}_{klm} (t)\Big)\bigg]\nonumber 
   \\
   &  +  \frac{4}{15 c^4}  \partial_{(i} \Big(\ddot{\mathcal{I}}^{(0)}_{j)ll}(t) /r\Big) - \frac{2}{c^5} \dddot{\mathcal{I}}^{(0)}_{\langle ij \rangle} -\frac{8}{c^4} \partial_{(i} \Big(Y_{j)}(t)/r \Big) +\cdots\,.\label{eq:2PNWZmetric2}
 \end{align}

\subsubsection{2PM mixed components}
Next, we want to compute the leading order contribution to the particular solution for $h_{it}^{[2]}$ using equation (\ref{eq:tau^2}) we find that 
\begin{align}
    S^{[2]}_{it} =& - \frac{8 \epsilon_{ikb} \mathcal{J}^{(0)}_b M n^k}{r^5c^4} + \frac{8Mn^k}{c^4 r^2} \partial_{il} \left( \frac{\dot{\mathcal{I}}^{(0)}_{kl}(u)}{r}\right) - \frac{6Mn^i}{c^4 r^2} \partial_{kl} \left( \frac{\dot{\mathcal{I}}^{(0)}_{kl}(u)}{r}\right) +\cdots\,.
\end{align}
In order to use the integral equation in (\ref{eq:W&PInt}) we decompose the source into irreducible representations
\begin{align}
    S^{[2]}_{it} =& - \frac{8 \epsilon_{ikb} \mathcal{J}_b^{(0)} M n^k}{r^5 c^4} + \frac{M}{c^4 r^5} \bigg[  6\left(\dot{\mathcal{I}}^{(0)}_{kl}(u) + \frac{r}{c} \ddot{\mathcal{I}}^{(0)}_{kl}(u) + \frac{1}{3} \frac{r^2}{c^2} \ddot{\mathcal{I}}^{(0)}_{kl}(u) \right) n^{\langle ilk \rangle}  \nonumber
    \\
    &+ \frac{n^i}{5} \left( 36\dot{\mathcal{I}}^{(0)}_{kk}(u) +  36 \frac{r}{c}\ddot{\mathcal{I}}^{(0)}_{kk}(u) + 2 \frac{r^2}{c^2}\dddot{\mathcal{I}}^{(0)}_{kk}(u) \right) \nonumber
    \\
    &- \frac{n^k}{5} \left( 28\dot{\mathcal{I}}^{(0)}_{ik}(u) + 28\frac{r}{c} \ddot{\mathcal{I}}^{(0)}_{ik}(u) 
    - 4 \frac{r^2}{c^2} \dddot{\mathcal{I}}^{(0)}_{ik}(u) \right) \bigg]+\cdots\,. \label{eq:S_it^2}
   \end{align}
We then apply the integral equation in (\ref{eq:W&PInt}) to each term individually, using integration by parts and dropping boundary terms that are expected to be cancelled by $B^{[2]}_{it}$, we find 
\begin{align}
    -\frac{1}{4\pi}\int_{\mathcal{E}} d^3 x'\frac{S_{it}^{[2]}(t-\vert x-x'\vert/c,x')}{\vert x-x'\vert}=&- \frac{2 \epsilon_{ikb}\mathcal{J}^{(0)}_b n^k}{c^4r^3} - \frac{M\left(\dot{\mathcal{I}}^{(0)}_{kl}(u) + r \ddot{\mathcal{I}}^{(0)}_{kl}(u)/c\right)}{c^4 r^3} n^{\langle ikl \rangle} \nonumber
    \\
    &+ \frac{9}{5} \frac{M\left(\dot{\mathcal{I}}^{(0)}_{kk}(u) + r \ddot{\mathcal{I}}^{(0)}_{kk}(u)/c\right)}{c^4 r^3} n^{i} \nonumber
    \\
    &- \frac{7}{5} \frac{M\left(\dot{\mathcal{I}}^{(0)}_{ik}(u) + r \ddot{\mathcal{I}}^{(0)}_{ik}(u)/c\right)}{c^4 r^3} n^{k} + \cdots\,.
\end{align}

Adding the homogeneous solution to this we get that the exterior zone metric is given by
\begin{align}
    g_{it}^\mathcal{E} =& \frac{4G}{c^2} \Bigg[ \frac{1}{2} \epsilon_{iab} \frac{n_a J_b}{r^2} + \frac{1}{2} \partial_l \Big( \frac{\dot{I}_{il}(u)}{r} \Big)  + \frac{1}{3} \epsilon_{iab} \partial_{la} \Big( \frac{J_{bl}(u)}{r} \Big) - \frac{1}{6} \partial_{kl} \Big( \frac{\dot{I}_{ikl}(u)}{r} \Big) \nonumber
    \\
    &\qquad+ \partial_i \bigg( \frac{W(u)}{r} \bigg) - \partial_{ik} \bigg( \frac{W_k(u)}{r} \bigg)+\frac{1}{4!} \partial_{klm} \bigg( \frac{\dot I_{iklm}(u)}{r}\bigg) - \frac{1}{8} \epsilon_{iab} \partial_{akl} \bigg( \frac{J_{bkl}(u)}{r}\bigg)  \nonumber
    \\
    &\qquad  +\frac{1}{2} \partial_{ikl} \bigg( \frac{W_{kl}(u)}{r}\bigg)  -\frac{1}{5!} \partial_{klmn} \bigg( \frac{\dot I_{iklmn}(u)}{r}\bigg) + \frac{1}{30} \epsilon_{iab} \partial_{aklm} \bigg( \frac{J_{bklm}(u)}{r}\bigg)\nonumber
    \\
    &\qquad   -\frac{1}{6} \partial_{iklm} \bigg( \frac{W_{klm}(u)}{r}\bigg) -\frac{1}{c^2} \partial_i \bigg( \frac{\dot X(u)}{r}\bigg) - \frac{1}{c^2}\frac{Y_i(u)}{r} + \frac{1}{2} \partial_a \left( \frac{\epsilon_{iab} \dot Z(u)}{r}\right)\Bigg] \nonumber
    \\
    & - \frac{2G^2 \epsilon_{ikb}\mathcal{J}_b^{(0)} n^k}{r^3} - \frac{G^2M\left(\dot{\mathcal{I}}^{(0)}_{kl}(u) + r \ddot{\mathcal{I}}^{(0)}_{kl}(u)/c\right)}{c^4 r^3} n^{\langle ikl \rangle} + \frac{9}{5} \frac{G^2M\left(\dot{\mathcal{I}}^{(0)}_{kk}(u) + r \ddot{\mathcal{I}}^{(0)}_{kk}(u)/c\right)}{c^4 r^3} n^{i} \nonumber
    \\
    &- \frac{7}{5} \frac{G^2M\left(\dot{\mathcal{I}}^{(0)}_{kk}(u) + r \ddot{\mathcal{I}}^{(0)}_{kk}(u)/c\right)}{c^4 r^3} n^{k}  +\cdots\,.
\end{align}
If we then go to the overlap region and $1/c$-expand the metric we get
\begin{align}
    \mathcal{C}\Big( g_{it}^\mathcal{E} \Big) =& \frac{4G}{c^2} \Bigg[ \frac{1}{2} \epsilon_{iab} \frac{n_a J_b}{r^2} + \frac{1}{2} \partial_l \Big( \frac{I_{il}(t)}{r} \Big)  + \frac{1}{3} \epsilon_{iab} \partial_{la} \Big( \frac{J_{bl}(t)}{r} \Big) - \frac{1}{6} \partial_{kl} \Big( \frac{\dot{I}_{ikl}(t)}{r} \Big)  \nonumber
    \\ 
    &\qquad + \frac{1}{4c^2} \partial_l \Big( r\dddot{I}_{il}(t) \Big)  + \frac{1}{6c^2} \epsilon_{iab} \partial_{la} \Big( r\ddot{J}_{bl}(t) \Big) - \frac{1}{12c^2} \partial_{kl} \Big( r\dddot{I}_{ikl}(t) \Big) \nonumber
    \\
    &\qquad -\frac{1}{6c^3} x^l \partial_t^4 I_{il}(t)+ \partial_i \bigg( \frac{W(t)}{r} \bigg) - \partial_{ik} \bigg( \frac{W_k(t)}{r} \bigg)  +\frac{1}{2} \partial_{i} \left( r \ddot W(t)\right)  \nonumber
    \\
    &\qquad - \frac{1}{2} \partial_{ik} \left( r \ddot W_k(t)\right) - \frac{1}{3} x^i \partial_t^3 W(t) + \partial_t^3 W_i(t) +\frac{1}{4!} \partial_{klm} \bigg( \frac{\dot I_{iklm}(t)}{r}\bigg) \nonumber
    \\
    &\qquad  - \frac{1}{8} \epsilon_{iab} \partial_{akl} \bigg( \frac{J_{bkl}(t)}{r}\bigg) +\frac{1}{2} \partial_{ikl} \bigg( \frac{W_{kl}(t)}{r}\bigg) -\frac{1}{5!} \partial_{klmn} \bigg( \frac{\dot I_{iklmn}(t)}{r}\bigg) \nonumber
    \\
    &\qquad  + \frac{1}{30} \epsilon_{iab} \partial_{aklm} \bigg( \frac{J_{bklm}(t)}{r}\bigg) -\frac{1}{6} \partial_{iklm} \bigg( \frac{W_{klm}(t)}{r}\bigg) -\frac{1}{c^2} \partial_i \bigg( \frac{\dot X(t)}{r}\bigg) \Bigg] \nonumber
        \\
    & - \frac{2G^2 \epsilon_{ikb}\mathcal{J}_b^{(0)} n^k}{c^4 r^3} - \frac{G^2 M \dot{\mathcal{I}}^{(0)}_{kl}(t) }{c^4 r^3} n^{\langle ikl \rangle} + \frac{9}{5} \frac{G^2M\dot{\mathcal{I}}^{(0)}_{kk}(t)}{c^4 r^3} n^{i}  \nonumber
    \\
    & - \frac{7}{5} \frac{G^2M\dot{\mathcal{I}}^{(0)}_{ik}(t)}{c^4 r^3} n^{k} +\cdots+ \mathcal{O} (c^{-6})\,. \label{eq:ExtZonmet(it)}
\end{align}

\subsubsection{3PM time components}
Finally, we move on to the $tt$-component.
Using equation (\ref{eq:tau^2}) we find 
\begin{align}
    S^{[2]}_{tt} =& - \frac{4M^2}{c^2r^4}  - \frac{4Mn^{\langle kl \rangle}}{c^2} \left[ \frac{9\mathcal{I}^{(0)}_{kl}}{r^6} + \frac{9 \dot{\mathcal{I}}^{(0)}_{kl}}{c r^5} - \frac{2 \ddot{\mathcal{I}}^{(0)}_{kl}}{c^2 r^4} - \frac{2 \dddot{\mathcal{I}}^{(0)}_{kl}}{c^3 r^3} \right]  - \frac{16M}{3c^2} \left[ \frac{ \ddot{\mathcal{I}}^{(0)}_{kk}}{c^2 r^4} +\frac{ \dddot{\mathcal{I}}^{(0)}_{kk}}{c^3 r^3}  \right] \nonumber 
 \\
 &-\frac{4Mn^{\langle klm \rangle}}{c^2} \left[ \frac{20\mathcal{I}^{(0)}_{klm}}{r^7} + \frac{20 \dot{\mathcal{I}}^{(0)}_{klm}}{c r^6} + \frac{3 \ddot{\mathcal{I}}^{(0)}_{klm}}{c^2 r^5} - \frac{11 \dddot{\mathcal{I}}^{(0)}_{klm}}{3c^3 r^4} \right] \nonumber
 \\
 &- \frac{ M n^{k}}{c^2} \left[ \frac{16}{5} \frac{ \ddot{\mathcal{I}}^{(0)}_{kll}}{c^2 r^5} + \frac{16}{5} \frac{ \dddot{\mathcal{I}}^{(0)}_{kll}}{c^3 r^4} \right] + \frac{16Mn^l \epsilon_{lmn} }{c^4 r^5} \left[ \dot{\mathcal{J}}^{(0)}_{mn}(u) +\frac{r}{c}\ddot{\mathcal{J}}^{(0)}_{mn}(u)  \right] +\cdots\,. \label{eq:S_tt^2}
\end{align}

Meanwhile, the source term for $h^{[3]}_{tt}$ is given below
\begin{align}
    S^{[3]}_{tt} =& \frac{12M^3}{c^4r^5} + \cdots\,. \label{eq:S_tt^3}
\end{align}
Solving these equation we find the following particular solution
\begin{align}
    \square^{-1}_{\text{ret}} S^{[3]}_{tt} =& \frac{2M^3}{c^4r^3}  \,,
    \\
    \square^{-1}_{\text{ret}} S^{[2]}_{tt} =& - 2 \frac{G^2I^2}{c^2 r^2} - \frac{4 M n^{k}(\ddot{\mathcal{I}}^{(0)}_{kll}(u) - \dddot{\mathcal{I}}^{(0)}_{kll}(u)r/c)}{5c^4 r^3} - \frac{8M (\ddot{\mathcal{I}}^{(0)}_{kk}(u) + \dddot{\mathcal{I}}^{(0)}_{kk}(u)r/c)}{3c^4 r^2} \nonumber 
    \\
    &- \frac{M n^{\langle kl \rangle}}{c^2 r} \left[ \frac{6 \mathcal{I}^{(0)}_{kl}(u)}{r^3} + \frac{6 \dot{\mathcal{I}}^{(0)}_{kl}(u)}{cr^2} +  \frac{8 \ddot{\mathcal{I}}^{(0)}_{kl}(u)}{c^2r} +  \frac{4 \dddot{\mathcal{I}}^{(0)}_{kl}(u)}{c^3}\right] \nonumber
    \\
    &-\frac{Mn^{\langle klm \rangle}}{rc^2} \left[ \frac{10\mathcal{I}^{(0)}_{klm}}{r^4} + \frac{10 \dot{\mathcal{I}}^{(0)}_{klm}}{c r^3} + \frac{8 \ddot{\mathcal{I}}^{(0)}_{klm}}{c^2 r^2} + \frac{14 \dddot{\mathcal{I}}^{(0)}_{klm}}{3c^3 r} \right] \nonumber
    \\
    &+ \frac{4M}{c^4} \frac{n^a \epsilon_{abl} (\dot{\mathcal{J}}^{(0)}_{bl} + \ddot{\mathcal{J}}^{(0)}_{bl}r/c)}{r^2} \,.
\end{align}

The final expression for the exterior zone metric is then given by
\begin{align}
    g^{\mathcal{E}}_{tt} =& -c^2 + \frac{2GI}{r} + \partial_{kl} \left(\frac{GI_{kl}(u)}{r} \right) - \frac{1}{3}\partial_{klm} \left(\frac{GI_{klm}(u)}{r} \right)+  \frac{1}{12} \partial_{klmn} \left(\frac{GI_{klmn}(u)}{r} \right) \nonumber
    \\
    &- \frac{1}{60} \partial_{klmnp} \left(\frac{GI_{klmnp}(u)}{r} \right)  + 8 \frac{ \dot W(u)}{c^2r} - \partial_k\left( 8 \frac{\dot W_{k}(u)}{c^2 r} \right) + \partial_{kl}\left( 4 \frac{\dot W_{kl}(u)}{c^2r} \right) \nonumber
    \\
    &- \partial_{klm}\left(  \frac{4 \dot W_{klm}(u)}{3c^2r} \right) - 2 \frac{G^2I^2}{c^2 r^2} - \frac{4 M n^{k}(\ddot{\mathcal{I}}^{(0)}_{kll}(u) - \dddot{\mathcal{I}}^{(0)}_{kll}(u)r/c)}{5c^4 r^3} \nonumber 
    \\
    &- \frac{M n^{\langle kl \rangle}}{c^2 r} \left[ \frac{6 \mathcal{I}^{(0)}_{kl}(u)}{r^3} + \frac{6 \dot{\mathcal{I}}^{(0)}_{kl}(u)}{cr^2} +  \frac{8 \ddot{\mathcal{I}}^{(0)}_{kl}(u)}{c^2r} +  \frac{4 \dddot{\mathcal{I}}^{(0)}_{kl}(u)}{c^3}\right] \nonumber
    \\
    &-\frac{Mn^{\langle klm \rangle}}{rc^2} \left[ \frac{10\mathcal{I}^{(0)}_{klm}}{r^4} + \frac{10 \dot{\mathcal{I}}^{(0)}_{klm}}{c r^3} + \frac{8 \ddot{\mathcal{I}}^{(0)}_{klm}}{c^2 r^2} + \frac{14 \dddot{\mathcal{I}}^{(0)}_{klm}}{3c^3 r} \right] \nonumber
    \\
    &- \frac{8M (\ddot{\mathcal{I}}^{(0)}_{kk}(u) + \dddot{\mathcal{I}}^{(0)}_{kk}(u)r/c)}{3c^4 r^2}  + \frac{4M}{c^4} \frac{n^a \epsilon_{abl} (\dot{\mathcal{J}}^{(0)}_{bl} + \ddot{\mathcal{J}}^{(0)}_{bl}r/c)}{r^2} + \frac{2 G^3 M^3}{c^4r^3} +\cdots\,.
\end{align}
We then $1/c$-expand in the overlap region to find
\begin{align}
    \mathcal{C}\big( g^{\mathcal{E}}_{tt} \big) =& -c^2 + \frac{2GI}{r} + \partial_{kl} \left(\frac{GI_{kl}(t)}{r} \right) - \frac{1}{3}\partial_{klm} \left(\frac{GI_{klm}(t)}{r} \right)+  \frac{1}{12} \partial_{klmn} \left(\frac{GI_{klmn}(t)}{r} \right) \nonumber
    \\
    &- \frac{1}{60} \partial_{klmnp} \left(\frac{GI_{klmnp}(t)}{r} \right) + \frac{1}{2c^2} \partial_{kl} \left( Gr \ddot I_{kl}(t) \right) - \frac{1}{6c^2}\partial_{klm} \left(Gr \ddot I_{klm}(t) \right) \nonumber
    \\
    &+  \frac{1}{24c^2} \partial_{klmn} \left(Gr \ddot I_{klmn}(t) \right) - \frac{1}{120c^2} \partial_{klmnp} \left(G r \ddot I_{klmnp}(t) \right) + \frac{1}{4!c^4} \partial_{kl} \left( Gr^3 \partial_t^4 I_{kl}(t) \right)   \nonumber
    \\
    & - \frac{1}{3\cdot 4!c^4}\partial_{klm} \left(Gr^3 \partial_t^4 I_{klm}(t) \right) + \frac{1}{12\cdot 4!c^4}\partial_{klmn} \left(Gr^3 \partial_t^4 I_{klmn}(t) \right) \nonumber
    \\
    &- \frac{1}{60\cdot 4!c^4}\partial_{klmnp} \left(Gr^3 \partial_t^4 I_{klmnp}(t) \right) -\frac{8}{5!c^5} x^{kl} \partial_t^5 I_{kl}(t)  + 8 \frac{ \dot W(t)}{c^2r} - \partial_k\left( 8 \frac{\dot W_{kl}(t)}{c^2 r} \right)  \nonumber
    \\
    &+ \partial_{kl}\left( 4 \frac{\dot W_{kl}(t)}{c^2r} \right) - \frac{8}{c^3} \ddot W(t) + \frac{4r}{c^4} \dddot W(t) -  \frac{4}{c^4} \partial_k (r \dddot W_k(t)) + \frac{2}{c^4} \partial_{kl} (r \dddot W_{kl}(t))\nonumber
    \\
    &- \frac{4}{3c^5} x^2 \partial_t^4 W(t) + \frac{8}{3c^5} x^k \partial_t^4 W_k(t) - \frac{4}{3c^5} \partial_t^4 W_{kk}(t) - 2 \frac{G^2M^2}{c^2 r^2}  \nonumber 
    \\
    &  - \frac{M n^{\langle kl \rangle}}{c^2 r} \left[ \frac{6 \mathcal{I}^{(0)}_{kl}(t)}{r^3}  +  \frac{5 \ddot{\mathcal{I}}^{(0)}_{kl}(t)}{c^2r} -  2\frac{ \dddot{\mathcal{I}}^{(0)}_{kl}(t)}{c^3}\right] - \frac{4 M n^{k} \ddot{\mathcal{I}}^{(0)}_{kll}(t)}{5c^4 r^3} - \frac{8M \ddot{\mathcal{I}}^{(0)}_{kk}(t) }{3c^4 r^2}\nonumber 
    \\
    &  -\frac{Mn^{\langle klm \rangle}}{rc^2} \left[ \frac{10\mathcal{I}^{(0)}_{klm}(t)}{r^4}  + \frac{3 \ddot{\mathcal{I}}^{(0)}_{klm}(t)}{c^2 r^2} \right] +\cdots+ \mathcal{O}(c^{-6})\,. \label{eq:Exteriorzonett2.5PN}
\end{align}
For convenience we have not explicitly expanded the multipole moments $I_L$ and $W_L$.

\subsection{Multipole expanding the near zone metric} \label{sec:MultNearZ}
In this subsection we will multipole expand the 2.5PN near zone metric. We start with some generalities. First of all, for integrals with some compact source term $\mu(x,t)$ that are of the following form
\begin{align}
     \int d^3x' \mu(t,x') |x-x'|^n\,,
\end{align}
we use that for $|x|>l_c$
\begin{align}
       |x-x'|^n = \sum_{l=0}^\infty \frac{(-)^l}{l!} \partial_L\left( r^n x'^L\right)\,. \label{eq:mult}
   \end{align}

The other type of term we will run into is the Poisson integral over a non-compact source term, $\sigma(t,x)$
\begin{align}
    \int d^3x' \frac{\sigma(t,x')}{|x-x'|}\,.
\end{align}
In this case we split the domain of integration in the integral over the interior and one over the exterior
\begin{align}
    \int d^3x' \frac{\sigma(t,x')}{|x-x'|} = \int_{\mathcal{I}^{(0)}n} d^3x' \frac{\sigma(t,x')}{|x-x'|} + \int_{\mathcal{E}} d^3x' \frac{\sigma(t,x')}{|x-x'|}\,, \qquad \mathcal{I}^{(0)}n = \{ x' \in \mathbb{R}^3 | r'<l_c\}\,.
\end{align}
The integration over the interior can be treated as a compact term and so we use what we learned in (\ref{eq:mult}). For the exterior zone integral, we use that the source term itself can be multipole expanded, so we find 
\begin{align}
    \sigma(t,x) = \frac{1}{4\pi} \sum_{m,l=0}^\infty \frac{\sigma_L^{\{m\}}(t) n^{\langle L \rangle}}{r^m}\,.
\end{align}
Each of these terms can be solved using a simpler version of equation (\ref{eq:W&PInt}), which can be derived in a similar fashion and results in
\begin{align}
    &\int_{\mathcal{E}} d^3x' \frac{\sigma(t,x')}{|x-x'|}= \sum_{m,l=0}^\infty \frac{n^{\langle L \rangle} \sigma_L^{\{m\}}(t) }{r} \bigg[\int_0^{l_c}  A(s,r) ds + \int_{l_c}^\infty B(s,r) ds \bigg]\,.
    \\
    &A(s,r) := \int_{l_c}^{r+s} dr' \frac{P_l(\xi)}{r'^{(m-1)}}\,, \qquad B(s,r) := \int_s^{r+s} dr' \frac{P_l(\xi)}{r'^{(m-1)}}\,.
\end{align}
This integration will naturally lead to terms that depend explicitly on $l_c$ but these will be cancelled by boundary terms from the integration of the interior.

For the integral over the interior one often makes use of the conserved currents in (\ref{eq:FluidConsCurnt}) as well as the associated identities in (\ref{eq:FluidIden}), which when integrated over will lead to the aforementioned boundary terms.

\subsubsection{Multipole expanding the spatial components}
We wish to perform the multipole expansion of the 2.5PN near zone metric. First we note that 
\begin{align}
    g^{\mathcal{N}}_{ij} =& h_{ij} + \frac{1}{c^2} h^{(2)}_{ij} + \frac{1}{c^4} h^{(4)}_{ij} + \frac{1}{c^5} h^{(5)}_{ij} +\mathcal{O}( \frac{1}{c^6})\,, \label{2PNNearzone}
\end{align}
where
\begin{align}
    h^{(2)}_{ij} =& 2 \delta_{ij} U\,,
    \\
    h^{(4)}_{ij} =& \delta_{ij} \bigg( 2 U^2 + \partial_t^2 X + 8\pi G P\big[ E_{(0)} -P_{(0)} +2 E_{(\sm 2)}U\big] \bigg) \nonumber
    \\
    &+ 16 \pi G P\big[  E_{(\sm 2)} v^i v^j\big] + 4 P\big[ \partial_i U \partial_jU\big] \,,
    \\
    h^{(5)}_{ij} =& \mathcal{H}^{(5)}_{ij}\,.
\end{align}

Using what we learned in the first part of this section, we see that the multipole expansion of the near zone metric in (\ref{2PNNearzone}) is given by
\begin{align}
    g^{\mathcal{N}}_{ij} 
      =&\delta_{ij} \bigg[ 1 +\frac{2(M + c^{\sm 2}M^{(2)})}{c^2r} + \partial_{kl} \Big( \frac{\mathcal{I}^{(0)}_{kl}(t)}{c^2 r} \Big) + \frac{M^2}{c^4r^2} + \frac{1}{2c^4} \partial_{kl} (r \ddot{\mathcal{I}}^{(0)}_{kl}(t)) - \frac{\ddot{\mathcal{I}}^{(0)}_{kk}}{c^4r} \bigg]\nonumber
     \\& +\frac{2\ddot{\mathcal{I}}^{(0)}_{ij}}{c^4 r} +\frac{M^2}{c^4r^2}n^i n^j - \frac{2}{3c^4} \partial_k \Big( \frac{\ddot{\mathcal{I}}^{(0)}_{ijk}}{r}\Big) + \frac{1}{c^4} \frac{8}{3} \partial_a \bigg( \frac{\epsilon_{ab(i} \dot{\mathcal{J}}^{(0)}_{|b|j)}(t)}{r}\bigg) \nonumber
     \\
     &+\delta_{ij} \bigg[ \frac{1}{3c^4} \partial_k \bigg( \frac{\ddot{\mathcal{I}}^{(0)}_{kll}(t)}{r}\bigg) - \frac{1}{3c^2}\partial_{klm} \Big( \frac{\mathcal{I}^{(0)}_{klm}(t)}{r} \Big) - \frac{4}{3c^4}  \partial_a \bigg( \frac{\epsilon_{abk} \dot{\mathcal{J}}^{(0)}_{bk}}{r}\bigg)\nonumber 
     \\
     &- \frac{1}{6c^4} \partial_{klm} \Big( r\ddot{\mathcal{I}}^{(0)}_{klm} (t)\Big)\bigg]  + \frac{1}{c^5} \mathcal{H}^{(5)}_{ij} +\cdots\,. \label{eq:2PNMNZmetric}
\end{align}
We see that the matching with (\ref{eq:2PNWZmetric2}) is consistent and fixes for us
\begin{align}
    \mathcal{H}^{(5)}_{ij}(t) &= -2 \dddot{\mathcal{I}}^{(0)}_{\langle ij \rangle }\,, 
    \qquad
    Y_j =  -\frac{1}{30} \ddot{\mathcal{I}}^{(0)}_{jll} - \frac{1}{6} \epsilon_{jkl} \partial_t^2 \mathcal{J}^{(0)}_{kl}\,.
\end{align}

\subsubsection{Multipole expanding the mixed components}
The $it$-components of the near zone metric up to 2.5PN order are given by
\begin{align}
    g_{it} = \frac{1}{c^2}g_{it}^{\text{1PN}} + \frac{1}{c^4} g_{it}^{\text{2PN}} + \frac{1}{c^5} g_{it}^{\text{2.5PN}}\,.
\end{align}
 We know that $g_{it}^{\text{2.5PN}}$ is just a harmonic function that we determined in section 7. So for the purpose of this appendix, $g_{it}^{\text{2.5PN}}$ is already fully matched and can be ignored. Meanwhile we know that
\begin{align}
    g_{it}^{\text{1PN}}=& 4U^i\,,
    \\
    g_{it}^{\text{2PN}}=& - 16\pi GP\left[-E_{(\sm2)} U_i +E_{(\sm2)}v^i_{(2)} + \left(\frac{1}{2}E_{(\sm2)}v^2+4E_{(\sm2)}U+E_{(0)} + P_{(0)}\right)v^i\right]\nonumber
    \\
    &-2\partial_t^2 X_i+4P\left[2 \partial_k U \partial_k  U_i -4 \partial_k U \partial_i  U_k - 3 \partial_i U \partial_t U\right]  + 4U^i U\,.
\end{align}
Multipole expanding these we find
\begin{align}
     \mathcal{M}(g_{ti}^{1\text{PN}}) =&  \frac{2G\epsilon_{ijk} n^j \mathcal{J}^{(0)}_k}{r^2} + \partial_k \Big( \frac{2G\dot{\mathcal{I}}^{(0)}_{\langle ik \rangle
    }(t)}{r}\Big) - \frac{2G}{3} \partial_{kl} \bigg(\frac{\dot{\mathcal{I}}^{(0)}_{ \langle ikl \rangle}(t)}{r} - 2 \epsilon_{ikm} \frac{\mathcal{J}_{ml}^{(0)}}{r}\bigg) \nonumber
    \\
    &+ \frac{G}{6} \partial_{klm}\left( \frac{\dot{\mathcal{I}}^{(0)}_{\langle iklm \rangle}}{r} - 3 \epsilon_{ika} \frac{\mathcal{J}_{alm}^{(0)}}{r} \right) -  \frac{G}{30} \partial_{klmn}\left( \frac{\dot{\mathcal{I}}^{(0)}_{\langle iklmn \rangle}}{r} - 4 \epsilon_{ika} \frac{\mathcal{J}_{almn}^{(0)}}{r} \right) \nonumber
    \\
    & + \frac{2}{3}  \partial_i \Big( \frac{G\dot{\mathcal{I}}^{(0)}_{ll
    }(t)}{r}\Big) - \frac{4}{15}   \partial_{ik} \Big( \frac{G\dot{\mathcal{I}}^{(0)}_{knn
    }(t)}{r}\Big)+ \frac{1}{14}   \partial_{ikl} \Big( \frac{G\dot{\mathcal{I}}^{(0)}_{klnn
    }(t)}{r}\Big) \nonumber
    \\
    &- \frac{2}{135}   \partial_{iklm} \Big( \frac{G\dot{\mathcal{I}}^{(0)}_{klmnn
    }(t)}{r}\Big) \,,
       \label{eq:PNmet0i}
 \end{align} 
\begin{align}
    \mathcal{M} \Big(g_{it}^{\text{2PN}} \Big)
        =&\frac{4G}{c^4}\Bigg[ \frac{1}{4}\partial_k (r \dddot{\mathcal{I}}^{(0)}_{\langle ik \rangle}) + \frac{1}{12} \partial_i (r \dddot{\mathcal{I}}^{(0)}_{kk}) - \frac{1}{12} \partial_{kl} \big( r \dddot{\mathcal{I}}^{(0)}_{\langle ikl \rangle}\big)- \frac{1}{30} \partial_{ik} \big( r \dddot{\mathcal{I}}^{(0)}_{knn}\big) \nonumber
    \\
    &\qquad 
    - \frac{1}{30} \frac{\dddot{\mathcal{I}}^{(0)}_{inn}}{r} + \frac{1}{6} \partial_{kl} \big( r \epsilon_{ikm} \ddot{\mathcal{J}}^{(0)}_{ml}\big) + \frac{1}{48} \partial_{klm} \big( r \dddot{\mathcal{I}}^{(0)}_{\langle iklm \rangle}\big) \nonumber\\
    &\quad
    + \frac{1}{112} \partial_{ikl}(r \dddot{\mathcal{I}}^{(0)}_{\langle kl \rangle nn})+ \frac{1}{56} \partial_k \left( \frac{\dddot{\mathcal{I}}^{(0)}_{\langle ik \rangle nn}}{r}\right)+\frac{1}{120} \partial_i \left( \frac{\dddot{\mathcal{I}}^{(0)}_{kkll}}{r}\right) \nonumber
    \\
    &\qquad  - \frac{1}{16} \partial_{klm} \big( r \epsilon_{ika} \ddot{\mathcal{J}}^{(0)}_{alm}\big)- \frac{1}{240} \partial_{klmn} \big( r \dddot{\mathcal{I}}^{(0)}_{\langle iklmn \rangle}\big) - \frac{1}{540} \partial_{iklm} \big( r \dddot{\mathcal{I}}^{(0)}_{\langle klm \rangle nn}\big)\nonumber
    \\
    &\qquad -\frac{1}{180} \partial_{kl} \left( \frac{\dddot{\mathcal{I}}^{(0)}_{\langle ikl \rangle nn}}{r}\right) - \frac{1}{350}\partial_{ik} \left( \frac{\dddot{\mathcal{I}}^{(0)}_{k ll nn}}{r}\right)  - \frac{1}{60} \partial_{klmn} \big( r \epsilon_{ika} \ddot{\mathcal{J}}^{(0)}_{almn}\big)\nonumber
    \\
    &\qquad+ \frac{1}{2} \epsilon_{iab} \frac{n_a \mathcal{J}_b^{(2)}}{r^2} + \frac{1}{2} \partial_l \bigg( \frac{\dot{\mathcal{I}}^{(2)}_{il}(t)}{r} \bigg) + \frac{1}{3} \epsilon_{iab} \partial_{la} \bigg( \frac{\mathcal{J}_{bl}^{(2)}(t)}{r} \bigg) 
    - \frac{1}{6} \partial_{kl} \bigg( \frac{\dot{\mathcal{I}}^{(2)}_{ikl}(t)}{r}\bigg) \Bigg]   \nonumber
    \\
    &- \frac{2G^2 \epsilon_{ikb}M\mathcal{J}_b^{(0)} n^k}{c^4 r^3} - \frac{G^2 M \dot{\mathcal{I}}^{(0)}_{kl}(t) }{c^4 r^3} n^{\langle ikl \rangle} + \frac{9}{5} \frac{G^2M\dot{\mathcal{I}}^{(0)}_{kk}(t)}{c^4 r^3} n^{i} - \frac{7}{5} \frac{G^2M\dot{\mathcal{I}}^{(0)}_{ik}(t)}{c^4 r^3} n^{k} \,.\label{eq:Mult2PNit}
\end{align}
We find that the matching with the metric in (\ref{eq:ExtZonmet(it)}) is consistent.

\subsubsection{Multipole expanding the time-time component}
The $tt$-component of the near zone metric up to 2.5PN order is given by
\begin{align}
    g_{tt} =-c^2 +2 U+ \frac{1}{c^2}g_{tt}^{\text{1PN}} + \frac{1}{c^2}g_{tt}^{\text{1.5PN}} + \frac{1}{c^4} g_{tt}^{\text{2PN}} + \frac{1}{c^5} g_{tt}^{\text{2.5PN}}\,,
\end{align}
where we know from previous sections that
\begin{align}
    g_{tt}^{\text{1PN}} =& 8\pi G  P[E_{(0)}+3P_{(0)}+2E_{(\sm2)}(v^2+U)] + \partial_t^2 X - 2 U^2 \,,
    \\
    g_{tt}^{\text{1.5PN}} =& -\frac{4}{3}\partial_t^3\mathcal{I}^{(0)}_{kk}\, ,
    \\
    g_{tt}^{\text{2PN}} =& - 2 \tau^{(6)}_t - 8\pi GU P[E_{(0)}+3P_{(0)}+2E_{(\sm2)}(v^2+U)]  - \partial_t^2 X   +U^3\, ,
    \\
    g_{tt}^{\text{2.5PN}} =& -\frac{2}{9}r^2 \partial_t^5 \mathcal{I}^{(0)}_{kk} +8 \pi G P[ E_{(3)}] +\frac{8}{3}U \partial_t^3\mathcal{I}^{(0)}_{kk}-2 \partial_t^3\mathcal{I}^{(0)}_{\langle ij \rangle} \partial_k\partial_l X -2 \mathcal{H}^{(7)}\, ,
\end{align}
with
\begin{align}
    \tau^{(6)}_t =&2U^3 -\frac{G}{2} \partial_t^2\int d^3x'\vert x-x'\vert\left(E_{(0)}+3P_{(0)}+2E_{(\sm2)}(v^2+U)\right)(t,x') \nonumber
    \\
    &-\frac{G}{24}\partial_t^4 \int d^3x' \vert x-x'\vert^3 E_{(\sm2)}(t,x') +P[ h_{kl}^{(4)} \partial_k \partial_l U] + 2 P[\partial_k U \partial_k \tau^{(4)}_{t}] \nonumber
    \\
    &-\frac{1}{2}P[\partial_t^2 U^2]  + 8 P[U_k \partial_t \partial_k U] +\frac{11}{2} P[\partial_k U \partial_k U^2]  +7 P[\partial_t U \partial_{t} U]\nonumber
    \\
    &+ 4P[U \partial_t \partial_t U] + 8P[\partial_j U_k \partial_j U_k - \partial_j U_k \partial_k U_j] \nonumber
    \\
     &-4 \pi G P\Big[ E_{(\sm 2)} \left(\tau^{(4)}_t + 4 v_{(2)}^k v^k + 2U(3v^2+2U) \right)  + E_{(0)}(U + 2v^2) \nonumber\\
    &\qquad \qquad+ P_{(0)} (3U + 2v^2) + 3P_{(2)} + E_{(2)} \Big]\,.
\end{align}
We then multipole expand, express the integrals in terms of conserved currents and apply the fluid identities of appendix \ref{app:conservation}. In the end we find
\begin{align}
     \mathcal{M}(g_{tt}^{1\text{PN}}) =& - \frac{2G^2M^2}{c^2r^2} - \frac{2G^2M}{r} \partial_{kl} \bigg( \frac{\mathcal{I}^{(0)}_{kl}(t)}{r} \bigg) + \frac{2G^2M}{3r} \partial_{klm} \bigg( \frac{\mathcal{I}^{(0)}_{klm}(t)}{r} \bigg) \nonumber
     \\
     &+\frac{2GM^{(2)}}{r} +\partial_{kl}\bigg( \frac{G\mathcal{I}_{kl}^{(2)}(t)}{r} \bigg) - \frac{1}{3} \partial_{klm} \bigg( \frac{G\mathcal{I}_{klm}^{(2)}(t)}{r} \bigg) +\frac{G}{2}\partial_{kl} (r \ddot{ \mathcal{I}}^{(0)}_{\langle kl\rangle}(t))  \nonumber
     \\
     & -  \frac{G}{6} \partial_{klm} (r \ddot{\mathcal{I}}^{(0)}_{\langle klm \rangle}(t)) +  \frac{G}{24} \partial_{klmn} (r \ddot{\mathcal{I}}^{(0)}_{\langle klmn \rangle}(t)) - \frac{G}{4!} \partial_{klmnp} (r \ddot{\mathcal{I}}^{(0)}_{\langle klmnp \rangle}(t)) \nonumber
    \\
    &  + \frac{4G\ddot{\mathcal{I}}^{(0)}_{ll}}{3c^2r} - \frac{8G}{15c^2} \partial_k \Big( \frac{\ddot{\mathcal{I}}^{(0)}_{kll}}{r}\Big)+ \frac{13G}{84c^2} \partial_{kl} \Big( \frac{\ddot{\mathcal{I}}^{(0)}_{klnn}}{r}\Big) -  \frac{19G}{540c^2} \partial_{klm} \Big( \frac{\ddot{\mathcal{I}}^{(0)}_{klmnn}}{r}\Big) \nonumber
     \\
     &+\frac{4G}{3c^2} \partial_k \Big( \frac{\epsilon_{kab} \dot{\mathcal{J}}^{(0)}_{ab}(t)}{r}\Big)- \frac{G}{2c^2} \partial_{kl} \Big( \frac{\epsilon_{kab} \dot{\mathcal{J}}^{(0)}_{abl}(t)}{r}\Big) + \frac{2G}{15c^2} \partial_{klm} \Big( \frac{\epsilon_{kab} \dot{\mathcal{J}}^{(0)}_{ablm}(t)}{r}\Big) \nonumber
     \\
     &   + \frac{16}{3} \partial_{kl} \left( \frac{ \mathcal{P}_{nnkl}^{(0)}}{r}\right) - \frac{4}{3} \partial_{klm} \left( \frac{ \mathcal{P}_{nnklm}^{(0)}}{r}\right) \,,
     \label{eq:PNmet00} 
 \end{align} 
\begin{align}
     \mathcal{M} \left( g^{2PN}_{tt} \right) 
    =& - \frac{5 n^{\langle kl \rangle} M \ddot{\mathcal{I}}^{(0)}_{kl}}{r^2} - \frac{3 n^{\langle klm \rangle} M \ddot{\mathcal{I}}^{(0)}_{klm}}{r^3} + \frac{2M^3}{r^3} \nonumber 
    \\
    &- \frac{4M M^{(2)}}{r^2} - \frac{8M \ddot{\mathcal{I}}^{(0)}_{kk}}{3r^2} -  \frac{M n^k}{r}  \left( \frac{4\ddot{\mathcal{I}}^{(0)}_{kll}}{5
    r} - 4 \frac{\epsilon_{klm} \dot{\mathcal{J}}^{(0)}_{lm}}{r} \right)  \nonumber
    \\
    &+\frac{2M^{(4)}}{r} + \frac{4\ddot{\mathcal{I}}_{kk}^{(2)}}{3r} - \frac{8}{15}\partial_k \left( \frac{\ddot{\mathcal{I}}_{kll}^{(2)}}{r}\right) + \frac{4}{3} \partial_k \left( \frac{\epsilon_{klm} \dot{\mathcal{J}}_{lm}^{(2)}}{r}\right) + \frac{1}{2} \partial_{kl} \left( r\ddot{\mathcal{I}}_{\langle kl \rangle}^{(2)} \right)\nonumber
    \\
    & - \frac{1}{6} \partial_{klm} \left( r\ddot{\mathcal{I}}_{\langle klm \rangle}^{(2)} \right) +\frac{1}{24} \partial_{ kl } \left( r^3 \partial_t^4\mathcal{I}^{(0)}_{\langle kl \rangle} \right) - \frac{1}{72} \partial_{klm} \left( r^3 \partial_t^4 \mathcal{I}^{(0)}_{\langle klm \rangle} \right)\nonumber
    \\
    &  + \frac{1}{12\cdot 4!} \partial_{klmn} \left( r^3 \partial_t^4 \mathcal{I}^{(0)}_{\langle klmn \rangle} \right) - \frac{1}{12 \cdot 5!} \partial_{klmnp} \left( r^3 \partial_t^4 \mathcal{I}^{(0)}_{\langle klmnp \rangle}\right) \nonumber
    \\
    & +\frac{2}{3} r \partial_t^4 \mathcal{I}^{(0)}_{kk} - \frac{4}{15} \partial_k \left( r \partial_t^4 \mathcal{I}^{(0)}_{knn} \right) + \frac{13}{168} \partial_{kl} \left( r \partial_t^4 \mathcal{I}^{(0)}_{\langle kl \rangle nn} \right) \nonumber
    \\
    & -  \frac{19}{1080} \partial_{klm} \left( r \partial_t^4 \mathcal{I}^{(0)}_{\langle klm \rangle nn} \right)+ \frac{2}{45} \frac{\partial_t^4 \mathcal{I}^{(0)}_{llnn}}{r} - \frac{3}{175} \partial_k \left( \frac{\partial_t^4 \mathcal{I}^{(0)}_{kllnn}}{r} \right)\nonumber
    \\
    & + \frac{2}{3} \partial_k \Big( r \epsilon_{kab} \dddot{\mathcal{J}}^{(0)}_{ab} \Big) - \frac{1}{4} \partial_{kl} \left( r \epsilon_{kab} \dddot{\mathcal{J}}^{(0)}_{abl} \right) + \frac{1}{15} \partial_{klm} \left( r\epsilon_{kab} \dddot{\mathcal{J}}^{(0)}_{ablm}\right) \nonumber
    \\
    &+ \frac{4}{3} \partial_{kl} \left( r \ddot{\mathcal{P}}_{mm\langle kl \rangle}^{(0)} \right) + \frac{8}{9} \frac{\ddot{\mathcal{P}}_{kkll}^{(0)}}{r}    - \frac{2}{3} \partial_{klm} \left( r \ddot{\mathcal{P}}_{nn\langle klm \rangle}^{(0)}\right) - \frac{4}{5} \partial_{k} \left( \frac{\ddot{\mathcal{P}}_{kllnn}^{(0)}}{r} \right) \,,\label{eq:Mult2PNtt}
    \\
    \mathcal{M} \left( g^{\text{2.5PN}}_{tt} \right) =& -\frac{2r^2}{9} \partial_t^5 {\mathcal{I}}^{(0)}_{kk}(t)+ \frac{2M^{(5)}}{r} +\frac{2 M\dddot{\mathcal{I}}^{(0)}_{\langle kl \rangle} n^{\langle kl \rangle }}{r}   + \frac{4M \dddot{\mathcal{I}}^{(0)}_{kk}}{3r} -2 \mathcal{H}^{(7)}\,.
\end{align}
We find that the matching with the exterior zone metric in (\ref{eq:Exteriorzonett2.5PN}) is consistent.

\addcontentsline{toc}{section}{References}
\bibliographystyle{JHEP}
\bibliography{Biblio}

\providecommand{\href}[2]{#2}\begingroup\raggedright\begin{thebibliography}{10}

\bibitem{Blanchet2014}
L.~Blanchet, \emph{{Gravitational Radiation from Post-Newtonian Sources and
  Inspiralling Compact Binaries}},
  \href{https://doi.org/10.12942/lrr-2014-2}{\emph{Living Rev. Rel.} {\bfseries
  17} (2014) 2} [\href{https://arxiv.org/abs/1310.1528}{{\ttfamily
  1310.1528}}].

\bibitem{poisson2014gravity}
E.~Poisson and C.~Will, \emph{Gravity: Newtonian, Post-Newtonian,
  Relativistic}, Cambridge University Press (2014).

\bibitem{Levi:2018nxp}
M.~Levi, \emph{{Effective Field Theories of Post-Newtonian Gravity: A
  comprehensive review}},
  \href{https://doi.org/10.1088/1361-6633/ab12bc}{\emph{Rept. Prog. Phys.}
  {\bfseries 83} (2020) 075901}
  [\href{https://arxiv.org/abs/1807.01699}{{\ttfamily 1807.01699}}].

\bibitem{Damour:1990pi}
T.~Damour, M.~Soffel and C.-m.~Xu, \emph{{General relativistic celestial
  mechanics. 1. Method and definition of reference systems}},
  \href{https://doi.org/10.1103/PhysRevD.43.3273}{\emph{Phys. Rev. D}
  {\bfseries 43} (1991) 3273}.

\bibitem{PhysRevD.57.7274}
P.~Jaranowski and G.~Sch\"afer, \emph{Third post-newtonian higher order adm
  hamilton dynamics for two-body point-mass systems},
  \href{https://doi.org/10.1103/PhysRevD.57.7274}{\emph{Phys. Rev. D}
  {\bfseries 57} (1998) 7274}.

\bibitem{Damour:2000kk}
T.~Damour, P.~Jaranowski and G.~Schaefer, \emph{{Poincare invariance in the ADM
  Hamiltonian approach to the general relativistic two-body problem}},
  \href{https://doi.org/10.1103/PhysRevD.62.021501}{\emph{Phys. Rev. D}
  {\bfseries 62} (2000) 021501}
  [\href{https://arxiv.org/abs/gr-qc/0003051}{{\ttfamily gr-qc/0003051}}].

\bibitem{Tichy:2011te}
W.~Tichy and E.E.~Flanagan, \emph{{Covariant formulation of the
  post-1-Newtonian approximation to General Relativity}},
  \href{https://doi.org/10.1103/PhysRevD.84.044038}{\emph{Phys. Rev. D}
  {\bfseries 84} (2011) 044038}
  [\href{https://arxiv.org/abs/1101.0588}{{\ttfamily 1101.0588}}].

\bibitem{VandenBleeken:2017rij}
D.~Van~den Bleeken, \emph{{Torsional Newton\textendash{}Cartan gravity from the
  large c expansion of general relativity}},
  \href{https://doi.org/10.1088/1361-6382/aa83d4}{\emph{Class. Quant. Grav.}
  {\bfseries 34} (2017) 185004}
  [\href{https://arxiv.org/abs/1703.03459}{{\ttfamily 1703.03459}}].

\bibitem{Hansen:2019pkl}
D.~Hansen, J.~Hartong and N.A.~Obers, \emph{{Action Principle for Newtonian
  Gravity}}, \href{https://doi.org/10.1103/PhysRevLett.122.061106}{\emph{Phys.
  Rev. Lett.} {\bfseries 122} (2019) 061106}
  [\href{https://arxiv.org/abs/1807.04765}{{\ttfamily 1807.04765}}].

\bibitem{VandenBleeken:2019gqa}
D.~Van~den Bleeken, \emph{{Torsional Newton-Cartan gravity and strong
  gravitational fields}},  in \emph{{15th Marcel Grossmann Meeting on Recent
  Developments in Theoretical and Experimental General Relativity,
  Astrophysics, and Relativistic Field Theories}}, 3, 2019
  [\href{https://arxiv.org/abs/1903.10682}{{\ttfamily 1903.10682}}].

\bibitem{HHO}
D.~Hansen, J.~Hartong and N.A.~Obers, \emph{{Non-Relativistic Gravity and its
  Coupling to Matter}},
  \href{https://doi.org/10.1007/JHEP06(2020)145}{\emph{JHEP} {\bfseries 06}
  (2020) 145} [\href{https://arxiv.org/abs/2001.10277}{{\ttfamily
  2001.10277}}].

\bibitem{Hartong:2023yxo}
J.~Hartong, E.~Have, N.A.~Obers and I.~Pikovski, \emph{{A coupling prescription
  for post-Newtonian corrections in Quantum Mechanics}},
  \href{https://arxiv.org/abs/2308.07373}{{\ttfamily 2308.07373}}.

\bibitem{Hartong:2022lsy}
J.~Hartong, N.A.~Obers and G.~Oling, \emph{{Review on Non-Relativistic
  Gravity}},  \href{https://arxiv.org/abs/2212.11309}{{\ttfamily 2212.11309}}.

\bibitem{Hansen:2019vqf}
D.~Hansen, J.~Hartong and N.A.~Obers, \emph{{Gravity between Newton and
  Einstein}}, \href{https://doi.org/10.1142/S0218271819440103}{\emph{Int. J.
  Mod. Phys. D} {\bfseries 28} (2019) 1944010}
  [\href{https://arxiv.org/abs/1904.05706}{{\ttfamily 1904.05706}}].

\bibitem{companionpaper}
J.~Hartong and J.~Musaeus, \emph{{Post-Newtonian Expansions in Transverse
  Gauge}}, {\emph{in preparation} }.

\bibitem{damour1983gravitational}
T.~Damour, \emph{Gravitational radiation and the motion of compact bodies.},
  {\emph{Lecture Notes in Physics, Berlin Springer Verlag} {\bfseries 124}
  (1983) 59}.

\bibitem{Blanchet:1985sp}
L.~Blanchet and T.~Damour, \emph{{Radiative gravitational fields in general
  relativity I. general structure of the field outside the source}},
  \href{https://doi.org/10.1098/rsta.1986.0125}{\emph{Phil. Trans. Roy. Soc.
  Lond. A} {\bfseries 320} (1986) 379}.

\bibitem{blanchet1987radiative}
L.~Blanchet, \emph{Radiative gravitational fields in general relativity ii.
  asymptotic behaviour at future null infinity}, {\emph{Proceedings of the
  Royal Society of London. A. Mathematical and Physical Sciences} {\bfseries
  409} (1987) 383}.

\bibitem{Blanchet:1989ki}
L.~Blanchet and T.~Damour, \emph{{Postnewtonian Generation of Gravitational
  Waves}}, {\emph{Ann. Inst. H. Poincare Phys. Theor.} {\bfseries 50} (1989)
  377}.

\bibitem{Blanchet:1992br}
L.~Blanchet and T.~Damour, \emph{{Hereditary effects in gravitational
  radiation}}, \href{https://doi.org/10.1103/PhysRevD.46.4304}{\emph{Phys. Rev.
  D} {\bfseries 46} (1992) 4304}.

\bibitem{PhysRevD.51.2559}
L.~Blanchet, \emph{Second-post-newtonian generation of gravitational
  radiation}, \href{https://doi.org/10.1103/PhysRevD.51.2559}{\emph{Phys. Rev.
  D} {\bfseries 51} (1995) 2559}.

\bibitem{LucBlanchet_1998}
L.~Blanchet, \emph{On the multipole expansion of the gravitational field},
  \href{https://doi.org/10.1088/0264-9381/15/7/013}{\emph{Classical and Quantum
  Gravity} {\bfseries 15} (1998) 1971}.

\bibitem{Damour:1990ji}
T.~Damour and B.R.~Iyer, \emph{{PostNewtonian generation of gravitational
  waves. 2. The Spin moments}}, {\emph{Ann. Inst. H. Poincare Phys. Theor.}
  {\bfseries 54} (1991) 115}.

\bibitem{PhysRevD.47.4392}
L.~Blanchet, \emph{Time-asymmetric structure of gravitational radiation},
  \href{https://doi.org/10.1103/PhysRevD.47.4392}{\emph{Phys. Rev. D}
  {\bfseries 47} (1993) 4392}.

\bibitem{PhysRevD.65.124020}
O.~Poujade and L.~Blanchet, \emph{Post-newtonian approximation for isolated
  systems calculated by matched asymptotic expansions},
  \href{https://doi.org/10.1103/PhysRevD.65.124020}{\emph{Phys. Rev. D}
  {\bfseries 65} (2002) 124020}.

\bibitem{PhysRevD.72.044024}
L.~Blanchet, G.~Faye and S.~Nissanke, \emph{Structure of the post-newtonian
  expansion in general relativity},
  \href{https://doi.org/10.1103/PhysRevD.72.044024}{\emph{Phys. Rev. D}
  {\bfseries 72} (2005) 044024}.

\bibitem{Will:1996zj}
C.M.~Will and A.G.~Wiseman, \emph{{Gravitational radiation from compact binary
  systems: Gravitational wave forms and energy loss to second postNewtonian
  order}}, \href{https://doi.org/10.1103/PhysRevD.54.4813}{\emph{Phys. Rev. D}
  {\bfseries 54} (1996) 4813}
  [\href{https://arxiv.org/abs/gr-qc/9608012}{{\ttfamily gr-qc/9608012}}].

\bibitem{Pati:2000vt}
M.E.~Pati and C.M.~Will, \emph{{PostNewtonian gravitational radiation and
  equations of motion via direct integration of the relaxed Einstein equations.
  1. Foundations}},
  \href{https://doi.org/10.1103/PhysRevD.62.124015}{\emph{Phys. Rev. D}
  {\bfseries 62} (2000) 124015}
  [\href{https://arxiv.org/abs/gr-qc/0007087}{{\ttfamily gr-qc/0007087}}].

\bibitem{Pati:2002ux}
M.E.~Pati and C.M.~Will, \emph{{Post-Newtonian gravitational radiation and
  equations of motion via direct integration of the relaxed Einstein equations.
  II. Two-body equations of motion to second post-Newtonian order, and
  radiation reaction to 3.5 post-Newtonian order }},
  \href{https://doi.org/10.1103/PhysRevD.65.104008}{\emph{Phys. Rev. D}
  {\bfseries 65} (2002) 104008}
  [\href{https://arxiv.org/abs/gr-qc/0201001}{{\ttfamily gr-qc/0201001}}].

\bibitem{Will:2005sn}
C.M.~Will, \emph{{Post-Newtonian gravitational radiation and equations of
  motion via direct integration of the relaxed Einstein equations. III.
  Radiation reaction for binary systems with spinning bodies}},
  \href{https://doi.org/10.1103/PhysRevD.71.084027}{\emph{Phys. Rev. D}
  {\bfseries 71} (2005) 084027}
  [\href{https://arxiv.org/abs/gr-qc/0502039}{{\ttfamily gr-qc/0502039}}].

\bibitem{Wang:2007ntb}
H.~Wang and C.M.~Will, \emph{{Post-Newtonian gravitational radiation and
  equations of motion via direct integration of the relaxed Einstein equations.
  IV. Radiation reaction for binary systems with spin-spin coupling}},
  \href{https://doi.org/10.1103/PhysRevD.75.064017}{\emph{Phys. Rev. D}
  {\bfseries 75} (2007) 064017}
  [\href{https://arxiv.org/abs/gr-qc/0701047}{{\ttfamily gr-qc/0701047}}].

\bibitem{Mitchell:2007ea}
T.~Mitchell and C.M.~Will, \emph{{Post-Newtonian gravitational radiation and
  equations of motion via direct integration of the relaxed Einstein equations.
  V. The Strong equivalence principle to second post-Newtonian order}},
  \href{https://doi.org/10.1103/PhysRevD.75.124025}{\emph{Phys. Rev. D}
  {\bfseries 75} (2007) 124025}
  [\href{https://arxiv.org/abs/0704.2243}{{\ttfamily 0704.2243}}].

\bibitem{bonnor1961transport}
W.~Bonnor and M.~Rotenberg, \emph{Transport of momentum by gravitational waves:
  the linear approximation}, {\emph{Proceedings of the Royal Society of London.
  Series A. Mathematical and Physical Sciences} {\bfseries 265} (1961) 109}.

\bibitem{bonnor1966gravitational}
W.~Bonnor and M.~Rotenberg, \emph{Gravitational waves from isolated sources},
  {\emph{Proceedings of the Royal Society of London. Series A. Mathematical and
  Physical Sciences} {\bfseries 289} (1966) 247}.

\bibitem{bonnor1959spherical}
W.~Bonnor, \emph{Spherical gravitational waves}, {\emph{Philosophical
  Transactions of the Royal Society of London. Series A, Mathematical and
  Physical Sciences} {\bfseries 251} (1959) 233}.

\bibitem{thorne1980multipole}
K.S.~Thorne, \emph{Multipole expansions of gravitational radiation},
  {\emph{Reviews of Modern Physics} {\bfseries 52} (1980) 299}.

\bibitem{10.1063/1.1665603}
W.L.~Burke, \emph{{Gravitational Radiation Damping of Slowly Moving Systems
  Calculated Using Matched Asymptotic Expansions}},
  \href{https://doi.org/10.1063/1.1665603}{\emph{Journal of Mathematical
  Physics} {\bfseries 12} (2003) 401}
  [\href{https://arxiv.org/abs/https://pubs.aip.org/aip/jmp/article-pdf/12/3/401/10951571/401\_1\_online.pdf}{{\ttfamily
  https://pubs.aip.org/aip/jmp/article-pdf/12/3/401/10951571/401\_1\_online.pdf}}].

\bibitem{hunter1968double}
A.~Hunter and M.~Rotenberg, \emph{The double-series approximation method in
  general relativity i. exact solution of the (24) approximation. ii.
  discussion of'wave tails' in the (2s) approximation}, {\emph{Journal of
  Physics A: General Physics} {\bfseries 2} (1968) 34}.

\bibitem{anderson1975equations}
J.L.~Anderson and T.C.~Decanio, \emph{Equations of hydrodynamics in general
  relativity in the slow motion approximation}, {\emph{General Relativity and
  Gravitation} {\bfseries 6} (1975) 197}.

\bibitem{PhysRevD.25.2038}
J.L.~Anderson, R.E.~Kates, L.S.~Kegeles and R.G.~Madonna, \emph{Divergent
  integrals of post-newtonian gravity: Nonanalytic terms in the near-zone
  expansion of a gravitationally radiating system found by matching},
  \href{https://doi.org/10.1103/PhysRevD.25.2038}{\emph{Phys. Rev. D}
  {\bfseries 25} (1982) 2038}.

\bibitem{Hansen:2019svu}
D.~Hansen, J.~Hartong and N.A.~Obers, \emph{{Non-relativistic expansion of the
  Einstein-Hilbert Lagrangian}},  in \emph{{15th Marcel Grossmann Meeting on
  Recent Developments in Theoretical and Experimental General Relativity,
  Astrophysics, and Relativistic Field Theories}}, 5, 2019
  [\href{https://arxiv.org/abs/1905.13723}{{\ttfamily 1905.13723}}].

\bibitem{Hansen:2020wqw}
D.~Hansen, J.~Hartong, N.A.~Obers and G.~Oling, \emph{{Galilean first-order
  formulation for the nonrelativistic expansion of general relativity}},
  \href{https://doi.org/10.1103/PhysRevD.104.L061501}{\emph{Phys. Rev. D}
  {\bfseries 104} (2021) L061501}
  [\href{https://arxiv.org/abs/2012.01518}{{\ttfamily 2012.01518}}].

\bibitem{Ergen:2020yop}
M.~Ergen, E.~Hamamci and D.~Van~den Bleeken, \emph{{Oddity in nonrelativistic,
  strong gravity}},
  \href{https://doi.org/10.1140/epjc/s10052-020-8112-6}{\emph{Eur. Phys. J. C}
  {\bfseries 80} (2020) 563}
  [\href{https://arxiv.org/abs/2002.02688}{{\ttfamily 2002.02688}}].

\bibitem{Wolf:2023xrv}
W.J.~Wolf, M.~Sanchioni and J.~Read, \emph{{Underdetermination in Classic and
  Modern Tests of General Relativity}},
  \href{https://arxiv.org/abs/2307.10074}{{\ttfamily 2307.10074}}.

\bibitem{Dautcourt:1996pm}
G.~Dautcourt, \emph{{PostNewtonian extension of the Newton-Cartan theory}},
  \href{https://doi.org/10.1088/0264-9381/14/1A/009}{\emph{Class. Quant. Grav.}
  {\bfseries 14} (1997) A109}
  [\href{https://arxiv.org/abs/gr-qc/9610036}{{\ttfamily gr-qc/9610036}}].

\bibitem{Kol:2010si}
B.~Kol and M.~Smolkin, \emph{{Einstein's action and the harmonic gauge in terms
  of Newtonian fields}},
  \href{https://doi.org/10.1103/PhysRevD.85.044029}{\emph{Phys. Rev. D}
  {\bfseries 85} (2012) 044029}
  [\href{https://arxiv.org/abs/1009.1876}{{\ttfamily 1009.1876}}].

\bibitem{Kol:2007bc}
B.~Kol and M.~Smolkin, \emph{{Non-Relativistic Gravitation: From Newton to
  Einstein and Back}},
  \href{https://doi.org/10.1088/0264-9381/25/14/145011}{\emph{Class. Quant.
  Grav.} {\bfseries 25} (2008) 145011}
  [\href{https://arxiv.org/abs/0712.4116}{{\ttfamily 0712.4116}}].

\bibitem{Elbistan:2022plu}
M.~Elbistan, E.~Hamamci, D.~Van~den Bleeken and U.~Zorba, \emph{{A 3+1
  formulation of the 1/c expansion of General Relativity}},
  \href{https://doi.org/10.1007/JHEP02(2023)108}{\emph{JHEP} {\bfseries 02}
  (2023) 108} [\href{https://arxiv.org/abs/2210.15440}{{\ttfamily
  2210.15440}}].

\bibitem{Kapustin:2021omc}
A.~Kapustin and M.~Touraev, \emph{{Non-relativistic Geometry and the
  Equivalence Principle}},
  \href{https://doi.org/10.1088/1361-6382/abfea5}{\emph{Class. Quant. Grav.}
  {\bfseries 38} (2021) 135003}
  [\href{https://arxiv.org/abs/2101.04153}{{\ttfamily 2101.04153}}].

\bibitem{Christensen:2013lma}
M.H.~Christensen, J.~Hartong, N.A.~Obers and B.~Rollier, \emph{{Torsional
  Newton-Cartan Geometry and Lifshitz Holography}},
  \href{https://doi.org/10.1103/PhysRevD.89.061901}{\emph{Phys. Rev. D}
  {\bfseries 89} (2014) 061901}
  [\href{https://arxiv.org/abs/1311.4794}{{\ttfamily 1311.4794}}].

\bibitem{Bergshoeff:2014uea}
E.A.~Bergshoeff, J.~Hartong and J.~Rosseel, \emph{{Torsional
  Newton\textendash{}Cartan geometry and the Schr\"odinger algebra}},
  \href{https://doi.org/10.1088/0264-9381/32/13/135017}{\emph{Class. Quant.
  Grav.} {\bfseries 32} (2015) 135017}
  [\href{https://arxiv.org/abs/1409.5555}{{\ttfamily 1409.5555}}].

\bibitem{Rendall}
A.D.~Rendall, \emph{On the definition of post-newtonian approximations},
  {\emph{Proceedings of the Royal Society of London. Series A: Mathematical and
  Physical Sciences} {\bfseries 438} (1992) 341}.

\bibitem{Fragkos:2022tbm}
V.~Fragkos, M.~Kopp and I.~Pikovski, \emph{{On inference of quantization from
  gravitationally induced entanglement}},
  \href{https://arxiv.org/abs/2206.00558}{{\ttfamily 2206.00558}}.

\bibitem{Smarr:1978dia}
L.~Smarr and J.W.~York, Jr., \emph{{Radiation gauge in general relativity}},
  \href{https://doi.org/10.1103/PhysRevD.17.1945}{\emph{Phys. Rev. D}
  {\bfseries 17} (1978) 1945}.

\bibitem{Dirac:1958jc}
P.A.M.~Dirac, \emph{{Fixation of coordinates in the Hamiltonian theory of
  gravitation}}, \href{https://doi.org/10.1103/PhysRev.114.924}{\emph{Phys.
  Rev.} {\bfseries 114} (1959) 924}.

\bibitem{Trautman:1958zdi}
A.~Trautman, \emph{{Radiation and Boundary Conditions in the Theory of
  Gravitation}}, {\emph{Bull. Acad. Pol. Sci. Ser. Sci. Math. Astron. Phys.}
  {\bfseries 6} (1958) 407} [\href{https://arxiv.org/abs/1604.03145}{{\ttfamily
  1604.03145}}].

\bibitem{dimopoulos2007testing}
S.~Dimopoulos, P.W.~Graham, J.M.~Hogan and M.A.~Kasevich, \emph{Testing general
  relativity with atom interferometry},
  \href{https://doi.org/https://doi.org/10.1103/PhysRevLett.98.111102}{\emph{Physical
  review letters} {\bfseries 98} (2007) 111102}.

\bibitem{zych2011quantum}
M.~Zych, F.~Costa, I.~Pikovski and {\v{C}}.~Brukner, \emph{Quantum
  interferometric visibility as a witness of general relativistic proper time},
  \href{https://doi.org/https://doi.org/10.1038/ncomms1498}{\emph{Nature
  communications} {\bfseries 2} (2011) 1}.

\bibitem{zych2012general}
M.~Zych, F.~Costa, I.~Pikovski, T.C.~Ralph and {\v{C}}.~Brukner, \emph{General
  relativistic effects in quantum interference of photons},
  \href{https://doi.org/10.1088/0264-9381/29/22/224010}{\emph{Classical and
  Quantum Gravity} {\bfseries 29} (2012) 224010}.

\bibitem{Pikovski:2013qwa}
I.~Pikovski, M.~Zych, F.~Costa and C.~Brukner, \emph{{Universal decoherence due
  to gravitational time dilation}},
  \href{https://doi.org/10.1038/nphys3366}{\emph{Nature Phys.} {\bfseries 11}
  (2015) 668} [\href{https://arxiv.org/abs/1311.1095}{{\ttfamily 1311.1095}}].

\bibitem{zych2016general}
M.~Zych, I.~Pikovski, F.~Costa and {\v{C}}.~Brukner, \emph{General relativistic
  effects in quantum interference of “clocks”},  in \emph{Journal of
  Physics: Conference Series}, 723, p.~012044, IOP Publishing, 2016,
  \href{https://doi.org/10.1088/1742-6596/723/1/012044}{DOI}.

\end{thebibliography}\endgroup

\end{document}